\documentclass[10pt]{article}

\usepackage{geometry}
\usepackage[citestyle=authoryear-comp, bibstyle=authoryear, uniquename=false, uniquelist=false, giveninits=true, maxbibnames=6, maxcitenames=2]{biblatex}
\bibliography{SPARmodel}
\AtBeginRefsection{\GenRefcontextData{sorting=ynt}}
\AtEveryCite{\localrefcontext[sorting=ynt]}

\usepackage{graphicx}
\graphicspath{ {./figures/} }
\usepackage{amsmath, amsfonts, amssymb}
\usepackage{bm}
\usepackage[normalem]{ulem}
\usepackage{url}
\usepackage{tabularx}
\usepackage{multirow}
\usepackage{makecell}
\usepackage{booktabs} 

\usepackage[hidelinks]{hyperref}
\hypersetup{colorlinks=true, allcolors=blue}

\usepackage{xcolor}

\usepackage{caption}
\usepackage{subcaption}

\usepackage[shortlabels]{enumitem}

\usepackage{mathtools} 

\usepackage{amsthm}
\newtheorem{theorem}{Theorem}[section]
\newtheorem{proposition}[theorem]{Proposition}
\newtheorem{lemma}[theorem]{Lemma}
\newtheorem{corollary}{Corollary}[theorem]

\theoremstyle{definition}
\newtheorem{definition}{Definition}

\newtheorem{example}{Example}

\usepackage{authblk}

\usepackage{placeins} 
\title{Modelling multivariate extremes through angular-radial decomposition of the density function}
\author[1]{Ed Mackay\thanks{email: e.mackay@exeter.ac.uk,\quad ORCID: 0000-0001-7121-4231}}
\author[2,3]{Philip Jonathan}
\affil[1]{University of Exeter, UK}
\affil[2]{Lancaster University, UK}
\affil[3]{Shell Research Ltd., London, UK}
\date{\today}

\begin{document}
\maketitle

\begin{abstract}
We present a new framework for modelling multivariate extremes, based on an angular-radial representation of the probability density function. Under this representation, the problem of modelling multivariate extremes is transformed to that of modelling an angular density and the tail of the radial variable, conditional on angle. Motivated by univariate theory, we assume that the tail of the conditional radial distribution converges to a generalised Pareto (GP) distribution. To simplify inference, we also assume that the angular density is continuous and finite and the GP parameter functions are continuous with angle. We refer to the resulting model as the semi-parametric angular-radial (SPAR) model for multivariate extremes. We consider the effect of the choice of polar coordinate system and introduce generalised concepts of angular-radial coordinate systems and generalised scalar angles in two dimensions. We show that under certain conditions, the choice of polar coordinate system does not affect the validity of the SPAR assumptions. However, some choices of coordinate system lead to simpler representations. In contrast, we show that the choice of margin does affect whether the model assumptions are satisfied. In particular, the use of Laplace margins results in a form of the density function for which the SPAR assumptions are satisfied for many common families of copula, with various dependence classes. We show that the SPAR model provides a more versatile framework for characterising multivariate extremes than provided by existing approaches, and that several commonly-used approaches are special cases of the SPAR model. 
\end{abstract}

\noindent%

\noindent%

\section{Introduction} \label{sec:intro}
\subsection{Motivation} \label{sec:motivation}
\begin{figure}[!t]
    \centering
    \begin{subfigure}[t]{0.45\textwidth}
        \centering
        \includegraphics[scale=0.6]{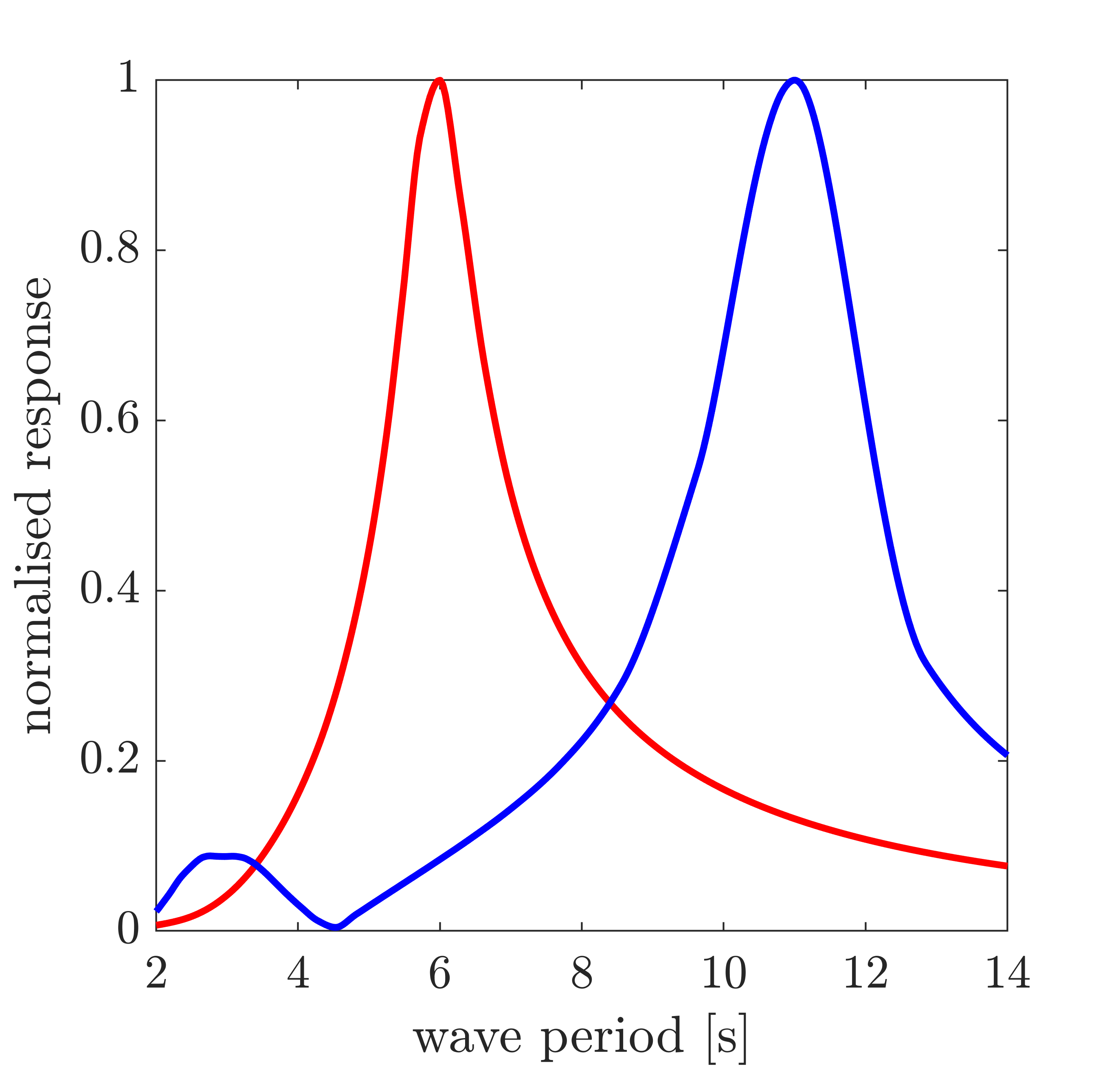}
        \caption{Indicative structural responses for two structures subject to wave loading, as a function of wave period. These responses could represent motions or a force on the structure.}
    \end{subfigure}
    \hskip2em
    \begin{subfigure}[t]{0.45\textwidth}
        \centering
        \includegraphics[scale=0.6]{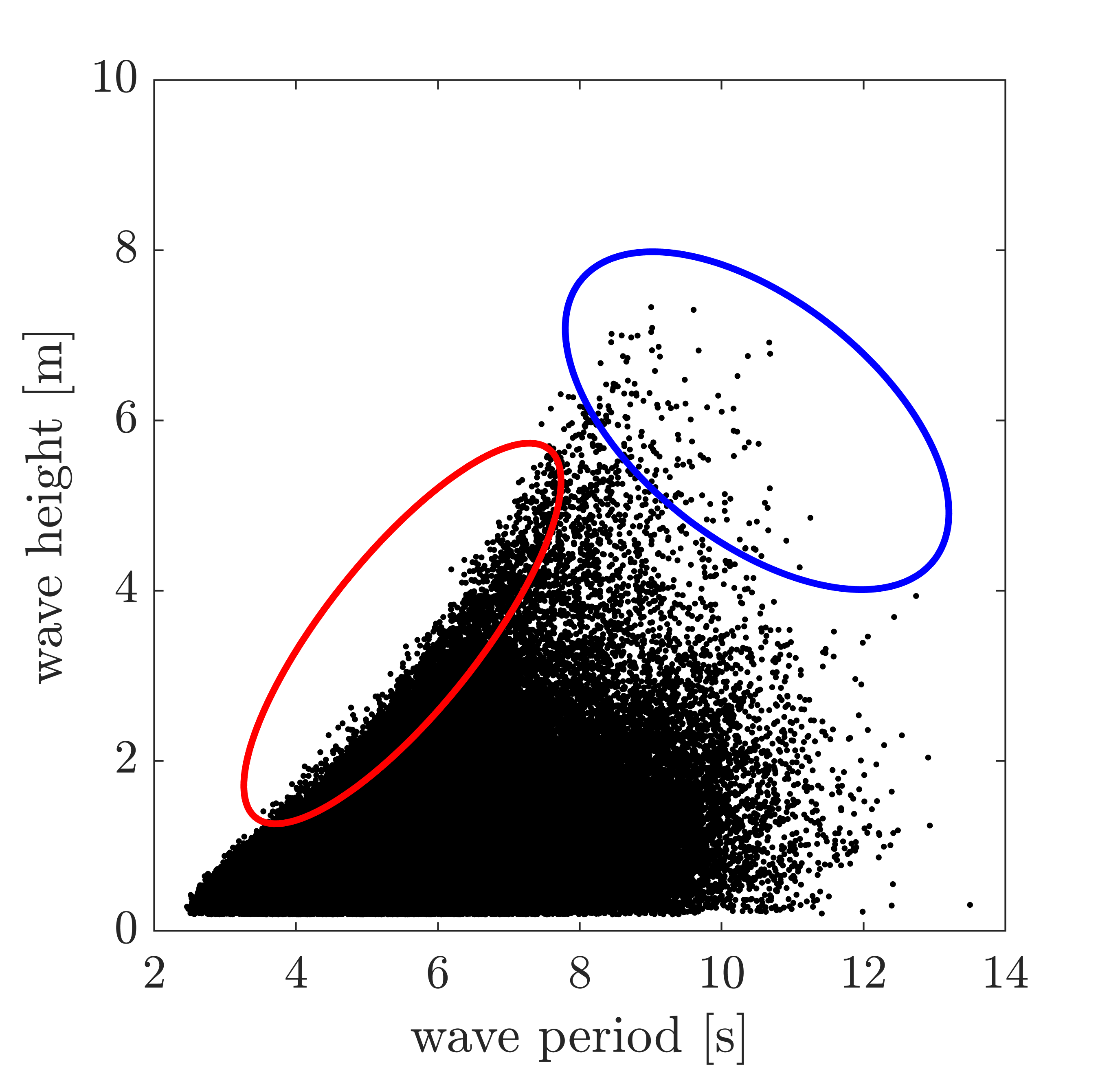}
        \caption{Hourly observations of wave height and wave period for a location off the US east coast.}
    \end{subfigure}\\
    \begin{subfigure}[t]{0.45\textwidth}
        \centering
        \includegraphics[scale=0.6]{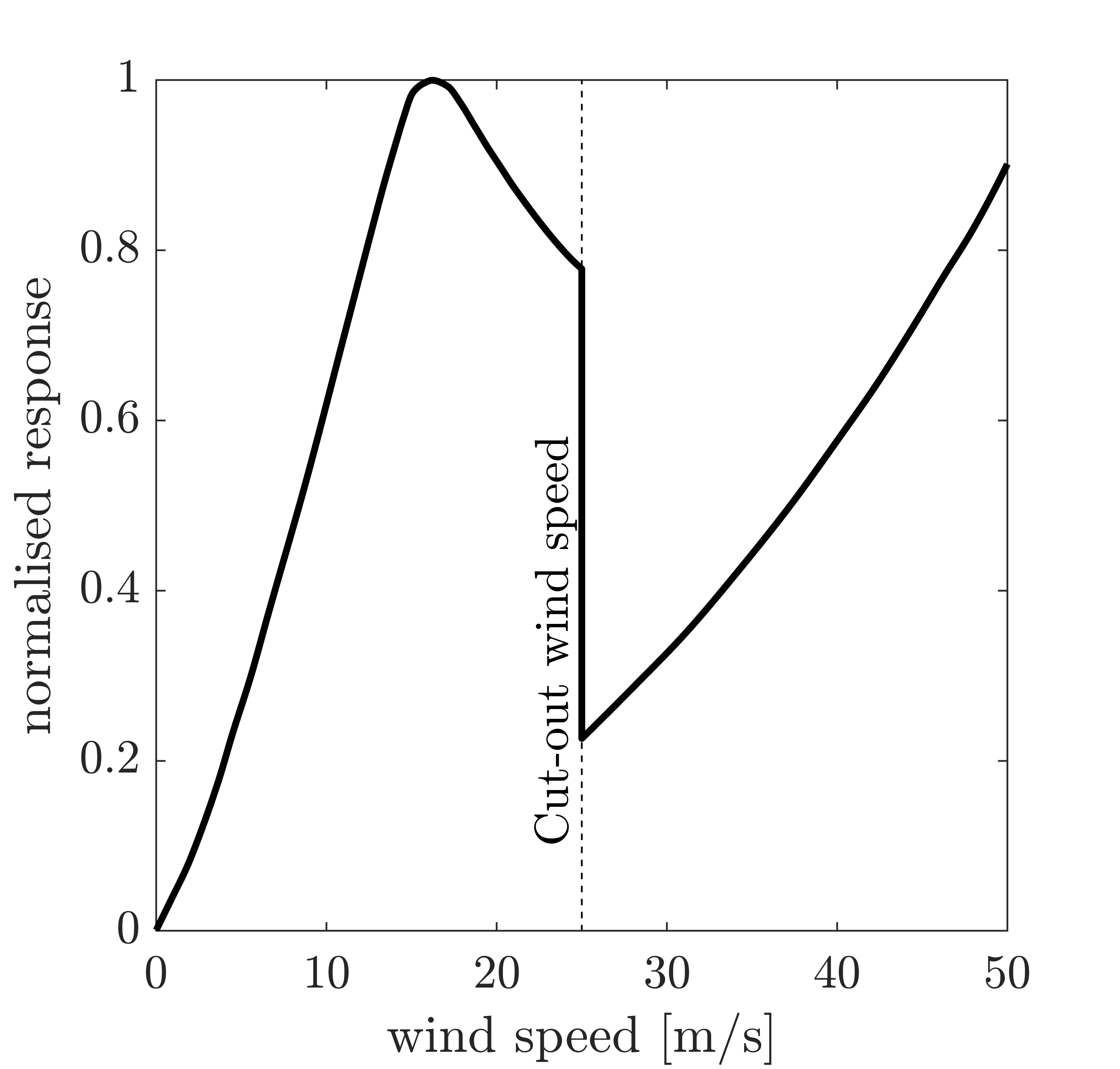}
        \caption{Indicative responses for an offshore wind turbine as a function of wind speed. This could represent various responses, such as bending moments in the turbine tower or blades.}
    \end{subfigure}
    \hskip2em
    \begin{subfigure}[t]{0.45\textwidth}
        \centering
        \includegraphics[scale=0.6]{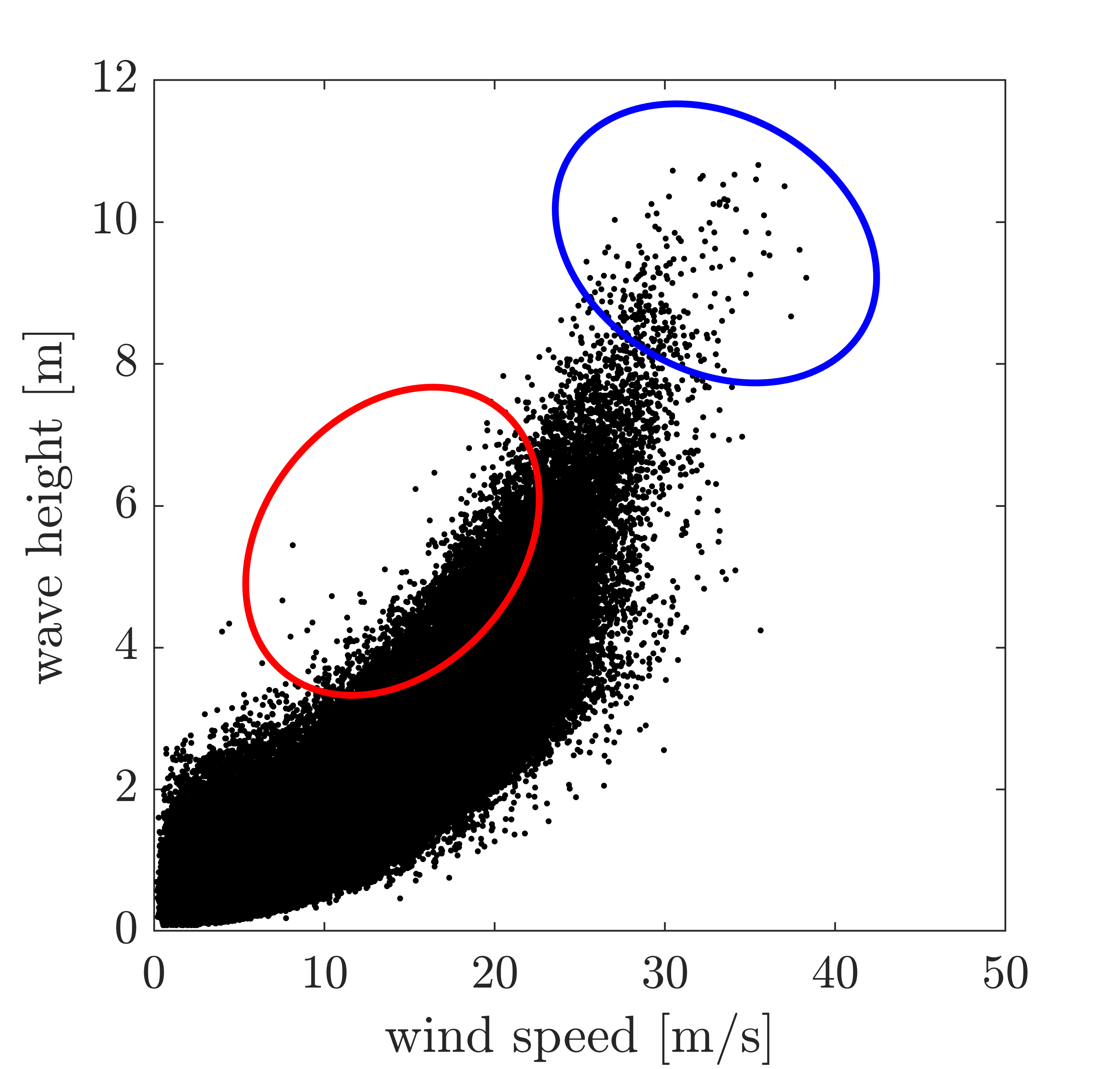}
        \caption{Hourly observations of wind speed and wave height for a location in the central North Sea.}
    \end{subfigure}
    \caption{Example datasets and structural responses for motivating problems. Ellipses on plots (b) and (d) indicate regions of variable space of where offshore structures may experience large responses. Regions in the red ellipses are not the largest values in either variable, whereas regions in the blue ellipses are extreme in at least one variable. The aim of the methodology proposed in this work is to characterise all `extreme regions' of a joint distribution using a single inference.}
    \label{fig:data_response}
\end{figure}

This work proposes a methodology which addresses two problems in multivariate extremes: (1) characterising extremes of random vectors in multiple orthants of $\mathbb{R}^d$ simultaneously; and (2) doing this for distributions with arbitrary dependence class. To motivate the first of these problems, consider the following problems in offshore engineering. Structures in the ocean must be designed to withstand the largest forces from waves and winds that are expected in their lifetime. To do this, we need a model for the joint distributions of environmental variables which affect loads on the structure. Wave-induced forces are dependent on both the height and period of the waves. The largest structural response will also depend on the resonant period of structure. \autoref{fig:data_response}(a) shows response curves for two structures as a function of wave period. These could represent a wide range of responses, such as vessel motions, bending moments in a structure, or tensions in a mooring line. Responses are assumed to increase with wave height at a given period. \autoref{fig:data_response}(b) shows 25 years of hourly observations of wave height and wave period for a location off the east coast of the US. The ellipses indicate regions of the variable space in which large structural loads may occur for different structures. For a structure with a resonant period around 5 seconds, the largest responses to wave loading may occur in the region indicated by the red ellipse. In this region neither variable is extreme, although the value of wave height, conditional on wave period can be viewed as extreme. In the blue region, large floating structures or vessels with resonant periods around 10 seconds may experience large responses. This region includes the largest observed wave heights, but not the largest or smallest observed values of wave period. 

A similar challenge arises in the design of offshore wind turbines. \autoref{fig:data_response}(c) shows a typical structural response curve as a function of wind speed, which could represent bending moments in the turbine tower or blades. Due to the way the turbine is controlled, the response to wind loading is non-monotonic. Wind turbines have a rated wind speed (usually around 10-15 m/s), at which they reach their maximum power output. For wind speeds above the rated speed, the turbines are controlled in a way that maintains constant power output with increased wind speed, but loads on the structure are reduced. To reduce loading in high wind speeds, the turbines are shut down at a certain wind speed, known as the cut-out speed, which results in a large reduction in the loading. The loads then increase with wind speed, due to passive drag loads on the structure. Loading on the structure also increases with wave height. This leads to two regions of the wind-wave variable space in which large loads are experienced. \autoref{fig:data_response}(d) shows a scatter plot of hourly values of wind speed and wave height over a 50-year period for a location in the North Sea. The ellipses indicate regions in which a turbine may experience large loads. In the red ellipse, neither variable is extreme, but the value of wave height is extreme conditional on the wind speed. In the blue ellipse both variables are extreme. 

In both example problems, the joint distribution of the variables could be characterized for the regions indicated by the ellipses in Figures \ref{fig:data_response}(b) and (d), using existing methods. In the red ellipses, a non-stationary univariate extreme value model could be used \parencite[e.g.][]{chavez2005, Randell2016, Youngman2019, Zanini2020, Barlow2023}. For the blue ellipse in \autoref{fig:data_response}(b), the conditional approach of \textcite{Heffernan2004} could be applied. And in the blue in ellipse in \autoref{fig:data_response}(d), one of the wide range of approaches in multivariate extreme value theory could be used, since both variables are extreme. However, ensuring consistency between the fitted models in the overlapping regions is difficult and it would be useful to characterise the joint distribution in all regions on the ‘outside’ of the sample in a single inference. The ‘outside’ of the sample is more formally defined as regions in which the radius of an observation is large relative to some origin within the body of the sample, where `large' is quantified locally, relative to other nearby observations. This motivates a transformation to some polar coordinate system and modelling of the distribution of large radii conditional on angle. 

\begin{figure}[!t]
    \centering
    \begin{subfigure}[t]{0.45\textwidth}
        \centering
        \includegraphics[scale=0.6]{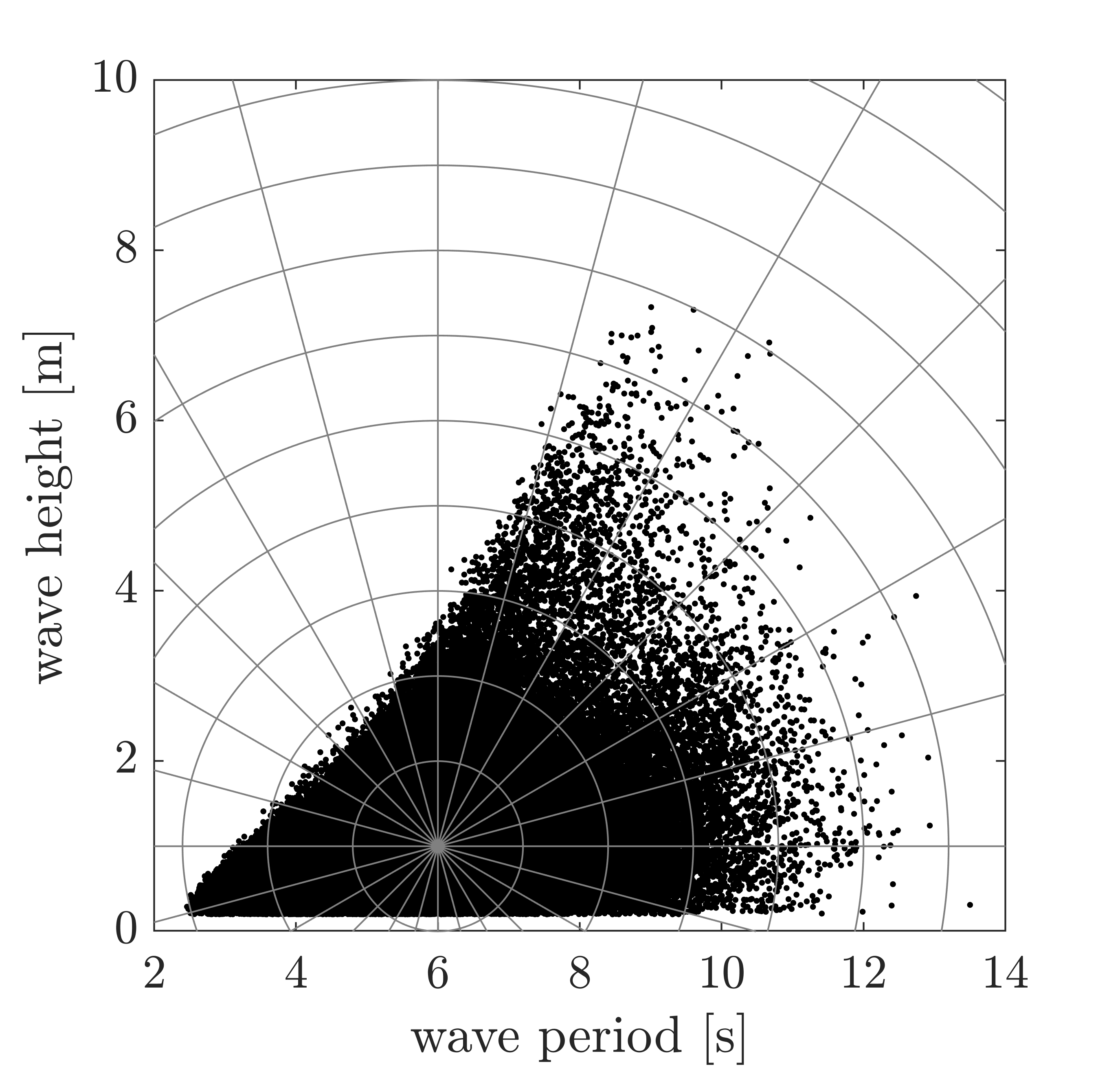}
        \caption{Wave height and period data with `standard' polar coordinate grid overlaid, with polar origin defined at $(x,y)=(6,1)$. `Standard' polar coordinates have radii defined in terms of the $L^2$ norm.}
    \end{subfigure}
    \hskip2em
    \begin{subfigure}[t]{0.45\textwidth}
        \centering
        \includegraphics[scale=0.6]{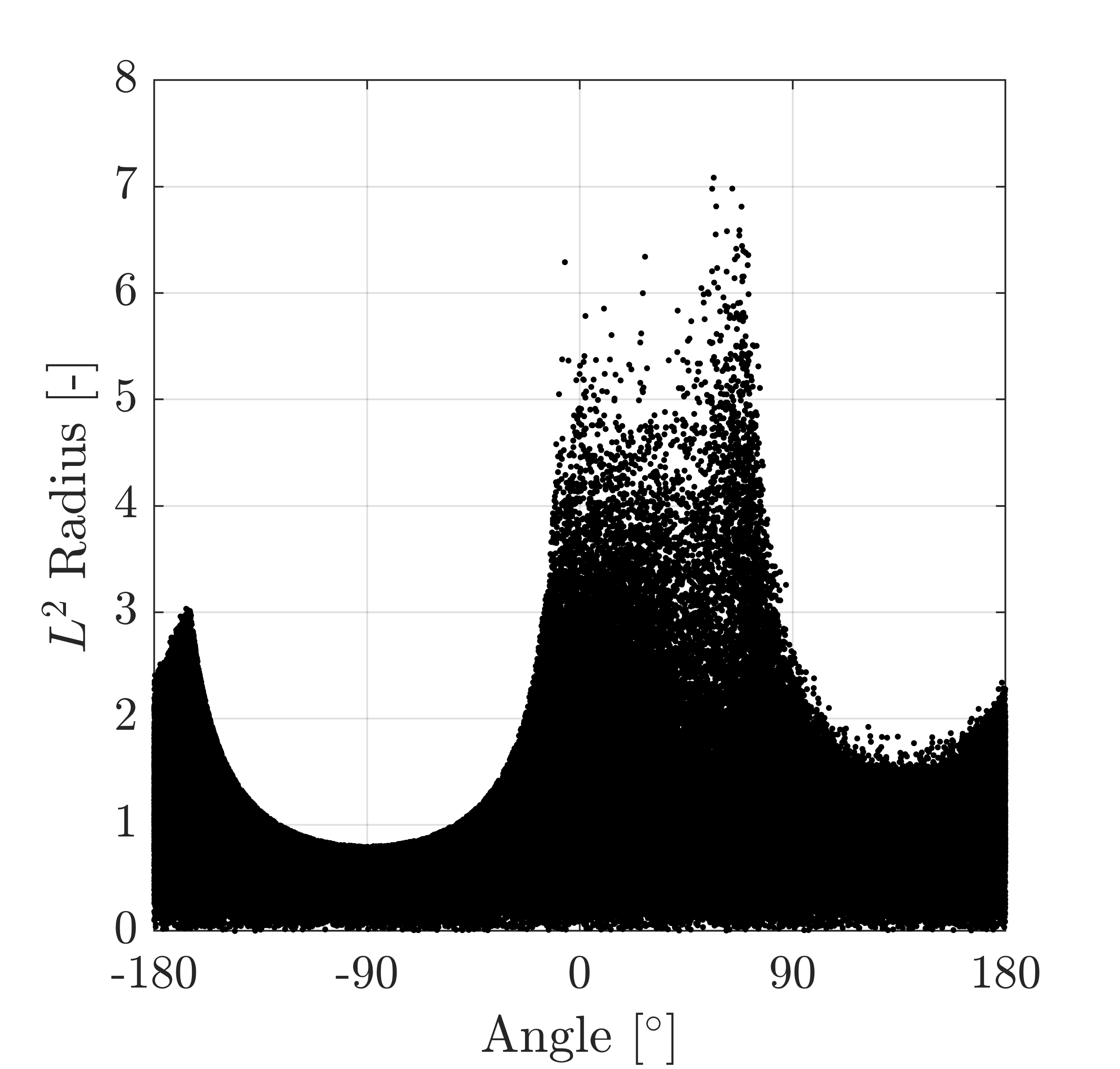}
        \caption{Wave height and period data shown in (a) transformed to standard polar coordinates.}
    \end{subfigure}\\
    \begin{subfigure}[t]{0.45\textwidth}
        \centering
        \includegraphics[scale=0.6]{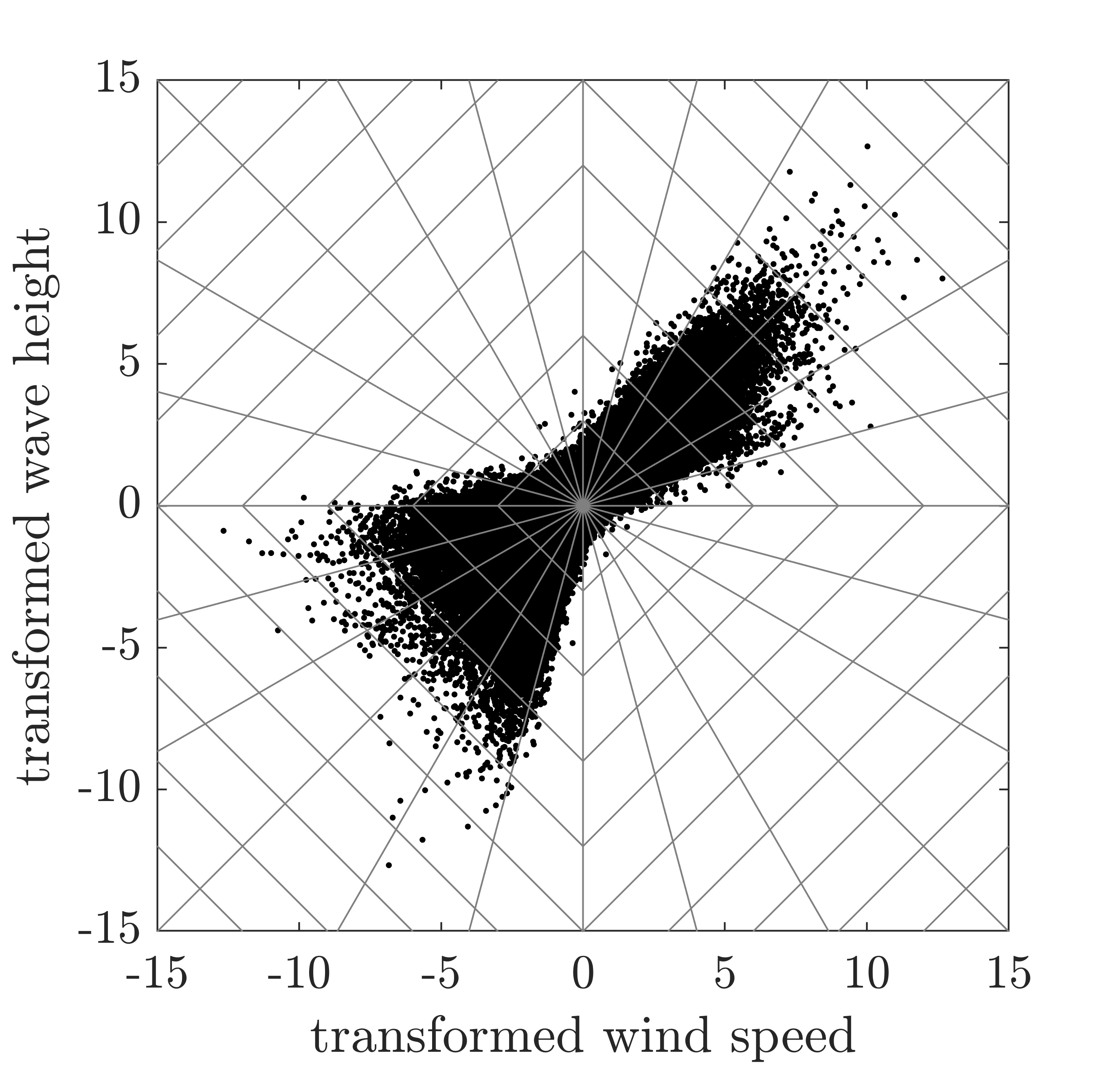}
        \caption{Wind and wave data transformed to Laplace margins with polar coordinate grid overlaid. Radii are defined in terms of the $L^1$ norm.}
    \end{subfigure}
    \hskip2em
    \begin{subfigure}[t]{0.45\textwidth}
        \centering
        \includegraphics[scale=0.6]{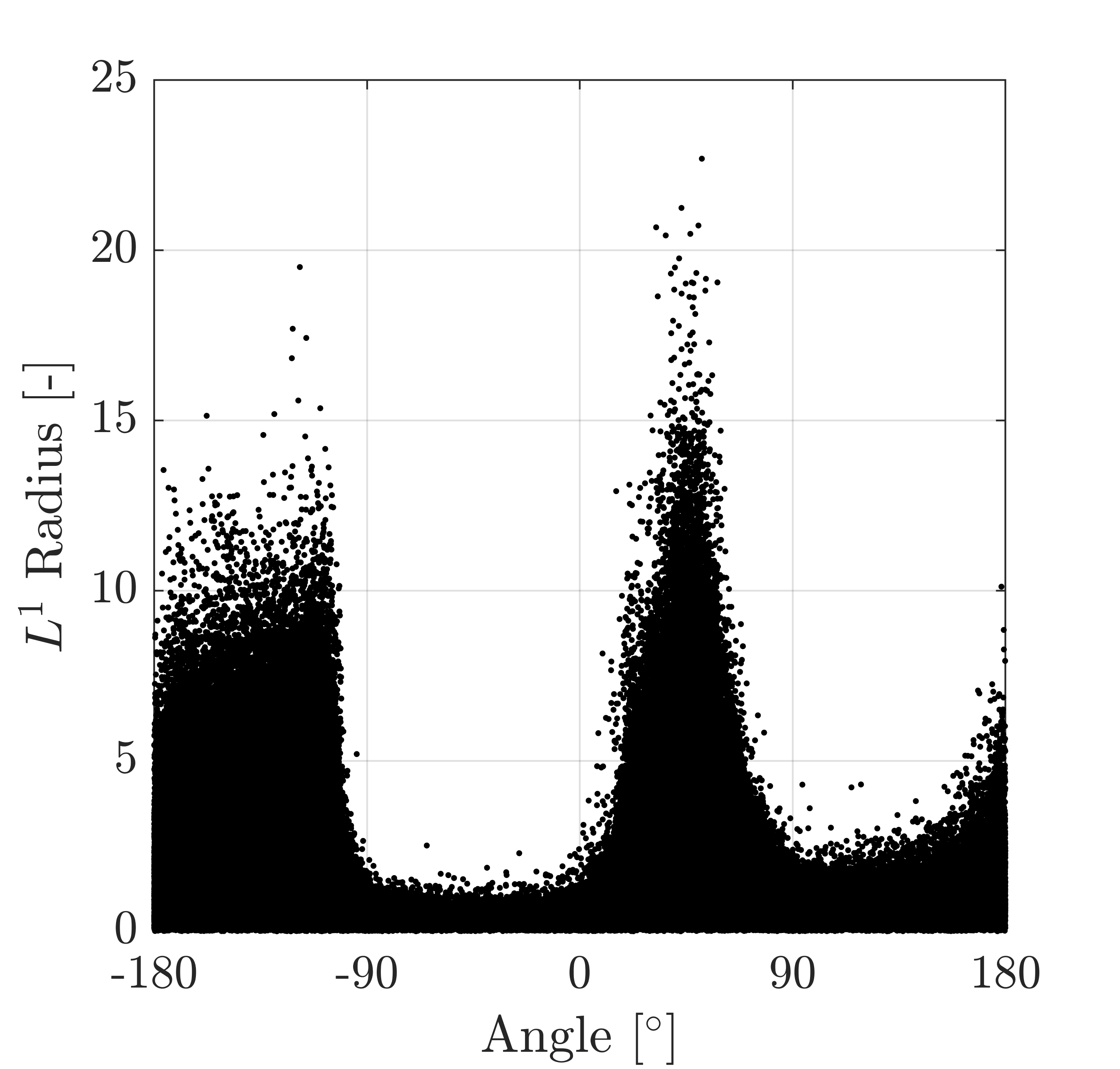}
        \caption{Wind and wave data shown in (c) transformed to polar coordinates, with radii define in terms of $L^1$ norm.}
    \end{subfigure}
    \caption{Examples of two possible transformations from Cartesian to polar coordinates, using data from the previous figure. After the polar coordinate transformation, the problem of modelling multivariate extremes is transformed to a problem of modelling univariate extremes conditional on angle.}
    \label{fig:data_transformation}
\end{figure}

Two possible transformations of the datasets shown in \autoref{fig:data_response} are shown in \autoref{fig:data_transformation}. For the wave height-period data, `standard' polar coordinates are used, where radii are defined in terms of the $L^2$ norm. For the wind-wave data, the observations have been transformed to Laplace margins before defining polar coordinates, and radii are defined in terms of the $L^1$ norm. After the coordinate transformation, the problem of modelling the joint distribution of extreme regions of the original sample is converted to a problem of modelling univariate extremes conditional on angle. We will show that this view of multivariate extremes is useful not only for the problems described here, but also for a much wider range of problems, including `standard' problems in multivariate extremes, where the interest lies only in the region where all variables are large. 

The coordinate transformations illustrated in \autoref{fig:data_transformation} are only two possibilities. This raises various questions such as: How should radii and angles be defined? Are inferences made using different polar coordinate systems equivalent? Should the margins be transformed to a standard scale before the angular-radial transformation is applied? Where should the origin of the polar coordinate system be defined? We will put aside these questions for now, and return to them later on. 

Next, consider the second problem mentioned above, of modelling the joint extremes of a distribution with an arbitrary dependence class. Suppose that random vector $\mathbf{X} \in \mathbb{R}^d$ has joint distribution function $F$, and continuous univariate marginal distribution functions $F_1, ..., F_d$, with corresponding joint and marginal survivor functions $\bar{F}$ and $\bar{F}_1, ..., \bar{F}_d$. The corresponding copula $C:[0,1]^d\to [0,1]$ and survival copula $\Hat{C}:[0,1]^d\to [0,1]$ are defined as 
\begin{align} 
    \begin{split}\label{eq:copula_defn}
        C(u_1,...,u_d) &= F(F_1^{-1}(u_1), ..., F_d^{-1}(u_d)),\\
        \Hat{C}(u_1,...,u_d) &= \bar{F}(\bar{F}_1^{-1}(u_1), ..., \bar{F}_d^{-1}(u_d)),
    \end{split}
\end{align} 
where $F_j^{-1}$ and $\bar{F}_j^{-1}$ are the inverse functions of $F_j$ and $\bar{F}_j$, $j=1,...,d$ \parencite[see e.g.][]{Joe2015}. Dependence in the lower and upper tail can be quantified in terms of the upper and lower tail dependence coefficients, $\chi_U,\chi_L\in[0,1]$, defined as \parencite{Joe2015}
\begin{align} 
    \begin{split}\label{eq:tail_dep_coef}
        \chi_L &= \lim_{t\to0^+} \frac{C(t,...,t)}{t},\\
        \chi_U &= \lim_{t\to0^+} \frac{\Hat{C}(t,...,t)}{t}.
    \end{split}
\end{align}
When $\chi_U>0$, the components of $\mathbf{X}$ are said to be \textit{asymptotically dependent} (AD) in the upper tail, and when $\chi_U=0$, they are said to be \textit{asymptotically independent} (AI). Dependence coefficients for joint tails in other regions can be defined analogously, as discussed in Section \ref{sec:copula_tail_model}.

Classical multivariate extreme value theory addresses the case where $\chi_U>0$, and has been widely studied -- see e.g. \textcite{Beirlant2004, haan2006extreme, Resnick2007} for reviews. In practice, we do not know the tail order a priori, so it is useful to adopt a modelling framework applicable to both AD and AI cases. \textcite{Ledford1996,Ledford1997} proposed a method to characterise multivariate extremes for distributions with arbitrary dependence class, in the region where all variables are large. However, when $\chi_U=0$, this may not be the region of most interest, since extremes of all variables may not occur simultaneously. An alternative approach was proposed by \textcite{Heffernan2004}, to estimate the joint distribution of variables conditional on at least one variable being large. However, inferences made using different conditioning variables are not necessarily consistent \parencite{Liu2014}. \textcite{Wadsworth2017} proposed a similar model to the one described in this work, which is capable of representing joint extremes of distributions with a range of dependence classes. The approach in \textcite{Wadsworth2017} involves a transformation to polar coordinates, and modelling the tail of the radial variable using a generalised Pareto (GP) distribution. However, the key difference to the approach proposed here is that Wadsworth \textit{et al.} assume that the angular and radial variables are AI at large radii, and the margins and polar coordinate system are chosen so that the assumption of AI angular and radial variables is a reasonable approximation. We will show that if margins and coordinates exist which satisfy the assumption of AI angular and radial variables, then the coordinates are, in general, non-standard and would be difficult to estimate in practice. This restricts the range of distributions that can be modelled using the approach proposed in \textcite{Wadsworth2017}. In contrast, we will show that the approach proposed here removes the need to select specific coordinate systems, simplifying the modelling approach, and making it applicable to a much wider range of cases.

\begin{figure}[!t]
    \centering
    (a) \includegraphics[scale=0.6]{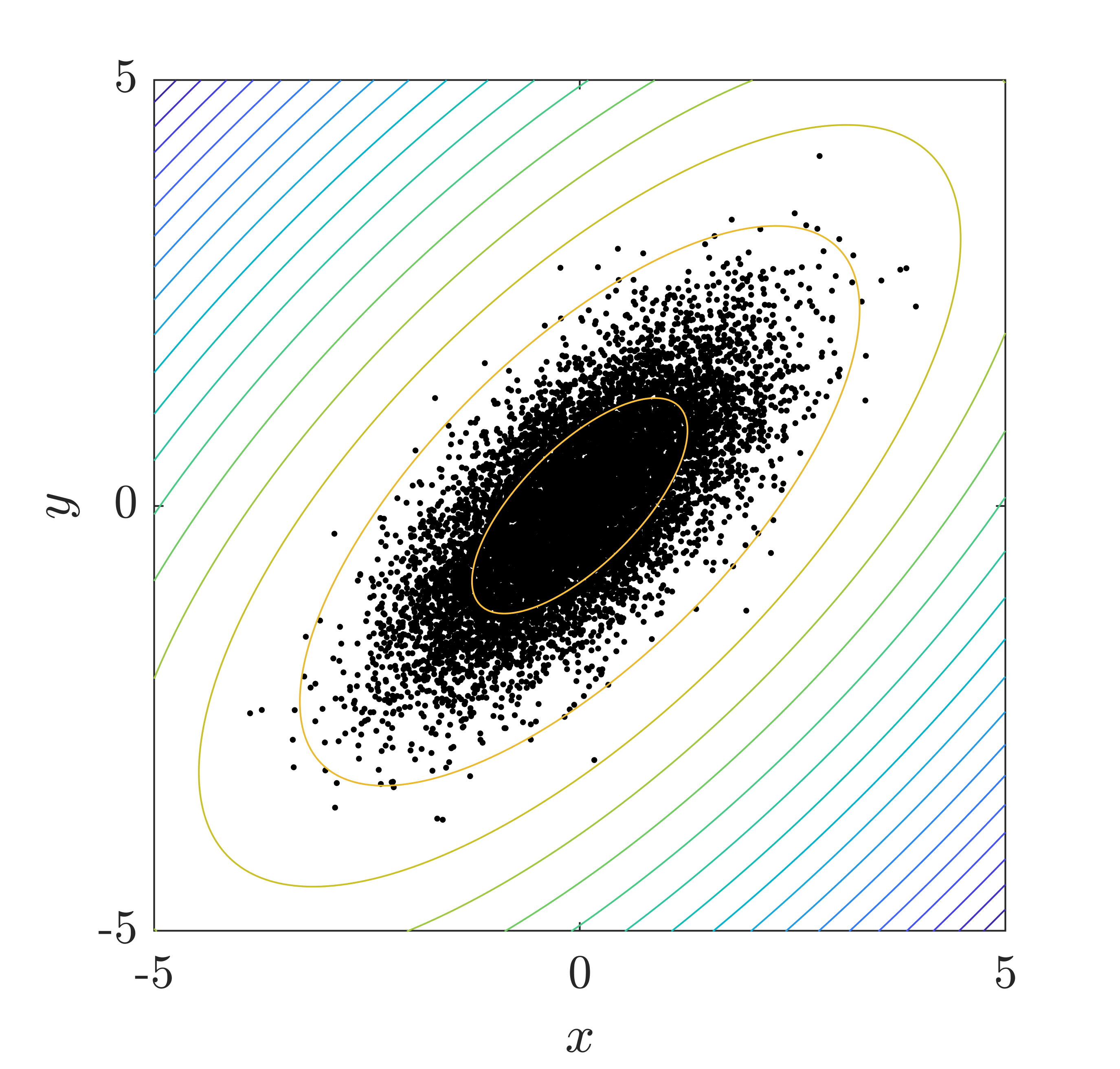}  (b) \includegraphics[scale=0.6]{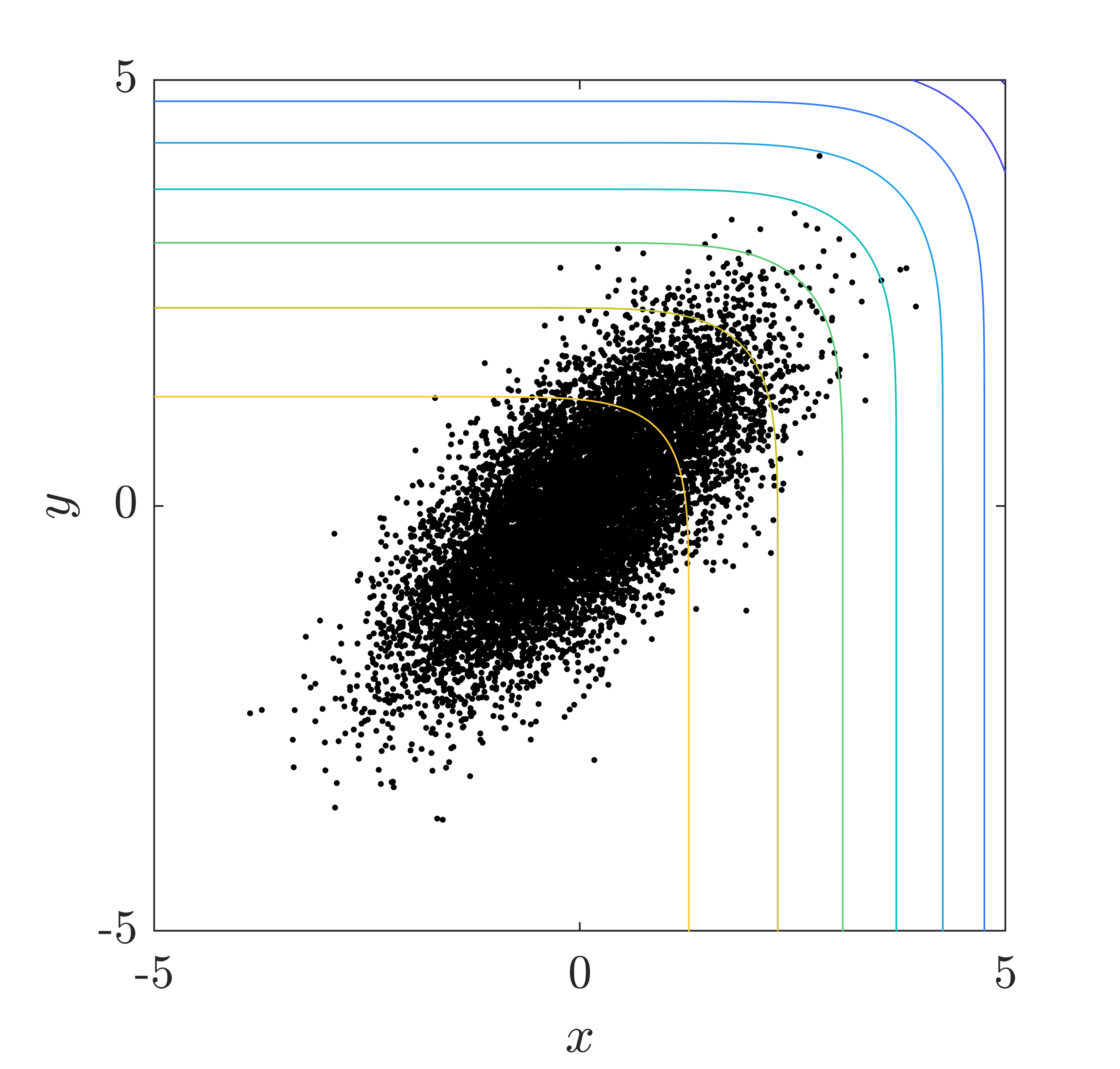}
    \caption{Black dots: Random sample of $10^4$ points from bivariate normal distribution with $\rho=0.7$. Coloured lines in (a) are isodensity contours at equal logarithmic increments. Coloured lines in (b) are level sets of the joint survivor function at equal logarithmic increments. In (b) an angular-radial description is only useful in the upper-right quadrant of the plane.}
    \label{fig:dens_vs_dist}
\end{figure}

To define our approach formally, we work with the joint density function, rather than the joint distribution or joint survival function. If we wish to describe the asymptotic behaviour of a random vector in all regions of the variable space, then the density function is a more natural quantity for analysis than the distribution function. Consider the example of the bivariate normal distribution, shown in \autoref{fig:dens_vs_dist}. The joint density function provides a description of where we are more or less likely to observe data points. In cases like this, where the density is unimodal and monotonically decreasing away from the mode in all directions, the isodensity contours can be viewed as describing locations which are equally `extreme', in the sense that they are equally rare. In contrast, isoprobability contours of the joint survival function only provide a useful description in the upper right quadrant of the plane. In the upper left and lower right quadrants, the contours at a given probability level asymptote toward the corresponding marginal quantile level. Visually, it appears that in this case the density function is likely to have a simple angular-radial description in all regions, whereas the survival function will not. We shall return to this example and the general case of elliptical distributions in Section \ref{sec:pdfs_in_polars}.

\subsection{Definition of the SPAR model} \label{sec:SPAR_def}
Consider a random vector $\mathbf{X}\in\mathbb{R}^d$, with joint density function $f_{\mathbf{X}}$. Assume there is a bijective map, $\mathcal{T}(\mathbf{X})=(R,\mathbf{W})$, from Cartesian coordinates to some polar coordinate system, where $R\in [0,\infty)$ is a radial variable and $\mathbf{W}\in\mathcal{S}$ is an angular variable. The set $\mathcal{S}\subset \mathbb{R}^d$ is a hypersurface, with the property that for each $\mathbf{w}\in \mathcal{S}$, the ray $\{r\mathbf{w}: r>0\}$ intersects $\mathcal{S}$ at a single point. Polar coordinate systems will be discussed in detail in Section \ref{sec:coords}. For now, we assume that a map to such a coordinate system exists. The obvious example is `standard' polar coordinates, where $\mathcal{S}$ is the unit hypersphere in $\mathbb{R}^d$. Unless otherwise stated, in this work we will assume that probability density functions of random vectors and variables exist and are integrable.

Returning to the definition of our model, suppose that $(R,\mathbf{W})$ has joint density function $f_{R,\mathbf{W}}$, and $\mathbf{W}$ has density $f_{\mathbf{W}}$. The angular-radial joint density function can be written in conditional form,
\begin{equation*}
	f_{R,\mathbf{W}}(r,\mathbf{w}) = f_{\mathbf{W}}(\mathbf{w}) f_{R|\mathbf{W}}(r|\mathbf{w}),
\end{equation*}
where $f_{R|\mathbf{W}}(r|\mathbf{w})$ is the density of $R$ conditional on $\mathbf{W}=\mathbf{w}$. The problem of modelling multivariate extremes then becomes a problem of modelling the angular density $f_{\mathbf{W}}(\mathbf{w})$ and the tail of the conditional density $f_{R|\mathbf{W}}(r|\mathbf{w})$. Define the conditional radial survivor function $\Bar{F}_{R|\mathbf{W}}(r|\mathbf{w}) = \int_r^{\infty} f_{R|\mathbf{W}}(s|\mathbf{w})\,ds$. For a given angle, $\Bar{F}_{R|\mathbf{W}}(r|\mathbf{w})$ is univariate. In univariate extremes, most methods for statistical inference begin with the assumption that the distribution is in the domain of attraction of an extreme value distribution. This suggests a possible assumption about the asymptotic behaviour of the tail of $\Bar{F}_{R|\mathbf{W}}(r|\mathbf{w})$, namely:
\begin{enumerate}[({A}1)]
    \item There exists functions $\sigma:(0,\infty)\times \mathcal{S}\to (0,\infty)$ and $\xi:\mathcal{S}\to \mathbb{R}$
    such that for each $\mathbf{w}\in\mathcal{S}$ and all $r>0$,
    \begin{equation} \label{eq:DOA_GP}
    \lim_{\mu\to r_F(\mathbf{w})} \sup_{r\in[0,r_F(\mathbf{w})-\mu]} \left| \frac{\Bar{F}_{R|\mathbf{W}}(\mu + r |\mathbf{w})}{\Bar{F}_{R|\mathbf{W}}(\mu|\mathbf{w})} - \Bar{F}_{GP}(r;\xi(\mathbf{w}),\sigma(\mu,\mathbf{w}))\right| = 0,
    \end{equation}
    where $r_F(\mathbf{w})$ is the upper endpoint of $R|(\mathbf{W}=\mathbf{w})$.
\end{enumerate}
$\Bar{F}_{GP}(r; \xi, \sigma)$ is the GP survivor function with shape parameter $\xi\in \mathbb{R}$ and scale parameter $\sigma>0$, given by
\begin{equation*}
	\Bar{F}_{GP}(r;\xi,\sigma) = 
	\begin{cases}
		\left(1 + \xi\dfrac{r}{\sigma} \right)^{-\tfrac{1}{\xi}}, & \xi\neq 0,\\[8pt]
		\exp\left(-\dfrac{r}{\sigma} \right), & \xi = 0.
	\end{cases}
\end{equation*}
The support of $F_{GP}$, the corresponding cumulative distribution function (cdf), is $0 \leq r < r_F$, where the upper end point is $r_F=\infty$ for $\xi\geq0$ and $r_F = - \sigma/\xi$ for $\xi<0$. Assumption (A1) is equivalent to the assumption that $F_{R|\mathbf{W}}(r|\mathbf{w})$ is in the domain of attraction of an extreme value distribution with shape parameter $\xi(\mathbf{w})$ \parencite{balkema1974,pickands1975}. 

In the univariate peaks-over-threshold (POT) method, assumption (A1) is used to motivate the approximation of the tail of a distribution above some high threshold by a GP distribution. In the multivariate context, assumption (A1) suggests a parametric model for the tail of the conditional radial distribution, but does not lead to a parametric form for the distribution of the angular variable. This motivates the so-called semi-parametric angular-radial (SPAR) model, introduced in \textcite{mackay2022imex}. A similar approach was also recently proposed in \textcite{papastathopoulos2023statistical}. The SPAR model assumes that the tail of $f_{R|\mathbf{W}}(r|\mathbf{w})$ above some threshold $\mu(\mathbf{w})$, can be represented by a GP density, with parameters conditional on $\mathbf{w}$. Let $\zeta(\mathbf{w}) \coloneqq \Pr(R > \mu(\mathbf{w}) \mid \mathbf{W} = \mathbf{w})$. Then the SPAR model for the joint angular-radial density can be written:
\begin{equation} \label{eq:SPAR_model}
	f_{R,\mathbf{W}}(r,\mathbf{w}) = \zeta(\mathbf{w})\, f_{\mathbf{W}}(\mathbf{w})\, f_{GP}(r-\mu(\mathbf{w});\xi(\mathbf{w}),\sigma(\mu,\mathbf{w})), \quad r>\mu(\mathbf{w}), 
\end{equation}
where $f_{GP}$ is the GP density function. For the purposes of inference, it is useful to make two further assumptions:
\begin{enumerate} [({A}1)]
    \setcounter{enumi}{1}
    \item There exist parameter functions $\sigma(\mu,\mathbf{w})$ and $\xi(\mathbf{w})$ that satisfy (A1) and are continuous, and $\sigma(\mu,\mathbf{w})$ is finite for finite $\mu$;
    \item The angular density $f_{\mathbf{W}}(\mathbf{w})$ is continuous and finite for all $\mathbf{w}$.
\end{enumerate}
Assumptions (A2) and (A3) are intended to simplify the inference procedure. Under these assumptions, the inference can be viewed as a non-stationary peaks-over-threshold analysis, for which there are many examples in the literature, \parencite[e.g.][]{chavez2005, Randell2016, Youngman2019, Zanini2020, Barlow2023}. As there are many potential methods for inference under the SPAR model assumptions, we defer an investigation of estimation methods to a separate work \parencite{MurphyBarltrop2023}. Instead, the focus of the present article is to consider conditions when assumptions (A1)-(A3) are valid. The intention is that these theoretical considerations will inform an approach to inference, that reframes multivariate extremes as a natural extension of univariate extremes with angular dependence.

We will consider cases where assumptions (A1)-(A3) hold either for all $\mathbf{w}\in \mathcal{S}$, or for all $\mathbf{w}\in\mathcal{S}^+$, where $\mathcal{S}^+=\mathcal{S} \cap [0,\infty)^d$ is the restriction of $\mathcal{S}$ to the (closed) non-negative orthant in $\mathbb{R}^d$. When these assumptions hold, we will say that $\mathbf{X}$ has a SPAR representation under map $\mathcal{T}$ for $\mathbf{w}\in \mathcal{S}$ or $\mathbf{w}\in \mathcal{S}^+$, respectively. Cases of other orthants in $\mathbb{R}^d$ can be treated analogously, by multiplying components of $\mathbf{X}$ by $-1$, so that the variable range of interest lies in the non-negative orthant of the transformed variable range. 

\subsection{Outline of the paper} \label{sec:outline}
The definition of the SPAR model requires a map from Cartesian to some angular-radial coordinates (or polar coordinates for short). As mentioned in Section \ref{sec:motivation}, there are various ways in which polar coordinates can be defined. This is discussed in detail in Section \ref{sec:coords}. We use more general types of polar coordinate system than those commonly-used for multivariate extremes, which are defined in terms of gauge functions for star-shaped sets. The more commonly-used polar coordinates, defined in terms of norms, are special cases of these generalised coordinate systems. Using these generalised polar coordinates is necessary for addressing the question of whether coordinate systems can be defined in which angular and radial variables are AI. As many of the examples in this work are two-dimensional, it is helpful to introduce a generalised concept of scalar angles in two dimensions, discussed in Section \ref{sec:pseudo_angle}. We show that using these generalised scalar angles leads to useful simplifications in some cases.

The effect of the choice of polar coordinate system is discussed in Section \ref{sec:pdfs_in_polars}. We consider which coordinate systems lead to a simple transformation of the probability density function from Cartesian to generalised polar coordinates, and if the choice of coordinates affects whether the SPAR assumptions are satisfied. Theorem \ref{thm:SPAR_indep} addresses the question of when it is possible to define coordinate systems in which the angular and radial components are AI. We show that when this is possible, the polar coordinate system for which this is true, is not necessarily a `standard' system, and may be difficult to estimate in practice. 

In order to make general statements about the impact of the choice of margins on whether a given copula has a SPAR representation, it is necessary to make some general assumptions about the asymptotic behaviour of copulas. Section \ref{sec:copula_tail_model} presents some generalised definitions of previous asymptotic models for copulas and copulas densities, as well as introducing some new definitions. We derive some links between these models, which are useful for understanding the limitations of angular-radial representations on various margins. 

In Section \ref{sec:margin}, we consider the effect of the choice of margins on whether the SPAR assumptions are met, and present some conditions on the copula which are sufficient for it to have a SPAR representation on certain margins. We show that using Laplace margins leads to forms of the density function in which the SPAR assumptions are satisfied for many commonly-used copulas, with various dependence classes. In contrast, the use of long-tailed or short-tailed margins imposes more restrictions on the types of copula that have SPAR representations. In Section \ref{sec:limit_set} we discuss the concept of limit sets of a random sample, introduced by \textcite{Davis1988}. Various existing representations for multivariate extremes can be related to the limit set, when it exists \parencite{Nolde2022}. We show that SPAR models on Laplace margins have a simple link to the corresponding limit set for the distribution, and provide a rigorous means of estimating the limit set, when it exists. 

Finally, a discussion and conclusions are presented in Section \ref{sec:discussion}. We return to some of the questions posed in Section \ref{sec:motivation}, regarding inference for the SPAR model, and discuss how the theoretical results derived in this work can inform the method used for inference. Proofs of results stated in the text are provided in Appendix \ref{app:proofs}. MATLAB code for calculating the numerical examples in this work is provided at \url{https://github.com/edmackay/SPAR}.

\section{Generalised angular-radial coordinate systems} \label{sec:coords}
In this section we define the generalised polar coordinate systems used throughout this work. Section \ref{sec:star_coords} defines polar coordinates in $\mathbb{R}^d$ in terms of gauge functions of star-shaped sets. This type of coordinate system has been studied by \textcite{Richter2014}, and we summarise the necessary information here. Then, Section \ref{sec:pseudo_angle} considers generalised scalar metrics of angle in $\mathbb{R}^2$. As far as we are aware, this type of generalised angle has not been defined previously, so these are considered in some detail. 

\subsection{Polar coordinate systems for star-shaped sets} \label{sec:star_coords}
We begin by presenting some preliminary definitions which can be found in reference texts \parencite[e.g.][]{Narici2010}. In the following discussion, we will associate various objects with norms and star-shaped sets. To refer to an arbitrary norm or set, we will use the notation $\|\cdot\|_*$ or $S_*$, replacing the asterisk with an appropriate label for definiteness as necessary. Throughout the article, we assume that the number of dimensions, $d$, is a natural number. 

\begin{definition}[Star-shaped set]
A star-shaped set, $G_*\subset\mathbb{R}^d$, is a set with the property that $\mathbf{x}\in G_*$ implies $t\mathbf{x}\in G_*$ for $0\leq t\leq 1$.
\end{definition}

\begin{definition}[Gauge function] \label{def:gaugefun}
Let $\mathcal{S}_*\subset\mathbb{R}^d$ be a surface which is the boundary to a compact star-shaped set containing the origin and assume $\mathbf{0}\notin\mathcal{S}_*$. The gauge function of $\mathcal{S}_*$, also referred to as the Minkowski functional, $\mathcal{R}_*: \mathbb{R}^d \to [0,\infty)$ is defined as $\mathcal{R}_*(\mathbf{x}) = \inf\{ r\in[0,\infty) : \mathbf{x} \in r\mathcal{S}_*\}$, where $r\mathcal{S}_* = \{(rw_1,...,rw_d) : (w_1,...,w_d)\in \mathcal{S}_*\}$.
\end{definition}

We use the notation $\mathcal{R}_*$ for the gauge function to connote that for each point $\mathbf{x}\in\mathbb{R}^d$, a gauge function defines a kind of radius of $\mathbf{x}$ from the origin. Informally, the gauge function defines how much we need to `inflate' or `contract' the surface, so that the point $\mathbf{x}$ lies on the scaled surface. Next, we consider norms and their unit spheres. 

\begin{definition}[Norm] \label{def:norm}
A norm on $\mathbb{R}^d$ is a function $\|\cdot\|_*:\mathbb{R}^d \to [0,\infty)$, for which the following properties hold for all $\mathbf{x},\mathbf{y}\in\mathbb{R}^d$ and $a\in\mathbb{R}$:
\begin{enumerate}
	\item Subadditivity: $\|\mathbf{x}+\mathbf{y}\|_* \leq \|\mathbf{x}\|_* + \|\mathbf{y}\|_*$.
	\item Absolute homogeneity: $\|a\mathbf{x}\|_* = |a|\|\mathbf{x}\|_*$.
	\item Positive definiteness: $\|\mathbf{x}\|_* = 0$ if and only if $\mathbf{x} = \mathbf{0}$.
\end{enumerate}
\end{definition}

\begin{definition}[Unit sphere of a norm]
The unit sphere with respect to the norm $\|\cdot\|_*$ is the set of points $\mathcal{U}_* = \{\mathbf{w}\in\mathbb{R}^d: \|\mathbf{w}\|_*=1\}$. In the case $d=2$, $\mathcal{U}_*$ is referred to as the unit circle.
\end{definition}

The subadditivity property of norms implies that the unit sphere with respect to any norm must be convex, and the absolute homogeneity property implies the unit sphere is centrally symmetric, i.e. $\mathbf{u}\in\mathcal{U}_*$ implies $-\mathbf{u}\in\mathcal{U}_*$. The unit sphere defines the boundary to a star-shaped set $G_*= \{ \mathbf{w} \in \mathbb{R}^d: \|\mathbf{w}\|_* \leq 1 \}$. The fact that $G_*$ is star-shaped can be seen from the absolute homogeneity property of norms: $\mathbf{x}\in G_*$ implies that for $0 \leq t\leq 1$ we have $\|t\mathbf{x}\|_* \leq \|\mathbf{x}\|_* \leq 1$, and hence $t\mathbf{x}\in G_*$. An equivalent way to define a norm would be in terms of the gauge function of its unit sphere or circle, i.e. 
$\|\mathbf{x}\|_* = \inf\{ r\in[0,\infty) : \mathbf{x} \in r\mathcal{U}_*\}$. Writing in this way, it is clear that norms are a special type of gauge function. In general, gauge functions differ from norms in that they do not necessarily satisfy subadditivity or absolute homogeneity. However, gauge functions are positive definite and positively homogeneous, with $\mathcal{R}_*(a\mathbf{x}) = a \mathcal{R}_*(\mathbf{x})$ for $a>0$. Moreover, the boundaries to star-shaped sets do not necessarily have the properties of the unit sphere for a norm. In particular, they need not be convex or centrally symmetric. Norms are usually given as explicit functions of the input variables, making their calculation straightforward. In contrast, evaluating the gauge function for an arbitrary surface $\mathcal{S}_*$ is less practical from a computational perspective, although clearly achievable in principle. We will see in Section \ref{sec:pdfs_in_polars} that defining polar coordinates in terms of this more general notion of gauge functions leads to simple forms of SPAR representations in which the angular and radial variables are independent in some cases. 

The most commonly-used norm used for defining angular-radial coordinates is the $L^p$ norm, defined below.

\begin{definition}[$L^p$ norm]
For a vector $\mathbf{x}=(x_1,\cdots,x_d)\in\mathbb{R}^d$ and real number $p\geq1$, the $L^p$ norm is defined as
\begin{equation} \label{eq:Lp}
	\|\mathbf{x}\|_p = \left(\sum_{j=1}^d |x_j|^p \right)^{1/p}.
	\end{equation}
In the case $p=\infty$, the $L^\infty$ norm is defined as $\|\mathbf{x}\|_\infty = \max\{|x_1|,\cdots,|x_d| \}$. The $L^p$ norms for $p=1,2,\infty$ are sometimes referred to as the sum norm, Euclidean norm and max norms, respectively.
\end{definition}

In the case $p\in(0,1)$, (\ref{eq:Lp}) does not define a norm, since it is not subadditive. However, $\|\mathbf{x}\|_p$ for $p\in(0,1)$ does define a gauge function. This is discussed further in Example \ref{ex:Lp_angle} below. Next, we come to the definition of generalised polar coordinates, defined in terms of the gauge function of a star-shaped set. Consider a vector $\mathbf{x} \in \mathbb{R}^d \setminus \{\mathbf{0}\}$. In the multivariate extremes literature, it is common to define angular-radial coordinates as 
\begin{equation*}
	r = \|\mathbf{x}\|_{rad}, \quad \mathbf{w} = \frac{\mathbf{x}}{\|\mathbf{x}\|_{ang}},
\end{equation*}
where $\|\cdot\|_{rad}$ and $\|\cdot\|_{ang}$ are two arbitrary norms, usually $L^p$ norms \parencite[see e.g.][p258]{Beirlant2004}. This can be generalised, by replacing the norms with gauge functions. To ensure that we can calculate the Jacobian of the transformation from Cartesian to polar coordinates, we require that the gauge functions used to define radii and angles correspond to star-shaped sets with continuous and piecewise-smooth boundaries. This will be discussed further in Section \ref{sec:pdfs_in_polars}.

\begin{definition}[Polar coordinates in $\mathbb{R}^d$]
Let $\mathbf{x} \in \mathbb{R}^d \setminus \{0\}$ and $\mathcal{R}_{rad}$, $\mathcal{R}_{ang}$, be a arbitrary gauge functions, corresponding to continuous and piecewise-smooth surfaces $\mathcal{S}_{rad}$ and $\mathcal{S}_{ang}$. Define the bijective map $\mathcal{T}:\mathbb{R}^d\setminus\{0\}\to(0,\infty) \times \mathcal{S}_{ang}$ as 
\begin{equation*} 
    \mathcal{T}(\mathbf{x}) = \left(\mathcal{R}_{rad}(\mathbf{x}), \frac{\mathbf{x}}{\mathcal{R}_{ang}(\mathbf{x})}\right).
\end{equation*}
The values $(r,\mathbf{w})=\mathcal{T}(\mathbf{x})$ are the polar coordinates of $\mathbf{x}$ with respect to gauge functions $\mathcal{R}_{rad}$ and $\mathcal{R}_{ang}$. The inverse map  $\mathcal{T}^{-1}:(0,\infty) \times \mathcal{S}_{ang} \to \mathbb{R}^d\setminus\{0\}$ from polar to Cartesian coordinates is given by
\begin{equation*}
    \mathcal{T}^{-1}(r,\mathbf{w}) = r \frac{\mathbf{w}}{\mathcal{R}_{rad}(\mathbf{w})}.
\end{equation*}
\end{definition}

Since $\mathcal{T}$ is dependent on the two gauge functions $\mathcal{R}_{rad}$, $\mathcal{R}_{ang}$, we could include this information and write $\mathcal{T}(\mathbf{x}; \mathcal{R}_{rad}, \mathcal{R}_{ang})$. However, as the gauge functions should be clear from the context, we omit this information to avoid overly-cumbersome notation. 

\subsection{Generalised angles in \texorpdfstring{$\mathbb{R}^2$}{R2}} \label{sec:pseudo_angle}
For the polar coordinate systems defined above, although the angular coordinate $\mathbf{w}\in\mathcal{S}_{ang}$ is $d$-dimensional, it only has $d-1$ degrees of freedom, due to the constraint that $\mathcal{R}_{ang}(\mathbf{w})=1$. For $d=2$, the angular variable $\mathbf{w}$ has only one degree of freedom, but is defined in terms of two coordinates, $\mathbf{w}=(w_1,w_2)$. For bivariate cases, it is useful to define a single variable to specify the angle. In many works on bivariate extremes, it is only the upper-right quadrant of the plane that is of interest. In this case, if there is a one-to-one relation between $w_1$ and $w_2$ in this quadrant, then the variable $w_1$ can be used as a surrogate for angle. However, if we wish to uniquely specify coordinates in all quadrants of the plane, then this definition of angle is ambiguous. For example, if angles are defined in terms of the $L^1$ norm, then for $(x,y)\in\mathbb{R}^2\setminus\{0\}$ we have $w_1=x/(|x|+|y|)$, which contains no information about the sign of $y$. 

Below, we will define a generalised scalar angle of a point, defined in terms of a gauge function. To motivate this, consider the definition of Euclidean angles, defined in terms of the $L^2$ norm. We have $(x,y) / \|(x,y)\|_2 = (\cos(\theta), \sin(\theta))$, where $\theta=\mathrm{atan2}(x,y)$, and $\mathrm{atan2}$ is the four-quadrant inverse tan function. In this case, the coordinates $(r,\theta)\in (0,\infty)\times(-\pi,\pi]$ uniquely specify a point in the punctured plane $\mathbb{R}^2\setminus\{0\}$. The Euclidean angle of $(x,y)$ can be defined as the distance around the circumference of the $L^2$ unit circle from $(1,0)$ to $(x,y)/\|(x,y)\|_2$, measured counter-clockwise, with negative values corresponding to clockwise distances. This definition can be generalised to arbitrary gauge functions as follows. To ensure that the angle is well-defined, we require that the boundary to the star-shaped set is continuous and piecewise-smooth. 

\begin{figure}[!t]
    \centering
    \includegraphics[scale=0.6]{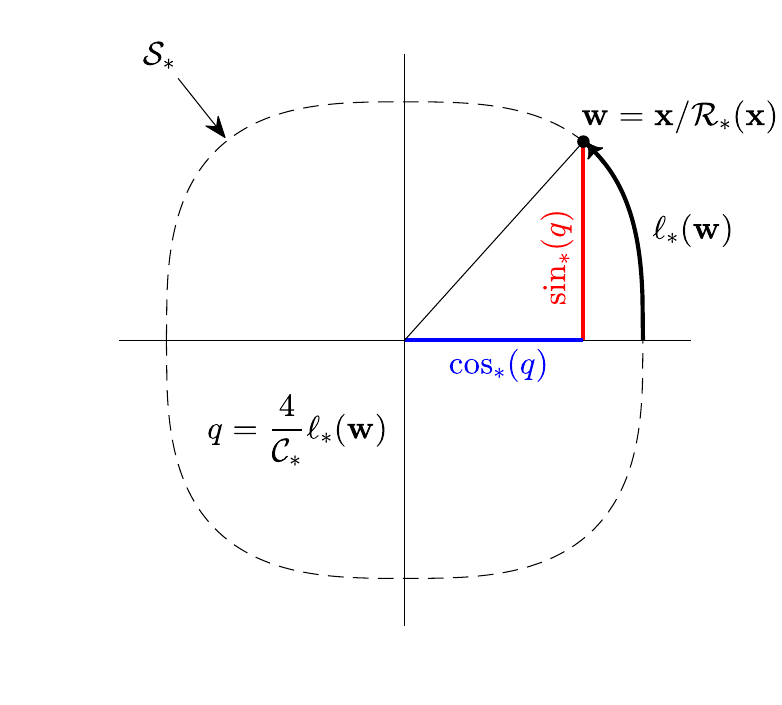}
    \vskip-30pt
    \caption{Illustration of definition of gauge function $\mathcal{R}_*$ and pseudo-angle, $q$, with respect to boundary $\mathcal{S}_*$. The circumference of the boundary is $\mathcal{C}_*$. Pseudo-trigonometric functions $\sin_*(q)$ and $\cos_*(q)$ relate the pseudo-angle with the corresponding x- and y-coordinates on the unit circle.}
    \label{fig:pseudo_angle}
\end{figure}

\begin{definition}[Arc length functions] \label{def:arclength}
Let $\mathcal{R}_*$ be a gauge function on $\mathbb{R}^2$ for a continuous and piecewise-smooth boundary $\mathcal{S}_*$. Denote the total length of $\mathcal{S}_*$ as $\mathcal{C}_*$ (in the case where $\mathcal{R}_*$ is a norm, this is the circumference of the unit circle). Define the arc length function $\ell_*: \mathcal{S}_* \to [0, \mathcal{C}_*)$, to be the distance along $\mathcal{S}_*$, from $(1,0)/\mathcal{R}_*(1,0)$ to $\mathbf{w}\in\mathcal{S}_*$, measured anti-clockwise (see \autoref{fig:pseudo_angle}). Define a set-valued function $\mathfrak{L}_*:\mathcal{S}_*\to \{ \{ d + n\mathcal{C}_* : n\in\mathbb{Z} \} : d \in [0, \mathcal{C}_*)\}$ as
\begin{equation*}
    \mathfrak{L}_*(\mathbf{w}) = \{\ell_*(\mathbf{w}) + n\mathcal{C}_* : n\in\mathbb{Z}\}.
\end{equation*}
The branches of $\mathfrak{L}_*$ for each value of $n\in\mathbb{Z}$ correspond to the number of loops around the unit circle, with $n<0$ corresponding to distances measured clockwise from the x-axis. We denote the restriction of the image of $\mathfrak{L}_*$ to a particular branch with values in the interval $I$, as $\mathfrak{L}_*^I$, where $I$ is a half open interval, either $(a,a+\mathcal{C}_*]$ or $[a,a+\mathcal{C}_*)$ for some $a\in\mathbb{R}$.
\end{definition}

The arc length function can be evaluated using integration. By the assumption that $\mathcal{S}_*$ is piecewise-smooth, for $(u,v)\in\mathcal{S}_*$ we can divide it into segments where either $dv/du$ or $du/dv$ (or both) is continuous and bounded. For segments where $dv/du$ is continuous and bounded, the arc length, $s$, between $u=a$ and $u=b$, can be calculated as
\begin{equation} \label{eq:ell_int}
    s = \int_a^b \left(1+\left(\frac{dv}{du}\right)^2 \right)^{1/2} du.
\end{equation}
In segments where $dv/du$ becomes infinite, we switch $u$ and $v$ in (\ref{eq:ell_int}). As the circumference $\mathcal{C}_*$ is dependent on $\mathcal{S}_*$, we define pseudo-angles in terms of a normalised distance around the boundary.

\begin{definition}[Pseudo-angles] \label{def:angle}
Let $\mathcal{R}_*$ be a gauge function on $\mathbb{R}^2$ for piecewise-smooth boundary $\mathcal{S}_*$. A set-valued pseudo-angle function $\mathcal{A}_*:\mathcal{S}_*\to \{\{q + 4n : n\in\mathbb{Z}\} : q\in[0,4)\}$, is defined as
\begin{equation*}
    \mathcal{A}_*(\mathbf{w}) \coloneqq \frac{4}{\mathcal{C}_*}\mathfrak{L}_*(\mathbf{w}).
\end{equation*}
The restriction of the image of $\mathcal{A}_*$ to a particular branch with values in the interval $I$, is denoted $\mathcal{A}_*^I$, where $I$ is a half-open interval, either $(a,a+4]$ or $[a,a+4)$ for some $a\in\mathbb{R}$. 
\end{definition}

The motivation for the inclusion of the factor $4$ in the definition of pseudo-angles is as follows. By definition, the set of angles $\{0+4n:n\in\mathbb{Z}\}$ corresponds to the direction of the positive x-axis for any gauge function, $\mathcal{R}_*$. When $\mathcal{R}_*$ is a norm, from the central symmetry property of the unit circle, the set of angles $\{2+4n:n\in\mathbb{Z}\}$ corresponds to the direction of the negative x-axis. Suppose that $\mathcal{S}_*$ is symmetric about the x- and y-axes (in the case of unit circles for norms, the absolute homogeneity property implies that if the unit circle is symmetric in one axis then it is symmetric in both axes). In this case, the arc length in each quadrant of the plane is equal. This symmetry property holds for $L^p$ norms. Under this symmetry, the sets of angles $\{1+4n:n\in\mathbb{Z}\}$ and $\{3+4n:n\in\mathbb{Z}\}$ correspond to the directions of the positive and negative y-axes, respectively. In these cases, the pseudo-angle, $q=\mathcal{A}_*^{[0,4)}(\mathbf{x}/\mathcal{R}_*(\mathbf{x}))$, can be interpreted as indicating the quadrant of the plane containing $\mathbf{x}$, i.e. $q\in(0,1)$ indicates $\mathbf{x}$ lies in the first quadrant, $q\in(1,2)$ indicates $\mathbf{x}$ lies in the second quadrant, etc., motivating the inclusion of the factor $4$ in the definition of pseudo-angle. The pseudo-angle could have been defined alternatively using the normalising factor $2\pi$, so that the definition coincides with the standard Euclidean definition for the $L^2$ norm. However, we have opted to normalise using the factor $4$, firstly for the interpretation in terms of quadrants of the plane and secondly to emphasise the difference between Euclidean angles and pseudo-angles.\footnote{The definition of pseudo-angle is given in terms of a relation between a vector and a point on the x-axis. More generally, we could define a pseudo-angle between any two vectors in the same way. However, apart from the case of the Euclidean norm (and norms which are scalar multiples of this), the pseudo-angle between two vectors is not invariant to rotation.}

To define polar coordinates on $\mathbb{R}^2$, we consider the case where the pseudo-angle and gauge functions are defined in terms of two arbitrary gauge functions. This can simplify the transformation of density functions from Cartesian to polar coordinates, discussed in the next section. When specifying polar coordinates, we will assume $I=(-2,2]$. This follows the convention for the angular variable $\theta$ used in standard polar coordinates, where it is usual to assume $\theta\in(-\pi,\pi]$. 

\begin{definition}[Polar coordinates on $\mathbb{R}^2$] \label{def:polarR2}
Let $\mathbf{x} = (x,y) \in \mathbb{R}^2 \setminus \{0\}$ and $\mathcal{R}_{rad}$, $\mathcal{R}_{ang}$, be arbitrary gauge functions, defined in terms of piecewise-smooth boundary $\mathcal{S}_{ang}$. Define the bijective map $\tilde{\mathcal{T}}:\mathbb{R}^2\setminus\{0\}\to(0,\infty)\times (-2,2]$ as 
\begin{equation*} 
    \tilde{\mathcal{T}}(\mathbf{x}) = \left(\mathcal{R}_{rad}(\mathbf{x}), \mathcal{A}_{ang}^{(-2,2]}\left(\frac{\mathbf{x}}{\mathcal{R}_{ang}(\mathbf{x})}\right)\right).
\end{equation*}
The values $(r,q)=\tilde{\mathcal{T}}(x,y)$ are the polar coordinates of $(x,y)$ with respect to gauge functions $\mathcal{R}_{rad}$ and $\mathcal{R}_{ang}$. The notation $\tilde{\mathcal{T}}$ is used for the two-dimensional map from Cartesian to polar coordinates, to emphasise that it differs from the map in $\mathcal{T}$ which uses vector angles. 
\end{definition}

The trigonometric functions sine and cosine provide a short-hand for the inverse map from standard polar coordinates to Cartesian coordinates, i.e. if $(r,\theta)$ are the standard polar coordinates of $(x,y)$ then $(x,y)=r(\cos(\theta),\sin(\theta))$. If gauge function $\mathcal{R}_*$ is used to define both radii and angles, then it is useful to define analogous pseudo-trigonometric functions $\cos_{*}(q)$ and $\sin_{*}(q)$ such that $\Tilde{\mathcal{T}}^{-1} (r,q) = r (\cos_{*}(q),\sin_{*}(q))$. In general, if different gauge functions are used to specify radii and angles, then the inverse map will be defined as
\begin{equation*}
    \Tilde{\mathcal{T}}^{-1} (r,q) = r \frac{(\cos_{ang}(q),\sin_{ang}(q))}{\mathcal{R}_{rad} (\cos_{ang}(q),\sin_{ang}(q))}.
\end{equation*}

\begin{definition}[Pseudo-trigonometric functions] \label{def:pseudotrig}
Let $\mathcal{A}_*^I$ be a pseudo-angle function for the boundary $\mathcal{S}_*$. Note that $\mathcal{A}_*^I$ is a bijection and denote the inverse map as $(\mathcal{A}_*^I)^{-1}: I\to\mathcal{S}_*$. Denote the extent of $\mathcal{S}_*$ in the x- and y-directions as:
\begin{align*}
	u_{min} &= \inf\{u: (u,v)\in\mathcal{S}_*, v\in\mathbb{R}\},\\
	u_{max} &= \sup\{u: (u,v)\in\mathcal{S}_*, v\in\mathbb{R}\},\\
	v_{min} &= \inf\{v: (u,v)\in\mathcal{S}_*, u\in\mathbb{R}\},\\
	v_{max} &= \sup\{v: (u,v)\in\mathcal{S}_*, u\in\mathbb{R}\}.
\end{align*}
The pseudo-trigonometric functions $\cos_*:\mathbb{R}\to[u_{min},u_{max}]$ and $\sin_*:\mathbb{R}\to [v_{min},v_{max}]$, are defined in terms of the inverse map as
\begin{equation} \label{eq:trigfun}
    (\cos_*(q), \sin_*(q)) = (\mathcal{A}_*^I)^{-1}(q).
\end{equation}
Although the domain on the RHS of (\ref{eq:trigfun}) is restricted to interval $I$, the definition extends to $q\in\mathbb{R}$, by noting that $I$ can be any half-open interval in $\mathbb{R}$ with length 4. 
\end{definition}

The pseudo-trigonometric functions relate the pseudo-angle to the corresponding x- and y-positions on the boundary $\mathcal{S}_*$ (see \autoref{fig:pseudo_angle}), in a way that is directly analogous to standard trigonometric functions. \textcite{Richter2007} also considered the definition of functions to relate the Euclidean angle $\theta$ with the corresponding point on the $L^p$ unit circle. The key difference with the functions defined here is that our pseudo-trigonometric functions take pseudo-angles as an input, rather than Euclidean angles. As far as we are aware, our definition of pseudo-angle has not been presented previously.

\begin{example}[Pseudo-angles for $L^p$ norms on $\mathbb{R}^2$] \label{ex:Lp_angle}
In this example we will consider the pseudo-angles and pseudo-trigonometric functions for the $L_p$ norm. As noted above, when $p\in(0,1)$, $\|\cdot\|_p$ is not a norm, but is a gauge function. We will include these cases for completeness. Examples of the unit circles for $L^p$ norms are shown in \autoref{fig:Pseudo_trig}. Let $(u,v)$ be points on the $L^p$ unit circle, so that $u = \pm (1-|v|^p)^{1/p}$. The absolute value of the gradient is given by
\begin{equation*}
    \left|\frac{du}{dv}\right| = \left(\frac{|v|^p}{1-|v|^p}\right)^{1-\tfrac{1}{p}}.
\end{equation*}
This can be substituted into (\ref{eq:ell_int}) to calculate the arc length $\ell_p(u,v)$ numerically (see Definition \ref{def:arclength}). Note that the point $u=v=2^{-1/p}$ corresponds to 1/8 of the circumference of the unit circle, $\mathcal{C}_p$, measured from the positive x- or y-axes. Therefore, we have $\mathcal{C}_p = 8\, \ell_p(2^{-1/p},2^{-1/p})$, so we need only calculate pseudo-angles for $u,v\in[0,2^{-1/p}]$, and the remaining values can be found by the symmetry of the $L^p$ unit circle. The pseudo-trigonometric functions $\cos_p(q)$ and $\sin_p(q)$ are shown in \autoref{fig:Pseudo_trig} for various values of $p$. For the special cases $p=1$ and 2 we can write explicit expressions for the pseudo-trigonometric and angle functions. For $p=1$ we have
\begin{align*}
    \cos_1(q) &= 1 - |q|, &  q&\in[-2,2]\\
    \mathcal{A}_1^{(-2,2]}(u,v) &= \varepsilon(v) (1 - u), & (u,v)&\in\mathcal{U}_1,
\end{align*}
where $\varepsilon$ is the generalised sign function, where $\varepsilon(v)=1$ for $v\geq0$ and $\varepsilon(v)=-1$ otherwise.\footnote{The difference is that $\text{sgn}(0)=0$ whereas $\varepsilon(0)=1$.} The domain of definition of $\cos_1$ can be extended to $q\in\mathbb{R}$ using the periodic property. We can then define $\sin_1(q) = \cos_1(q-1)$ for $q\in\mathbb{R}$. For $p=2$, we have $\cos_2(q) = \cos(\pi q/2)$, $\sin_2(q) = \sin(\pi q/2)$ for $q\in\mathbb{R}$ and $\mathcal{A}_2^{(-2,2]}(u,v) = (2/\pi) \text{atan2}(u,v)$ for $(u,v)\in\mathcal{U}_2$.\hfill $\blacksquare$

\begin{figure}[!t]
    \centering
    \includegraphics[scale=0.5]{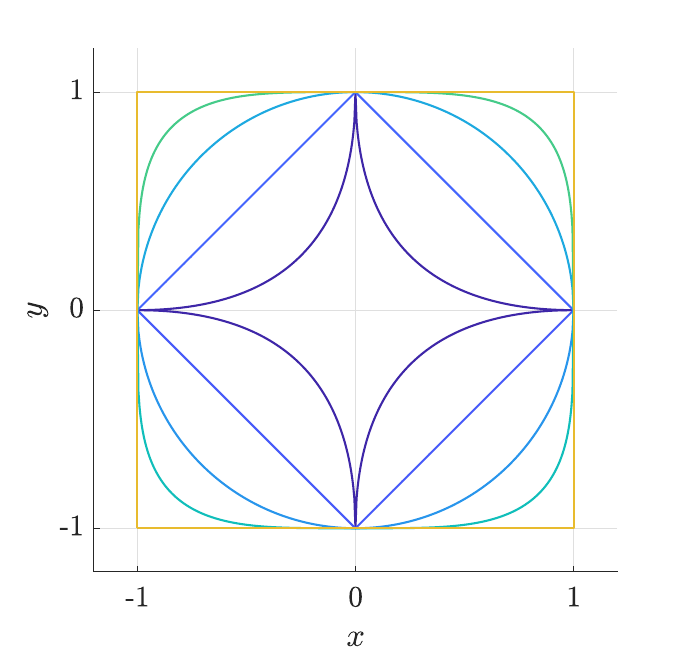}
    \includegraphics[scale=0.5]{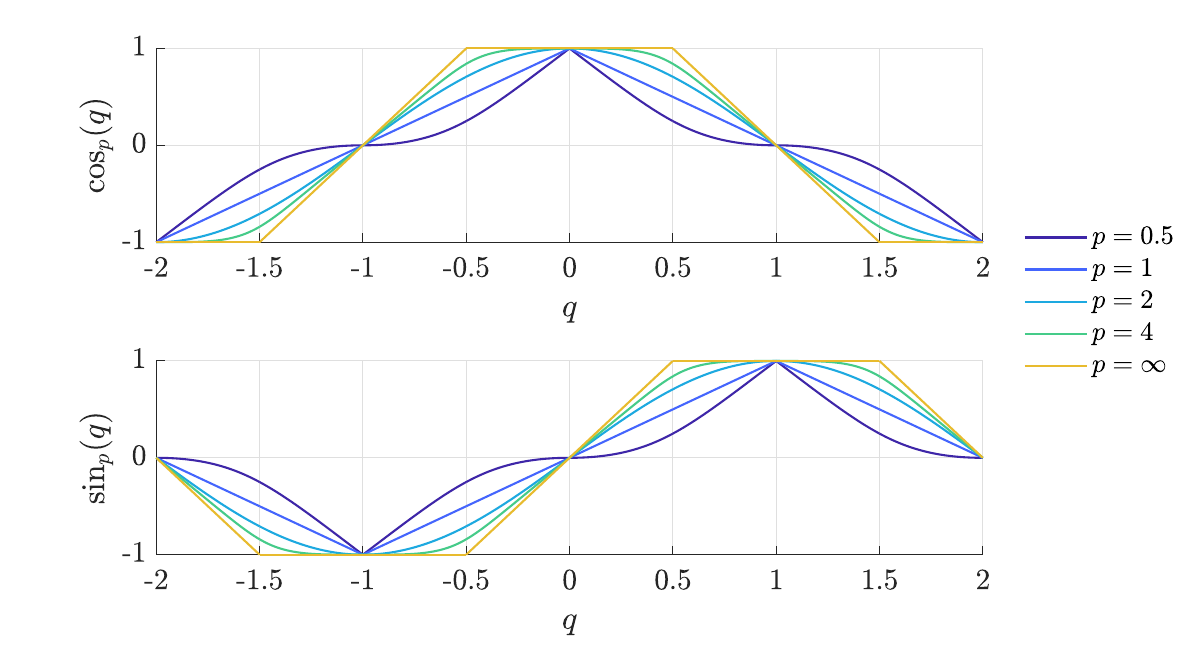}
    \caption{Left: Unit circles for the $L^p$ norm for various values of $p$. When $0<p<1$, $\|\cdot\|_p$ is not a norm, but is a gauge function. Right: Corresponding pseudo-trigonometric functions.}
    \label{fig:Pseudo_trig}
\end{figure}
\end{example}

\section{Effects of the choice of coordinate system} \label{sec:pdfs_in_polars}
The SPAR model requires a transformation of density functions from Cartesian to polar coordinates. In this section we consider the effect of the choice of polar coordinate system on SPAR representations for a given joint density. To illustrate the general principles, we start by considering bivariate cases in Section \ref{sec:pdf_coords_bivar}, with polar coordinates specified in terms of scalar pseudo-angles. We consider the Jacobian of the transformation from Cartesian to generalised polar coordinates, with focus on pseudo-angles defined in terms of $L^p$ norms. We then consider the effect of changing the gauge functions used to define radii and angles, and how this affects whether the SPAR assumptions are satisfied. Finally, these results are used to address the question of when coordinate systems can be defined in which angular and radial variables are asymptotically independent. In Section \ref{sec:pdf_coords_multivar} we go on to consider the general case of density functions in $\mathbb{R}^d$, and show that similar results apply. The results in this section are mostly based on calculating Jacobians of various transformations of coordinate systems. However, as the coordinate systems involved are non-standard in some cases, we illustrate the results using simple examples. 

\subsection{Bivariate density functions} \label{sec:pdf_coords_bivar}
Suppose that random vector $(X,Y)$ has probability density function $f_{X,Y}(x,y)$. Define the random vector $(R,Q)=\tilde{\mathcal{T}}(X,Y)$ for some gauge functions $\mathcal{R}_{rad}$ and $\mathcal{R}_{ang}$ (see Definition \ref{def:polarR2}). Then $(R,Q)$ has density function
\begin{equation} \label{eq:fRQ_Jacobian}
    f_{R,Q}(r,q)=f_{X,Y}\left(r \frac{(\cos_{ang}(q),\sin_{ang}(q))}{\mathcal{R}_{rad}(\cos_{ang}(q),\sin_{ang}(q))}\right) |\det(\mathbf{J}(r,q))|,
\end{equation}
where $\mathbf{J}$ is the Jacobian matrix of $\tilde{\mathcal{T}}^{-1}(r,q)$, so that
\begin{equation}  \label{eq:detJ}
    |\det(\mathbf{J}(r,q))| = \frac{r}{(\mathcal{R}_{rad}(\cos_{ang}(q),\sin_{ang}(q)))^2} \left| \cos_{ang}(q)\sin_{ang}'(q) - \cos_{ang}'(q)\sin_{ang}(q) \right|,
\end{equation}
where the prime indicates differentiation with respect to the argument. As the surface $\mathcal{S}_{ang}$, used to define the pseudo-angle, is only required to be piecewise smooth, the functions $\cos_{ang}$ and $\sin_{ang}$ may not be differentiable everywhere, but we can divide $\mathcal{S}_{ang}$ into a finite number of disjoint segments where the derivatives exist. Finding explicit expressions for the pseudo-trigonometric functions and their derivatives is not always easy. The following proposition considers the case for $L^p$ norms. [Proofs of results stated in the text are provided in Appendix \ref{app:proofs}].

\begin{proposition} \label{prop:LpJacobian}
Let $\cos_p$ and $\sin_p$ be pseudo-trigonometric functions for the $L^p$ norm. Then for $q\in\mathbb{R}$,
\begin{equation} \label{eq:Jq}
	J_p(q) \coloneqq \left| \cos_p(q)\sin_p'(q) - \cos_p'(q)\sin_p(q) \right| = \frac{\mathcal{C}_p}{4} \left( \cos_p^{2(p-1)}(q) + \sin_p^{2(p-1)}(q) \right)^{-1/2}.
\end{equation}
When $p=1$, 2 or $\infty$, $J_p(q)$ is a constant and we have
\begin{equation*}
	J_1(q) = 1, \qquad
	J_2(q) = \frac{\pi}{2}, \qquad
	J_\infty(q) = 2, \qquad q\in\mathbb{R}.
\end{equation*}
For $p\in(1,2)\cup(2,\infty)$, $J_p(q)$ is not constant.
\end{proposition}

Proposition \ref{prop:LpJacobian} shows that using pseudo-angles defined in terms of the $L^p$ norm for $p=1,2,\infty$ results in a simple transformation of density functions from Cartesian to polar coordinates. For other values of $p$, the Jacobian of the transformation is non-trivial to calculate. However, since the $L^\infty$ unit circle is a scaled and rotated copy of the $L^1$ unit circle (see \autoref{fig:Pseudo_trig}), pseudo-angles defined in terms of the $L^\infty$ norm correspond to a phase-lagged version of pseudo-angles defined in terms of the $L_1$ norm, so we will not consider these pseudo-angles further. For the case of the $L^2$ norm, it is more natural to use the usual Euclidean angle. As above, suppose that random vector $(X,Y)$ has probability density function $f_{X,Y}(x,y)$, and define the radial variable $R=\mathcal{R}_*(X,Y)$, for some gauge function $\mathcal{R}_*$, and angular variables  $\Theta=\mathrm{atan2}(X,Y)$ and $Q_1 = \mathcal{A}_1^{(-2,2]}((X,Y)/\|(X,Y)\|_1)$. Then we have
\begin{align} 
\label{eq:fxy2frq}	f_{R,Q_1}(r,q) &= \frac{r}{(\mathcal{R}_*(\cos_1(q),\sin_1(q)))^2} f_{X,Y}\left(r \frac{(\cos_1(q),\sin_1(q))}{\mathcal{R}_*(\cos_1(q),\sin_1(q))}\right), \\
\label{eq:fxy2frtheta}	f_{R,\Theta}(r,\theta) &= \frac{r}{(\mathcal{R}_*(\cos(\theta),\sin(\theta)))^2} f_{X,Y}\left(r \frac{(\cos(\theta),\sin(\theta))}{\mathcal{R}_*(\cos(\theta),\sin(\theta))}\right).
\end{align}
The variable $\Theta$ is used here instead of $Q_2 = \mathcal{A}_2^{(-2,2]}((X,Y)/\|(X,Y)\|_2)$, for which the joint density of $(R,Q_2)$ differs by a factor of $\pi/2$ from the joint density of $(R,\Theta)$. In both cases, the calculation of the angular-radial density from the Cartesian density is straightforward. For the remainder of this work, we will work with either Euclidean angles or $L^1$ pseudo-angles. When working in $L^1$ coordinates, we will denote the angular variable $Q$, dropping the subscript. The next example illustrates that the most convenient coordinate system to use depends on the application. 

\begin{example}[Independence on normal and Laplace margins] \label{ex:indep}
Suppose that $(X,Y)$ is a random vector on $\mathbb{R}^2$ and define angular-radial random vectors
\begin{align*}
    (R_1,Q)&=\left(\|(X,Y)\|_1, \mathcal{A}_1^{(-2,2]} \left(\frac{(X,Y)}{\|(X,Y)\|_1}\right)\right),\\
    (R_2,\Theta) &= \left(\|(X,Y)\|_2, \mathrm{atan2}(X,Y) \right).
\end{align*}
First, suppose that $X$ and $Y$ are independent standard normal variables. The corresponding joint density functions are 
\begin{align*}
    f_{X,Y}(x,y) &= \frac{1}{2\pi} \exp \left( -\tfrac{1}{2}(x^2+y^2)\right),\\
    f_{R_1,Q}(r,q) &= \frac{r}{2\pi} \exp \left( -\frac{r^2}{2} (\cos_1^2(q) + \sin_1^2(q))\right),\\
    f_{R_2,\Theta}(r,\theta) &= \frac{r}{2\pi} \exp \left( -\frac{r^2}{2}\right).
\end{align*}
So $f_{R_1,Q}$ has angular dependence, whereas $f_{R_2,\Theta}$ is independent of $\Theta$. Now suppose that $X$ and $Y$ are independent standard Laplace variables. In this case, the corresponding joint densities are
\begin{align*}
    f_{X,Y}(x,y) &= \tfrac{1}{4} \exp(-(|x|+|y|)),\\
    f_{R_1,Q}(r,q) &= \frac{r}{4} \exp(-r),\\
    f_{R_2,\Theta}(r,\theta) &= \frac{r}{4} \exp(-r (|\cos(\theta)| + |\sin(\theta)|)).
\end{align*}
In this case $f_{R_2,\Theta}$ has angular dependence, whereas $f_{R_1,Q}$ does not. \hfill $\blacksquare$ 
\end{example}

In the cases considered in Example \ref{ex:indep}, the joint densities in Cartesian coordinates are given in terms of a norm of the coordinates, so using that particular norm to define the angular-radial coordinates is a natural choice, and leads to a simple representation. The next example illustrates this for the case of elliptical distributions. 

\begin{example}[SPAR models for elliptical distributions] \label{ex:elliptic}
Suppose that random vector $(X,Y)$ follows an elliptical distribution, and that both $X$ and $Y$ have zero median, unit variance and Pearson correlation coefficient $\rho\in(-1,1)$. Then the joint density function can be written $f_{X,Y}(x,y)=f_0(\|(x,y)\|_{e,\rho}^2)$, where $f_0$ is a generator function, and $\|\cdot\|_{e,\rho}$ is the elliptical norm given by \parencite{Cambanis1981}
\begin{equation*}
    \|(x,y)\|_{e,\rho} = \left(\frac{x^2 + y^2 - 2\rho xy}{1-\rho^2}\right)^{1/2}.
\end{equation*}
For example, the bivariate normal distribution is given by setting $f_0(z) = \gamma \exp(-z/2)$, where $\gamma=1/(2\pi\sqrt{1-\rho^2})$. Similarly, the bivariate t distribution with $\nu$ degrees of freedom, is given by setting $f_0(z) = \gamma \, (1 + z/\nu)^{-1-\tfrac{\nu}{2}}$. Define the random variables $R_e=\|(X,Y)\|_{e,\rho}$, $\Theta=\mathrm{atan2}(X,Y)$. From (\ref{eq:fxy2frtheta}), the random vector $(R_e,\Theta)$ has joint density \begin{align*}
    f_{R_e,\Theta}(r,\theta) &= \frac{r}{\|(\cos(\theta),\sin(\theta))\|_{e,\rho}^2} f_{X,Y}\left(r \frac{(\cos(\theta),\sin(\theta))}{\|(\cos(\theta),\sin(\theta)\|_{e,\rho}}\right)
    = 
    \alpha^2(\theta) r f_0(r^2),
\end{align*}
where
\begin{equation*}
	\alpha(\theta) = \|(\cos(\theta), \sin(\theta))\|_{e,\rho}^{-1}= \left(\frac{1-\rho^2}{1 - \rho \sin(2\theta)} \right)^{1/2}.
\end{equation*}
Under this transformation, the density factorises to independent radial and angular components, where the angular component is invariant to the type of elliptical distribution. This choice of radial variable was proposed by \textcite{Wadsworth2017} in their Example 3. Provided that $r f_0(r^2)$ is in the domain of attraction of an extreme value distribution and $\int_0^\infty r f_0(r^2) dr$ is finite, then $(X,Y)$ has a SPAR representation for $\theta\in(-\pi,\pi]$. 

If the generator function has antiderivative $F_0$ (i.e. $F'_0=f_0$), then $\partial F_0(r^2)/\partial r = 2r f_0(r^2)$. We can then write explicit expressions for the angular density and conditional radial distribution in terms of $F_0$,
\begin{align*}
	f_{\Theta}(\theta) &= \int_0^\infty f_{R,\Theta}(r,\theta)\,dr = \alpha^2(\theta) \int_0^\infty r\,f_0(r^2)\,dr = \frac{\alpha^2(\theta)}{2} \left[ F_0(r^2) \right]^{r=\infty}_{r=0},\\
	\Bar{F}_{R|\Theta}(r|\theta) &= \frac{\int_r^\infty f_{R,\Theta}(s,\theta)\,ds}{f_{\Theta}(\theta)} =  \frac{[F_0(s^2)]_{s=r}^{s=\infty}}{[F_0(s^2)]_{s=0}^{s=\infty}}.
\end{align*}
For a bivariate normal distribution, $F_0(z)=-2\gamma \exp(-z/2)$, and we have
\begin{align*}
	f_{\Theta}(\theta) &= \frac{\sqrt{1-\rho^2}}{2\pi(1 - \rho \sin(2\theta))},\\
	\Bar{F}_{R|\Theta}(r|\theta) &= \exp\left(-\frac{r^2}{2}\right).
\end{align*}
So the angular density is continuous and finite, satisfying assumption (A3). In this case, the conditional radial distribution is Rayleigh, with unit scale parameter. This satisfies assumptions (A1) and (A2) with $\xi(\theta)=0$ and $\sigma(\mu,\theta) = 1/\mu$. However, the convergence to the asymptotic form is slow. This is directly analogous to the univariate case, in the sense that a normal distribution is in the domain of attraction of a Gumbel distribution, but converges slowly. For a bivariate t distribution, $F_0(z) = -2\gamma \, (1 + z/\nu)^{-\tfrac{\nu}{2}}$, and the angular density is identical to that for the bivariate normal distribution. The survivor function of the conditional radial distribution is 
\begin{equation*}
	\Bar{F}_{R|\Theta}(r|\theta) = \left(1+\frac{r^2}{\nu}\right)^{-\nu/2}.
\end{equation*}
The quantile $\mu$ at exceedance probability $\zeta$ is therefore $\mu^2 = \nu (\zeta^{-2/\nu}-1)$. Moreover, for all $r>0$ we have
\begin{equation} \label{eq:Student_Fgp}
    \frac{\Bar{F}_{R|\Theta}(\mu + r|\theta)}{\Bar{F}_{R|\Theta}(\mu|\theta)} 
    =
    \left(\frac{1+\dfrac{(\mu+r)^2}{\nu}}{1+\dfrac{\mu^2}{\nu}}\right)^{-\nu/2}
    \sim
    \left(1+\frac{r}{\mu}\right)^{-\nu}
    =
    \bar{F}_{GP}(r;1/\nu,\mu/\nu), \quad \mu\to\infty.
\end{equation}
[Throughout this paper we use the notation $f(x)\sim g(x)$ as $x\to\infty$ to mean $\lim_{x\to\infty} f(x)/g(x)=1$.] Hence assumptions (A1) and (A2) are satisfied with $\xi(\theta)=1/\nu$ and $\sigma(\mu,\theta)=\mu/\nu$. An example of a SPAR approximation to a bivariate t distribution with $\nu=2$ and $\rho=0.6$ is shown in Figure \ref{fig:SPAR_student}. In this example, the threshold exceedance probability has been set at $\zeta=0.05$. The isodensity contours of the SPAR approximation, closely follow those of the true distribution. The angular density is also shown. As it is smoothly-varying and finite it is readily amenable to non-parametric estimation. \hfill $\blacksquare$

\begin{figure}[!t]
	\centering
	\includegraphics[scale=0.5]{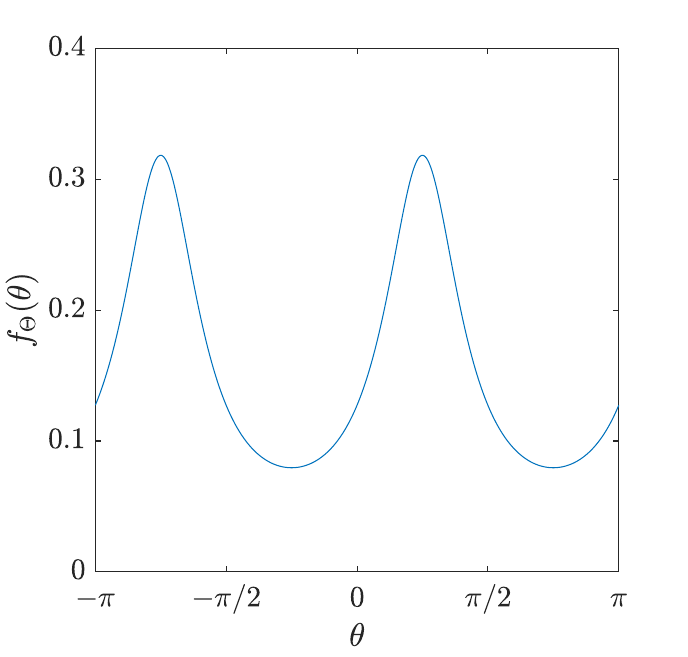}
	\includegraphics[scale=0.5]{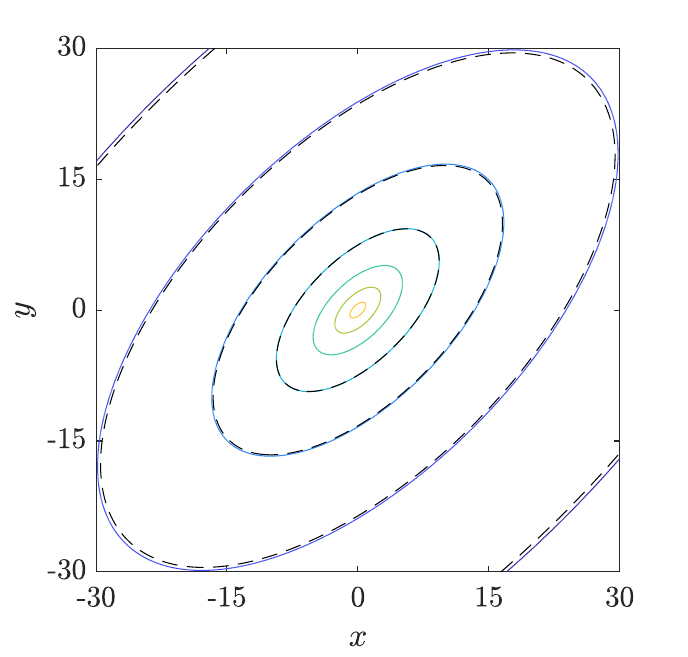}
	\caption{SPAR representation for bivariate t distribution with $\rho=0.6$ and $\nu=2$. Left: Angular density function. Right: Isodensity contours of joint pdf at equal logarithmic increments from $10^{-1}$ to $10^{-7}$. Coloured lines are exact values, and dashed lines are density from SPAR model with threshold exceedance probability $\zeta=0.05$.}
	\label{fig:SPAR_student}
\end{figure}
\end{example}

Example \ref{ex:elliptic} illustrates that for elliptical distributions, a judicious choice of coordinate system can lead to SPAR representations in which the angular and radial components are independent. However, the coordinate system used here is non-standard and may be difficult to estimate in practice. Example \ref{ex:indep} showed that changing the angular-radial coordinate system leads to a change in the angular dependence. In the remainder of the section, we consider the effect of changing the gauge functions used to define angular and radial variables.

\begin{lemma}[Change of radial variable] \label{lem:radius_change}
For random vector $(X,Y)$ define radial variables $R_a=\mathcal{R}_a(X,Y)$ and $R_b=\mathcal{R}_b(X,Y)$, for some gauge functions $\mathcal{R}_a$ and $\mathcal{R}_b$. Define the angular variable  $Q = \mathcal{A}_*^{(-2,2]}((X,Y)/\mathcal{R}_*(X,Y))$, for arbitrary gauge function $\mathcal{R}_*$. Suppose that $(R_a,Q)$ has joint density $f_{R_a,Q}$, then $(R_b,Q)$ has joint density $f_{R_b,Q}$, with
\begin{align*}
	f_{R_b,Q}(r_b,q) &= \frac{\mathcal{R}_a(\mathbf{w})}{\mathcal{R}_b(\mathbf{w})} f_{R_a,Q}\left(r_b \frac{\mathcal{R}_a(\mathbf{w})}{\mathcal{R}_b(\mathbf{w})},q\right),
\end{align*}
where $\mathbf{w}=(\cos_*(q),\sin_*(q))$.
\end{lemma}

When the star-shaped sets used to define the gauge functions have a continuous boundary, then the Jacobian of the transformation, $\mathcal{R}_a(\mathbf{w})/\mathcal{R}_b(\mathbf{w})$, is also continuous. In this case, if $(R_a,Q)$ has a SPAR representation, then $(R_b,Q)$ will also have a SPAR representation. This is stated in the following corollary. 

\begin{corollary} \label{cor:SPAR_radius_change}
In addition to the assumptions of Lemma \ref{lem:radius_change}, suppose that gauge functions $\mathcal{R}_a$ and $\mathcal{R}_b$ are defined in terms of star-shaped sets with continuous boundaries. Suppose that $f_{R_a,Q}$ satisfies assumptions (A1)-(A3) for $q\in I\subseteq(-2,2]$ with angular density $f_Q$ and GP parameter functions $\xi(q)$, $\sigma_a(\mu,q)$. Then $f_{R_b,Q}$  satisfies SPAR assumptions (A1)-(A3) for $q\in I\subseteq(-2,2]$ with the same angular density and GP parameter functions $\xi(q)$, $\sigma_b(\mu,q)$. Define the conditional quantile functions $\mu_a(\zeta,q) = \Bar{F}_{R_a|Q}^{-1} (\zeta|q)$ and $\mu_b(\zeta,q) = \Bar{F}_{R_b|Q}^{-1} (\zeta|q)$ for $\zeta\in[0,1]$. Then the conditional quantile functions and GP scale functions are related by
\begin{align*}
    \mu_b(\zeta,q) &= \frac{\mathcal{R}_b(\mathbf{w})}{\mathcal{R}_a(\mathbf{w})} \mu_a(\zeta,q)\\
	\sigma_b(\mu_b(\zeta,q),q) &= \frac{\mathcal{R}_b(\mathbf{w})}{\mathcal{R}_a(\mathbf{w})} \sigma_a \left(\mu_a(\zeta,q), q \right).
\end{align*}
\end{corollary}

These results can be applied to the case of elliptical distributions, to give SPAR representations in terms of standard polar coordinates.

\begin{example}[Elliptical distributions in standard polar coordinates] \label{ex:elliptic_standard}
Continuing from Example \ref{ex:elliptic}, if we work in standard polar coordinates and define $R=\|(X,Y)\|_2$, then from Lemma \ref{lem:radius_change} we have 
\begin{align*}
	f_{R,\Theta}(r,\theta) = \frac{\|(\cos(\theta),\sin(\theta)\|_{e,\rho}}{\|(\cos(\theta),\sin(\theta)\|_2} f_{R_e,Q}\left(r \frac{\|(\cos(\theta),\sin(\theta)\|_{e,\rho}}{\|(\cos(\theta),\sin(\theta)\|_2},\theta\right)
     = r f_0 \left(\left(\frac{r}{\alpha(\theta)} \right)^2\right).
\end{align*}
This agrees with the density obtained using (\ref{eq:fxy2frtheta}) directly. As there is no change in the angular variable from the previous example, the angular density is unchanged in this coordinate system. From Corollary \ref{cor:SPAR_radius_change}, whenever the generator $f_0$ is such that the SPAR assumptions are satisfied in the elliptic coordinate system (see Example \ref{ex:elliptic}), the SPAR assumptions will also be satisfied in the standard polar coordinate system. In these cases, the conditional radial quantiles and GP scale parameter function will be scaled by $\|(\cos(\theta),\sin(\theta)\|_{e,\rho}^{-1} = \alpha(\theta)$, when expressed in standard polar coordinates. \hfill $\blacksquare$ 
\end{example}

Examples \ref{ex:indep} and \ref{ex:elliptic} showed that, in some cases, coordinates could be chosen for which angular and radial components are independent. The question of when the SPAR assumptions can be satisfied with a GP approximation to the tail of the radial variable that is independent of angle can be addressed using Corollary \ref{cor:SPAR_radius_change}. This is stated in the following theorem.

\begin{theorem}[Coordinate systems for asymptotically independent angular and radial components] \label{thm:SPAR_indep}
Suppose that $(X,Y)$ is a random vector on $\mathbb{R}^2$. Then the following statements are equivalent.
\begin{enumerate}[(i)]
	\item There exists a map $\tilde{\mathcal{T}}^*$ to a polar coordinate system in which $(S,Q)=\tilde{\mathcal{T}}^*(X,Y)$ satisfies SPAR assumptions (A1)-(A3) for all $q\in I \subseteq (-2,2]$, with $I$ an interval, and GP parameter functions that are independent of angle.
	\item There exists a map $\tilde{\mathcal{T}}$ to a polar coordinate system with the same angle function as $\tilde{\mathcal{T}}^*$, for which the variables $(R,Q)=\tilde{\mathcal{T}}(X,Y)$ satisfy SPAR assumptions (A1)-(A3) for all $q\in I$, with constant GP shape parameter and scale parameter function $\sigma(\mu,q) \sim \alpha(q) h(\mu)$ as $\mu\to r_F(q)$, where $r_F(q)$ is the upper end point of $R|(Q=q)$, for some continuous functions $\alpha(q)>0$ and $g(\mu)>0$.
\end{enumerate}
When (ii) is satisfied, the map $\tilde{\mathcal{T}}^*$ can be defined in terms of a radial gauge function $\mathcal{R}_\alpha (x,y) = r/\alpha(q)$, where $(r,q) = \tilde{\mathcal{T}}(x,y)$.
\end{theorem}

Referring back to Example \ref{ex:elliptic_standard}, we can see immediately, that for elliptical distributions $\alpha(\theta) =  \|(\cos(\theta), \sin(\theta))\|_{e,\rho}^{-1}$, and hence $\mathcal{R}_\alpha (x,y) = \|(x,y)\|_{e,\rho}$. Example \ref{ex:elliptic} showed that using this gauge function to define the radial variable does indeed lead to independent angular and radial variables. Theorem \ref{thm:SPAR_indep} provides necessary and sufficient conditions to satisfy the model assumptions of \textcite{Wadsworth2017}. However, the coordinate systems required to satisfy the model assumptions do not always exist, and can be non-standard when they do exist. This will be illustrated in Sections \ref{sec:SPAR_Laplace}-\ref{sec:SPAR_shorttail}, which presents examples of the functions $\sigma(\mu,q)$ on various margins. The requirement for AI angular and radial variables in the Wadsworth \textit{et al.} model, limits the types of distribution which can be represented using this approach. In contrast, the SPAR approach relaxes this restriction, enabling a representation for a wider range of distributions.

To conclude this section, we consider the effect of changing the angular variable. 

\begin{lemma}[Change of angular variable] \label{lem:angle_change}
For random vector $(X,Y)$ define the radial variable $R=\mathcal{R}_*(X,Y)$, for some gauge function $\mathcal{R}_*$, and angular variables
\begin{equation*}
    Q_a=\mathcal{A}_a^{(-2,2]}\left(\frac{(X,Y)}{\mathcal{R}_a(X,Y)}\right), \qquad Q_b=\mathcal{A}_b^{(-2,2]}\left(\frac{(X,Y)}{\mathcal{R}_b(X,Y)}\right),
\end{equation*}
for gauge functions $\mathcal{R}_a$ and $\mathcal{R}_b$. Suppose that $(R,Q_a)$ has joint density $f_{R,Q_a}$ then $(R,Q_b)$ has joint density $f_{R,Q_b}$, given by
\begin{align*}
    f_{R,Q_b}(r,q) &= \frac{d q_{a,b}(q)}{d q} f_{R,Q_a}\left(r, q_{a,b}(q)\right), \qquad q_{a,b}(q) = \mathcal{A}_a^{(-2,2]} \left(\frac{(\cos_b(q),\sin_b(q))}{\mathcal{R}_a(\cos_b(q),\sin_b(q))}\right).
\end{align*}
In the case that $\mathcal{R}_a=\|\cdot\|_1$ and $\mathcal{R}_b=\|\cdot\|_2$ we have
\begin{align*}
    q_{1,2}(q) 
    &= \varepsilon(\sin_2(q))\left(1-\frac{\cos_2(q)}{\|(\cos_2(q), \sin_2(q)) \|_1}\right), 
    & 
    \frac{d q_{1,2}(q)}{d q}
    &=\frac{\pi}{2} \|(\cos_2(q), \sin_2(q)) \|_1^{-2}\\
    q_{2,1}(q) 
    &= \frac{2}{\pi}\mathrm{atan2}(\cos_1(q),\sin_1(q)),  
    & 
    \frac{d q_{2,1}(q)}{d q} &= \frac{2}{\pi}\|(\cos_1(q),\sin_1(q))\|_2^{-2},
\end{align*}
where $\varepsilon$ is the generalised sign function, defined in Example \ref{ex:Lp_angle}.
\end{lemma}

Lemma \ref{lem:angle_change} considers the Jacobians involved when changing between two coordinate systems defined in terms of pseudo-angles. Since Euclidean angles differ from $L^2$ pseudo-angles by a factor of $\pi/2$, if we change between coordinates defined in terms of $L^1$ pseudo-angles and Euclidean angles, then the factors of $\pi/2$ in the Jacobians in Lemma \ref{lem:angle_change} vanish. The Jacobians of the transformation between Euclidean and $L^1$ angles are continuous. However, although the pseudo-angle is a continuous function of the Cartesian coordinates, the Jacobian $d q_{a,b}(q)/dq$ is not guaranteed to be continuous. This is illustrated in the following example.

\begin{example}
Suppose that $\mathcal{S}_a$ is the curve illustrated in \autoref{fig:badcurve}, and define corresponding gauge and angle functions $\mathcal{R}_a$ and $\mathcal{A}_a$. The circumference of $\mathcal{S}_a$ is $\mathcal{C}_a=2(1+\sqrt{2})$. For $(w_1,w_2)\in\mathcal{S}_a\cap[0,1]^2$ we have 
\begin{equation*}
     \mathcal{A}_a^{(-2,2]}(w_1,w_2) = \frac{1}{1+\sqrt{2}}
    \begin{cases}
        2w_2, & w_2<1/2,\\
        1 + \sqrt{2}(2w_2-1), & w_2\ge 1/2.
    \end{cases}
\end{equation*}
The $L^1$ angle corresponding to $\mathcal{S}_a$ angle $q$ is therefore
\begin{equation*}
     q_{1,a}(q) = \mathcal{A}_1 \left(\frac{(\cos_a(q),\sin_a(q))}{\cos_a(q)+\sin_a(q)}\right) =
    \begin{cases}
        \dfrac{(1+\sqrt{2})q}{1+(1+\sqrt{2})q}, & q\in[0,(1+\sqrt{2})^{-1}],\\
        \dfrac{(1+\sqrt{2})q-1+\sqrt{2}}{2\sqrt{2}}, & q\in((1+\sqrt{2})^{-1},1].
    \end{cases}
\end{equation*}
Consider the case of independent standard exponential variables, $(X,Y)$. In $L^1$ polar coordinates $(R_1,Q_1)=(X+Y,Y)$, the joint density is $f_{R_1,Q_1}(r,q) = r \exp(-r)$ for $(r,q)\in[0,\infty)\times[0,1]$, and angular density is $f_{Q_1}(q) = 1$ for $q\in[0,1]$. If we define $Q_a=\mathcal{A}_a^{(-2,2]}((X,Y)/\mathcal{R}_a(X,Y))$ then from Lemma \ref{lem:angle_change}, the joint density of $(R_1,Q_a)$ is $f_{R_1,Q_a}(r,q) = (dq_{1,a}(q)/dq) r \exp(-r)$ for $(r,q)\in[0,\infty)\times[0,1]$, and the corresponding angular density is $f_{Q_a}(q) = (dq_{1,a}(q)/dq)$ for $q\in[0,1]$. Since $dq_{1,a}(q)/dq$ is discontinuous in $q$, so is $f_{Q_a}(q)$. \hfill $\blacksquare$
\begin{figure}[!t]
    \centering
    \includegraphics[scale=0.5]{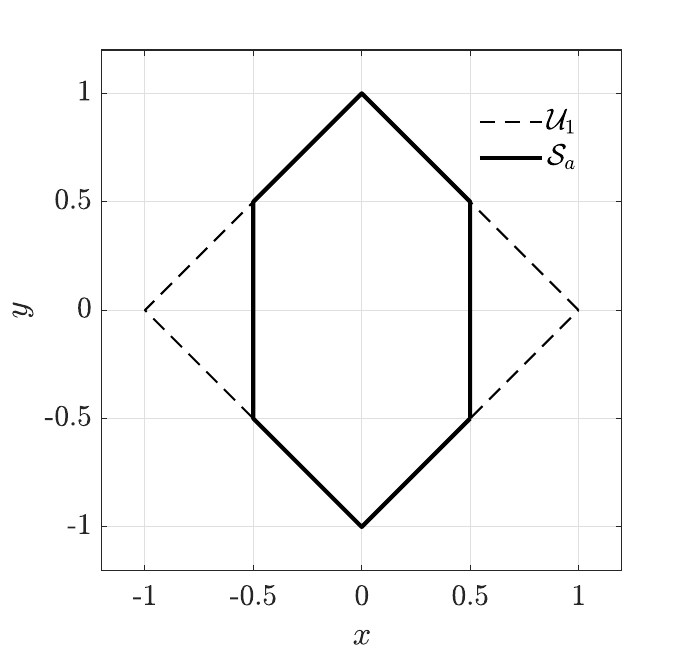}
    \includegraphics[scale=0.5]{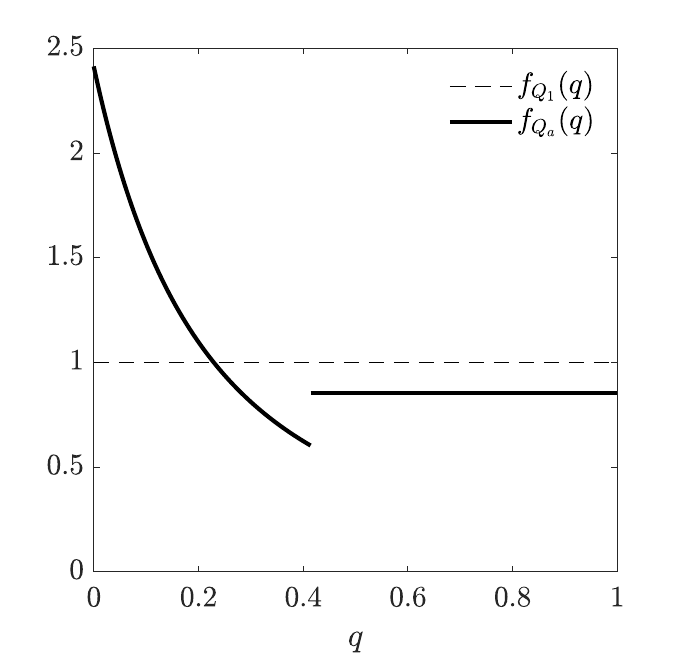}
    \caption{Left: Curve $\mathcal{S}_a$ used to define angular variable $Q_a$, and unit circle for $L^1$ norm, $\mathcal{U}_1$, used to define angular variable $Q_1$. Right: Angular density of independent standard exponential variables in terms of angular variables $Q_1$ and $Q_a$. Note that $f_{Q_1}(q)$ is continuous, whereas $f_{Q_a}(q)$ is not.}
    \label{fig:badcurve}
\end{figure}
\end{example}

Therefore, to ensure that a change of angle does not affect whether the SPAR assumptions are satisfied, we need to specify that the Jacobian of the transformation is also continuous. This is stated in the following corollary to Lemma \ref{lem:angle_change}.

\begin{corollary} \label{cor:SPAR_angle_change}
Under the assumptions of Lemma \ref{lem:angle_change}, suppose that the joint density $f_{R,Q_a}$ satisfies SPAR assumptions (A1)-(A3) for all $q\in [s,t] \subseteq (-2,2]$, with parameter functions $\xi_a(q)$ and $\sigma_a(\mu,q)$ and angular density $f_{Q_a}(q)$. If the Jacobian $J(q) = d q_{a,b}(q)/d q$ is continuous for $q\in[q_{b,a}(s),q_{b,a}(t)]$, then the joint density $f_{R,Q_b}$ satisfies assumptions (A1)-(A3) for $q\in[q_{b,a}(s),q_{b,a}(t)] \subseteq (-2,2]$ with parameter functions $\xi_b(a) = \xi_a(q_{a,b}(q))$, $\sigma_b(\mu,\theta) = \sigma_a(\mu,q_{a,b}(q))$ and angular angular density  
\begin{align*}
f_{Q_b}(q) &= \frac{d q_{a,b}(q)}{d q} f_{Q_a}\left(q_{a,b}(q)\right).
\end{align*}
\end{corollary}

\subsection{Multivariate density functions} \label{sec:pdf_coords_multivar}
For the general multivariate case, we start by considering the situation where both radii and angles are defined in terms of the same gauge function, $\mathcal{R}_*$, with corresponding piecewise-smooth surface $\mathcal{S}_* \subset \mathbb{R}^d$. As will become apparent below, piecewise-smoothness is required so that partial gradients of the surface can be calculated. Suppose that $\mathbf{w}=(w_1,...,w_d) \in \mathcal{S}_*$, and that $\mathcal{S}_*$ can be partitioned into a finite number of pairwise disjoint segments $\mathcal{S}_{*,j}$, such that $\mathcal{S}_*=\bigcup_j \mathcal{S}_{*,j}$, and in each segment $\mathcal{S}_{*,j}$ is smooth and $w_d$ is uniquely determined by the values $w_1,...,w_{d-1}$. In this case, we can define a function $g_j$ for each segment such that $w_d=g_j(w_1,...,w_{d-1})$. For example, consider the case of the $L^2$ unit circle in $\mathbb{R}^2$. In this case, we would partition the circle in segments in the upper and lower halves of the plane. Then for $(w_1,w_2)\in\mathcal{U}_2$, we can write $w_2=\sqrt{1-w_1^2}$ in the upper half of the plane and $w_2=-\sqrt{1-w_1^2}$ in the lower half. Obviously, the general case will be more complicated, since the surface $\mathcal{S}_*$ will not necessarily have any symmetries. For $\mathbf{w}\in \mathcal{S}_{*,j}$, the absolute value of the Jacobian of the transformation from generalised polar to Cartesian coordinates can then be written \parencite[][Lemma 1]{Richter2014}
\begin{equation*}
    \left|\det(J(r,\mathbf{w}))\right| = r^{d-1} \left|g_j(w_1,...,w_{d-1}) - \sum_{i=1}^{d-1} w_i \frac{\partial g_j(w_1,...,w_{d-1})}{\partial w_i} \right|.
\end{equation*}

In the case where both radii and angles are defined in terms of the $L^1$ norm, we can partition the unit sphere $\mathcal{U}_1$ into the sections falling in each orthant of $\mathbb{R}^d$. In each orthant $g_j(w_1,...,w_{d-1})=\pm\left(1-\sum_{i=1}^{d-1}|w_i|\right)$, where the sign is dependent on the orthant. It is then straightforward to confirm that $\left| \det(J(r,\mathbf{w})) \right| = r^{d-1}$ in any orthant. Therefore, for random vector $\mathbf{X}\in\mathbb{R}^d$, with density $f_{\mathbf{X}}$, if we define $R=\|\mathbf{X}\|_1$ and $\mathbf{W} = \mathbf{X}/R$, then $(R,\mathbf{W})$ has joint density
\begin{equation} \label{eq:fx2frw_L1}
    f_{R,\mathbf{W}} (r,\mathbf{w}) = r^{d-1} f_{\mathbf{X}}(r\mathbf{w}).
\end{equation}
We can then consider how changing the gauge function used for the radius affects the joint density. Suppose that  $R_* = \mathcal{R}_* (\mathbf{X})$ for some gauge function $\mathcal{R}_*$, then we have
\begin{equation*}
    \mathbf{X} = R \mathbf{W} = \frac{R_*}{\mathcal{R}_* (\mathbf{W})} \mathbf{W}.
\end{equation*}
Hence if $(R,\mathbf{W})$ has joint density $f_{R,\mathbf{W}} (r,\mathbf{w})$ then $(R_*,\mathbf{W})$ has joint density 
\begin{equation*}
    f_{R_*,\mathbf{W}} (r,\mathbf{w}) = \frac{1}{\mathcal{R}_* (\mathbf{w})} f_{R,\mathbf{W}} \left(\frac{r}{\mathcal{R}_* (\mathbf{w})},\mathbf{w}\right).
\end{equation*}
An analogous result to Theorem \ref{thm:SPAR_indep} can then be derived for the multivariate case, showing that in certain cases we can define generalised polar coordinate systems in which $R$ and $\mathbf{W}$ are AI, but that these coordinate systems are dependent on the distribution. Since a change of coordinate system does not affect whether the SPAR assumptions are satisfied, we shall restrict our attention to the case where both radial and angular variables are defined in terms of the $L^1$ norm, so that the angular-radial density has the simple form given in (\ref{eq:fx2frw_L1}).  

Working in $L^1$ polar coordinates also has the advantage that we can switch easily between vector and scalar angles in $\mathbb{R}^2$, since for $(w_1,w_2)=(\cos_1(q),\sin_1(q))$ we have $|dw_1/dq|=1$. Therefore, joint angular-radial densities defined in terms of scalar angular variable $Q$ have unit Jacobian when transforming from those defined in terms of vector angular variable $\mathbf{W}$. That is, $f_{R,Q}(r,q) = f_{R,\mathbf{W}}(r,(\cos_1(q),\sin_1(q)))$. Similarly, for the angular density we have $f_{Q}(q) = f_{\mathbf{W}}(\cos_1(q),\sin_1(q))$. The ability to switch between vector and scalar angles does not apply for standard polar coordinates, where for $(w_1,w_2)=(\cos(\theta),\sin(\theta))$ we have $|dw_1/d\theta|=|\sin(\theta)|$.

\section{Angular-radial models for copulas} \label{sec:copula_tail_model}
The joint extremal behaviour of random variables is determined by the asymptotic behaviour of their copula. In order to make general statements about the effect of the choice of margins, it is necessary to start by introducing general assumptions about the asymptotic behaviour of copulas and copula densities. The copula, $C$, of a random vector $\mathbf{X} \in \mathbb{R}^d$ was defined in (\ref{eq:copula_defn}). The corresponding copula density $c:[0,1]^d\to[0,\infty)$ is defined as $c(u_1,...,u_d) = \partial^d C(u_1,...,u_d)/ (\partial u_1 \cdots \partial u_d) $, when this derivative exists. Although our primary interest in this work is in the density, it is useful to also consider the behaviour of the copula, since various measures of dependence, such as the tail dependence coefficients (\ref{eq:tail_dep_coef}) are defined in terms of the copula. We will also show that the asymptotic models for the copula considered here are limited in terms of the range of distributions they can distinguish between, in that a wide range of copulas have the same asymptotic representation. In contrast, we will show that a certain type of asymptotic model for the copula density can distinguish between a wider range of distributions with various dependence classes.

The asymptotic models for copula and copula density can be stated as models for $C(\mathbf{u})$ and $c(\mathbf{u})$ as $\mathbf{u}\to(0,...,0)\coloneqq \mathbf{0}_d$ along different paths. Section \ref{sec:ARL_model} considers models where $\mathbf{0}_d$ is approached at the same rate in each variable, and Section \ref{sec:ARE_model} considers models where $\mathbf{0}_d$ is approached at a different rate in each variable. These two types of models can be viewed as angular-radial models for the behaviour of a copula or copula density, using different polar coordinate systems. However, in contrast to Sections \ref{sec:coords} and \ref{sec:pdfs_in_polars}, the polar coordinate systems are nonlinear, in a sense that will be discussed in Section \ref{sec:ARL_ARE_relation}. This view of the asymptotic models for copulas is instructive for understanding the relations between the models and the limitations of what can be described by each model. In Section \ref{sec:margin}, we discuss how the limitations of the models can be related to whether a copula has a SPAR representation on a given margin. Finally, in Section \ref{sec:delta_model}, we introduce a more general angular-radial description of copula densities, and discuss how it is related to the models defined in Sections \ref{sec:ARL_model} and \ref{sec:ARE_model}. Examples of the various models discussed in this section are presented in Appendix \ref{app:copula_calcs}. As with other sections, proofs are presented in Appendix \ref{app:proofs}.

The models we consider here can be viewed as making various types of assumptions about regular variation of the copula and copula density. We start by defining regularly-varying functions and listing some properties (see e.g. \cite{bingham1987}).

\begin{definition}[Regularly-varying function]
A measurable function $f:[0,\infty) \to [0,\infty)$ is regularly-varying at $r_0$ with index $\alpha\in\mathbb{R}$, denoted $f\in RV_\alpha (r_0)$, if for any $x > 0$,
\begin{equation}
	\lim_{r\to r_0} \frac{f(r x)}{f(r)} = x^\alpha.
\end{equation}
A regularly-varying function with index $\alpha=0$ is said to be slowly-varying at $r_0$. A regularly-varying function with index $\alpha=\pm\infty$ is said to be rapidly-varying at $r_0$. Any $f\in RV_\alpha(r_0)$ can be written as $f(r) = \mathcal{L}(r) r^\alpha$, where $\mathcal{L}$ is slowly-varying at $r_0$. Also, note that $f(r)\in RV_\alpha (\infty)$ if and only if $f(1/r)\in RV_\alpha (0^+)$.
\end{definition}

Previous asymptotic models for copulas and copula densities have considered behaviour of $C(\mathbf{u})$ and $c(\mathbf{u})$ for limits $\mathbf{u}\to \mathbf{0}_d= (0,...,0)$ or $\mathbf{u}\to \mathbf{1}_d= (1,...,1)$, which govern the behaviour of a random vector when all components are small or all components are large. For our purposes, it is useful to generalise this to any corner of the copula. To do this, we start by introducing some notation.

\begin{definition}
Suppose $\mathbf{u}_0=(u_{0,1},\dots,u_{0,d})$ where $u_{0,j}\in\{0,1\}$, $j=1,...,d$. For $d$-dimensional copula $C$ with corresponding density $c$ define 
\begin{align*}
	c_{\mathbf{u}_0} (u_1,\dots,u_d) &= c\left(\left(u_{0,1} + (-1)^{u_{0,1}} u_1\right),\, \dots\, , \left(u_{0,d} + (-1)^{u_{0,d}} u_d\right)\right),\\ 
	C_{\mathbf{u}_0} (u_1,\dots,u_d) &=  \int\displaylimits_{0}^{u_1} \cdots \int\displaylimits_{0}^{u_d} c_{\mathbf{u}_0} (s_1,\dots,s_d) \, ds_1 \cdots ds_d.
\end{align*}
Note that $C_{\mathbf{u}_0}$ and $c_{\mathbf{u}_0}$ are also a copula and copula density, and that, by definition, $c_{\mathbf{0}_d}\equiv c$ and $C_{\mathbf{0}_d}\equiv C$. In words, $C_{\mathbf{u}_0}$ gives the value of copula $C$ in coordinates relative to the corner $\mathbf{u}_0$, and $c_{\mathbf{u}_0}$ is the corresponding density function. We refer to $C_{\mathbf{u}_0}$ and $c_{\mathbf{u}_0}$ as the copula and copula density of $C$ with respect to corner $\mathbf{u}_0$
\end{definition}

An equivalent way to define $C_{\mathbf{u}_0}$ and $c_{\mathbf{u}_0}$ is in terms of `reflections' \parencite{taylor2007}. Suppose that the random vector with uniform margins $(U_1,...,U_d)$ has copula $C$. For $j=1,...,d$, define $\tilde{U}_j=U_j$ when $u_{0,j}=0$ and $\tilde{U}_j=1-U_j$ when $u_{0,j}=1$. Then $(\Tilde{U}_1,...,\Tilde{U}_d)$ has copula $C_{\mathbf{u}_0}$ and corresponding density $c_{\mathbf{u}_0}$ (when it exists). For bivariate copula $C$ we have $C_{(0,0)}(u,v) = C(u,v)$, $C_{(1,0)}(u,v) = v - C(1-u,v)$, $C_{(0,1)}(u,v) = u - C(u,1-v)$, and $C_{(1,1)}(u,v) = u + v - 1 + C(1-u,1-v) = \hat{C}(u,v)$. Similar but more complex relations can be derived in higher dimensions.  

\subsection{Constant scaling order} \label{sec:ARL_model}
The models defined by \textcite{Hua2011}, based on the concepts introduced in \textcite{Ledford1996, Ledford1997}, assume that the corner of the copula is approached at a constant rate in each variable. In the original presentation, \textcite{Hua2011}, considered corners $\mathbf{u}_0 = \mathbf{0}_d = (0,...,0)$ and $\mathbf{u}_0 = \mathbf{1}_d = (1,...,1)$, and referred to the indices of regular variation in these corners as the upper and lower tail orders. The definition below generalises this to any corner of the copula.

\begin{definition}[Tail order]\label{def:tail_order}
Suppose that $C$ is a $d$-dimensional copula and for corner $\mathbf{u}_0$ there exists some $\kappa_{\mathbf{u}_0} >0$ and $\mathcal{L}_{\mathbf{u}_0} (t)\in RV_0(0^+)$, such that
\begin{equation} \label{eq:tail_order}
	C_{\mathbf{u}_0} (t\mathbf{1}_d) \sim \mathcal{L}_{\mathbf{u}_0}(t) t^{\kappa_{\mathbf{u}_0}}, \quad t\to0^+.
\end{equation} 
Then $\kappa_{\mathbf{u}_0}$ is the tail order of $C$ in corner $\mathbf{u}_0$. Note that when $\kappa_{\mathbf{u}_0}$ is defined we have $\kappa_{\mathbf{u}_0} \geq 1$, since $C_{\mathbf{u}_0} (t\mathbf{1}_d) \leq t$.
\end{definition}

Clearly, $\mathcal{L}_{\mathbf{u}_0}$ and $\kappa_{\mathbf{u}_0}$ are also dependent on the copula $C$, so we could write $\mathcal{L}_{\mathbf{u}_0}(t;C)$ and $\kappa_{\mathbf{u}_0}(C)$ to make this explicit. However, the copula in question should be clear from the context, so we omit this information for brevity. This convention is also applied to other coefficients and functions of the copula, defined below.

Definition \ref{def:tail_order} states that if copula $C$ is regularly-varying in corner $\mathbf{u}_0$, when approached on the diagonal, then the index of regular variation is referred to as the tail order. Proposition 0.8(i) in \textcite{Resnick1987} states that if $f\in RV_\alpha(0^+)$ then $\lim_{x\to0^+} \log(f(x))/\log(x) = \alpha$. Therefore, if (\ref{eq:tail_order}) is satisfied,
\begin{equation*}
    \kappa_{\mathbf{u}_0} = \lim_{t\to0^+} \frac{\log(C_{\mathbf{u}_0} (t\mathbf{1}_d))}{\log(t) }.
\end{equation*}
The upper and lower tail dependence coefficients in \eqref{eq:tail_dep_coef} can be generalised analogously. The tail dependence coefficient in corner $\mathbf{u_0}$ is defined as
\begin{equation*}
    \chi_{\mathbf{u}_0} = \lim_{t\to0^+} \frac{C_{\mathbf{u}_0} (t\mathbf{1}_d)}{t},
\end{equation*}
when this limit exists. Moreover, if we define $\Upsilon_{\mathbf{u}_0} = \lim_{t\to0^+} \mathcal{L}_{\mathbf{u}_0}(t)$, then when $\kappa_{\mathbf{u}_0}=1$, we can see that $\Upsilon_{\mathbf{u}_0 } = \chi_{\mathbf{u}_0}$. \textcite{Joe2015} makes the following classification of tail behaviour:
\begin{itemize}
    \item \textit{Strong tail dependence}: $\kappa_{\mathbf{u}_0}=1$ and $\Upsilon_{\mathbf{u}_0}>0$. 
    \item \textit{Intermediate tail dependence}: $1<\kappa_{\mathbf{u}_0}<d$ or $\kappa_{\mathbf{u}_0}=1$ and $\Upsilon_{\mathbf{u}_0}=0$.
    \item \textit{Tail orthant independence}: $\kappa_{\mathbf{u}_0} = d$ and $\Upsilon_{\mathbf{u}_0} \in (0,\infty)$.
    \item \textit{Negative tail dependence}: $\kappa_{\mathbf{u}_0}>d$.
\end{itemize}
The notion of strong tail dependence corresponds to the usual definition of asymptotic dependence, while the remaining categories distinguish between different types of asymptotic independence. The tail order describes the behaviour of the copula as the corner is approached along the diagonal. The tail order function generalises this to other linear paths toward the corner.

\begin{definition}[Tail order function and ARL model]
Suppose that $C$ is a $d$-dimensional copula and has tail order $\kappa_{\mathbf{u}_0}$ in corner $\mathbf{u}_0$ with corresponding slowly-varying function $\mathcal{L}_{\mathbf{u}_0} (t)$. Suppose also that there exists a function $B_{\mathbf{u}_0} : [0,\infty)^d \to [0,\infty)$ such that 
\begin{equation} \label{eq:ARL_dist}
C_{\mathbf{u}_0} (t \mathbf{z}) \sim B_{\mathbf{u}_0} (\mathbf{z}) \mathcal{L}_{\mathbf{u}_0}(t) t^{\kappa_{\mathbf{u}_0}}, \qquad t\to0^+.
\end{equation}
Then $B_{\mathbf{u}_0}$ is the tail order function of $C$ in corner $\mathbf{u}_0$. We refer to the asymptotic assumption (\ref{eq:ARL_dist}) as the Angular-Radial model for the copula in Linear coordinates (ARL model).
\end{definition}

\textcite{Hua2011} derived various properties of tail order functions. In particular, they showed that $B_{\mathbf{u}_0} (\mathbf{z})$ is homogeneous of order $\kappa_{\mathbf{u}_0}$, i.e. $B_{\mathbf{u}_0} (t\mathbf{z}) = t^{\kappa_{\mathbf{u}_0}} B_{\mathbf{u}_0} (\mathbf{z})$. Moreover, \textcite{Joe2010} showed for strong tail dependence, the function $\Upsilon_{\mathbf{u}_0} B_{\mathbf{u}_0}(\mathbf{z})$, can be related to the stable tail dependence function (e.g. \cite[p257]{Beirlant2004}), for the corresponding lower extreme value limit of $C_{\mathbf{u}_0}$. An analogous model for the copula density can also be defined \parencite{Hua2011, Li2013a, Li2015a}.

\begin{definition}[Tail density function and ARL density model] \label{def:tail_density}
Suppose that $C$ is a $d$-dimensional copula with density function $c$, and that $C$ has tail order $\kappa_{\mathbf{u}_0}$ in corner $\mathbf{u}_0$, with corresponding slowly-varying function $\mathcal{L}_{\mathbf{u}_0} (t)$. Suppose also that there exists a function $b_{\mathbf{u}_0} : [0,\infty)^d \to [0,\infty)$ such that
\begin{equation} \label{eq:ARL_dens}
	c_{\mathbf{u}_0} (t \mathbf{z}) \sim b_{\mathbf{u}_0} (\mathbf{z}) \mathcal{L}_{\mathbf{u}_0}(t) t^{\kappa_{\mathbf{u}_0}-d}, \qquad t\to0^+.
\end{equation}
Then $b_{\mathbf{u}_0}$ is the tail density function of $c$ in corner $\mathbf{u}_0$. We refer to (\ref{eq:ARL_dens}) as the ARL model for the copula density.
\end{definition}

Under some mild constraints, the tail density function can be derived from the tail order function. Suppose that copula $C$ has a tail order function in corner $\mathbf{u}_0$, and the partial derivatives $\partial C_{\mathbf{u}_0}/\partial u_j$ are ultimately monotone as $u_j\to0^+$. In this case, the operations of taking limits and partial differentiation can be interchanged. Note that for $u_j = tz_j$, $\partial/\partial u_j = t^{-1} \, \partial/\partial z_j$. Therefore, for $t\to0^+$, we can write
\begin{equation} \label{eq:tail_dens_pdif}
	c_{\mathbf{u}_0}(t\mathbf{z}) = \frac{\partial^d}{\partial u_1 \cdots \partial u_d} C_{\mathbf{u}_0}(t\mathbf{z}) \sim \mathcal{L}_{\mathbf{u}_0} (t) \, t^{\kappa_{\mathbf{u}_0}-d} \frac{\partial^d}{\partial z_1 \cdots \partial z_d} B_{\mathbf{u}_0}(\mathbf{z}) = \mathcal{L}_{\mathbf{u}_0} (t) \, t^{\kappa_{\mathbf{u}_0}-d} b_{\mathbf{u}_0}(\mathbf{z}).
\end{equation}

When the tail density function exists it is homogeneous of order $\kappa_{\mathbf{u}_0}-d$. For the case $\kappa_{\mathbf{u}_0}<d$ (i.e. intermediate or strong tail dependence) the copula density becomes infinite in corner $\mathbf{u}_0$. Many families of copulas and copula densities have asymptotic forms (\ref{eq:ARL_dist}) and (\ref{eq:ARL_dens}), with examples presented in \parencite{Hua2011, Li2013a, Li2015a}, and further examples are given in Appendix \ref{app:copula_calcs}. However, it is important to note that the ARL models for the copula and copula density assume that $t\to0^+$ with $\mathbf{z}=(z_1,...z_d)$ fixed. In some cases, the limit does not apply when $z_j\to 0^+$, for some $j=1,...,d$. In such cases the ARL models do not describe the behaviour close to the margins. This is discussed further in Section \ref{sec:ARL_ARE_relation} and the implications for SPAR models on long-tailed margins are discussed in Section \ref{sec:SPAR_longtail}. 

\subsection{Different scaling orders} \label{sec:ARE_model}
An alternative assumption about multivariate tail probabilities was presented by \textcite{Wadsworth2013}, who considered the joint tail region for variables with standard Pareto or exponential margins.  This assumption can be expressed in terms of an assumption about the asymptotic form of the copula in the corners. The Wadsworth-Tawn model considers the behaviour of the copula as the corner is approached at different rates in each variable.

\begin{definition}[Copula exponent function and ARE model]
Suppose that $C$ is a $d$-dimensional copula and there exists a function $\Lambda_{\mathbf{u}_0} : [0,\infty)^d \to [0,\infty)$ such that
\begin{equation} \label{eq:AREmod}
	C_{\mathbf{u}_0} (t^{\mathbf{z}}) \sim \mathcal{L}_{\mathbf{u}_0}(t; \mathbf{z}) t^{\Lambda_{\mathbf{u}_0}(\mathbf{z})}, \quad t\to0^+,
\end{equation} 
where $\mathbf{z}=(z_1,...,z_d)$, $t^{\mathbf{z}} = (t^{z_1},...,t^{z_d})$, and $\mathcal{L}_{\mathbf{u}_0} (t; \mathbf{z})$ is slowly-varying in $t$ at $0^+$. Then $\Lambda_{\mathbf{u}_0}$ is the exponent function of copula $C$ in corner $\mathbf{u}_0$. We refer to assumption (\ref{eq:AREmod}) as the Angular-Radial model for the copula in Exponential coordinates (ARE model). 
\end{definition}

\textcite{Wadsworth2013} refer to $\Lambda_{\mathbf{u}_0}$ as the angular dependence function. However, as we are considering various types of angular-radial model, we use the term `copula exponent function' to avoid confusion. \textcite{Wadsworth2013} showed that $\Lambda_{\mathbf{u}_0}$ has properties analogous to those of the stable tail dependence function for EV copulas. In particular, $\Lambda_{\mathbf{u}_0}$ is homogeneous of order 1, and by comparing (\ref{eq:tail_order}), and (\ref{eq:AREmod}) we see that when both these representations are valid, $\Lambda_{\mathbf{u}_0}(\mathbf{1}_d) = \kappa_{\mathbf{u}_0}$. Moreover, when $\kappa_{\mathbf{u}_0}=1$, we have $\Lambda_{\mathbf{u}_0}(\mathbf{u}_0)(z_1,...,z_d)= \max(z_1,...,z_d)$.\footnote{This can be seen as follows. First note that since $C_{\mathbf{u}_0}(u_1,...,u_d)\leq \min(u_1,...,u_d)$, we have $\Lambda_{\mathbf{u}_0}(\mathbf{u}_0)(z_1,...,z_d)\geq \max(z_1,...,z_d)$. Secondly, since $C(r^{aw_1},r^{w_2},...,r^{w_d})\leq C(r^{w_1},r^{w_2},...,r^{w_d})$ for $a>1$, the exponent function $\Lambda_{\mathbf{u}_0}$ is non-decreasing in the first argument, and similarly for the other arguments. In the case that $\Lambda_{\mathbf{u}_0}(\mathbf{1}_d) = \kappa_{\mathbf{u}_0}=1$, combining these observations gives the result.} We can also introduce an analogous definition for the density function.

\begin{definition}[Copula density exponent function and ARE density model]
Suppose that $c$ is a $d$-dimensional copula density and there exists a function $\lambda_{\mathbf{u}_0} : (0,\infty)^d \to [0,\infty)$  such that
\begin{equation} \label{eq:AREmod_dens}
	c_{\mathbf{u}_0} (t^{\mathbf{z}}) \sim \mathcal{M}_{\mathbf{u}_0}(t; \mathbf{z}) t^{\lambda_{\mathbf{u}_0}(\mathbf{z})-S_{\mathbf{z}}}, \quad t\to0^+,
\end{equation} 
where $S_{\mathbf{z}} = \sum_{j=1}^d z_j$ and $\mathcal{M}_{\mathbf{u}_0} (t; \mathbf{z})$ is slowly-varying in $t$ at $0^+$. Then $\lambda_{\mathbf{u}_0}$ is the exponent function of copula density $c$ in corner $\mathbf{u}_0$. We refer to assumption (\ref{eq:AREmod_dens}) as the Angular-Radial model for the copula density in Exponential coordinates (ARE model for the copula density). The reason for including the term $S_{\mathbf{z}}$ in the index will be discussed further below. 
\end{definition}

As with the copula exponent function, $\lambda_{\mathbf{u}_0}$ is also homogeneous of order 1. However, in contrast to the copula exponent function, the copula density exponent function is only defined for $\mathbf{z}=(z_1,...,z_d)$ such that $z_j>0$ for $j=1,...,d$. In the case that $z_j=0$, the path $t^{\mathbf{z}}$ does not terminate in corner $\mathbf{u}_0$ as $t\to0^+$, and hence the variation of the density along this path is not representative of the behaviour in corner $\mathbf{u}_0$. The relation between $\Lambda_{\mathbf{u}_0}$ and $\lambda_{\mathbf{u}_0}$ is more complicated than that between $B_{\mathbf{u}_0}$ and $b_{\mathbf{u}_0}$, but can be inferred from Proposition 3.3 in \textcite{Nolde2022}.\footnote{Specifically, on exponential margins for $\mathbf{z}\in\mathcal{U}_1^+$, $f_{\mathbf{X}}(r\mathbf{z}) = \exp(-r) c_{(\mathbf{1}_d)} (\exp(-r\mathbf{z})) \sim \mathcal{M}_{\mathbf{1}_d}(\exp(-r); \mathbf{z}) \exp(-r  \lambda_{\mathbf{1}_d} (\mathbf{z}))$. Proposition 2.6(i) in \textcite{Resnick2007} shows that $\lim_{r\to\infty} [- \log(f_{\mathbf{X}}(r\mathbf{z})) /r ] = \lambda_{\mathbf{1}_d}(\mathbf{z})$. Proposition 2.2 in \textcite{Nolde2022} shows that this is a sufficient condition for $f_{\mathbf{X}}$ to have a limit set with gauge function $\lambda_{\mathbf{1}_d}(\mathbf{z})$. The assumptions of Proposition 3.3 in \textcite{Nolde2022} then apply, and provide a relation between $\Lambda_{\mathbf{1}_d}(\mathbf{z})$ and $\lambda_{\mathbf{1}_d}(\mathbf{z})$.} We can also relate $\lambda_{\mathbf{u}_0}$ to the tail order. Comparing (\ref{eq:ARL_dens}) and (\ref{eq:AREmod_dens}) we see that when both these representations are valid, we have $\lambda_{\mathbf{u}_0}(\mathbf{1}_d) = \kappa_{\mathbf{u}_0}$.

The lower tails of EV copulas represent canonical cases of copula exponent functions, where $\mathcal{L}_{\mathbf{0}_d}(t,\mathbf{z}) = 1$ and $\mathcal{M}_{\mathbf{0}_d}(t,\mathbf{z})$ is a function of $\mathbf{z}$ only. Moreover for EV copulas, $\Lambda_{\mathbf{0}_d}(\mathbf{z}) = \lambda_{\mathbf{0}_d}(\mathbf{z}) = A(\mathbf{z})$, where $A(\mathbf{z})$ is the stable tail dependence function (see Appendix \ref{app:EVcopula}). This is, in part, the motivation for including the term $S_{\mathbf{z}}$ in the definition of angular density functions. The other motivation is related to expressions for the density on exponential margins, discussed in Section \ref{sec:delta_model}.

\subsection{Relations between models for constant and variable scaling orders} \label{sec:ARL_ARE_relation}
\begin{figure}[!t]
	\centering
    \begin{subfigure}[t]{0.35\textwidth}
        \centering
        \includegraphics[scale=0.5]{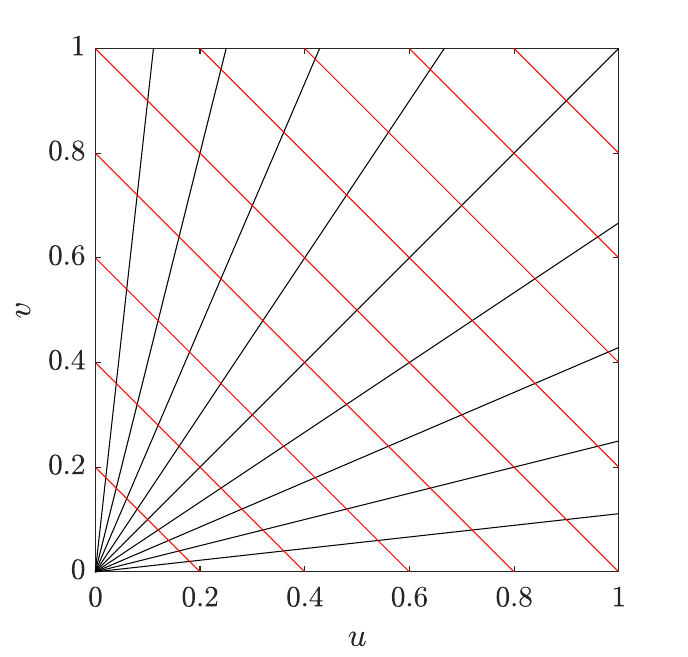}
        \caption{Constant scaling order / ARL model (linear scale).}
    \end{subfigure}
    \hskip2em
    \begin{subfigure}[t]{0.35\textwidth}
        \centering
        \includegraphics[scale=0.5]{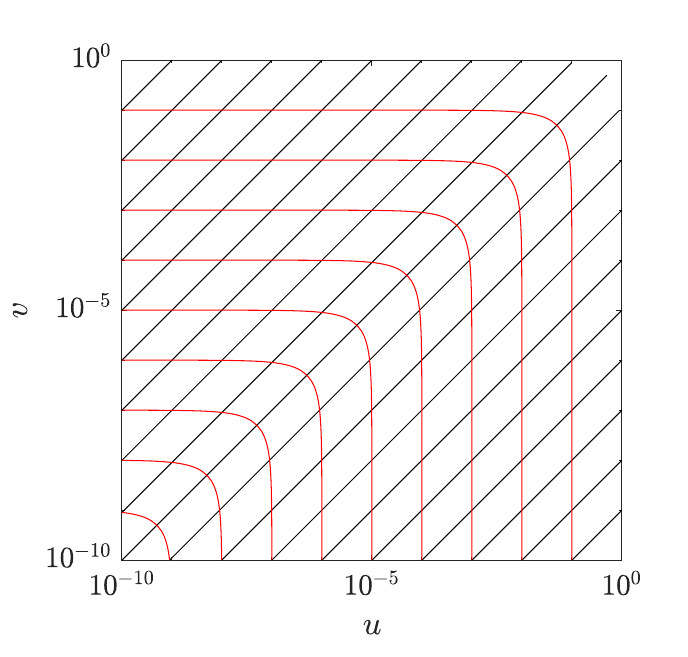}
        \caption{Constant scaling order / ARL model (log-log scale).}
    \end{subfigure}\\
    \begin{subfigure}[t]{0.35\textwidth}
        \centering
        \includegraphics[scale=0.5]{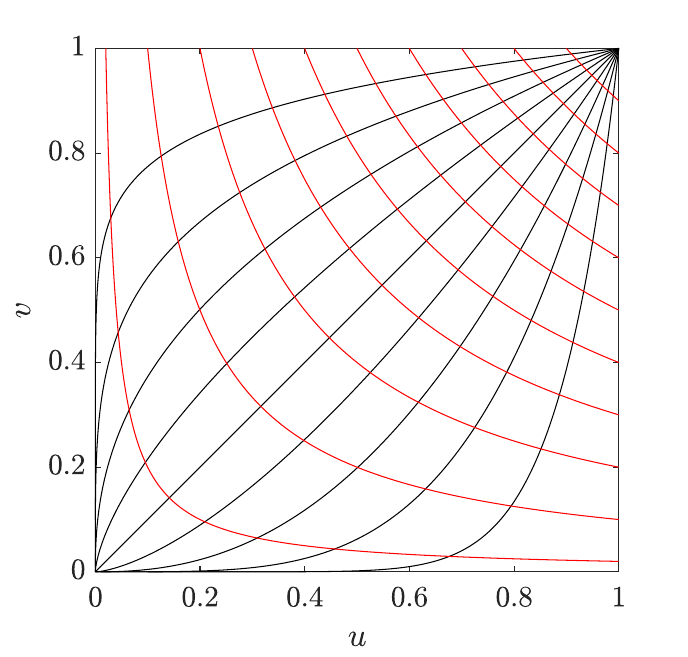}
        \caption{Different scaling orders / ARE model (linear scale).}
    \end{subfigure}
    \hskip2em
    \begin{subfigure}[t]{0.35\textwidth}
        \centering
        \includegraphics[scale=0.5]{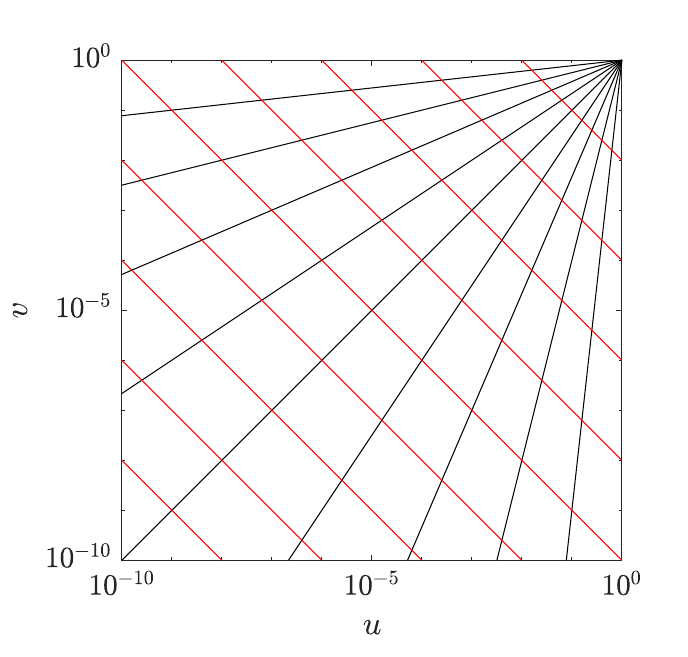}
        \caption{Different scaling orders / ARE model (log-log scale).}
    \end{subfigure}
    \caption{Polar coordinate grids used for asymptotic models for copulas. Top: ARL models / constant scaling order. Bottom: ARE models / different scaling orders. Red lines: constant radius. Black lines: constant angle.}
    \label{fig:copula_coords}
\end{figure}

The ARL and ARE models for the copula and copula density were presented in terms of assumptions about the scaling order of each variable. However, we can also view them as describing the asymptotic behaviour of the copula in different angular-radial coordinate systems. To illustrate the points of interest, it will suffice to consider two-dimensional cases. The ARL models consider the variation of copula and copula density in terms of $L^1$ polar coordinates $(u,v)=(t(1-z),tz)$.\footnote{Strictly, the definition was presented for $\mathbf{u}=t\mathbf{z}$ and $\mathbf{z}\in[0,\infty)^d$, but the homogeneity property of the tail order and tail density functions allows us to restrict the domain of $\mathbf{z}$ to lie on the unit simplex in the non-negative orthant.} Similarly, the ARE models describe the variation of the copula and copula density in terms of coordinates $(u,v)=(r^{1-w},r^w)$, for $r\to 0^+$ and $w\in(0,1)$. In this case the coordinates $(r,w)$ define a nonlinear polar coordinate system, as shown in \autoref{fig:copula_coords}. 

When considering the behaviour of the copula density for $(u,v)\to(0,0)$, it is useful to use a log-log scale for visualisations. Consider a two-dimensional case with Cartesian coordinates $(u,v)=(t(1-z),tz)$. A line of constant angle $z$ has equation by $v=uz/(1-z)$. When viewed on a log-log scale, the equation becomes $\log(v)= \log(u)+ \log(z/(1-z))$, which has unit gradient and the value of $z$ determines the y-intercept (see \autoref{fig:copula_coords}). In contrast, for Cartesian coordinates $(u,v)=(r^{1-w},r^w)$, a line of constant $w$ is given by $v=u^{w/(1-w)}$, whereas on the log-log scale the line has equation $\log(v)=(w/(1-w))\log(u)$. So in this case, lines of constant $w$ correspond to linear rays on the log-log scale, with gradient $w/(1-w)$ (see \autoref{fig:copula_coords}). Or to put it another way, $(\log(r),w)=(\log(uv),\log(v)/\log(uv))$, so that $(\log(r),w)$ are the $L^1$ polar coordinates of $(\log(u),\log(v))$.

For the polar coordinate system used in ARL models, on the log-log scale, all rays with fixed $z$ are parallel to the ray $w=1/2$ for the coordinate system used in the ARE model. Informally speaking, any information about how the copula or copula density vary with $z\in(0,1)$ at small values of $t$ which is contained in the ARL model, is collapsed onto the ray $w=1/2$ for the ARE model. Conversely, the information about the angular variation of the density, described by exponent functions for the copula and copula density must relate to information about the behaviour of the tail order and tail density functions for $z\to0^+$ or $z\to1^-$. This is made precise in the following propositions. To express the tail order functions in terms of the exponent functions, we make the additional assumption that $\Lambda_{\mathbf{u}_0}(1-w,w)$ and $\lambda_{\mathbf{u}_0}(1-w,w)$ have a Taylor series expansion about $w=1/2$, and that the first derivatives are equal at this point. In the case that $\lambda_{\mathbf{u}_0}(1-w,w)$ is a convex function, and is differentiable at $w=1/2$ the equality of the first derivatives is implied by Proposition 3.3 in \textcite{Nolde2022}. This this assumption is satisfied for the examples considered in Appendix \ref{app:copula_calcs}.

\begin{proposition}[Tail functions in terms of exponent functions] \label{prop:ARE2ARL}
Suppose that bivariate copula $C$ and corresponding density $c$, satisfy the ARE model assumptions in corner $\mathbf{u}_0$. Suppose also that and $\Lambda_{\mathbf{u}_0}(1-w,w)$ and $\lambda_{\mathbf{u}_0}(1-w,w)$ both have a Taylor series expansion about $w=1/2$ with $\beta \coloneqq [\tfrac{d}{dw}\Lambda_{\mathbf{u}_0}(1-w,w)]_{w=1/2} = [\tfrac{d}{dw}\lambda_{\mathbf{u}_0}(1-w,w)]_{w=1/2}$. Suppose that $C$ and $c$ also satisfy the ARL model assumptions in corner $\mathbf{u}_0$. Then the tail order and tail density functions are given by
\begin{align*}
    B_{\mathbf{u}_0}(z_1,z_2) &= \gamma_1 z_1^{\tfrac{1}{2}(\kappa_{\mathbf{u}_0}-\beta)} z_2^{\tfrac{1}{2}(\kappa_{\mathbf{u}_0}+\beta)},\\
    b_{\mathbf{u}_0}(z_1,z_2) &= \gamma_2 z_1^{\tfrac{1}{2}(\kappa_{\mathbf{u}_0}-\beta) -1} z_2^{\tfrac{1}{2}(\kappa_{\mathbf{u}_0}+\beta) -1},
\end{align*}
for $(z_1,z_2)\in(0,\infty)^2$ and $\kappa_{\mathbf{u}_0}=\Lambda_{\mathbf{u}_0}(1,1)=\lambda_{\mathbf{u}_0}(1,1)$ and constants $\gamma_1, \gamma_2>0$. When $\kappa_{\mathbf{u}_0}<2$ and $\beta=0$, the ARL model assumptions are not valid for $z_1\to0^+$ or $z_2\to0^+$.
\end{proposition}

When the assumptions of Proposition \ref{prop:ARE2ARL} are satisfied, the tail order and tail density functions are defined completely by properties of the exponent functions at $w=1/2$. It was noted in Section \ref{sec:ARE_model} that when $\kappa_{\mathbf{u}_0}=1$, we have $\Lambda_{\mathbf{u}_0}(z_1,z_2)=\max(z_1,z_2)$, so $\Lambda_{\mathbf{u}_0}$ is not differentiable along the ray $z_1=z_2$. Moreover, it will be shown in Section \ref{sec:limit_set}, that when $\kappa_{\mathbf{u}_0}=1$, the function $\lambda_{\mathbf{u}_0}(1-w,w)$ is not differentiable at $w=1/2$. So Proposition \ref{prop:ARE2ARL} does not apply in the case $\kappa_{\mathbf{u}_0}=1$. In cases of intermediate tail dependence (i.e. $\kappa_{\mathbf{u}_0}\in(1,2)$), Proposition \ref{prop:ARE2ARL} tells us that the ARL models cannot be valid for $z_1\to0^+$ or $z_2\to0^+$ whenever the exponent functions are symmetric (and hence $\beta=0$). In Section \ref{sec:SPAR_longtail}, we will see that this implies that the SPAR representation of these types of copula on long-tailed margins is only valid in the region where both variables are large.

A trivial example where the assumptions of Proposition \ref{prop:ARE2ARL} are satisfied is the independence copula, where $\Lambda_{\mathbf{u}_0}(\mathbf{z})=\lambda_{\mathbf{u}_0}(\mathbf{z})=1$ and $B_{\mathbf{u}_0}(z_1,z_2)=z_1 z_2$, $b_{\mathbf{u}_0}(\mathbf{z})=1$. Non-trivial examples where the proposition applies are the Gaussian copula and the lower tail of EV copulas (see Appendix \ref{app:copula_calcs}). \autoref{fig:copula_models}(a) shows contours of $C(u,v)$ for the lower tail of an EV copula with symmetric logistic dependence, together with contours of the ARL model (\ref{eq:ARL_dist}). The ARL model is a good approximation close to the line $u=v$, but deviates from the true contours away from this region. For EV copulas, the ARE model is exact for the lower tail (see Appendix \ref{app:EVcopula}). Informally, when $\Lambda_{\mathbf{u}_0}$ is differentiable along the ray $u=v$, the level sets of $C$ asymptote to straight lines in the region of the ray $u=v$. Due to the coordinate system used in the ARL model, the tail order function is an asymptotic representation of this section of the copula exponent function, resulting in differences from the true values in other regions. 

\begin{figure}[!t]
	\centering
    \begin{subfigure}[t]{0.35\textwidth}
        \centering
        \includegraphics[scale=0.5]{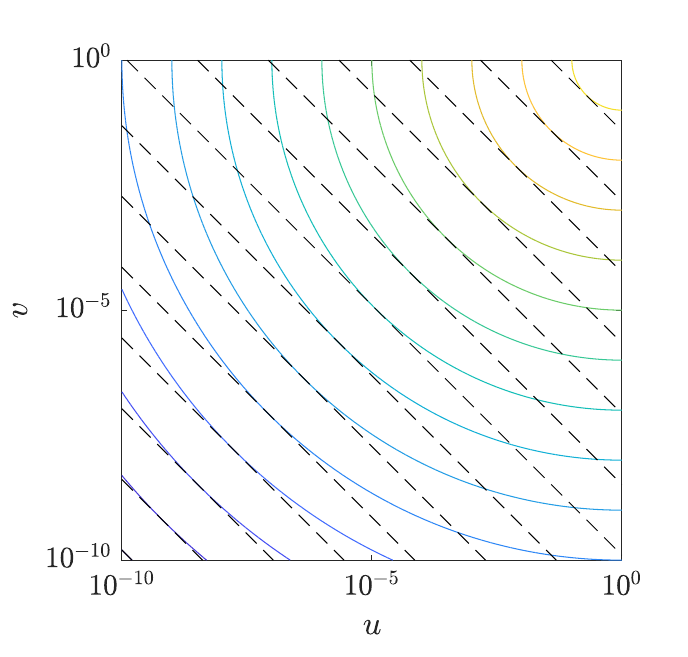}
        \caption{Contours of $C_{(0,0)}$ (coloured lines) and ARL model (\ref{eq:ARL_dist}) (dashed lines).}
    \end{subfigure}
    \hskip2em
    \begin{subfigure}[t]{0.35\textwidth}
        \centering
        \includegraphics[scale=0.5]{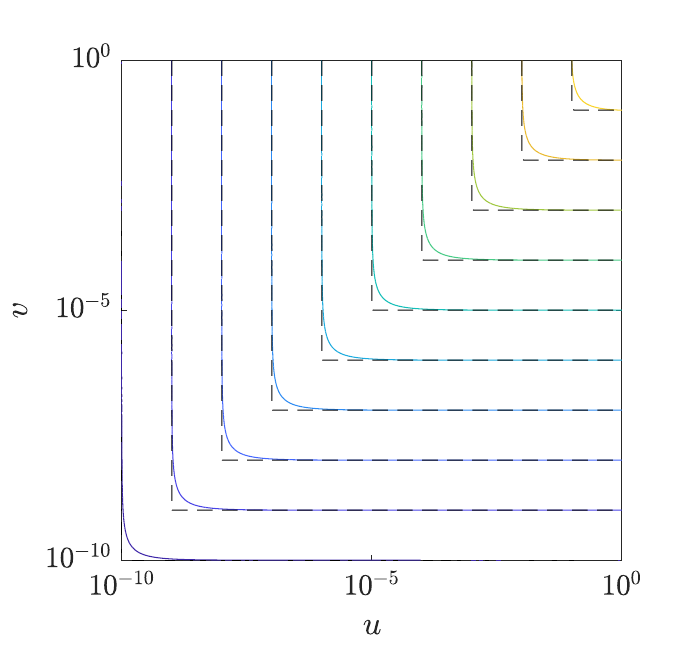}
        \caption{Contours of $C_{(1,1)}$ (coloured lines) and ARE model (\ref{eq:AREmod}) (dashed lines).}
    \end{subfigure}
    \caption{Illustration of limitations of ARL and ARE models for lower and upper tails of EV copula, $C$, with symmetric logistic dependence and parameter $\alpha=2$. For copulas with intermediate tail dependence in a given corner, meeting the assumptions of Proposition \ref{prop:ARE2ARL} with symmetric exponent function, contours of ARL model appear as straight lines on a log-log scale, as shown on panel (a). For copulas with strong tail dependence in a given corner, the ARE model always has the form shown in panel (b).}
    \label{fig:copula_models}
\end{figure}

In contrast, contours of the upper corner $C_{(1,1)}$ are shown in \autoref{fig:copula_models}(b), together with the corresponding ARE model. In this case, since $\kappa_{(1,1)}=1$, the copula exponent function is $\Lambda_{(1,1)}(z,1-z)=\max(z,1-z)$. On the log-log scale of the axes, the convergence to the asymptotic model is evident. However, it is clear that the ARE model does capture the small portion (on this scale) of the contour that is curved, close to the line $u=v$. This information about the variation of the copula is described by ARL model, but is lost in the description provided by the ARE model. This is made precise in the converse result to Proposition \ref{prop:ARE2ARL}, stated below. Since we know that the ARE model for the copula always takes the same form when $\kappa_{\mathbf{u}_0}=1$, and that for intermediate tail dependence, ARL models have the limitations described in Proposition \ref{prop:ARE2ARL}, we focus on the relation between the ARL and ARE models for the copula density here.

\begin{proposition}[Copula density exponent function in terms of tail density function] \label{prop:ARL2ARE}
Suppose that bivariate copula density $c$ satisfies ARL model assumptions in corner $\mathbf{u}_0$, and that $c_{\mathbf{u}_0}(t(1-z),tz) \sim b_{\mathbf{u}_0}(1-z,z) \mathcal{L}_{\mathbf{u}_0} (t)\, t^{\kappa_{\mathbf{u}_0}-2}$ for $t\to0^+$ and $z\in[0,1]$. Suppose also that $b_{\mathbf{u}_0}(1-z,z)\in RV_{\beta_1}(0^+)$ and $b_{\mathbf{u}_0}(1-z,z)\in RV_{\beta_2}(1^-)$ for some $\beta_1,\beta_2\in(0,\infty)$. Then $c$ also satisfies the ARE model assumptions in corner $\mathbf{u}_0$, with copula density exponent function given by
\begin{equation*}
    \lambda_{\mathbf{u}_0} (1-w,w) = \frac{\kappa_{\mathbf{u}_0}}{2} + 
    \begin{cases}
        (2(1+\beta_2)-\kappa_{\mathbf{u}_0}) \, \left|w-\tfrac{1}{2} \right|, & w\in(0,1/2],\\
        (2(1+\beta_1)-\kappa_{\mathbf{u}_0}) \, \left|w-\tfrac{1}{2} \right|, & w\in(1/2,1).
    \end{cases}
\end{equation*}
\end{proposition}

The form of $\lambda_{\mathbf{u}_0}$ given in Proposition \ref{prop:ARL2ARE} consists of two linear segments, with $\lambda_{\mathbf{u}_0}(1^-,0^+)=1+\beta_2$, $\lambda_{\mathbf{u}_0}(\tfrac{1}{2},\tfrac{1}{2})=\kappa_{\mathbf{u}_0}/2$ and $\lambda_{\mathbf{u}_0}(0^+,1^-)=1+\beta_1$. When the assumptions of the proposition hold, $\lambda_{\mathbf{u}_0}$ only contains information about the tail order $\kappa_{\mathbf{u}_0}$ and the indices of regular variation for the tails of $b_{\mathbf{u}_0}(1-z,z)$. Information about the variation of $b_{\mathbf{u}_0}(1-z,z)$ for values of $z$ away from $0$ and $1$ is lost. The assumptions of Proposition \ref{prop:ARL2ARE} hold trivially for the case of the independence copula. Non-trivial examples with $\kappa_{\mathbf{u}_0}=1$ where the assumptions of Proposition \ref{prop:ARL2ARE} hold with $\beta_1=\beta_2$, include the upper tail of the bivariate EV copula with symmetric logistic dependence (see Appendix \ref{app:copula_calcs}) and any copulas that are in the domain of attraction of this -- this includes all Archimedean copulas with strong tail dependence \parencite{charpentier2009}, and all corners of the t-copula (see Appendix \ref{app:t_copula}). Asymmetric examples with $\beta_1\neq\beta_2$ include the upper tails of other EV copulas, with stable tail dependence function given by the Dirichlet model \parencite{Coles1991} or the polynomial model \parencite{kluppelberg2006}. The assumptions of the proposition do not hold for EV copulas with the H{\"u}sler-Reiss dependence model \parencite{Husler1989}, as the function $b_{(1,1)}$ is rapidly-varying in the tails, i.e. regularly-varying with index $+\infty$ (see Appendix \ref{app:EVcopula} and Example \ref{ex:EV_laplace} below).

In summary, ARL models for the copula and copula density do not apply close to the margins in cases of intermediate tail dependence. Copulas with strong tail dependence all have the same form of ARE model, making this form less useful for describing multivariate extremes. However, the ARE model for the copula density is more versatile in the types of distribution it can describe, over a range of dependence classes. In Section \ref{sec:SPAR_Laplace} we will see that the ARE model for the copula density is closely linked to SPAR representations of distributions on Laplace margins, and their corresponding limit sets.

\FloatBarrier
\subsection{Generalised angular-radial dependence functions} \label{sec:delta_model}
As noted in Section \ref{sec:ARE_model}, the copula exponent function was originally defined in terms of angular-radial representations of a joint distribution function on exponential margins. Similarly, as discussed below, the tail order and tail density functions are closely related to asymptotic models for the angular-radial behaviour of the survivor and density functions for variables on long-tailed margins. We can generalise this idea and define functions to represent the variation of the copula density for angular-radial coordinates defined on arbitrary margins. 

\begin{definition}
Consider a random vector $\mathbf{X}_* \in \mathbb{R}^d$ with copula density $c$ and common marginal distribution $F_*$ and marginal density $f_*$. For $(r,\mathbf{w})\in\{(t, \mathbf{z})\in[0,\infty)\times\mathcal{U}_1 : t\mathbf{z}\in\mathrm{supp}(\mathbf{X}_*)\}$ define
\begin{align*}
    \delta_*(r,\mathbf{w};c) &= c(F_*(rw_1),\cdots,F_*(rw_d)),\\
    m_*(r,\mathbf{w}) &= \prod_{i=1}^d f_*(rw_i).
\end{align*}
We refer to $\delta_*$ as the angular-radial representation of the copula density (abbreviated to \textit{AR copula function}) on margins $F_*$, and $m_*$ as the marginal product function. The joint density of $\mathbf{X}_*$ can then be expressed as $f_{\mathbf{X}_*} (r\mathbf{w}) = m_*(r,\mathbf{w}) \delta_*(r,\mathbf{w};c)$.
\end{definition}

In the notation defined above we have written $\delta_*(r,\mathbf{w};c)$ to emphasise that $\delta_*$ is dependent on the copula density, whereas $m_*$ is not. As with other functions and coefficients of the copula, we will omit this information when $c$ is clear from the context, and write $\delta_*(r,\mathbf{w})$. In the case of the independence copula, $c(\mathbf{u})\equiv 1$, we have $\delta_*(r,\mathbf{w})=1$ for any choice of margin, and $f_{\mathbf{X}_*} (r\mathbf{w},c) = m_*(r,\mathbf{w})$. For arbitrary copula density $c$, the function $\delta_*(r,\mathbf{w},c)$ therefore describes how the joint density of independent random variables with common margins $F_*$ is modified to obtain the density of dependent random variables with copula $c$.

Using this notation, we can relate the asymptotic form of the AR copula function on exponential margins, in terms of the copula density exponent function. If we define $r=-\log(t)$, then we can write
\begin{align*}
    \delta_E(r,\mathbf{w}) = c_{\mathbf{1}_d}(\exp(-r\mathbf{w})) 
    \sim \mathcal{M}_{\mathbf{1}_d}(\exp(-r); \mathbf{w}) \exp(-r (\lambda_{\mathbf{1}_d}(\mathbf{w})-1)), \quad r\to\infty, \mathbf{z}\in\mathcal{U}_1\cap(0,1]^d.
\end{align*}
Similarly, the copula density exponent function can also be related to the asymptotic form of AR copula function on Laplace margins. 

\begin{proposition} \label{prop:delta_laplace_ARE}
Suppose that $d$-dimensional copula density $c$ satisfies the ARE model assumptions in corner $\mathbf{u_0}=(u_{0,1},...,u_{d,1})$, with copula density exponent function $\lambda_{\mathbf{u}_0}(\mathbf{z})$ that is continuous with bounded partial derivatives everywhere apart from the ray $\mathbf{z}=\mathbf{1}_d$. Then for all $\mathbf{w} \in \{(w_1,...,w_d)\in\mathcal{U}_1 : (-1)^{1+u_{d,j}} w_j>0, \, j=1,...,d\}$ there exists a function $\mathcal{M}_{\mathbf{u_0}}(t,\mathbf{w})$ that is slowly-varying in $t$ at $0^+$, such that 
\begin{equation} \label{eq:delta_laplace_ARE}
    \delta_L(r,\mathbf{w}) \sim \mathcal{M}_{\mathbf{u_0}} (\exp(-r),\mathbf{w}) \exp(-r (\lambda_{\mathbf{u}_0} (\mathbf{w}) - 1)), \quad r\to\infty.
\end{equation}
\end{proposition}

In the case that $\mathbf{w}=(w_1,...,w_d)$ has $w_j=0$ for one or more $j\in\{1,...,d\}$, the path through the copula, $\mathbf{u} = F_L(r\mathbf{w})$, does not terminate in a corner as $r\to\infty$, so the copula density exponent functions do not provide information about the asymptotic behaviour of $\delta_L(r,\mathbf{w})$ for these values of $\mathbf{w}$. However, in these cases, for the examples considered in Appendix \ref{app:copula_calcs}, we find that $\delta_L(r,\mathbf{w}) \sim \mathcal{M}_{\mathbf{u_0}} (\exp(-r), \mathbf{w}) \exp(-r (\lim_{\mathbf{z} \to \mathbf{w}} \lambda_{\mathbf{u}_0} (\mathbf{z}) - 1))$, for $r\to\infty$. In Section \ref{sec:SPAR_Laplace}, we will consider more general examples of asymptotic forms for $\delta_L(r,\mathbf{w})$ than that given in Proposition \ref{prop:delta_laplace_ARE}. However, in the examples presented in Section \ref{sec:Laplace_examples}, we will show that this form is applicable for many families of copulas.

We can also relate the ARL model for the copula density to the asymptotic form of the AR copula function on GP margins with positive shape parameter.

\begin{proposition} \label{prop:delta_GP_ARL}
Suppose that $d$-dimensional copula density $c$ satisfies the ARL model assumptions in the upper tail, i.e. for $z\in(0,\infty)^d$ we have $c_{\mathbf{1}_d} (t\mathbf{z}) \sim b_{\mathbf{1}_d} (\mathbf{z}) \mathcal{L}_{\mathbf{1}_d} (t) t^{\kappa_{\mathbf{1}_d}-d}$ as $t\to0^+$, for some $\mathcal{L}_{\mathbf{1}_d} \in RV_0(0^+)$ and $b_{\mathbf{1}_d} (\mathbf{z})>0$. Then for GP margins with shape parameter $\xi_m>0$ and $\mathbf{w} \in \mathcal{U}_1 \cap (0,1]^d$, we have
\begin{equation} \label{eq:delta_GP_ARL}
    \delta_{GP}(r,\mathbf{w}) \sim s_{\mathbf{w}}^{\kappa_{\mathbf{1}_d}-d} b_{\mathbf{1}_d} \left(s_{\mathbf{w}}^{-1} \mathbf{w}^{-1/\xi_m} \right) \mathcal{L}_{\mathbf{1}_d} \left(r^{-1/\xi_m} \right) r^{\tfrac{d-\kappa_{\mathbf{1}_d}}{\xi_m}}, \quad r\to\infty,
\end{equation}
where $s_{\mathbf{w}} = \xi_m^{-1/\xi_m} \sum_{j=1}^d w_j^{-1/\xi_m}$. 
\end{proposition}

Propositions \ref{prop:delta_laplace_ARE} and \ref{prop:delta_GP_ARL} relate a description of the marginal-scale angular-radial behaviour to a description of the copula. In Section \ref{sec:margin}, we will see how this can be used to make some general statements about how the choice of margin affects whether a given copula has a SPAR representation.

\section{Effects of the choice of margin} \label{sec:margin}
In this section we consider the effect of the choice of margins on whether a given copula has a SPAR representation, i.e. whether assumptions (A1)-(A3) are satisfied. Assumptions (A1) and (A2) are related to the tail of the conditional radial distribution, whereas assumption (A3) is related to the angular density, which is an integral over the entire conditional radial distribution. We start by discussing assumption (A3) in Section \ref{sec:ang_dens}, and consider under what choice of margins the angular density is finite and continuous for the types of copula introduced in Sections \ref{sec:ARL_model} and \ref{sec:ARE_model}. To examine whether assumption (A3) is satisfied, we need to specify the entire marginal distribution, rather than just the properties of the tail. We show that, in some cases, for two choices of marginal distribution functions, $F$, $G$, which are asymptotically equivalent in the tail (i.e. $\Bar{F}(x)\sim \Bar{G}(x)$ as $x\to\infty$), using margin $F$ satisfies assumption (A3), whereas using margin $G$ does not. We show that using Laplace margins leads to forms of the joint density where (A3) is satisfied for a wide range of copulas, whereas for other choices of margins the angular density is not finite in some cases. In particular, using two-sided Laplace margins has distinct advantages to using one-sided exponential margins.

In the remainder of the section, we consider particular choices of margin in more detail, in relation to whether assumption (A2) is satisfied (i.e. whether the convergence to a GP tail in assumption (A1) is satisfied with parameter functions that are finite and continuous with angle). We consider three cases, where the margins are in the domain of attraction of a generalised extreme value (GEV) distribution with positive, zero, or negative shape parameter. Section \ref{sec:SPAR_Laplace} considers the case of Laplace margins. We show that using Laplace margins leads to forms of the joint density which satisfy (A2) for a large family of copulas, including many commonly-used models. In Section \ref{sec:limit_set} we demonstrate that there is a useful relation between SPAR representations on Laplace margins and the corresponding limit set, when they exist. This provides links between SPAR and other representations for multivariate extremes, which can be related to properties of the limit set \parencite{Nolde2022}. The link between SPAR representations and limit sets also provides a means of estimating limit sets. We also show that some copulas which do not have limit sets on Laplace margins, do have SPAR representations.

Section \ref{sec:SPAR_longtail} considers SPAR representations on long-tailed margins (i.e. margins in the domain of attraction of an extreme value distribution with positive shape parameter). We show that copulas with asymptotic dependence in the upper tail have a convenient representation on long-tailed margins. However, there are copulas with asymptotic dependence in the upper tail for which the SPAR representation on long-tailed margins is only valid in the region where all variables are large, whereas the SPAR representation on Laplace margins is valid in all `extreme regions'. Moreover, we show that for a large class of copulas which are asymptotically independent in the upper tail, the SPAR representations on long-tailed margins are only valid in the region where all variables are large. We also demonstrate that SPAR representations on GP margins with $\xi_m=1$ are equivalent to the representations proposed by \textcite{Coles1991} for the case of asymptotic dependence, and the representation of \textcite{Ledford1996} for the case of asymptotic independence. 

Finally, Section \ref{sec:SPAR_shorttail} considers SPAR representations on short-tailed margins (i.e. margins in the domain of attraction of a extreme value distribution with negative shape parameter). We show that, in general, there are significant restrictions on the types of copula which have SPAR representations on these margins. However, there are certain types of copula, which do not have support on the whole of $(0,1)^d$, that do have SPAR representations on short-tailed margins. We discuss the implications this finding has for modelling real-world datasets, such as those used in the motivating examples in Section \ref{sec:motivation}. 

Given the results of Section \ref{sec:pdfs_in_polars}, without loss of generality, we can work in $L^1$ polar coordinates, and for random vector $\mathbf{X}\in \mathbb{R}^d$ we define $R = \|\mathbf{X}\|_1$ and $\mathbf{W}= \mathbf{X}/R$. When $d=2$ we define $Q=\mathcal{A}_1^{(-2,2]}(\mathbf{W})$. As noted at the end of Section \ref{sec:pdfs_in_polars}, this allows us to switch between the use of vector and scalar angles in $\mathbb{R}^2$, with unit Jacobian. As with other sections, proofs of results stated in the text are provided in Appendix \ref{app:proofs}.

\subsection{Angular density} \label{sec:ang_dens}
To understand how the choice of margins affects assumption (A3), we need to consider the behaviour of the copula density along rays of constant angle, defined on various margins. Due to the wide range of possibilities for choices of margin, we restrict our interest to two families of distributions. For cases where we are only interested in extremes in the non-negative orthant of $\mathbb{R}^d$, a natural choice is GP margins with unit scale and shape parameter $\xi_m$ (we denote the shape parameter of the margins as $\xi_m$, to distinguish it from the shape parameter of the tail of the conditional radial distribution). The three canonical cases are $\xi_m=-1$ (uniform margins), $\xi_m=0$ (exponential margins) and $\xi_m=1$ (asymptotically equivalent to standard Fr\'echet or Pareto margins). When interest is in extremes in all orthants of $\mathbb{R}^d$, it is beneficial to use symmetric margins. We define the symmetric GP (SGP) density function to be $f_{SGP}(x;\xi_m) = \tfrac{1}{2}f_{GP}(|x|;\xi_m,1)$ for $x\in(-r_F,r_F)$, where $f_{GP}$ is the usual GP density function, and $r_F=\infty$ for $\xi_m\geq0$ and $r_F=-1/\xi_m$ for $\xi_m<0$. The three cases above now correspond to a uniform distribution on $[-1,1]$ when $\xi_m=-1$, the standard Laplace distribution when $\xi_m=0$, and a \textit{`double Pareto'} distribution when $\xi_m=1$. 

\begin{figure}[!t]
	\centering
    \begin{subfigure}[t]{0.32\textwidth}
        \centering
         \includegraphics[scale=0.5]{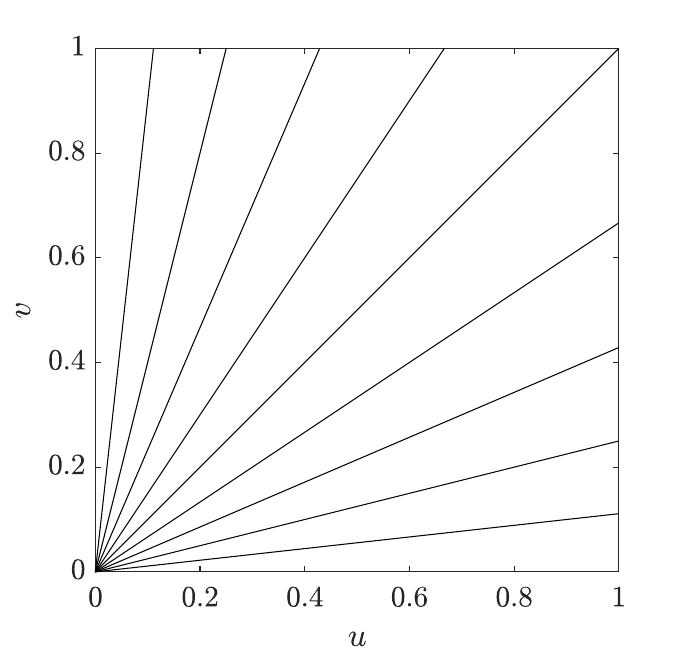}
         \caption{Uniform (GP with $\xi_m=-1$)}
     \end{subfigure}
     \hfill
     \begin{subfigure}[t]{0.32\textwidth}
         \centering
         \includegraphics[scale=0.5]{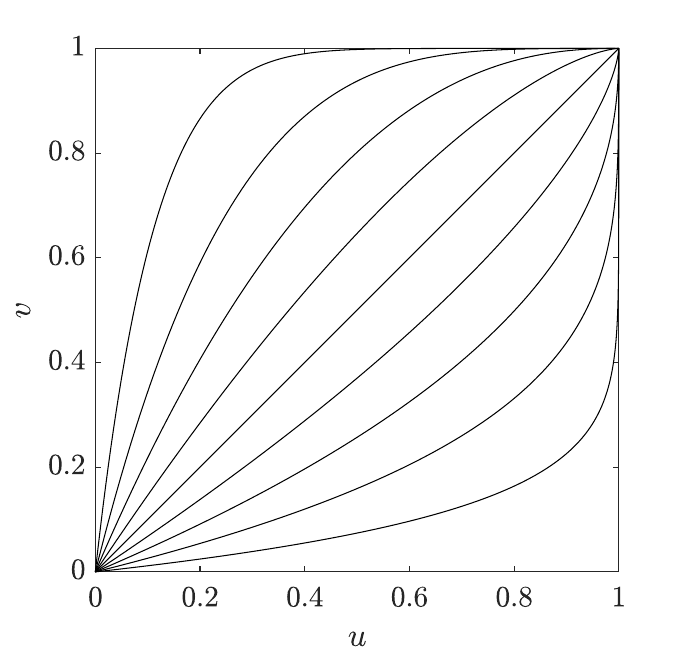}
         \caption{Exponential (GP with $\xi_m=0$)}
     \end{subfigure}
     \hfill
     \begin{subfigure}[t]{0.32\textwidth}
         \centering
         \includegraphics[scale=0.5]{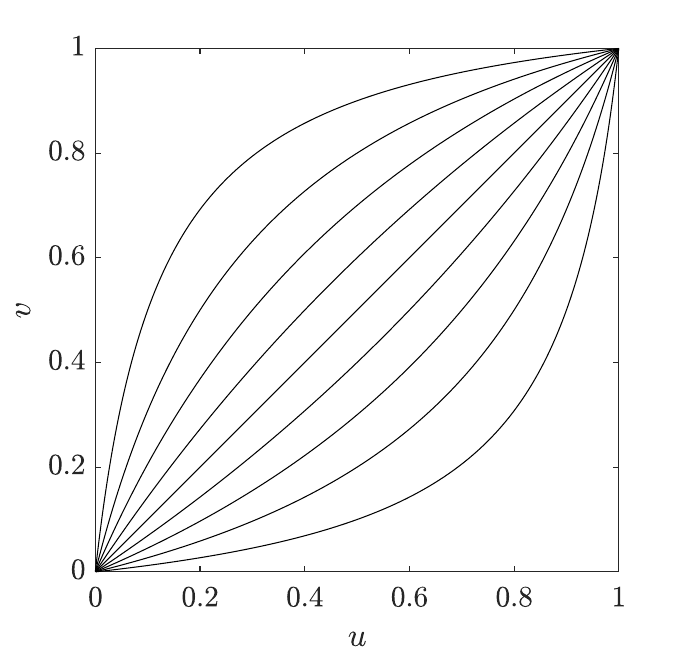}
         \caption{GP with $\xi_m=1$}
     \end{subfigure}\\
     \begin{subfigure}[t]{0.32\textwidth}
        \centering
         \includegraphics[scale=0.5]{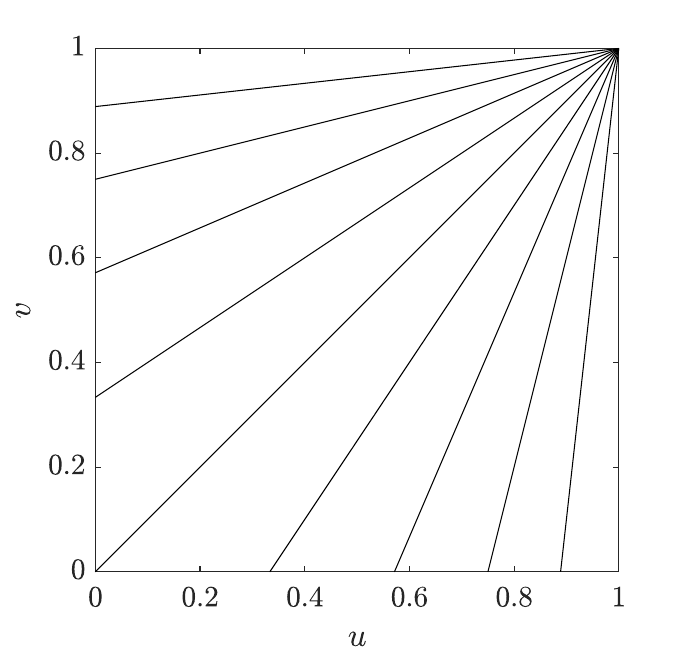}
         \caption{Standard Pareto}
     \end{subfigure}
     \hfill
     \begin{subfigure}[t]{0.32\textwidth}
         \centering
         \includegraphics[scale=0.5]{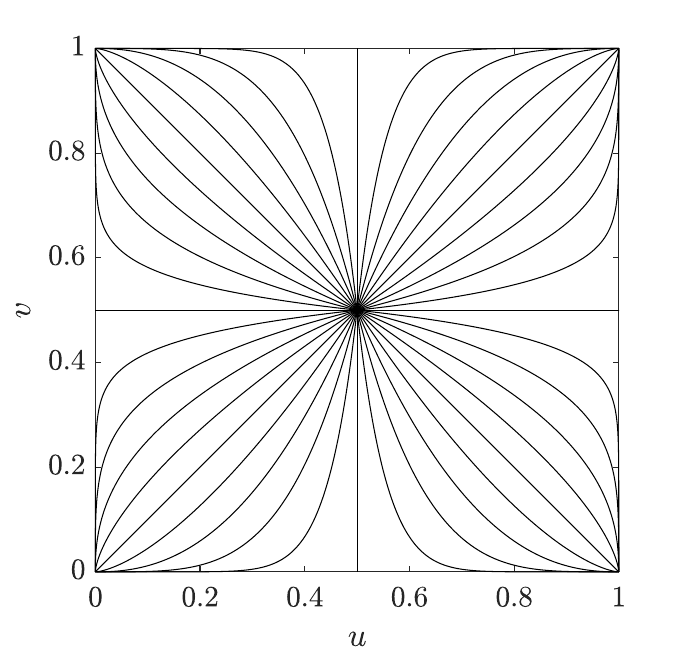}
         \caption{Laplace, (SGP with $\xi_m=0$)}
     \end{subfigure}
     \hfill
     \begin{subfigure}[t]{0.32\textwidth}
         \centering
         \includegraphics[scale=0.5]{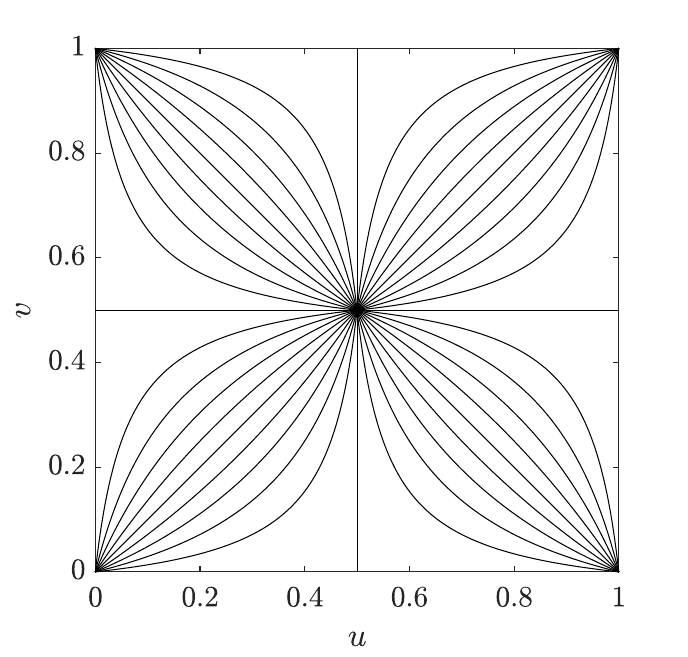}
         \caption{SGP with $\xi_m=1$}
     \end{subfigure}
	\caption{Paths through the copula corresponding to rays of constant angle on different margins.}
	\label{fig:Copula_paths}
    \bigskip
	\centering
    \captionsetup[subfigure]{justification=centering}
    \begin{subfigure}[t]{0.32\textwidth}
        \centering
        \includegraphics[scale=0.5]{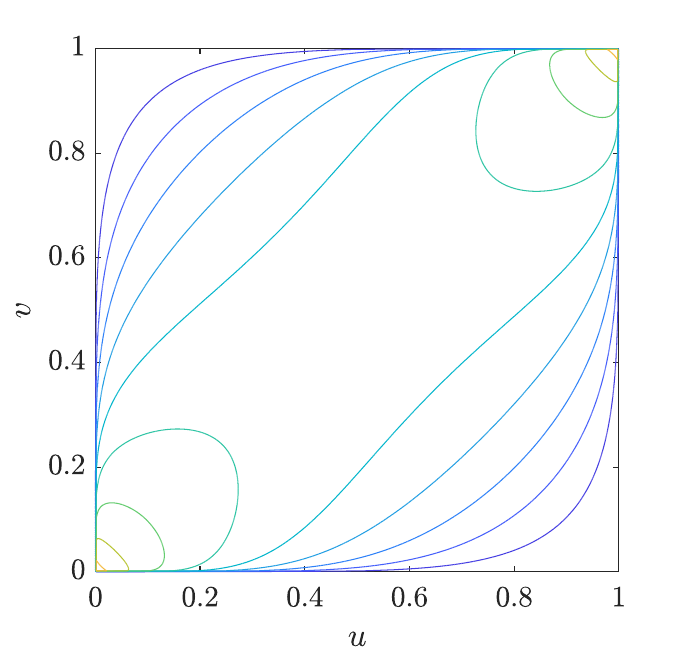}
        \caption{Gaussian copula, $\rho=0.6$ }
    \end{subfigure}
    \hfill
    \begin{subfigure}[t]{0.32\textwidth}
        \centering
        \includegraphics[scale=0.5]{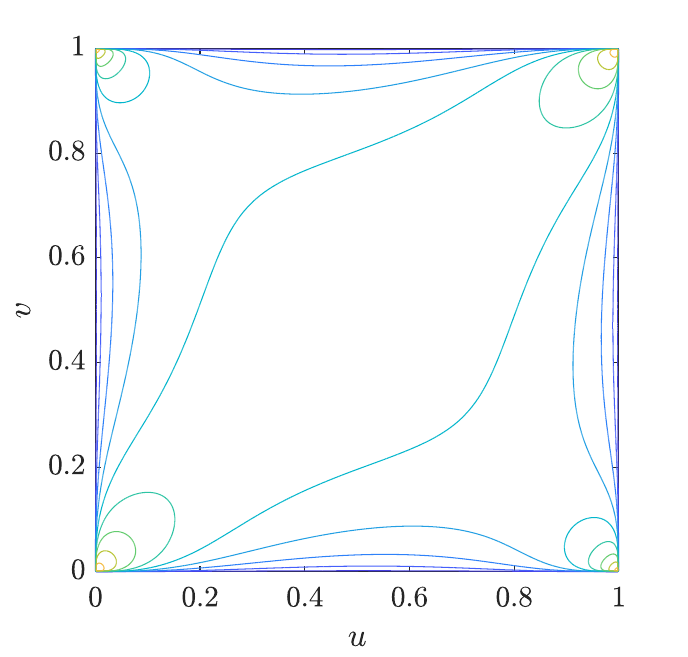}
        \caption{t copula, $\rho=0.2$, $\nu=2$ }
    \end{subfigure}
    \hfill
    \begin{subfigure}[t]{0.32\textwidth}
        \centering
        \includegraphics[scale=0.5]{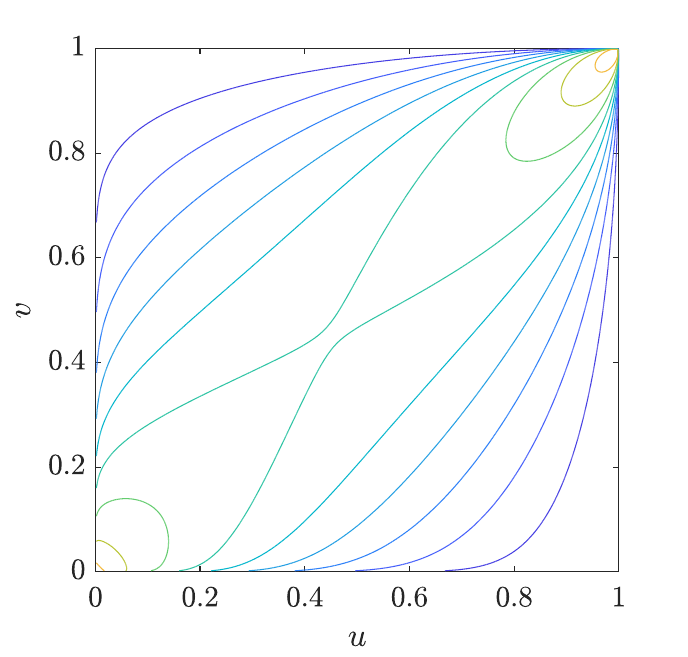}
        \caption{EV copula with symmetric logistic dependence, $\alpha=2$}
    \end{subfigure}
    \centering 
	\caption{Contours of scaled copula density $c_s(u,v) = c(u,v)/(1+c(u,v))$ for various two-dimensional copulas, with contour levels from 0.1 (dark blue) to 0.9 (yellow) at steps of 0.1. Note that $c_s(u,v) \to 1^-$ as $c(u,v)\to\infty$.}
	\label{fig:Copula_density}
\end{figure}

Suppose that $(X_{GP},Y_{GP})$, $(X_{SGP},Y_{SGP})$ and $(X_P,Y_P)$ are random vectors with copula density $c$, and GP, SGP and standard Pareto margins (i.e. $\Pr(X_P>x)=1/x$, $x\ge1$), respectively. Even though GP margins with $\xi_m=1$ are asymptotically equivalent to standard Pareto margins, the angular density can differ in important ways, described below. Define corresponding $L^1$ polar variables $R_* = \|(X_*,Y_*)\|_1$, $Q_* = \mathcal{A}_1^{(-2,2]} ((X_*,Y_*)/R_*)$, where $*$ denotes the respective marginal distribution. \autoref{fig:Copula_paths} shows the paths through the copula corresponding to rays of constant $Q_*$. For GP margins, the rays all emanate from $(u,v)=(0,0)$ on the copula scale, whereas for SGP margins, the rays emanate from $(u,v)=(\tfrac{1}{2}, \tfrac{1}{2})$. The case of Pareto margins forms an interesting contrast to uniform margins. The cdf of the standard Pareto distribution is $F_P(x)=1-x^{-1}$, $x\geq1$. For $q\in(0,1)$ and $r\geq\max(q^{-1},(1-q)^{-1})$ the angular-radial representation of the copula density on Pareto margins is
\begin{align*}
    \delta_P(r,(1-q,q)) = c_{(1,1)}((r(1-q))^{-1}, (rq)^{-1}) = c_{(1,1)} (t(1-w),tw),
\end{align*}
where $t=(rq(1-q))^{-1}$ and $w=1-q$. That is, rays $Q_p=q$ correspond to straight lines on the copula scale, emanating from the upper right corner.

From \autoref{fig:Copula_paths}, it is evident that on symmetric margins the paths only encounter one corner, whereas on one-sided margins the paths can asymptote close to the lower left / upper right corner as $q\to 0^+$ or $1^-$. In these cases, if the copula density tends to infinity in the lower right or upper left corners (this corresponds to $\kappa_{(0,1)}<2$ or $\kappa_{(1,0)}<2$), then in order to consider whether the angular density is finite, we need to consider how the joint density $f_{R_*,Q_*}(r,q)$ behaves when $\mathbf{u}=(F_*(r(1-q)),F_*(r,q))$ is close to these corners. \autoref{fig:Copula_density} shows contours of scaled copula density $c_s(u,v) = c(u,v)/(1+c(u,v))$ for various copulas. The scaling is used so that $c_s(u,v)\in[0,1]$, with $c_s(u,v) \to 1^-$ as $c(u,v)\to\infty$. For the Gaussian copula with $\rho\in(-1,1)$ the tail order is $2/(1+\rho)$ in the upper right and lower left corners, and $2/(1-\rho)$ in the upper left and lower right corners (see Appendix \ref{app:Gaussian} for details). So when $\rho<0$, the copula density is infinite in the lower right and upper left corners. For the t-copula, there is strong tail dependence in all corners, and hence the density is infinite at each corner (see Appendix \ref{app:t_copula}). For 2-dimensional EV copulas there is strong tail dependence in the upper right corner and intermediate tail dependence in the lower left corner (see Appendix \ref{app:EVcopula}).

The following propositions consider the angular density $f_{Q_*}(q)$ for various choices of marginal distribution, when the copula density has either ARL or ARE form in the corners. Not all copulas have these forms, but they are applicable for many widely-used families of copula. In particular, the cases considered highlight that, for a given copula, $f_{Q_*}(q)$ may be finite for one choice of margin, but not for other choices.

\begin{proposition}[Angular densities on different margins] \label{prop:fq_convergence_HJ}
Suppose that two-dimensional copula density $c$ has ARL form in each corner, and is continuous and finite away from the corners. Suppose that angular and radial variables on various margins are defined as above. Then the following results hold for the angular densities.\par
\noindent\textbf{GP margins:}
\begin{enumerate}[(a)]
\item If $-\kappa_{(1,1)}<\xi_m$ then $f_{Q_{GP}}(q)$ is finite for $q\in(0,1)$.
\item Suppose there is strong tail dependence in the lower right corner, i.e. $\kappa_{(1,0)}=1$.
\begin{enumerate}[(i)]
	\item If $\xi_m\geq 0$ then $f_{Q_{GP}}(q)\to\infty$ for $q\to0^+$. 
	\item If $\xi_m<0$ and there exist $\beta_0,\beta_1>0$ such that $b_{(1,0)}(1-z,z) \in RV_{\beta_0}(0^+)$ and $b_{(1,0)}(1-z,z) \in RV_{\beta_1}(1^-)$, then $f_{Q_{GP}}(q)$ is finite for $q\to0^+$. 
\end{enumerate}
\end{enumerate} 
\textbf{SGP margins:}
\begin{enumerate}[(a)]
	\setcounter{enumi}{2}
	\item If $\xi_m<1$ then $f_{Q_{SGP}}(q)$ is finite for $q\in \mathbb{R}$. 
	\item If $\xi_m\geq1$ and $c$ is positive and finite on the edges, then $f_{Q_{SGP}}(q)$ is infinite for $q\in\mathbb{Z}$.
	\item Suppose that $c(1-t,1/2)\in RV_{\alpha}(0^+)$ for some $\alpha>0$, then $f_{Q_{SGP}}(0)$ is finite for $\xi_m<1+\alpha$, with similar constraints for other values of $q\in\mathbb{Z}$.
\end{enumerate}
\textbf{Pareto margins:}
\begin{enumerate}[(f)]
\item If there is strong tail dependence in the lower left corner ($\kappa_{(0,0)}=1$), then $f_{Q_{P}}(1/2) = \infty$.
\end{enumerate} 
\end{proposition}

Proposition \ref{prop:fq_convergence_HJ} shows that there is some advantage to using two-sided SGP margins over one-side GP margins, even when interest is only in the upper right quadrant of the plane, since $f_{Q_{SGP}}(q)$ is finite for all $q\in\mathbb{R}$ when $\xi_m<1$, whereas and $f_{Q_{GP}}(q)$ can tend to infinity for $q\to 0^+$ or $1^-$. This is because the rays of constant $Q_{SGP}$ pass close to at most one corner of the copula, whereas rays of constant $Q_{GP}$ pass close to the lower right and upper left corners when $q\to 0^+$ or $1^-$. Proposition \ref{prop:fq_convergence_HJ} considers the case of strong tail dependence in the lower right corner. However, this is not a necessary condition for $f_{Q_{GP}}(q)\to\infty$ for $q\to0^+$. For example, for the case of the independence copula and GP margins with $\xi_m=1$, we can calculate directly $f_{Q_{GP}}(q) = (\log((1+z)/(1-z))-2z) \,z^{-3}$, where $z=1-2q$, and hence $f_{Q_{GP}}(q)\to\infty$ as $q\to 0^+$ or $q\to1^-$.


When one-sided margins are used, Proposition \ref{prop:fq_convergence_HJ} shows that GP margins have some advantage over Pareto margins, in that $f_{Q_P}(1/2)=\infty$ when there is strong lower tail dependence, whereas $f_{Q_{GP}}(q)$ is finite for $q\in(0,1)$. However, note that when $\xi_m=1$, $(X_{GP},Y_{GP}) = (X_P - 1, Y_P - 1)$. So, defining angular-radial variables on GP margins with the origin at $(X_{GP},Y_{GP}) = (0,0)$ is equivalent to defining angular-radial variables on Pareto margins with origin at $(X_P,Y_P)=(1,1)$. When plotting the joint density on a log-log scale, it is more convenient to work with Pareto margins, since the lower end point of the margins is $1=10^0$, rather than $0=10^{-\infty}$ for GP margins. Therefore, in section \ref{sec:SPAR_longtail}, we will work with Pareto rather than GP margins, and define the origin of the polar coordinates at $(1,1)$. 

In many cases, copulas have both ARL and ARE asymptotic forms in the corners. Moreover, in certain cases the ARE representation can be defined in terms of the ARL representation (Proposition \ref{prop:ARL2ARE}). However, since it is not always possible to calculate the ARE form in terms of the ARL form, it is useful to consider the behaviour of the angular density when the copula has ARE form in the corners. Since Proposition \ref{prop:fq_convergence_HJ} established that Laplace margins have some advantages over other margins in two-dimensional cases, we only consider Laplace margins here. We also extend to the general multivariate case. 

\begin{proposition}[Angular density on Laplace margins] \label{prop:fw_convergence_WT}
Suppose that $d$-dimensional copula density $c$ satisfies the ARE model assumptions in each corner, and is continuous and finite away from the corners. Then on Laplace margins, $f_{\mathbf{W}}(\mathbf{w})$ is finite for $\mathbf{w}\in \mathcal{U}_1 = \{\mathbf{x}\in\mathbb{R}^d: \|\mathbf{x}\|_1=1\}$.
\end{proposition}

So, on Laplace margins, under the assumptions of Proposition \ref{prop:fw_convergence_WT}, the angular density is finite for all angles. SPAR assumption (A3) also requires the angular density to be continuous. To complete our consideration of assumption (A3), we make the following observations about the continuity of $f_{\mathbf{W}}$.

\begin{proposition}[Continuity of angular density] \label{prop:fw_continuity}
Suppose $f_{R,\mathbf{W}}$ is continuous and bounded for all $(r,\mathbf{w})$ in the domain, and that $f_{\mathbf{W}}(\mathbf{w})$ is finite for all $\mathbf{w}$ in the domain. Then $f_{\mathbf{W}}(\mathbf{w})$ is also continuous and hence satisfies SPAR assumption (A3).
\end{proposition}

Note that Proposition \ref{prop:fw_continuity} applies whenever $f_{R,\mathbf{W}}$ is continuous and bounded, not just for the case of Laplace margins. The next lemma shows that the assumptions of Proposition \ref{prop:fw_convergence_WT} are consistent with those of Proposition \ref{prop:fw_continuity}, and hence also imply continuity of angular density.

\begin{lemma} \label{lemma:frw_conditions}
Under the assumptions of Proposition \ref{prop:fw_convergence_WT}, $f_{R,\mathbf{W}}$ is continuous and bounded for all $(r,\mathbf{w})\in [0,\infty)\times \mathcal{U}_1$, and hence $f_{\mathbf{W}}$ satisfies SPAR assumption (A3).
\end{lemma}

\subsection{SPAR models on Laplace margins} \label{sec:SPAR_Laplace}
The previous section showed that under certain conditions, the angular density on Laplace margins satisfies SPAR assumption (A3). In this section we consider assumptions (A1) and (A2), regarding the tail of the conditional radial distribution, for the case of Laplace margins. The standard Laplace density function is $f_L(x)=\tfrac{1}{2}\exp(-|x|)$ for $x\in\mathbb{R}$. For $\mathbf{w}=(w_1,...,w_d)\in\mathcal{U}_1$ and $r\in[0,\infty)$, the marginal product function is
\begin{equation*}
    m_{L}(r,\mathbf{w}) = 2^{-d} \exp\left( -r \sum_{i=1}^d |w_i|\right) = 2^{-d} \exp( -r).
\end{equation*}
In the examples below, we will show that many commonly-used copulas have asymptotic form (\ref{eq:delta_laplace_ARE}). In this case, for $(r,\mathbf{w})\in [0,\infty)\times\mathcal{U}_1$, the angular-radial joint density has asymptotic form
\begin{equation} \label{eq:frw_Laplace}
	f_{R,\mathbf{W}}(r,\mathbf{w}) = r^{d-1} m_L(r,\mathbf{w}) \delta_L(r,\mathbf{w}) \sim 2^{-d} r^{d-1} \mathcal{M}(\exp(-r),\mathbf{w}) \exp( -r \lambda(\mathbf{w})), \quad r\to\infty,
\end{equation}
for some function $\mathcal{M}(t,\mathbf{w})>0$, which is slowly-varying in $t$ at $0^+$, and $\lambda(\mathbf{w}) > 0$, which is continuous in $\mathbf{w}$. From Propositions \ref{prop:fw_convergence_WT} and \ref{prop:fw_continuity}, we know that this asymptotic form together with the copula being continuous and bounded away from the corners is sufficient for the angular density to satisfy assumption (A3). The next proposition gives sufficient conditions for assumptions (A1)-(A2) to be satisfied. 

\begin{proposition} \label{prop:SPAR_Laplace}
Suppose that angular-radial density $f_{R,\mathbf{W}}$ has asymptotic form (\ref{eq:frw_Laplace}), and is continuous and bounded, and that $f_{R,\mathbf{W}}(r,\mathbf{w})$ is ultimately monotone in $r$ for each $\mathbf{w}\in \mathcal{U}_1$. Then joint density $f_{R,\mathbf{W}}$ satisfies SPAR assumptions (A1) and (A2) for all $\mathbf{w}\in\mathcal{U}_1$, with GP shape parameter $\xi(\mathbf{w})=0$ and scale parameter $\sigma(\mu,\mathbf{w}) = 1/\lambda(\mathbf{w})$.
\end{proposition}

Returning briefly to the discussion of coordinate systems, angular-radial densities of the form given in Proposition \ref{prop:SPAR_Laplace} also satisfy the assumptions of Theorem \ref{thm:SPAR_indep}(ii), with $\sigma(\mu,\mathbf{w}) = h(\mu)\alpha(\mathbf{w})$, where $h(\mu)=1$ and $\alpha(\mathbf{w}) = 1 / \lambda(\mathbf{w})$. We can therefore define alternative angular-radial coordinates so that the angular and radial variables are AI. However, as we will show in the examples in Section \ref{sec:Laplace_examples}, the general forms of $\lambda(\mathbf{w})$ do not result in standard coordinate systems. The following example considers the case for the independence copula on Laplace margins, where angular and radial variables are independent using $L^1$ polar coordinates.

\begin{example}[Independence on Laplace margins] \label{ex:Indep_laplace}
The independence copula $c(\mathbf{u})=1$ for $\mathbf{u}\in[0,1]^d$, has asymptotic form (\ref{eq:delta_laplace_ARE}) on Laplace margins with $\mathcal{M}(r,\mathbf{w})=1$ and $\lambda(\mathbf{w})=1$, for $(r,\mathbf{w})\in[0,\infty)\times\mathcal{U}_1$. The angular-radial joint density is $f_{R,\mathbf{W}}(r,\mathbf{w}) = 2^{-d} r^{d-1} \exp(-r)$. The assumptions of Proposition \ref{prop:SPAR_Laplace} are therefore satisfied, so $c$ has a SPAR representation on Laplace margins. In this case, the angular density is constant, $f_{\mathbf{W}}(\mathbf{w}) = 2^{-d} (d-1)!$. \hfill $\blacksquare$
\end{example}

\subsubsection{Alternative approximation for Laplace margins}
In the SPAR model (\ref{eq:SPAR_model}), we assume that the tail of the conditional radial density $f_{R|\mathbf{W}}(r|\mathbf{w})$ can be approximated by a GP distribution. This is a direct analogue to the univariate case. In the multivariate setting, we have
\begin{equation*}
    f_{R|\mathbf{W}}(r|\mathbf{w}) = r^{d-1} m_*(r,\mathbf{w}) \delta_*(r,\mathbf{w}).
\end{equation*}
In the multivariate case, the assumption about the form of the tail could either be applied to $f_{R|\mathbf{W}}(r|\mathbf{w})$ itself, or it could be applied to the tail of the angular-radial dependence function $\delta_*(r,\mathbf{w})$. In the examples below, we will show that for many families of copulas, the angular-radial dependence function on Laplace margins takes a specific asymptotic form, where the slowly-varying function in (\ref{eq:delta_laplace_ARE}) is given by $\mathcal{M}(\exp(-r),\mathbf{w}) = g(\mathbf{w}) r^{\beta(\mathbf{w})}$, for bounded functions $g(\mathbf{w})>0$ and $\beta(\mathbf{w})$. That is, we have
\begin{equation} \label{eq:delta_Laplace}
	\delta_L(r,\mathbf{w}) \sim  g(\mathbf{w}) r^{\beta(\mathbf{w})} \exp( - r (\lambda(\mathbf{w})-1)), \quad r\to\infty,
\end{equation}
The independence copula is a trivial example, with $g(\mathbf{w})=1$ and $\beta(\mathbf{w})=0$. Similar forms were also considered by \textcite{Wadsworth2022} for exponential margins. 

If $\delta_L(r,\mathbf{w})$ has asymptotic form (\ref{eq:delta_Laplace}) with $\beta(\mathbf{w})=0$, then $f_{R|\mathbf{W}}(r|\mathbf{w}) \sim \tilde{g}(\mathbf{w}) r^{d-1} \exp( -r \lambda(\mathbf{w}))$ for $r\to\infty$, where $\tilde{g}(\mathbf{w}) = 2^{-d} (g(\mathbf{w})/f_{\mathbf{W}}(\mathbf{w}))$. In this case it is more accurate to apply the assumption of a GP-type (exponential) tail to $\delta_{L} (r,\mathbf{w})$, rather than to the conditional radial density. This motivates an alternative POT assumption for the conditional radial density on Laplace margins, that 
\begin{equation} \label{eq:frgw_Laplace}
    f_{R|\mathbf{W}}(r|\mathbf{w}) \sim \frac{r^{d-1}}{A(\mathbf{w})}\, \exp\left(-\frac{r}{\sigma(\mathbf{w})} \right), \quad r\to\infty,
\end{equation}
where $\sigma(\mathbf{w}) = 1 / \lambda(\mathbf{w})$ and $A(\mathbf{w})$ is a normalisation constant. The RHS of (\ref{eq:frgw_Laplace}) is a gamma density function with shape $d$ and scale $\sigma(\mathbf{w})$, and hence $A(\mathbf{w}) = (d-1)! \, (\sigma(\mathbf{w}))^d$. The difference between model (\ref{eq:SPAR_model}) and assumption (\ref{eq:frgw_Laplace}) becomes larger as the number of dimensions $d$ increases. Therefore, in many cases assumption (\ref{eq:frgw_Laplace}) will be a more accurate basis for SPAR models on Laplace margins than model (\ref{eq:SPAR_model}), where the tail of $f_{R|\mathbf{W}}$ is approximated with a GP distribution. If assumption (\ref{eq:frgw_Laplace}) is assumed to hold above some threshold $\mu(\mathbf{w})$ with corresponding exceedance probability $\zeta(\mathbf{w})$, then a modified version of SPAR model (\ref{eq:SPAR_model}) for Laplace margins can be written as
\begin{equation} \label{eq:SPAR_laplace}
    f_{R,\mathbf{W}}(r,\mathbf{w}) = \frac{\zeta(\mathbf{w})f_{\mathbf{W}}(\mathbf{w})}{A(\mathbf{w})}  r^{d-1}\, \exp\left(-\frac{r}{\sigma(\mathbf{w})} \right), \quad r\geq \mu(\mathbf{w}),
\end{equation}
where the normalisation constant becomes
\begin{align*}
    A(\mathbf{w}) &= \int_{\mu(\mathbf{w})}^{\infty} r^{d-1}\, \exp\left(-\frac{r}{\sigma(\mathbf{w})} \right) dr = (\sigma(\mathbf{w}))^d \Gamma\left(d,\frac{\mu(\mathbf{w})}{\sigma(\mathbf{w})}\right),
\end{align*}
and $\Gamma(\cdot,\cdot)$ is the upper incomplete gamma function. In the case $d=2$ we have
\begin{align*}
    A(\mathbf{w}) &= \sigma(\mathbf{w})(\sigma(\mathbf{w})+\mu(\mathbf{w})) \exp\left(-\frac{\mu(\mathbf{w})}{\sigma(\mathbf{w})}\right).
\end{align*}

\subsubsection{Relation to limit sets} \label{sec:limit_set}
It is useful to introduce the concept of limit sets and show how these are related to SPAR models on Laplace margins. 

\begin{definition}[Limit set]
Let $\mathbf{X}_1,\dots, \mathbf{X}_n$ be a sequence of independent and identically distributed random vectors in $\mathbb{R}^d$. Suppose that for some non-random compact set $G\in\mathbb{R}^d$, there is a sequence of real constants $r_n>0$, with $r_n\to\infty$ as $n\to\infty$, such that as $n\to\infty$,
\begin{equation*}
    G_n = \{\mathbf{X}_1/r_n, \cdots, \mathbf{X}_n/r_n\} \xrightarrow{P} G,
\end{equation*}
with respect to the Hausdorff metric, and $\xrightarrow{P}$ denotes convergence in probability. Then $G$ is referred to as the limit set of the scaled sample cloud \parencite{Davis1988}.    
\end{definition}

\textcite{Kinoshita1991} showed that when the limit set exists, it is compact and star-shaped. The limit set can be used to describe various extremal dependence properties of $\mathbf{X}_j$ \parencite{Balkema2010, Nolde2014}. Moreover, \textcite{Nolde2022} showed how various dependence coefficients and representations for multivariate extremes could be related to properties of the limit set, including those of \textcite{Ledford1996, Ledford1997}, \textcite{Heffernan2004}, and \textcite{Wadsworth2013}. We will not repeat the details here, but note that SPAR representations can also be related to these methods, through the link to the limit set, described below. 

Limit sets can be viewed as a multivariate analogue to the scaling of univariate maxima. Consider a variable $X_E$ which follows an exponential distribution with scale parameter $\sigma$, with cdf $F_{X_E}(x)=1-\exp(-x/\sigma)$, $x\geq 0$. Define $M_n$ to be the maximum of the maximum of $n$ independent variables with cdf $F_{X_E}$. As $n\to\infty$, the distribution of $M_n$ converges to a Gumbel distribution with scale $\sigma$ and location $\sigma\log(n)$, i.e. $[F_{X_E}(x)]^n \sim \exp(-\exp(-(x-\sigma\log(n))/\sigma))$, as $n\to\infty$. Hence the distribution of scaled maxima $M_n/\log(n)$ converges to a Gumbel distribution with scale $\sigma/\log(n)$ and location $\sigma$. Since $\sigma/\log(n) \to 0$ as $n\to\infty$, the limit distribution is 
degenerate and we have $\lim_{n\to\infty}\Pr(|M_n/\log(n)-\sigma| > \varepsilon) = 0$ for all $\varepsilon>0$, and hence $M_n/\log(n)$ converges in probability to $\sigma$. In the multivariate setting, the boundary of the limit set can be viewed as the radius of scaled maxima from the conditional radial distribution.\footnote{In the multivariate context, the observations are distributed over angle. For a small $\epsilon$-neighbourhood around a particular angle $\mathbf{w}$, the mean number falling within this neighbourhood converges to $m = n\epsilon V_d f_{\mathbf{W}}(\mathbf{w})$ as $\epsilon\to0^+$, where $V_d$ is the volume of the $d$-dimensional $L^1$ sphere. However, if we set $\epsilon=1/\log(n)$ then $\log(m)/\log(n)\to 1$ as $n\to\infty$, and we obtain a similar convergence of scaled maxima at any particular angle.} The following proposition, which is an adaptation of Proposition 2.2 in \textcite{Nolde2022}, makes this explicit. 


\begin{proposition} \label{prop:limitset}
Suppose that the random vector $\mathbf{X}\in\mathbb{R}^d$ has standard Laplace margins and has joint density function $f_{\mathbf{X}}$ that satisfies
\begin{equation} \label{eq:fx_limitset}
    f_{\mathbf{X}}(r\mathbf{x}) \sim \mathcal{M}(\exp(-r),\mathbf{x}) \exp\left(- \frac{r}{\sigma (\mathbf{x})}\right), \quad r\to\infty,
\end{equation}
for $\mathbf{x} \in \mathbb{R}^d\setminus\{\mathbf{0}\}$, where $\mathcal{M}(t,\mathbf{w})$ is slowly-varying in $t$ at $0^+$ and the function $\sigma$ is positive, continuous and homogeneous of order $-1$. Define $G_n = \{\mathbf{X}_1/\log(n), \dots, \mathbf{X}_n/\log(n)\}$ to be a sequence of scaled random samples from $f_{\mathbf{X}}$. Then the following results hold.
\begin{enumerate}[(i)]
    \item The scaled sample cloud, $G_n$, has a limit set, $G$, with boundary $\partial G = \{\mathbf{w} \, \sigma(\mathbf{w}):  \mathbf{w}\in\mathcal{U}_1 \}$.
    \item The limit set is bounded by the unit box, $G \subseteq \{\mathbf{x}\in\mathbb{R}^d : \|\mathbf{x}\|_{\infty}=1\}$, and this bound is reached in each dimension, i.e. for $j=1,...,d$, 
    \begin{align*}
        \max\{x_j : (x_1,...,x_d)\in G\} &= 1,\\
        \min\{x_j : (x_1,...,x_d)\in G\} &= -1.
    \end{align*}
\end{enumerate}
\end{proposition}

Clearly, both SPAR model (\ref{eq:SPAR_model}) and modified SPAR model (\ref{eq:SPAR_laplace}), which have constant shape parameter $\xi(\mathbf{w})=0$ and scale function $\sigma(\mathbf{w})$, satisfy the assumptions of Proposition \ref{prop:limitset}. Therefore, the SPAR framework provides a rigorous means of estimating limit sets when they exist, since the GP scale parameter is equal to the radius of the limit set. A similar approach was proposed by \textcite{Simpson2022}, where a large quantile of the conditional radial distribution was used as a proxy for the limit set, estimated by fitting a GP distribution to the tail of the conditional radial distribution (see also \cite{majumder2023semiparametric}). However, part (i) of Proposition \ref{prop:limitset} shows explicitly that the $L^1$ radius of points on the boundary of limit set is described by the scale function of the SPAR model, since $\|\mathbf{w} \sigma(\mathbf{w})\|_1 = \sigma(\mathbf{w})$ for $\mathbf{w} \in \mathcal{U}_1$. Part (ii) provides bounds on the scale function in these cases, namely that for $\mathbf{w} = (w_1,\dots,w_d) \in \mathcal{U}_1$ we have $0< |w_j| \sigma(\mathbf{w})\leq 1$ for $j=1,\dots,d$ and that the upper bound is achieved for some $|w_j|\leq 1/2$. The upper and lower bounds in part (ii) are simply a consequence on the choice of margins, for which scaled maxima and minima must converge to one, as described above. 

In Section \ref{sec:ARL_ARE_relation} it was mentioned that for two-dimensional copula densities which have both an ARL and ARE representation, when $\kappa_{(u_0,v_0)}=1$ the function $\lambda_{(u_0,v_0)}(1-w,w)$ is not differentiable at $w=1/2$. The reason for this is now evident. Namely, when $\kappa_{(u_0,v_0)}=1$ we have $\lambda_{(u_0,v_0)}(1/2,1/2)=1/2$ and hence the $L^1$ radius of the limit set is $1/\lambda_{(u_0,v_0)}(1/2,1/2)=2$ in the direction $((-1)^{u_0+1},(-1)^{v_0+1})$. Since the limit set is bounded by the unit box, and the upper bound is achieved in the corner $((-1)^{u_0+1},(-1)^{v_0+1})$, the function $\lambda_{(u_0,v_0)}(1-w,w)$ must have a cusp at $w=1/2$, and hence is not differentiable at this point.

Again returning to the question of coordinate systems discussed in Theorem \ref{thm:SPAR_indep}, we can see that when a joint density on Laplace margins has asymptotic form (\ref{eq:fx_limitset}), the coordinate system required to obtain AI angular and radial variables can be defined in terms of the gauge function of the limit set. 

Finally, we note that there are copulas for which the limit set on Laplace margins is degenerate in certain regions, i.e. the boundary has zero radius at certain angles. Examples of this type are EV copulas with H{\"u}sler-Reiss dependence and the bivariate exponential copula, discussed in Examples \ref{ex:EV_laplace} and \ref{ex:biv_exp} below. We will show that in these cases, the copula still has a SPAR representation, despite the limit set being degenerate.

\subsubsection{Bivariate examples} \label{sec:Laplace_examples}
In the bivariate examples below, we will use the notation $\mathbf{w}=(w_1,w_2)=(\cos_1(q),\sin_1(q))$ and $\sigma(q)=1/\lambda(\mathbf{w})$ for $q\in(-2,2]$. We will demonstrate that SPAR models on Laplace margins can represent a range of tail dependence levels. From the discussion in Section \ref{sec:ARE_model}, when $\delta_L(r,\mathbf{w})$ has asymptotic form (\ref{eq:delta_Laplace}) and also has an ARL representation, the tail orders in each quadrant are given by $\kappa_{(u_0,v_0)} = \lambda((-1)^{u_0+1},(-1)^{v_0+1})$. Throughout this section we use the modified SPAR model (\ref{eq:SPAR_laplace}) to approximate the joint density. In the examples below, the angular density and thresholds have been calculated using numerical integration in most cases. Example \ref{ex:elliptic} considered SPAR models for bivariate Gaussian and t distributions on their own respective margins. In examples \ref{ex:Gauss_Laplace} and \ref{ex:Student_Laplace}, we revisit these examples and show that Gaussian and t copulas also have SPAR representations on Laplace margins. Extreme value copulas have interesting properties so we consider this family of copulas in some detail. Expressions for the copulas and copula densities presented here can be found in reference works, \parencite[e.g.][]{Joe2015}.

\begin{figure}[!t]
	\centering
	\includegraphics[scale=0.5]{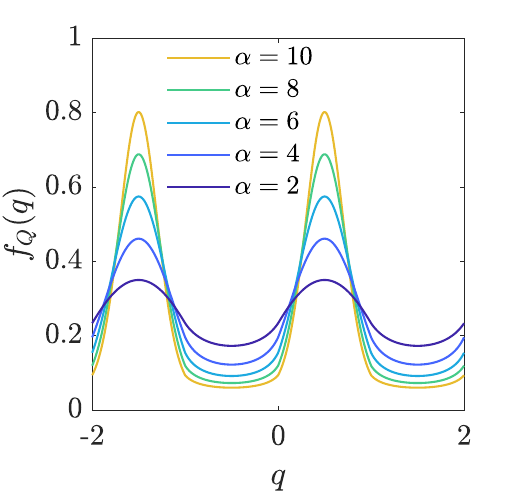}
    \includegraphics[scale=0.5]{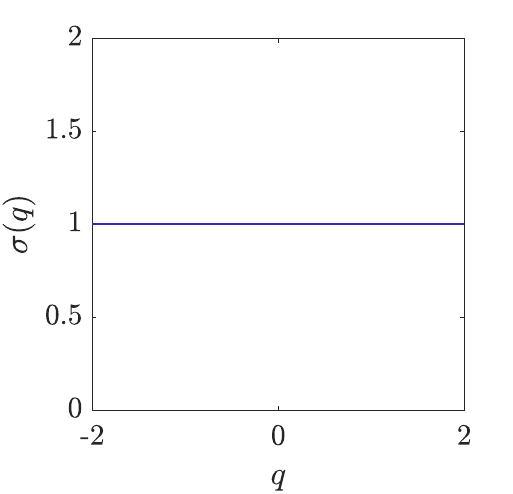}
    \includegraphics[scale=0.5]{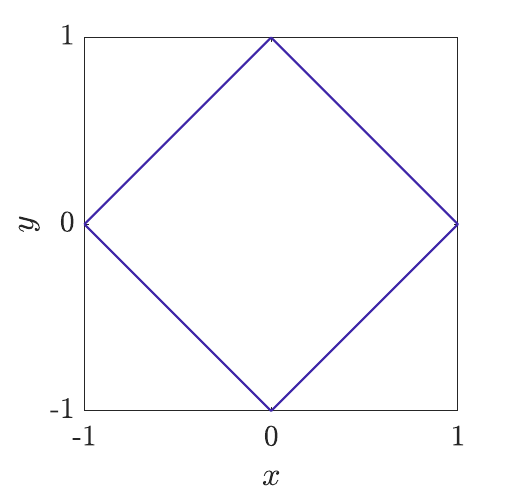}
    \includegraphics[scale=0.5]{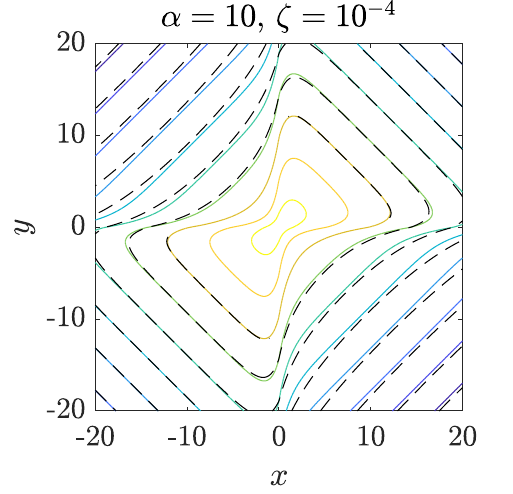}\\
    \includegraphics[scale=0.5]{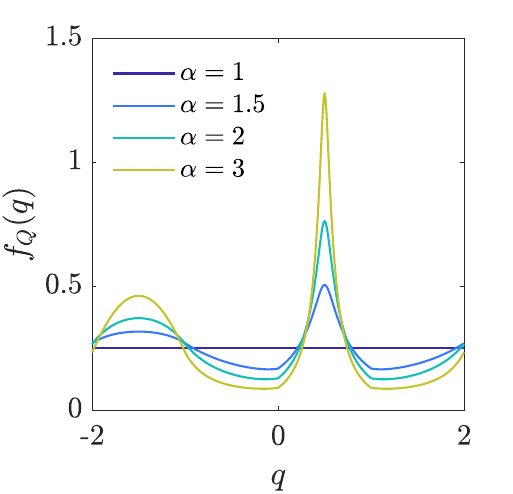}
    \includegraphics[scale=0.5]{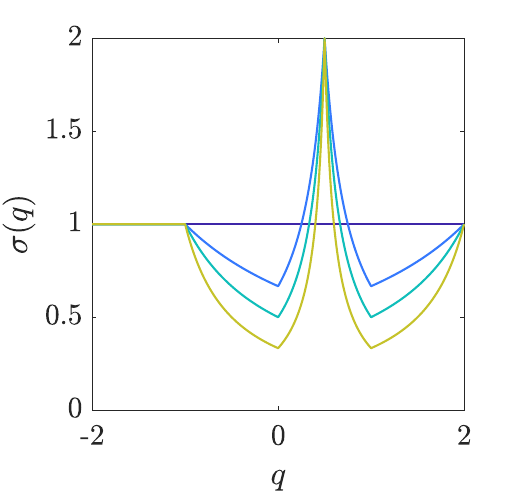}
    \includegraphics[scale=0.5]{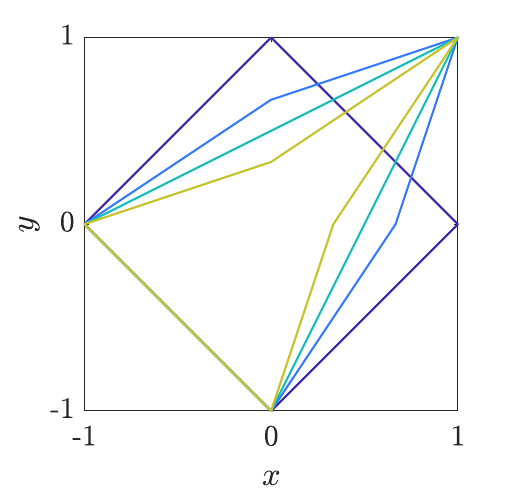}
    \includegraphics[scale=0.5]{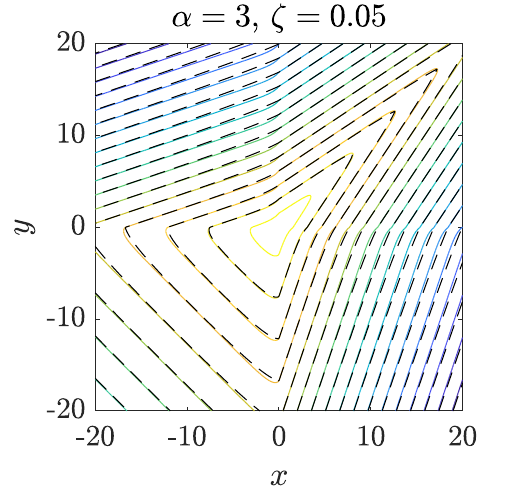}\\
    \includegraphics[scale=0.5]{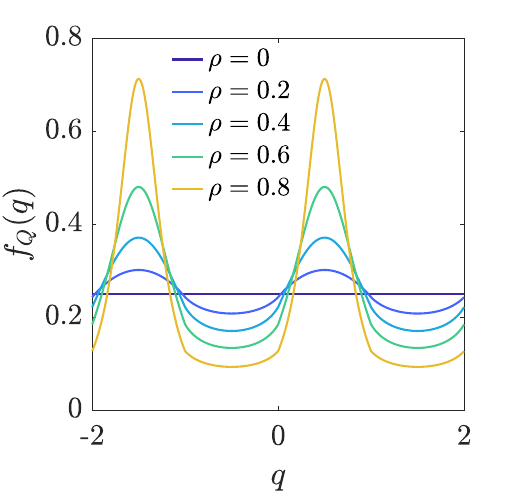}
    \includegraphics[scale=0.5]{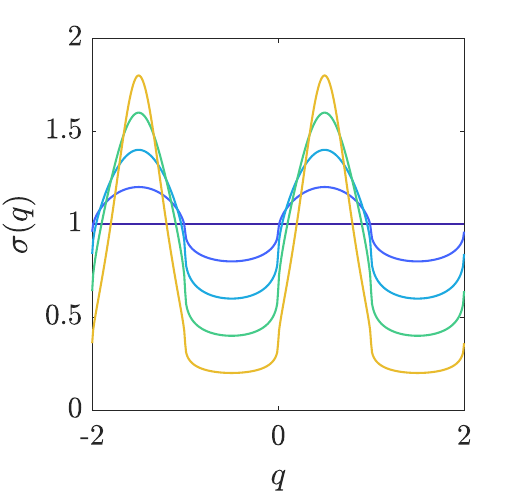}
    \includegraphics[scale=0.5]{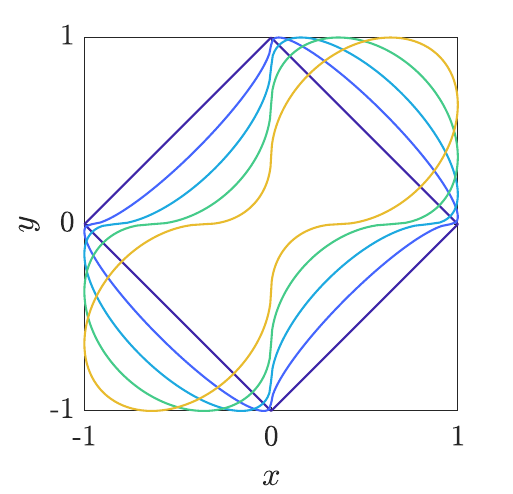}
    \includegraphics[scale=0.5]{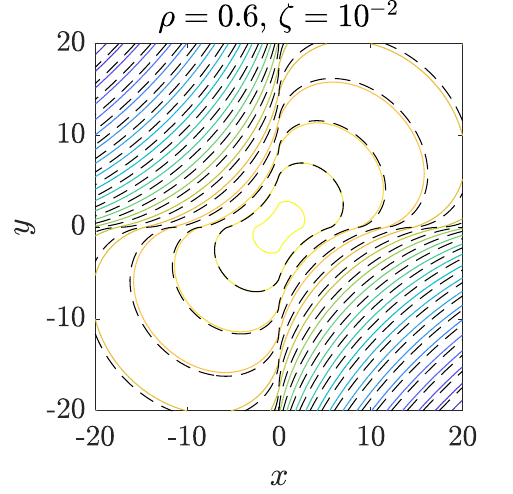}\\
    \includegraphics[scale=0.5]{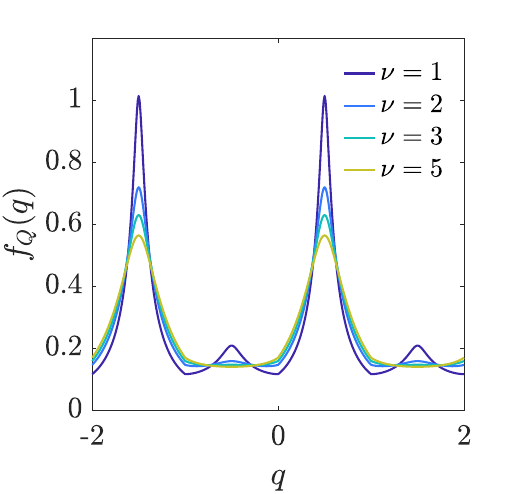}
    \includegraphics[scale=0.5]{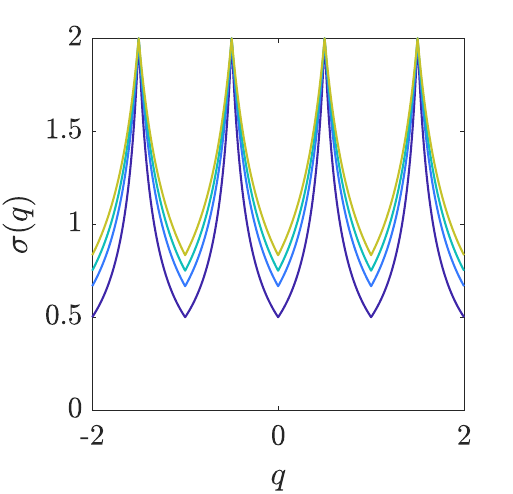}
    \includegraphics[scale=0.5]{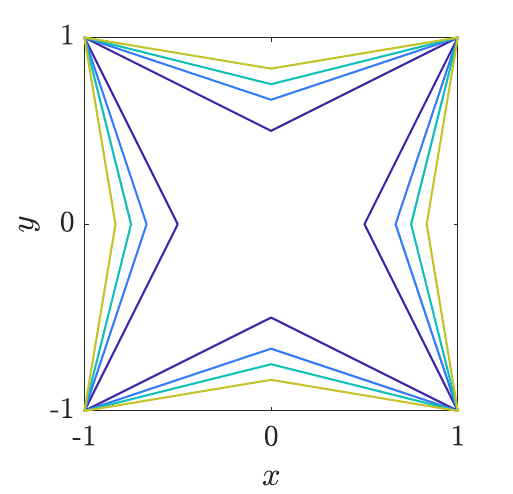}
    \includegraphics[scale=0.5]{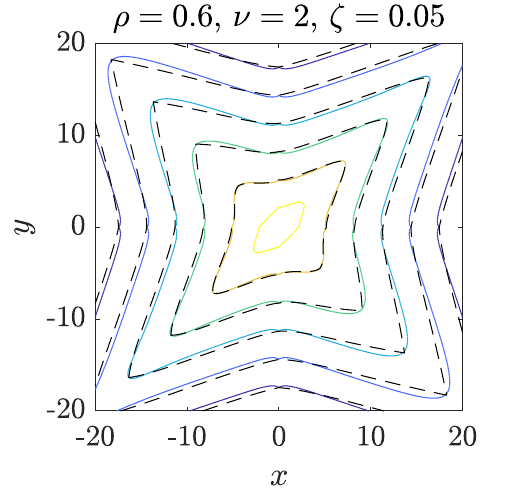}
	\caption{SPAR models for Frank (top row), Joe (2nd row), Gaussian (3rd row) and t (bottom row) copulas on Laplace margins. Left column: angular density for various values of dependence parameters $\alpha$, $\rho$ and $\nu$. Second column: Gamma scale parameter. Third column: Limit sets. Fourth column: Isodensity contours for joint density (coloured lines) and SPAR approximation (black dashed lines). Contours are at density levels $10^{-2}$, $10^{-4}$, $10^{-6}$, etc.}
	\label{fig:SPAR_Laplace}
\end{figure}

\begin{example}[Frank copula]
The Frank copula density is given by 
\begin{align*}
     c(u,v) &= -\alpha h(1) \frac{1+h(u+v)}{(h(u)h(v)+h(1))^2},
\end{align*}
for $(u,v)\in[0,1]^2$ and $\alpha \in \mathbb{R} \setminus \{0\}$, where $h(z) = e^{-\alpha z} - 1$. It is straightforward to show that $c(u,v)\in(0,\infty)$ for $(u,v)\in[0,1]^2$, and hence $\delta_L(r,\mathbf{w})$ is asymptotically a function of $\mathbf{w}$ only, and therefore of form (\ref{eq:delta_Laplace}) with $\beta(\mathbf{w})=0$, $\lambda(\mathbf{w})=1$. This corresponds to asymptotic independence in each quadrant. The angular density for various values of $\alpha$ is shown in \autoref{fig:SPAR_Laplace} (top row). A comparison of the joint density from the SPAR model (dashed black lines) with the true joint density (solid coloured lines) is shown for a case with $\alpha=10$ and threshold exceedance probability $\zeta = 10^{-4}$. In this case, the modified SPAR model (\ref{eq:SPAR_laplace}) is asymptotically exact. However, a low threshold exceedance probability is required to give good agreement between the SPAR density and true density, due to slow convergence of $\delta_L(r,\mathbf{w})$ to the limit in the 2nd and 4th quadrants of the plane. Any copula with continuous density which is positive along the edges, has a SPAR representation on Laplace margins, determined solely by the angular density $f_Q(q)$. Other examples include Plackett, Ali-Mikhail-Haq, and Farlie-Gumbel-Morgenstern copulas. Note that for these cases the limit sets are all the same, but the SPAR models differ, due to the different angular densities. \hfill $\blacksquare$
\end{example}

\begin{example}[Joe copula]
The Joe copula density is given by
\begin{align*}
c(1-u,1-v) &= (uv)^{\alpha-1} z^{\tfrac{1}{\alpha}-2} \left(z + \alpha-1 \right),
\end{align*}
where $z=u^\alpha+v^\alpha-(uv)^\alpha$, $(u,v)\in[0,1]^2$ and $\alpha\geq1$. The Joe copula with $\alpha=1$ is equal to the independence copula. The dependence function has asymptotic form (\ref{eq:delta_Laplace}) with $\beta(\mathbf{w})=0$ and for $\alpha>1$ we have
\begin{equation*}
\lambda(\mathbf{w}) = 
\begin{cases}
	1, & q\in(-2,-1]\\
	1 + (\alpha - 1) |w_1|, & q\in(-1,0]\\
	1 - \alpha + (2\alpha - 1) \max(|w_1|,|w_2|), & q\in(0,1]\\
	1 + (\alpha - 1) |w_2| & q\in(1,2].
\end{cases}
\end{equation*}
The angular density, scale parameter and corresponding limit sets are shown in \autoref{fig:SPAR_Laplace} for various values of $\alpha$. When $\alpha>1$ the distribution is asymptotically dependent in the 1st quadrant and negatively dependent in the 2nd and 4th quadrants. The distribution is asymptotically quadrant independent in the 3rd quadrant for all values of $\alpha$. A comparison of the joint density with the SPAR approximation is shown in \autoref{fig:SPAR_Laplace} for $\alpha=3$ and threshold exceedance probability $\zeta=0.05$. The agreement is good in all quadrants.\hfill $\blacksquare$ 
\end{example}

\begin{example}[Gaussian copula] \label{ex:Gauss_Laplace}
The Gaussian copula density, with Pearson correlation $\rho\in(-1,1)$, is given by
\begin{equation*}
c(u,v) = \frac{1}{\sqrt{1-\rho^2}} \exp\left(-\frac{\rho^2 (x^2 + y^2) - 2\rho xy}{2(1-\rho^2)}\right),
\end{equation*}
for $(u,v)\in[0,1]^2$, where $x=\Phi^{-1}(u)$ and $y=\Phi^{-1}(v)$, and $\Phi^{-1}$ is the inverse of the standard normal cdf. $\delta_L(r,\mathbf{w})$ has asymptotic form (\ref{eq:delta_Laplace}) with 
\begin{equation*}
    \lambda(\mathbf{w}) = \frac{1 - 2 \rho\, \mathrm{sgn}(w_1w_2) \sqrt{|w_1w_2|}}{1-\rho^2},
\end{equation*}
and $\beta(\mathbf{w})=(\lambda(\mathbf{w})-1)/2$ (see Appendix \ref{app:Gaussian}). The modified SPAR model for the Gaussian copula on Laplace margins is therefore not asymptotically exact. However, an approximation using $\beta(\mathbf{w})=0$ leads to a reasonably good agreement, as shown in \autoref{fig:SPAR_Laplace}. In the example shown, $\rho=0.6$ and $\zeta=0.01$. The angular density, scale parameter and corresponding limit sets are also shown in \autoref{fig:SPAR_Laplace} for various values of $\rho$. \hfill $\blacksquare$ 
\end{example}

\begin{example}[t copula] \label{ex:Student_Laplace}
The t copula density, with Pearson correlation $\rho\in(-1,1)$ and $\nu>0$ degrees of freedom is given by
\begin{equation*}
c(u,v) = \left(\frac{\Gamma(\nu/2)}{\Gamma((\nu+1)/2)}\right)^2 \, \frac{\nu}{2\sqrt{1-\rho^2}} \, \left(1+\frac{x^2 +y^2 - 2\rho xy}{\nu(1-\rho^2)}\right)^{-\nu/2-1} \, \left(\left(1+\frac{x^2}{\nu}\right) \left(1+\frac{y^2}{\nu}\right)\right)^{(\nu+1)/2} , 
\end{equation*}
for $(u,v)\in[0,1]^2$, where $x=F_t^{-1}(u;\nu)$, $y=F_t^{-1}(v;\nu)$ and $F_t^{-1}(\cdot;\nu)$ is the inverse cdf of the univariate t distribution on $\nu$ degrees of freedom. $\delta_L(r,\mathbf{w})$ has asymptotic form (\ref{eq:delta_Laplace}) with $\beta(\mathbf{w})=0$ and (see Appendix \ref{app:t_copula})
\begin{equation*}
    \lambda(\mathbf{w}) = \max(|w_1|,|w_2|) + \frac{1}{\nu} ||w_1|-|w_2||.
\end{equation*}
In this case, the SPAR model is asymptotically exact. The scale parameter and corresponding limit sets are independent of the correlation coefficient $\rho$. However, the angular density is a function of both $\nu$ and $\rho$. The example shown in \autoref{fig:SPAR_Laplace} for the joint density has $\rho=0.6$, $\nu=2$ and threshold exceedance probability $\zeta=0.05$, the same case as considered in Example \ref{ex:elliptic} and shown in \autoref{fig:SPAR_student}. On Laplace margins, the agreement between the SPAR model and the true density is slightly better in the first and third quadrants, than in the 2nd and 4th. \hfill $\blacksquare$ 
\end{example}

\begin{example}[Extreme value copulas] \label{ex:EV_laplace}
As mentioned above, EV copulas have ARE form in the lower tail and hence from Propositions \ref{prop:delta_laplace_ARE} and \ref{prop:SPAR_Laplace}, they satisfy the SPAR model assumptions for the negative orthant of $\mathbb{R}^d$. In other orthants, it is useful to consider some specific examples to illustrate various possibilities. Appendix \ref{app:EVcopula} considers two-dimensional cases where the stable tail dependence function follows either the symmetric logistic, asymmetric logistic, or H{\"u}sler-Reiss form. Here, we summarise results that are of general interest and refer to Appendix \ref{app:EVcopula} for details.

For the case of symmetric logistic dependence with parameter $\alpha>1$, we find that $\delta_L(r,\mathbf{w})$ has asymptotic form (\ref{eq:delta_Laplace}), with $\beta(\mathbf{w})=0$ for $q\in(-2,-1)\cup[0,1]$ and $\beta(\mathbf{w})=1-\alpha$ for $q\in[-1,0)\cup(1,2]$. The angular density, GP scale parameter and limit sets are shown in \autoref{fig:SPAR_Laplace_EV} for various values of $\alpha$. A comparison of the SPAR approximation with the true density is also shown, for the case $\alpha=2$ and $\zeta=0.05$. The agreement is good in the first and third quadrants, but slightly worse in the second and fourth due to the non-zero value of $\beta$ for this angular range.

For the case of the asymmetric logistic model \parencite{Tawn1988}, we find that $\delta_L(r,\mathbf{w})$ has asymptotic form (\ref{eq:delta_Laplace}), with $\beta(\mathbf{w})=0$ for all $q$. The limit set is symmetric about the line $q=1/2$. However, at finite levels, the corresponding isodensity contours are not symmetric about $q=1/2$, as illustrated in \autoref{fig:SPAR_Laplace_EV}, for a case with $\alpha=5$, $\gamma_1=0.9$ and $\gamma_2=0.1$. To improve the agreement between the SPAR approximation and true joint density at finite levels, we can define the angular-radial coordinate system relative to a different origin. A finite shift in the origin does not affect the asymptotic behaviour of $\delta_L(r,\mathbf{w})$, and hence the GP scale and shape parameters for the SPAR approximation, but can improve the rate of convergence of the true density to the SPAR approximation. The SPAR approximation for a case with parameters $\alpha=5$, $\gamma_1=0.9$ and $\gamma_2=0.1$ is shown in \autoref{fig:SPAR_Laplace_EV}, with the origin placed at $(x_0,y_0)=(-0.859,-3.3)$. The rationale for this choice of origin is discussed in Appendix \ref{app:EVcopula}, where it is shown to improve the speed of convergence of $\delta_L(r,\mathbf{w})$ to its asymptotic form. In the first quadrant, the threshold has been set as $\mu(q) = \max \{ 10, |x_0/\cos_1(q)|, |y_0/\sin_1(q)|\}$, to ensure good agreement between the asymptotic SPAR approximation and the true density. The values $|x_0/\cos_1(q)|$ and $|y_0/\sin_1(q)|$ correspond to the positive x- and y-axes of the original Cartesian coordinate system, and the minimum value of 10 has been chosen somewhat arbitrarily through trial and error so that the asymptotic approximation is reasonable. In the third quadrant, the density is strictly decreasing for all $q\in[-2,-1]$ and the threshold is set at a constant value $\mu(q)=10$. Similarly, we set $\mu(q) = \max \{ 10, |y_0/\sin_1(q)|\}$ in the second quadrant, and $\mu(q) = \max \{ 10, |x_0/\cos_1(q)|\}$ in the fourth quadrant. The corresponding threshold exceedance probabilities are then calculated using numerical integration. In the regions where the SPAR model is defined, the agreement with the true density contours is good, including the region along the higher-density `finger' in the first quadrant. 

Finally, for the case of the H{\"u}sler-Reiss dependence model \parencite{Husler1989}, with parameter $\alpha>0$, we find that $\delta_L(r,\mathbf{w})$ does not have asymptotic form (\ref{eq:delta_Laplace}) or a continuous limit set. Instead, we find that $\delta_L(r,\mathbf{w}) \propto \exp\left(-\tfrac{1}{2} \left[\left(\tfrac{\alpha}{2}(w_1-w_2) r\right)^2 - r \right] \right)$ for $q\in(0,1)$. However, it is shown in Appendix \ref{app:EVcopula} that the SPAR assumptions (A1)-(A3) are nevertheless satisfied for this copula. This illustrates that the conditions of Proposition \ref{prop:SPAR_Laplace} are sufficient but not necessary for a copula to have a SPAR representation on Laplace margins. Moreover, this example shows that the SPAR approach provides a more general framework for representing multivariate extremes than using limit sets alone.

In summary, we find that the lower tail of EV copulas have a `natural' representation on Laplace margins, but that the case for other orthants is more complicated. In Section \ref{sec:SPAR_longtail} it will be shown that the upper tails of EV copulas have a natural representation on long-tailed margins, but that there are severe limitations on the types of copula that have SPAR representations on these margins.\hfill $\blacksquare$ 

\begin{figure}[!t]
	\centering
    \includegraphics[scale=0.5]{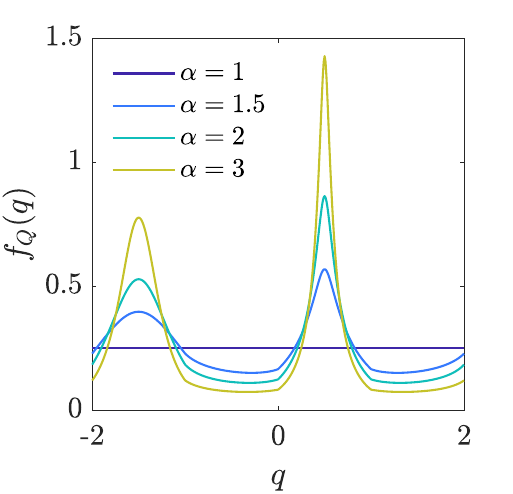}
    \includegraphics[scale=0.5]{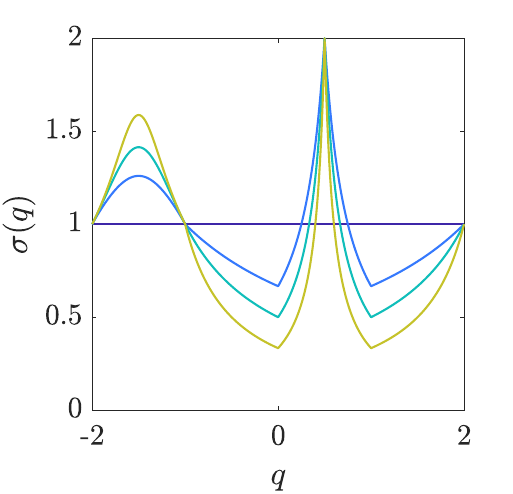}
    \includegraphics[scale=0.5]{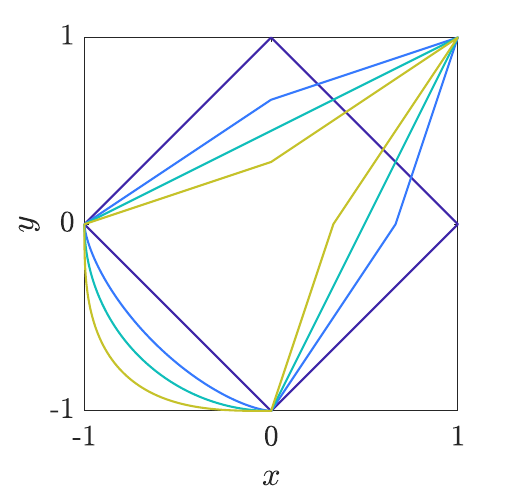}
    \includegraphics[scale=0.5]{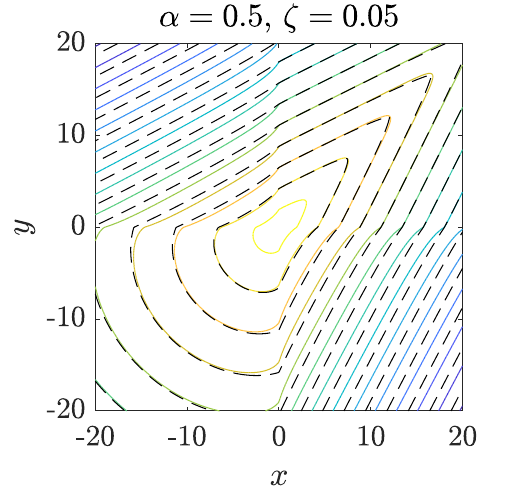}\\
    \includegraphics[scale=0.5]{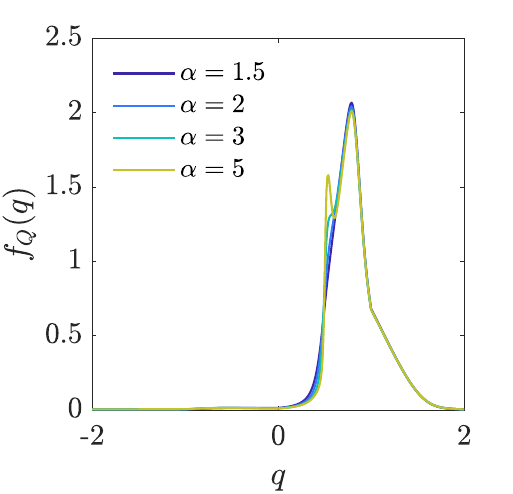}
    \includegraphics[scale=0.5]{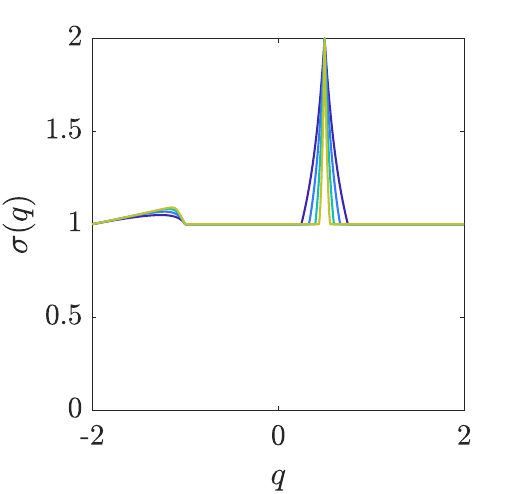}
    \includegraphics[scale=0.5]{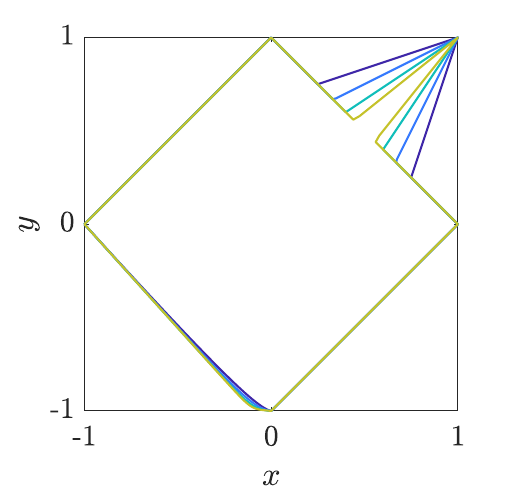}
    \includegraphics[scale=0.5]{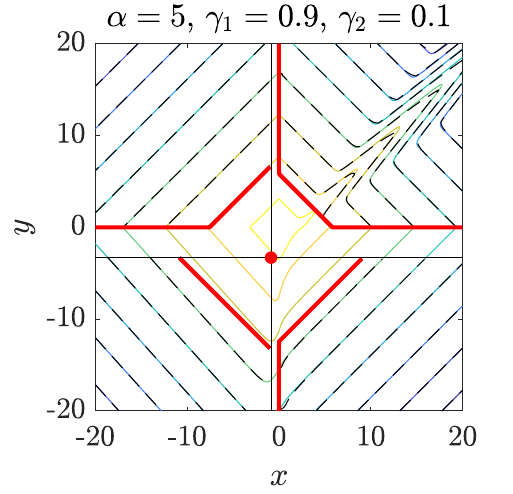}
	\caption{As previous figure, but for extreme value copulas with symmetric logistic dependence (top row) and asymmetric logistic dependence (bottom row) on Laplace margins. For EV asymmetric logistic copula, red dot indicates origin of angular-radial coordinates, red line indicates SPAR threshold.}
	\label{fig:SPAR_Laplace_EV}
\end{figure}
\end{example}

\begin{example}[Bivariate exponential copula] \label{ex:biv_exp}
This example illustrates that although the tail order is infinite in the lower left corner, and the limit set is degenerate in this quadrant, the bivariate exponential copula does have a SPAR representation on Laplace margins. The bivariate exponential copula with parameter $\alpha\in[0,1]$ is given by
\begin{equation*}
    C(u,v) = uv \exp(-\alpha\log(u)\log(v)),
\end{equation*}
for $(u,v)\in[0,1]^2$, with corresponding density function
\begin{equation*}
    c(u,v) = \exp(-\alpha \log(u) \log(v)) \left(\alpha^2\log(u)\log(v) - \alpha\log(uv)-\alpha+1\right).
\end{equation*}
The case $\alpha=0$ corresponds to the independence copula. For $\alpha\in(0,1]$ and $x>0$, in the lower left corner we have
\begin{align*}
    \lim_{t\to0^+} \frac{C(x(t,t))}{C(t,t)} = \lim_{t\to0^+} \frac{(xt)^{2-\alpha\log(xt)}}{t^{2-\alpha\log(t)}} = \begin{cases}
        0, & x<1,\\
        1, & x=1,\\
        \infty, & x>1.
    \end{cases}
\end{align*}
Hence the lower tail order is $\kappa_{(0,0)}=\infty$. For $\alpha\in(0,1)$, the angular-radial representation of the copula density has asymptotic form
\begin{align*}
    \delta_L(r,\mathbf{w}) &\sim 
    \begin{cases}
        2^{-\alpha \log(2)} \alpha r \left[\alpha w_1 w_2 r +  \alpha \log(2) + 1 \right] \exp(-\alpha [r^2w_1w_2+r\log(2)]), & q\in[-2,-1], \\
        \alpha |w_2| r, & q\in(-1,0),\\
        1 - \alpha + \alpha \log(2), & q=0,1,\\
        1 - \alpha, & q\in(0,1),\\
        \alpha |w_1| r, & q\in(1,2).
    \end{cases}
\end{align*}
This does not have asymptotic form (\ref{eq:delta_Laplace}) in the lower left quadrant. However, assumptions (A1) and (A2) are satisfied, taking $\xi(q)=0$ and $\sigma(\mu,q)=1$ for $q\in[-1,2]$ and $\sigma(\mu,q)=(1+2\alpha w_1 w_2 \mu)^{-1}$ for $q\in(-1,-2)$. When $\alpha=1$, the asymptotic form of $\delta_L(r,\mathbf{w})$ is the same for $q\in (-2,0]\cup[1,2]$, whereas for $q\in(0,1)$ we have $\delta_L(r,\mathbf{w})\sim g(\mathbf{w}) \exp(-r\min(w_1,w_2))$, where $g(\mathbf{w})=1$ for $q=\tfrac{1}{2}$ and $g(\mathbf{w})=\tfrac{1}{2}$ otherwise. For any value of $\alpha\in[0,1]$, the angular density is continuous and finite and hence satisfies assumption (A3).\hfill $\blacksquare$ 
\end{example}

\subsection{SPAR models on long-tailed margins} \label{sec:SPAR_longtail}
When considering joint extremes on long-tailed margins (i.e. margins in the domain of attraction of an extreme value distribution with positive shape parameter), it is common to use the framework of multivariate regular variation \parencite[MRV,][]{Resnick1987} and hidden regular variation \parencite[HRV,][]{Resnick2002}. However, \textcite{Li2013a} showed that an MRV distribution can be constructed from a copula with upper tail order $\kappa_{\mathbf{1}_d}=1$, coupled with long-tailed margins. Similarly, \textcite{Li2015a} showed that an HRV distribution can be constructed from a copula with upper tail order $\kappa_{\mathbf{1}_d}\in(1,d)$, coupled with long-tailed margins. Therefore, for consistency with previous sections, we continue to use the framework of tail order functions and tail density functions and the ARL model. 

Proposition \ref{prop:delta_GP_ARL} showed that the AR copula function on GP margins can be derived in terms of the ARL representation of the copula density in the upper corner, when this exist. Since the relation in Proposition \ref{prop:delta_GP_ARL} is valid for any $\xi_m>0$, for simplicity we will assume $\xi_m=1$ in the following. Moreover, as discussed in Section \ref{sec:ang_dens}, the use of Pareto margins with the polar origin defined at $\mathbf{x}=\mathbf{1}_d$ is equivalent to using GP margins with the polar origin defined at $\mathbf{x}=\mathbf{0}_d$, but Pareto margins are more convenient for visualisations on a log-log scale. Therefore, throughout this section, we will work with Pareto margins with the polar origin defined at $\mathbf{x}=\mathbf{1}_d$.

\begin{proposition} \label{prop:SPAR_Pareto}
Suppose that $\mathbf{X}\in [1,\infty)^d$ is a random vector with standard Pareto margins and copula density $c$. Suppose also that $c$ satisfies the ARL model assumptions in the upper and lower tails and is continuous and finite away from the corners. Define $R=\|\mathbf{X} - \mathbf{1}_d\|_1$ and $\mathbf{W}=(\mathbf{X} - \mathbf{1}_d)/R$. Then 
\begin{equation*}
    f_{R,\mathbf{W}}(r,\mathbf{w}) \sim \Tilde{b}(\mathbf{w}) \mathcal{L}_{\mathbf{1}_d} \left(r^{-1} \right) r^{-1-\kappa_{\mathbf{1}_d}}, \quad r\to\infty,
\end{equation*}
where $\Tilde{b}(\mathbf{w}) = \left[\prod_{j=1}^d w_j^{-2} \right] s_{\mathbf{w}}^{\kappa_{\mathbf{1}_d}-d} b_{\mathbf{1}_d} \left((s_{\mathbf{w}} \mathbf{w})^{-1} \right)$ and $s_{\mathbf{w}} = \sum_{j=1}^d w_j^{-1}$. Moreover, the joint density $f_{R,\mathbf{W}}(r,\mathbf{w})$ satisfies assumptions (A1)-(A3) for $\mathbf{w} \in \mathcal{U}_1 \cap (0,1]^d$, with $\xi(\mathbf{w}) = 1/\kappa_{\mathbf{1}_d}$ and $\sigma(\mu,\mathbf{w})=\mu/\kappa_{\mathbf{1}_d}$.
\end{proposition}

The expression for $f_{R,\mathbf{W}}$ on Pareto margins, given in Proposition \ref{prop:SPAR_Pareto} is equivalent to that obtained from the \textcite{Ledford1996, Ledford1997} model in Cartesian coordinates. This is a direct consequence of the formulation of the ARL model for the copula, discussed by \textcite{Hua2011}. Proposition \ref{prop:SPAR_Pareto} therefore gives a relation between the Ledford-Tawn model and SPAR models on Pareto margins. However, Proposition \ref{prop:SPAR_Pareto} only tells us that the SPAR assumptions are satisfied in the joint exceedance region, where all variables are large, since we have assumed that $\mathbf{w}=(w_1,...,w_d) \in \mathcal{U}_1 \cap (0,1]^d$. In the remainder of the section, we will examine the circumstances in which the SPAR assumptions are satisfied for $w_j\to 0^+$ for one or more $j\in\{1,...,d\}$.

Before we consider this further, there is another important point to note regarding the asymptotic GP scale parameters. Proposition \ref{prop:SPAR_Pareto} gives a scaling function $\sigma(\mu,\mathbf{w})$ for the asymptotic GP distribution that is independent of $\mathbf{w}$. However, it is important to note that whilst the conditional radial tail distribution is asymptotically equivalent to $\Bar{F}_{GP} (r; 1/\kappa_{\mathbf{1}_d} , \mu/\kappa_{\mathbf{1}_d})$, this is not necessarily the best approximation at finite levels. Consider the case of a GP distribution, with scale $\sigma_0$ and shape $\xi>0$. In this case we have
\begin{equation*}
    \frac{\Bar{F}_{GP} (\mu+r;\xi,\sigma_0)}{\Bar{F}_{GP} (\mu;\xi,\sigma_0)} = \frac{\left(1+\xi \dfrac{\mu+r}{\sigma_0} \right)^{-1/\xi}}{\left(1+\xi \dfrac{\mu}{\sigma_0} \right)^{-1/\xi}} = \left(1+\xi \frac{r}{\sigma_0+\xi\mu} \right)^{-1/\xi} = \Bar{F}_{GP} (r;\xi,\sigma_0+\xi\mu).
\end{equation*}
That is, the tail distribution of a GP distribution at threshold level $\mu$ is also a GP distribution, with the same shape parameter and scale parameter $\sigma_0+\xi\mu$. This is the well-known threshold stability property of the GP distribution. However, $\Bar{F}_{GP} (r;\xi,\sigma_0+\xi\mu) \sim \Bar{F}_{GP} (r;\xi,\xi\mu)$ as $\mu\to\infty$. So, the Pickands-Balkema-de Haan theorem would be satisfied taking $\sigma(\mu)=\mu\xi$, whereas a more accurate approximation to the tail (an equality in this case) is obtained by taking $\sigma(\mu)=\sigma_0+\xi\mu$. When $\sigma_0$ is large relative to $\xi\mu$, neglecting this term can lead to large errors in the GP approximation at finite levels. The next example illustrates that in some bivariate cases $\sigma_0\to\infty$ as $\mathbf{w}\to(0,1)$.

\begin{example}[EV copulas on Pareto margins]
For EV copulas we have $c_{\mathbf{1}_d}(t\mathbf{w}) \sim t^{1-d} \left|A^{(1,...,1)}(\mathbf{w})\right|$, where $A$ is the stable tail dependence function (see Appendix \ref{app:EVcopula}). In the two-dimensional case, for random vector $(X,Y)\in[1,\infty)^2$, we define $R=(X-1)+(Y-1)$ and $Q=(Y-1)/R$. Recalling that in two-dimensions $f_{R,\mathbf{W}}(r,(1-q,q))=f_{R,Q}(r,q)$ (see discussion in Section \ref{sec:pdf_coords_multivar}), from Proposition \ref{prop:SPAR_Pareto} we have
\begin{equation} \label{eq:EV_pareto}
    f_{R,Q}(r,q) \sim \Tilde{b}(q) r^{-2}, \quad r\to\infty,
\end{equation}
where $\Tilde{b}(q) = - (q(1-q))^{-1} A^{(1,1)}(q,1-q)$. Note that $f_{GP}(r;1,\sigma) \sim \sigma r^{-2}$ for $r\to\infty$. Hence, for the conditional radial density we have
\begin{align*}
    f_{R|Q}(r|q) &\sim f_{GP}\left(r; 1, \sigma_0(q) \right), \quad r\to\infty,
\end{align*}
where $\sigma_0(q) = \Tilde{b}(q) /f_Q(q)$. Hence, from the discussion above a more accurate approximation to the tail of the radial distribution at threshold $\mu$, than given in Proposition \ref{prop:SPAR_Pareto}, is a GP distribution with shape $\xi(q)=1$ and scale $\sigma(\mu,q) = \mu + \sigma_0(q)$. To investigate the behaviour of $\sigma_0(q)$ for $q\to0^+$, we revisit the three models for the stable tail dependence function which were considered in Example \ref{ex:EV_laplace} (see Appendix \ref{app:EVcopula} for details).

For the case of symmetric logistic dependence with parameter $\alpha\geq1$, we find that $f_Q(q)\to 0^+$ as $q\to0^+$ for all $\alpha>1$. When $\alpha=1$, the copula is equal to the independence copula and hence $f_Q(q)\to \infty$ as $q\to0^+$. When $\alpha>1$, $\Tilde{b}(q) = (\alpha - 1) (q(1-q))^{\alpha - 2} (q^\alpha + (1-q)^\alpha)^{\tfrac{1}{\alpha} - 2}$ and hence 
\begin{equation*}
   \Tilde{b}(q) \to 
   \begin{cases}
        \infty, & \alpha<2,\\
        1, & \alpha=2,\\
       0, & \alpha>2,
   \end{cases}
   \qquad q\to0^+.
\end{equation*}
Therefore $\sigma_0(q)\to\infty$ as $q\to0^+$ when $\alpha\leq 2$. So, in this case the assumption of a finite scale parameter function (A2) does not hold for $q\to0^+$. When $\alpha> 2$, numerical calculations indicate that $\sigma_0(q)\to0^+$ as $q\to0^+$, so for these cases the copula appears to have a valid SPAR representation for all $q\in[0,1]$. Examples of the angular density and GP scale parameter as a function of $q$ are shown in \autoref{fig:SPAR_Pareto_EV} for various values of $\alpha$. Contour plots of the joint densities $f_{R,Q}$ and $f_{X,Y}$ are also shown for the case $\alpha=3$ together with the SPAR approximations. For the Cartesian density, the agreement is good both in the joint exceedance region and close to the margins. 

For the case of asymmetric logistic dependence, the angular density goes to infinity as $q\to0^+$ or $q\to1^-$ for any $\alpha\geq1$ and $\gamma_1,\gamma_2\in(0,1)$. Hence the SPAR model assumptions are only valid in the joint exceedance region $0<q<1$. Examples of the joint densities $f_{R,Q}$ and $f_{X,Y}$ are shown in \autoref{fig:SPAR_Pareto_EV} for the case $\alpha=5$, $\gamma_1=0.1$, $\gamma_2=0.9$ (the same case considered on Laplace margins), together with the SPAR approximations. For the Cartesian density, the location of the `finger' is captured by the SPAR approximation, without the need for a judicious choice of origin, required on Laplace margins. The reason for this is that the asymmetric behaviour of the joint density is contained in the partial derivative of the stable tail dependence function $A^{(1,1)}$, and hence this information is also represented in $\sigma_0(q)$. In contrast, $\delta_L(r,\mathbf{w})$ is symmetric, so the asymmetric behaviour at finite levels must be accounted for using a shift of origin on Laplace margins. However, on Pareto margins, the density close to the axes is not captured by the SPAR approximation. In the angular-radial space, the SPAR approximation appears much better, since for any fixed $q\in(0,1)$ the SPAR approximation converges to the true density as $r\to\infty$. 

Finally, for the H{\"u}sler-Reiss dependence model, we find that $f_Q(q)\to0^+$ and $\sigma_0(q)\to0^+$ for $q\to0^+$ or $1^-$ for any $\alpha>0$ -- see \autoref{fig:SPAR_Pareto_EV}. So the SPAR assumptions are valid close to the margins as well. An example of the SPAR approximation to the joint density for the case $\alpha=1$ is shown in \autoref{fig:SPAR_Pareto_EV}. The SPAR approximation is in good agreement with the true density both in the joint exceedance region and close to the margins. 

\begin{figure}[!t]
	\centering
    \includegraphics[scale=0.5]{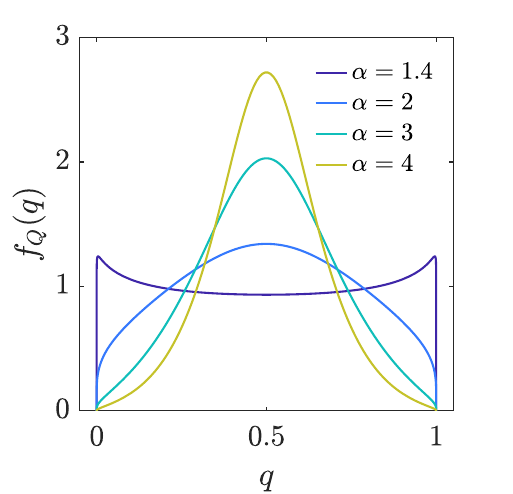}
    \includegraphics[scale=0.5]{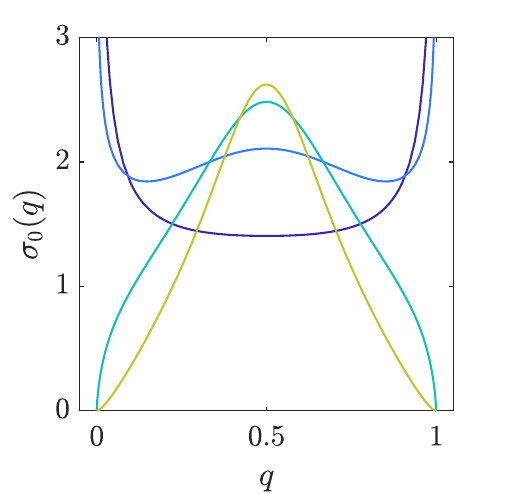}
    \includegraphics[scale=0.5]{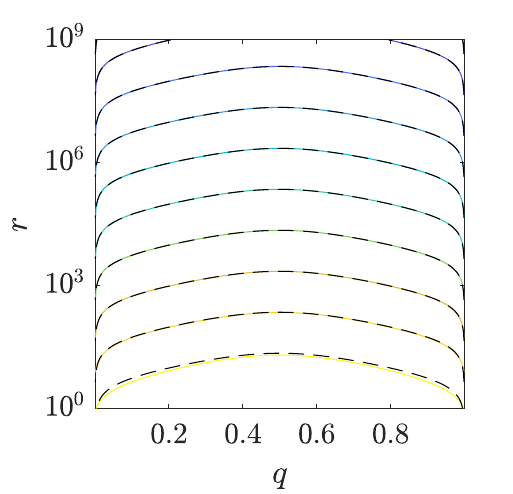}
    \includegraphics[scale=0.5]{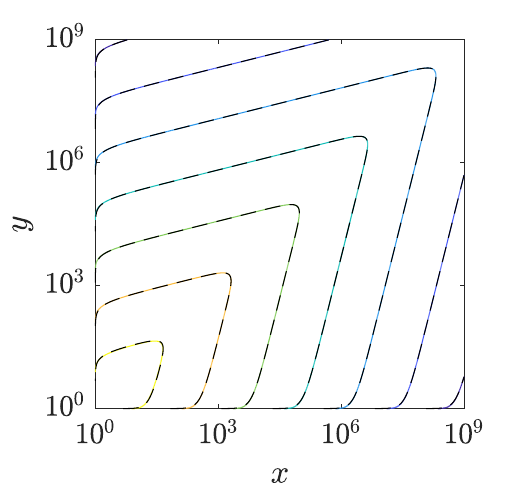}\\
    \includegraphics[scale=0.5]{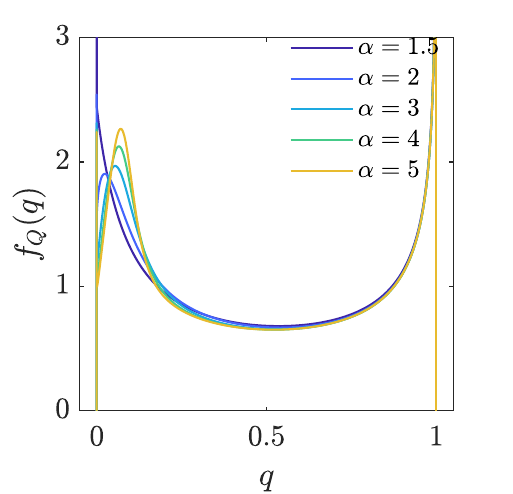}
    \includegraphics[scale=0.5]{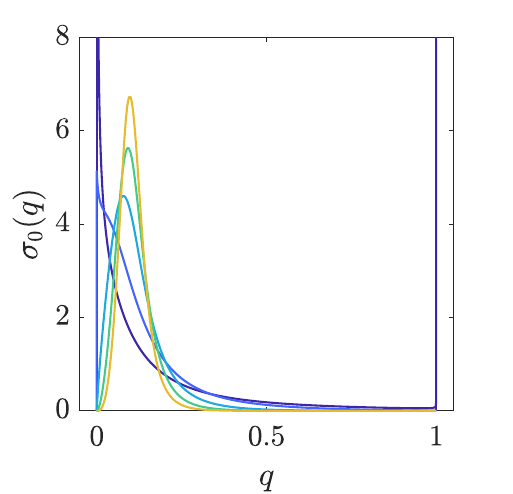}
    \includegraphics[scale=0.5]{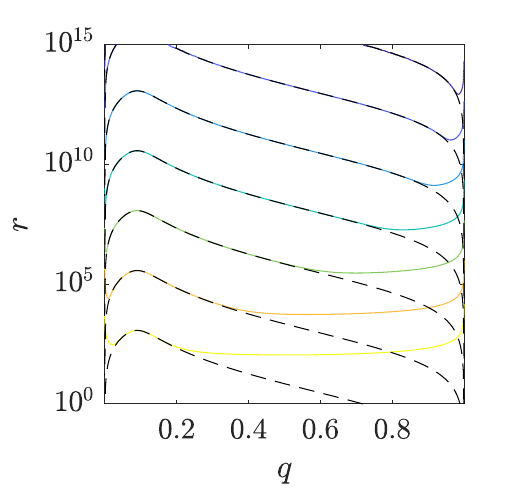}
    \includegraphics[scale=0.5]{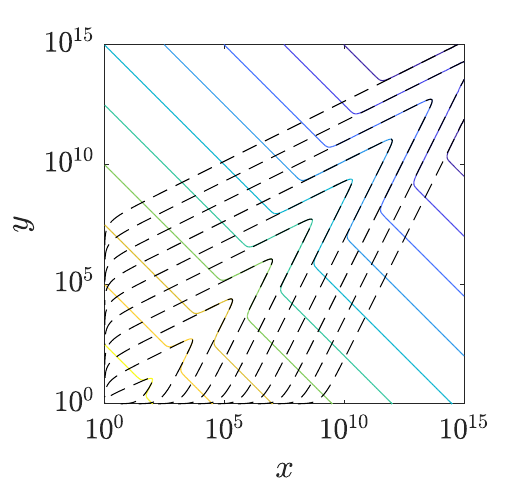}\\
    \includegraphics[scale=0.5]{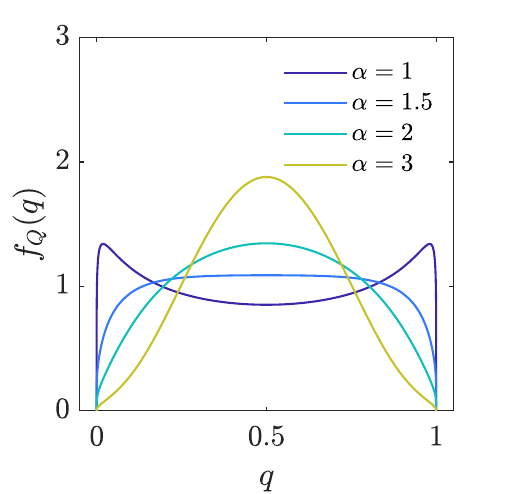}
    \includegraphics[scale=0.5]{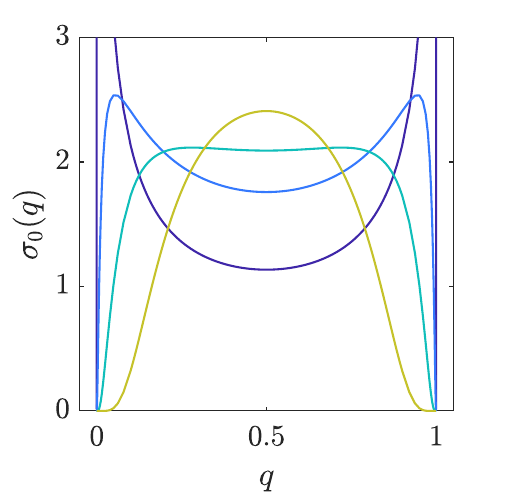}
    \includegraphics[scale=0.5]{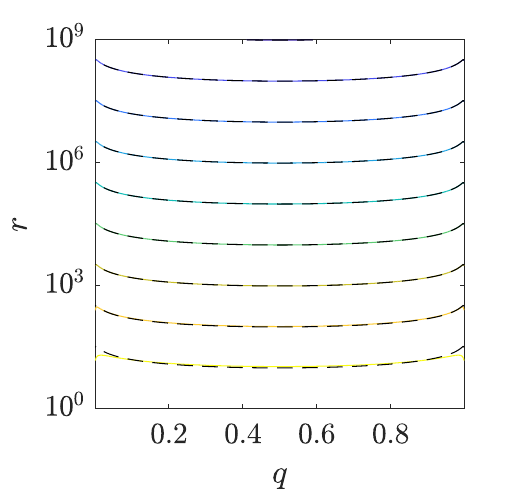}
    \includegraphics[scale=0.5]{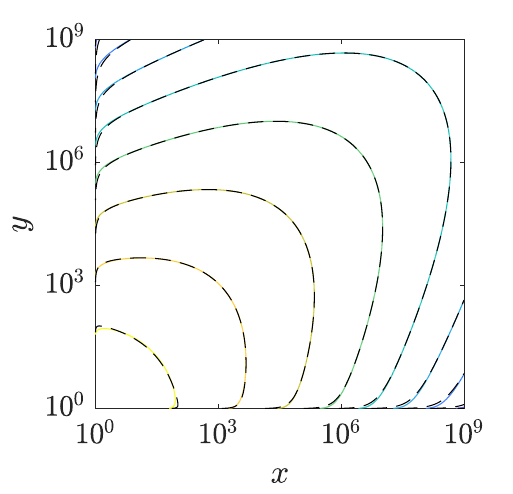}
	\caption{SPAR models for EV copulas on standard Pareto margins, with polar origin at $(x,y)=(1,1)$, for symmetric logistic dependence (top row), asymmetric logistic dependence (middle row) and H{\"u}sler-Reiss dependence model (bottom row). First and second columns show angular density and GP scale parameter for various dependence parameter values, $\alpha$. Third column shows logarithmically-spaced contours of the true angular-radial joint density (coloured lines) and SPAR approximation (black dashed lines). Right hand column shows logarithmically-spaced contours of the true Cartesian joint density (coloured lines) and SPAR approximation (black dashed lines).}
	\label{fig:SPAR_Pareto_EV}
\end{figure}

These examples illustrate that while SPAR models provide a good approximation to the upper tail of EV copulas on Pareto margins in the joint exceedance region, they do not necessarily provide a good approximation close to the axes, since either the angular density or scale parameter can become unbounded for $q\to0^+$ or $1^-$. The angular-radial representation of the joint density in (\ref{eq:EV_pareto}), is the same as that derived by \textcite{Coles1991}. So from a probabilistic standpoint, the SPAR model for EV copulas on Pareto margins is equivalent to the Coles-Tawn method. However, from a statistical perspective the SPAR approach is less parsimonious in this application, since it requires estimation of both the angular density and GP scale parameter, whereas $\Tilde{b}(q)$ is estimated directly by fitting parametric models in the Coles-Tawn method.\hfill $\blacksquare$ 
\end{example}

Proposition \ref{prop:ARE2ARL} showed that in certain cases with $\kappa_{\mathbf{1}_d}>1$, the ARL representation for the copula was only valid in a limited region. The next proposition makes use of this result together with Proposition \ref{prop:SPAR_Pareto}, to show that for these cases, SPAR models on Pareto margins are only valid in the joint exceedance region. 

\begin{proposition} \label{prop:SPAR_Pareto_AI}
Suppose that random vector $(X,Y)\in[1,\infty)^2$ has standard Pareto margins and copula density $c$, and define $R=(X-1)+(Y-1)$ and $Q=(Y-1)/R$. Suppose also that $c$ satisfies the assumptions of Proposition \ref{prop:ARE2ARL} in the upper right corner, with $\beta=0$ and $\kappa_{(1,1)}>1$. Then for $q\in(0,1)$ the angular-radial joint density is asymptotically
\begin{equation*}
    f_{R,Q}(r,q) \sim (q(1-q))^{-1 - \tfrac{\kappa_{(1,1)}}{2}} \mathcal{L}(r) r^{-1-\kappa_{(1,1)}}, \qquad r\to\infty.
\end{equation*}
for some $\mathcal{L}\in RV_0(\infty)$.
\end{proposition}

Under the assumptions of Proposition \ref{prop:SPAR_Pareto_AI}, the angular variation of the density is dependent only on the tail order. For fixed $r$, the approximation for $f_{R,Q}(r,q)$ given in Proposition \ref{prop:SPAR_Pareto_AI} tends to infinity when $q\to0^+$ or $1^-$, and hence the SPAR representation is only valid in the joint exceedance region. Examples where the assumptions of Proposition \ref{prop:SPAR_Pareto_AI} are satisfied include the case where $c(u,v)=c_{EV}(1-u,1-v)$ where $c_{EV}$ is an extreme value copula density, with symmetric stable tail dependence function, such as the symmetric logistic model or H{\"u}sler-Reiss model (see Appendix \ref{app:EVcopula}), and when $c$ is the Gaussian copula (see Appendix \ref{app:Gaussian}). \autoref{fig:SPAR_Pareto_AI} illustrates these examples, and shows contours of the true joint density together with the SPAR approximations, both in the angular-radial space and Cartesian space. It is evident that the asymptotic expression for the angular-radial joint density given in Proposition \ref{prop:SPAR_Pareto_AI} is valid for any fixed $q\in(0,1)$ as $r\to\infty$. However, in the Cartesian space with a log-log scale used for the margins, the approximation is only valid in the region where both variables are large, and the approximations tend to infinity on the margins.

\begin{figure}[!t]
	\centering
    \includegraphics[scale=0.5]{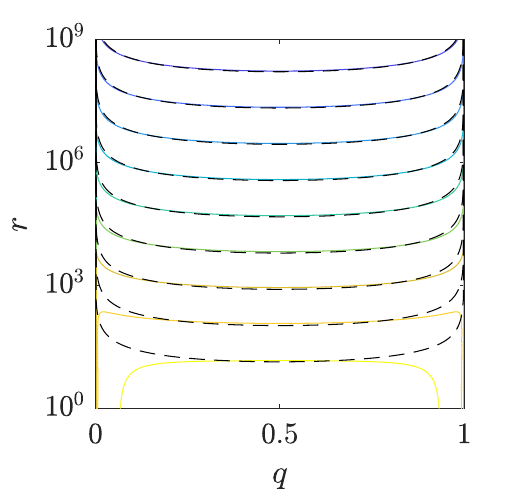}
    \includegraphics[scale=0.5]{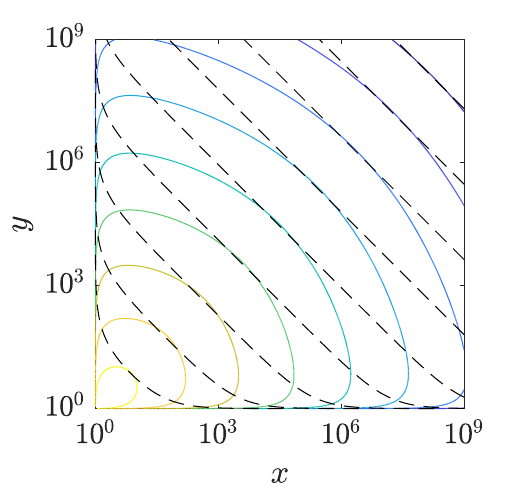}\\
    \includegraphics[scale=0.5]{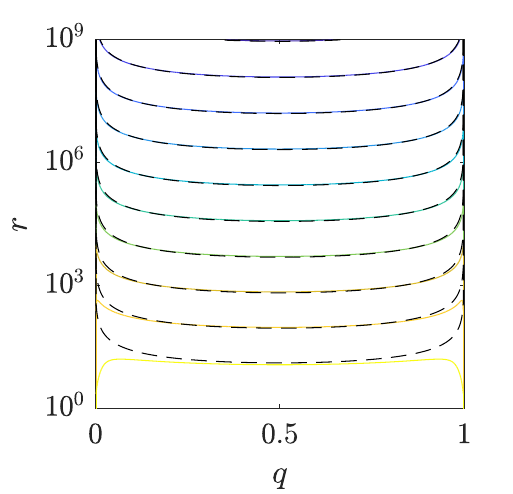}
    \includegraphics[scale=0.5]{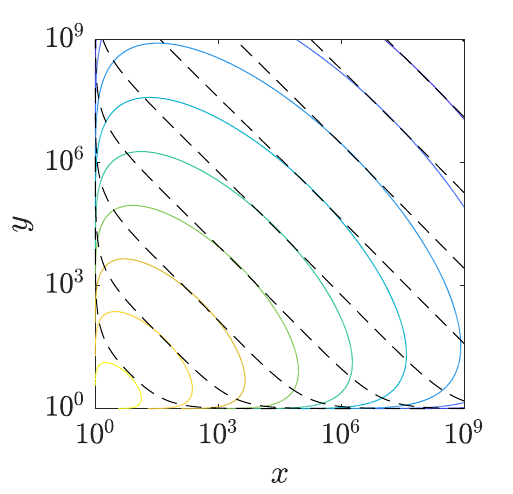}
    \caption{As previous figure, but for lower tail of EV copula with symmetric logistic dependence with $\alpha=3$ (top) and Gaussian copula with $\rho=0.6$ (bottom). Solid lines are contours of true joint densities at logarithmically-spaced increments, and dashed lines are SPAR approximations.}
	\label{fig:SPAR_Pareto_AI}
\end{figure}

In summary, copulas with strong tail dependence have a natural SPAR representation on long-tailed margins in the joint exceedance region. However, this representation is not always valid close to the margins. Moreover, for many copulas without strong tail dependence, the SPAR representation on long-tailed margins is only valid in the joint exceedance region. In general, long-tailed margins are less useful than Laplace margins for SPAR representations.

Finally, we note that joint densities on long-tailed margins do not have limit sets. Consider the case of GP margins with shape $\xi_m>0$ and scale $\sigma=1$. The distribution of $M_n$, the maximum of $n$ samples from the marginal distribution, converges to a generalised extreme value (GEV) distribution with shape $\xi$, scale $n^\xi$ and location $(n^\xi-1)/\xi$, as $n\to\infty$. Therefore, the distribution of scaled maxima $M_n/n^\xi$, converges to a GEV distribution with shape $\xi$, unit scale and location $1/\xi$. This limit is non-degenerate, and therefore the scaled sample has no upper bound. Since the scaled margins do not converge to a finite interval, scaled samples from the joint distribution do not converge onto a finite set.

\subsection{SPAR models on short-tailed margins} \label{sec:SPAR_shorttail}
To illustrate some features of SPAR models on short-tailed margins, it is useful to start by considering the case of the independence copula. Throughout this section, we will consider two-dimensional examples and assume that $(X,Y)\in\mathcal{D}\subset \mathbb{R}^2$ has GP margins with unit scale parameter and shape parameter $\xi_m<0$. Angular and radial variables are defined as $R=X+Y$ and $Q=Y/R$. The upper end point of $X$ and $Y$ is $-1/\xi_m$. The upper end point of $R$ is dependent on $Q$ and the copula of $(X,Y)$, and will be denoted $r_F(q)$ at $Q=q$. 

\begin{example}[Independence on GP margins with negative shape] \label{ex:indep_GPneg}
Suppose that $(X,Y)$ and $(R,Q)$ are random vectors defined as above, and that $X$ and $Y$ are independent. The upper end point of $R$ at $Q=q$ is given by $r_F(q) = -(\xi_m \max(q,1-q))^{-1}$. The angular-radial joint density is given by $f_{R,Q}(r,q) = r\, m_{GP}(r,(1-q,q))$, where $m_{GP}$ is the marginal product function, given below. If we substitute $r=r_F(q)-s$ with $s\in[0,r_F(q)]$, then for $r\in[0,r_F(q)]$ and $q\in[0,1]$ we have
\begin{align*}
m_{GP}(r,(1-q,q)) &= f_{GP}(r(1-q); \xi_m,1) f_{GP}(rq; \xi_m,1)\\
&=
\left[(1+\xi_m r(1-q))(1+\xi_m rq)\right]^{-1-\tfrac{1}{\xi_m}}\\
& = 
\begin{cases}
\left[\left(1 - \dfrac{\min(q,1-q)}{\max(q,1-q)} -\dfrac{s}{r_F(q)} \right)\left(\dfrac{s}{r_F(q)}\right) \right]^{-1-\tfrac{1}{\xi_m}}, & q\in[0,1/2)\cup(1/2,1]\\
\left(\dfrac{s}{r_F(q)}\right)^{-2-\tfrac{2}{\xi_m}}, & q=1/2.
\end{cases}
\end{align*}
The angular density, $f_Q(q)$, is finite for $q\in[0,1]$ for all $\xi_m<0$. Moreover, assumption (A1) is satisfied, with $\xi(q) = \xi_m$ for $q\in[0,1/2)\cup(1/2,1]$ and $\xi(1/2)=(\xi_m+2)/\xi_m$. Hence, unless $\xi_m=-1$, there is a discontinuity in the radial shape parameter and hence (A2) is not satisfied. In the case that $\xi_m=-1$, we have $m_{GP}(r,(1-q,q))=1$ and $f_{R,Q}(r,q)=r$. So although assumptions (A1)-(A3) are satisfied, the approximation of the tail of $f_{R|Q}(r|q)$ with a GP distribution with $\xi=-1$ (i.e. a uniform distribution) is not a good approximation, since $f_{R|Q}(r|q)$ is linearly increasing in $r$. \hfill $\blacksquare$
\end{example}

Example \ref{ex:indep_GPneg} shows that, for the case of the independence copula, there is a discontinuity in radial shape parameter $\xi(q)$ at $q=1/2$. This arises because for $q<1/2$ the variable $Y$ does not reach its upper end point as $r\to r_F(q)$ and hence the density $f_{GP}(rq; \xi_m,1)$ converges to a non-zero value as $r\to r_F(q)$. Similarly, for $q>1/2$ the variable $X$ does not reach its upper end point as $r\to r_F(q)$ (see \autoref{fig:Copula_paths}). So for assumption (A2) to be satisfied on GP margins with negative shape, either a copula must eliminate this discontinuity, or the support of the copula must be such that $\delta_{GP}(r,(1-q,q))=0$ for $r<-(\xi_m\max(q,1-q))^{-1}$, so that $m_{GP}(r,(1-q,q))$ is non-zero everywhere within the support of $f_{R,Q}$, and the domain of attraction of $f_{R|Q}(r|q)$ is controlled by the behaviour of $\delta_{GP}(r,(1-q,q))$ at the edge of its support. 

For the copulas considered so far in this work, $c(u,v)>0$ for $(u,v)\in(0,1)^2$, resulting in tail behaviour of $f_{R|Q}$ that is governed by both $m_{GP}$ and $\delta_{GP}$. Archimedean copulas with non-strict generators (see e.g. \cite{nelsen2006}) provide an example where the Hausdorff distance between the support of $c$ and $[0,1]^2$ can be greater than zero, and the SPAR assumptions are satisfied on short-tailed margins in some cases. Two examples are given below. In both examples, we establish the parameters of asymptotic GP distribution by reparameterising the density as follows. The GP density $f_{GP}(r; \xi, \sigma)$ with $\xi<0$, has upper end point $r_F = -\sigma/\xi$. If we substitute $r=r_F-s$, then 
\begin{equation} \label{eq:GPdens_negxi}
    f_{GP}(r_F - s; \xi, \sigma) = \sigma^{\tfrac{1}{\xi}} (-\xi s)^{-1-\tfrac{1}{\xi}}, \quad s\in[0,r_F].
\end{equation}
Now suppose that the conditional radial density has asymptotic form 
\begin{equation*}
    f_{R|Q}(r_F(q)-s|q) \sim g(q) s^\beta, \quad s\to0^+,
\end{equation*}
for some continuous function $g(q)>0$ and $\beta>-1$, where $r_F(q)$ is the upper end point of $R$ at $Q=q$. Equating the last two expressions gives $f_{R|Q}(r|q) \sim f_{GP}(r; \xi, \sigma(q))$ as $r\to r_F(q)=-\sigma(q)/\xi$, with $\xi=-1/(1+\beta)$ and $\sigma(q) = (g(q))^{\xi} (-\xi)^{1+\xi}$.

\begin{figure}[!t]
	\centering
    \includegraphics[scale=0.5]{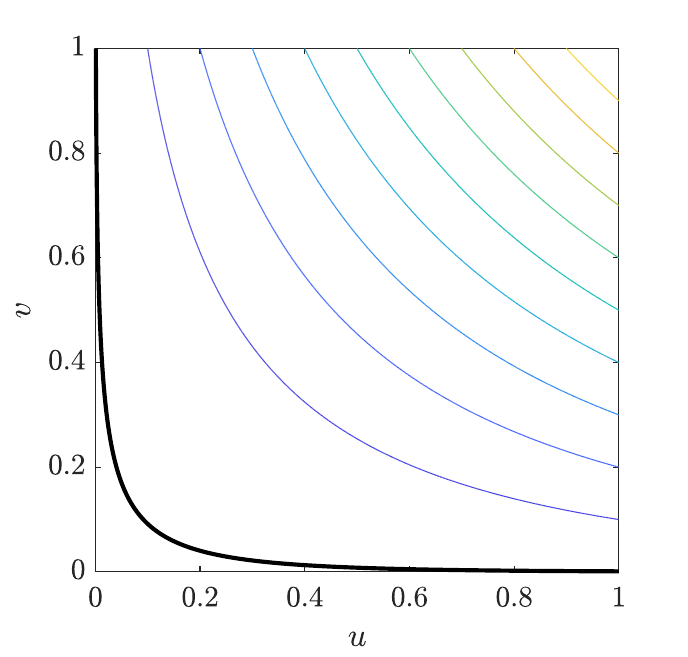}
    \includegraphics[scale=0.5]{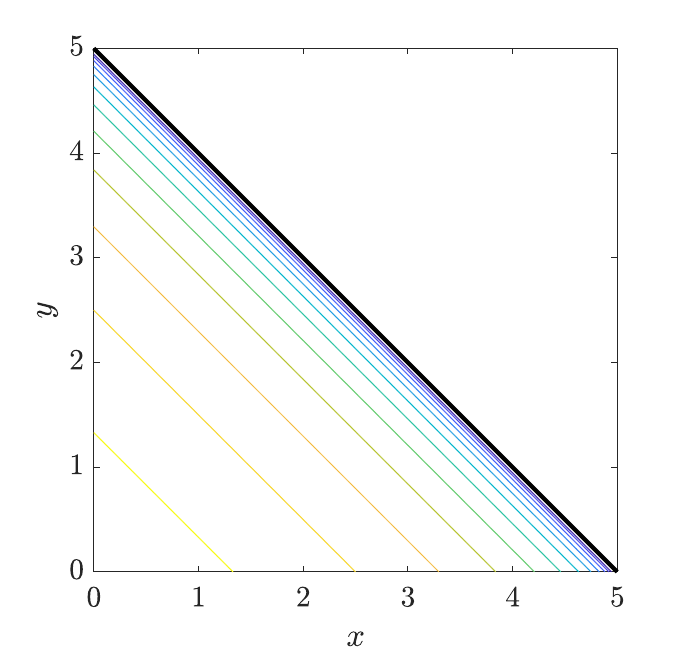}
    \includegraphics[scale=0.5]{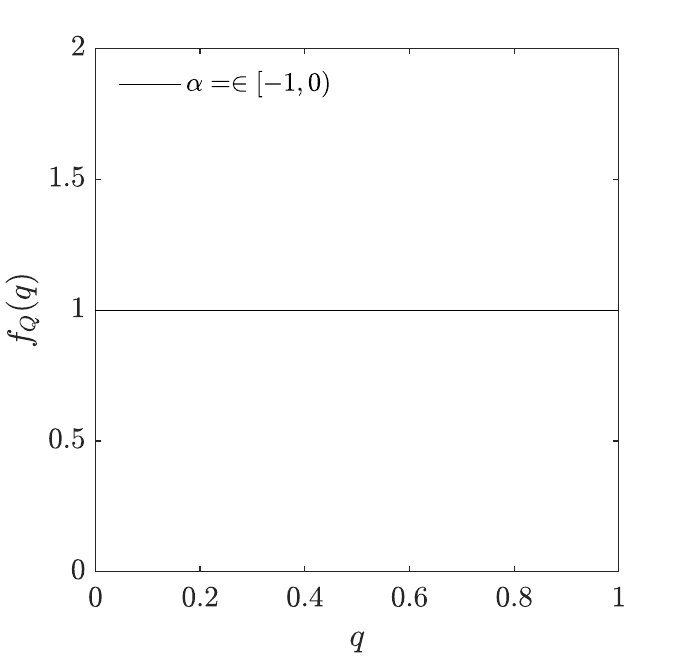}\\
    \includegraphics[scale=0.5]{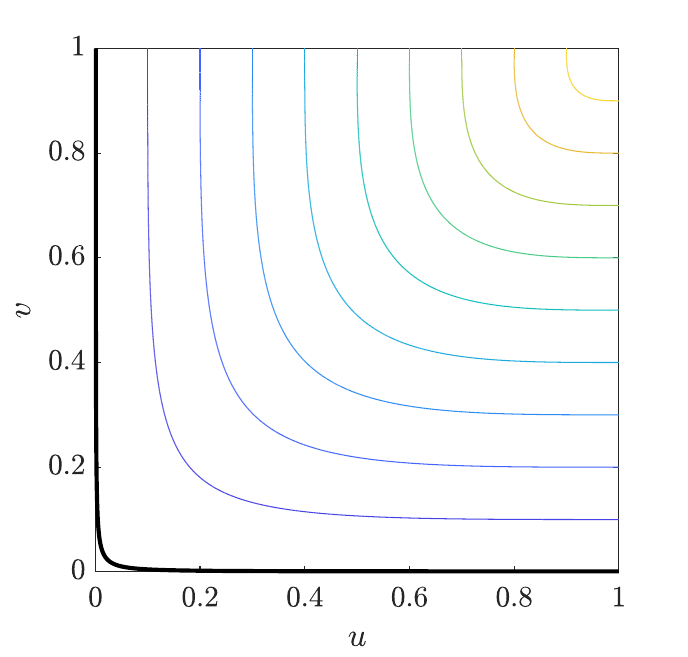}
    \includegraphics[scale=0.5]{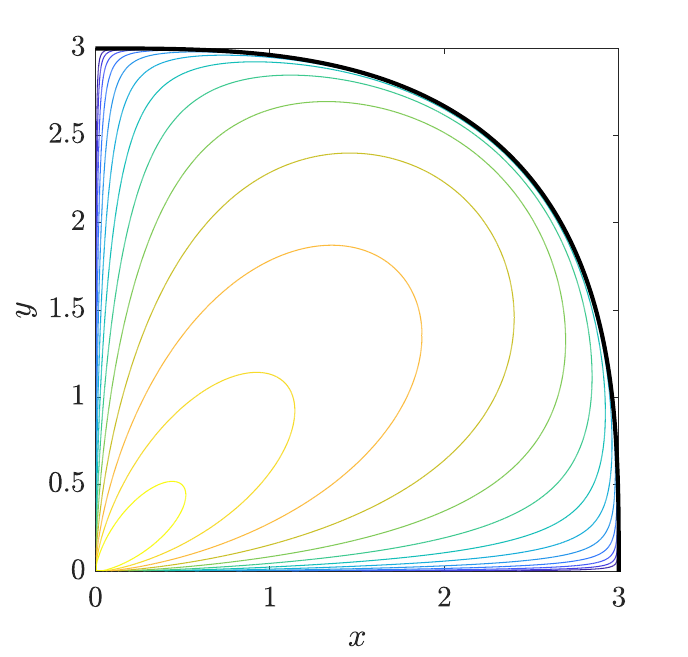}
    \includegraphics[scale=0.5]{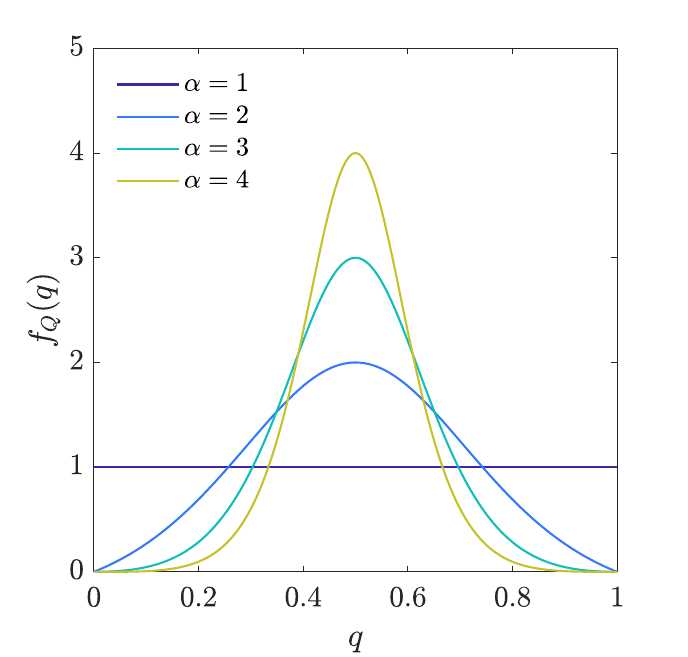}
    \caption{Examples of Archimedean copulas on GP margins with negative shape parameter, for Clayton copula with $\alpha=-0.2$ (top row) and Nelsen copula 4.2.15 with $\alpha=3$ (bottom row). Left column: contours of the copula levels from 0 to 0.9 at increments of 0.1. Thick black line indicates level set $C(u,v)=0$. Middle column: contours of joint density $f_{X,Y}$ for corresponding survival copulas on GP margins, with contour levels at logarithmic increments. Thick black line indicates level set $f_{X,Y}(x,y)=0$. Right column: angular density for various copula parameter $\alpha$.}
	\label{fig:SPAR_shorttail}
\end{figure}

\begin{example}[Clayton copula] \label{ex:Clayton_GPneg}
The Clayton copula with parameter $\alpha \in [-1,\infty) \setminus \{0\}$ is given by $C(u,v) = z^{-1/\alpha}$, where $z=\max(u^{-\alpha}+v^{-\alpha}-1,0)$ and $(u,v)\in[0,1]^2$. When $\alpha<0$, we have $C(t,t)=0$ for $t\in[0, 2^{1/\alpha}]$. When $\alpha>0$, we have $C(t,t)>0$ for $t>0$. An example contour plot of the copula is shown in \autoref{fig:SPAR_shorttail} for the case $\alpha=-0.2$. In the remainder of this example, we assume that $\alpha\in[-1,0)$. The corresponding density is 
\begin{equation*}
    c(u,v) = 
    \begin{cases}
        (1-\alpha)(uv)^{-\alpha-1}z^{-\tfrac{1}{\alpha}-2}, & (u,v)\in\{(a,b)\in[0,1]^2: a^{-\alpha}+b^{-\alpha}\geq 1\},\\
        0, & \text{otherwise.}
    \end{cases}
\end{equation*}
Since the support of $c$ is limited in the lower left corner, we work with the density of the survival copula instead, $c_{(1,1)}(u,v) = c(1-u,1-v)$. On GP margins with $\xi_m<0$, the upper end point of $\delta_{GP}(r,(1-q,q);c_{(1,1)})$ occurs when $z=0$, or equivalently $(1+\xi_m r(1-q))^{\alpha/\xi_m} + (1+\xi_m rq)^{\alpha/\xi_m} = 1$. Setting $\xi_m=\alpha$ leads to a simple solution, with the upper end point of $\delta_{GP}$ given by $r_F(q)=-1/\alpha$ for $q\in[0,1]$. Substituting $r=r_F(q)-s$, we find that the angular-radial joint density is independent of $q$ and given by
\begin{align*}
    f_{R,Q}(r,q) &= (1+\alpha)r(1+\alpha r)^{-\tfrac{1}{\alpha}-2}\\
    &= -(1+\alpha)\left(\frac{1}{\alpha} + s\right)(-\alpha s)^{-\tfrac{1}{\alpha}-2}.\\
    &\sim -\left(\frac{1}{\alpha}+1\right) (-\alpha s)^{-\tfrac{1}{\alpha}-2}, \quad s\to 0^+\\
    &= f_{GP}(r_F-s; \xi, \sigma),
\end{align*}
where $\xi=\alpha/(1+\alpha)$ and $\sigma = (-\xi)^{-\xi} (1-\xi)^{\xi+1}$. 
Since $f_{R,Q}(r,q)$ is independent of $q$, $f_{Q}(q)=1$ and $f_{R,Q}(r,q)=f_{R|Q}(r|q)$. Hence SPAR model assumptions (A1)-(A3) are satisfied. The Cartesian joint density $f_{X,Y}(x,y)$ for the case $\alpha=1/5$ is shown in \autoref{fig:SPAR_shorttail}. In this case, the corresponding SPAR approximation is not shown, as the contours are closely spaced. However, the discussion above shows that it is asymptotically exact. \hfill $\blacksquare$
\end{example}

\begin{example}[Nelsen copula 4.2.15] \label{ex:Nelsen_GPneg}
Table 4.2 in \textcite{nelsen2006} presents a list of one-parameter families of Archimedean copulas. The copula labelled 4.2.15, with parameter $\alpha\geq1$ is given by $C(u,v) = z^{\alpha}$ for $(u,v)\in[0,1]^2$, where 
\begin{equation*}
    z=\max\left(1- \|(x,y)\|_\alpha, 0\right),
\end{equation*}
and $x=1-u^{1/\alpha}$, $y=1-u^{1/\alpha}$, and $\|\cdot\|_\alpha$ is the $L^\alpha$ norm. We have $C(t,t)=0$ for $t\leq (1-2^{-1/\alpha})^\alpha$. An example contour plot of the copula is shown in \autoref{fig:SPAR_shorttail} for the case $\alpha=3$. The corresponding density is 
\begin{equation*}
    c(u,v) = 
    \begin{cases}
        \left(1-\tfrac{1}{\alpha}\right) (uv)^{\tfrac{1}{\alpha}-1} (xy)^{\alpha-1} \left(x^{\alpha} + y^{\alpha}\right)^{\tfrac{1}{\alpha} -2} z^{\alpha-2}, & (u,v)\in\{(a,b)\in[0,1]^2: \|(1-a^{1/\alpha},1-b^{1/\alpha})\|_\alpha\leq 1\}\\
        0, & \text{otherwise}.
    \end{cases}
\end{equation*}
As in the previous example, we will work with the density of the survival copula, $c_{(1,1)}(u,v) = c(1-u,1-v)$. Using GP margins with $\xi_m=-1/\alpha$, the upper end point of $\delta_{GP}(r,(1-q,q);c_{(1,1)})$ occurs when $z=0$. Solving for $r$ shows that the upper end point is given by $r_F(q)=\alpha/\|(1-q,q)\|_\alpha$. Making the substitution $r=r_F(q)-s$, we find that the angular-radial joint density is given by
\begin{equation*}
    f_{R,Q}(r_F(q)-s,q) = \alpha^{2-\alpha} (\alpha-1) (q(1-q))^{\alpha-1} ((1-q)^\alpha + q^\alpha )^{-\tfrac{1}{\alpha}-1} s^{\alpha-2}.
\end{equation*}
Integrating this gives the angular density and conditional radial density as
\begin{align*}
    f_{Q}(q) &= \alpha (q(1-q))^{\alpha-1} ((1-q)^\alpha + q^\alpha )^{-2},\\
    f_{R|Q}(r_F(q)-s|q) &= \alpha^{1-\alpha} (\alpha-1) ((1-q)^\alpha + q^\alpha )^{1-\tfrac{1}{\alpha}} s^{\alpha-2}.
\end{align*}
The conditional radial density $f_{R|Q}$ is exactly of the form (\ref{eq:GPdens_negxi}) and hence is a GP density, with parameters $\xi=-1/(\alpha-1)$ and $\sigma(q) = (g(q))^{\xi} (-\xi)^{1+\xi}$, with $g(q)=\alpha^{1-\alpha} (\alpha-1) ((1-q)^\alpha + q^\alpha )^{1-\tfrac{1}{\alpha}}$. The Cartesian joint density $f_{X,Y}(x,y)$ for the case $\alpha=3$ is shown in \autoref{fig:SPAR_shorttail}, together with the angular density for various values of $\alpha$. Since the SPAR representation is exact everywhere in the domain, this is not shown.  \hfill $\blacksquare$
\end{example}

The examples presented above are for cases where the angular-radial density assumes a simple form for a judicious choice of shape parameter for the margins. Some Archimedean copulas with non-strict generators have infinite density along the boundary of the level set $C(u,v)=0$ (see e.g. Table 4.2 in \cite{nelsen2006}). In these cases, the SPAR assumptions are not satisfied on GP margins. 

The use of short-tailed margins and copulas whose support does not cover the whole of $(0,1)^2$ has natural applications to some environmental variables, such as winds, waves, temperatures, etc., discussed in Section \ref{sec:motivation}. It is reasonable to expect that the distribution of an environmental variable has a finite upper end point. Moreover, joint domains of variables may be restricted by physical constraints. In the case of ocean surface gravity waves, combinations of wave heights, $H$, and wavelengths, $L$, that are physically possible are constrained by the wave steepness, defined as the ratio $H/L$.\footnote{In deep water, the wavelength is proportional to the square of the wave period -- see \autoref{fig:data_transformation}(a).} When the steepness exceeds a certain value, the wave breaks, limiting the wave height as a function of wavelength. Therefore, the range of possible values of the random vector $(H,L)$ is bounded by a line of the form $H=sL$, for some limiting steepness $s$. The joint density of the Clayton survival copula on GP margins with negative shape, illustrated in \autoref{fig:SPAR_shorttail}, behaves analogously, with the support bounded by the line $x+y=-1/\alpha$. However, when modelling environmental variables which are assumed to have bounded support, it may be more appropriate not to use prescribed margins, but instead estimate the SPAR model on the original margins (after scaling and centering). This is discussed further in the following section. 

In summary, Example \ref{ex:indep_GPneg} shows that the independence copula does not have a SPAR representation on GP margins with negative shape parameter, and illustrates that there are constraints on the types of copula that have SPAR representations on these margins. However, Examples \ref{ex:Clayton_GPneg} and \ref{ex:Nelsen_GPneg} show that some copulas whose support does not cover the whole of $(0,1)^2$ do have SPAR representations on short-tailed margins.

\section{Discussion and conclusions} \label{sec:discussion}
\subsection{Main findings}
This work presents a new and general framework for modelling multivariate extremes, based on an angular-radial representation of the density function. The framework requires a transformation of variables from Cartesian to polar coordinates. In Section \ref{sec:pdfs_in_polars} it was shown that the choice of generalised polar coordinates, introduced in Section \ref{sec:coords}, does not affect whether the SPAR model assumptions are satisfied, but that certain coordinate systems lead to simpler representations in which the angular and radial variables are asymptotically independent. The coordinate systems for which this is true depend on both the copula and margins. In general, using polar coordinates defined in terms of the $L^1$ norm is sufficient, and gives a coordinate system which is straightforward to work with in higher dimensions, and has a simple transformation between Cartesian and polar density functions. 

For the theoretical examples considered in this work, the SPAR model assumption (A1), that the conditional radial distribution converges to a GP distribution was always satisfied. However, the more restrictive assumptions (A2) and (A3), regarding continuity and finite values for the angular density and GP parameter functions, were not always satisfied. For a given copula, the choice of margin affects whether the SPAR model assumptions are satisfied. It was shown that the use of Laplace margins leads to density functions where the SPAR model assumptions are satisfied for many commonly-used copulas. Using exponential margins leads to similar asymptotic representations in the non-negative orthant of $\mathbb{R}^d$. However, in certain cases the angular density can become unbounded on exponential margins, whereas it remains finite on Laplace margins. So there is some advantage to using Laplace margins over exponential margins, even when the interest is only in extremes in the non-negative orthant of $\mathbb{R}^d$. 

SPAR models on Laplace margins also provide a useful link to the limit set of the scaled sample cloud, when it exists. The scale parameter function of a SPAR model on Laplace margins describes the radius of the boundary of the limit set as a function of angle. However, it is important to note that for some copulas, the limit set is degenerate in some regions (i.e. the boundary has zero radius from the origin at some angles). For the examples considered where this was the case, SPAR representations on Laplace margins were nevertheless possible. While limit sets are of interest from a theoretical point of view, estimation of the density in extreme regions (or the corresponding probability of observations falling within an extreme set) is usually of greater practical importance. Characterising the density in extreme regions of variable space is the main aim of the SPAR approach.

SPAR models on Laplace margins can represent a wide range of copulas. However, for copulas with strong dependence in the upper tail, it was shown that the joint exceedance region in the upper tail has a more natural representation on Pareto margins. From a probabilistic perspective, SPAR representations for these cases are equivalent to the representation derived by \textcite{Coles1991}. Similarly, for copulas with tail order $\kappa_{\mathbf{1}_d}>1$ (\ref{eq:tail_order}), the SPAR representation on Pareto margins is equivalent to the \textcite{Ledford1996,Ledford1997} model. However, in both cases $\kappa_{\mathbf{1}_d}=1$ and $\kappa_{\mathbf{1}_d}>1$, there are copulas for which the SPAR representation on Pareto margins is valid only in the region where all variables are large, and is not valid when the value of one or more variables is small. In contrast, in the examples presented, the use of Laplace margins for such copulas leads to SPAR representations which are valid for all angles. 

With regard to how the SPAR framework relates to existing approaches, we can summarise that the models of \textcite{Coles1991, Ledford1996, Ledford1997} and \textcite{Wadsworth2017} are all special cases of SPAR representations, and that SPAR models can represent extremal properties of a wider range of distributions than these models. Moreover, the decomposition of the angular-radial joint density function, into an angular density and conditional radial density, provides a simple intuitive motivation for the SPAR model. Multivariate extremes can be seen as a simple extension of univariate extremes, with angular dependence.  

\subsection{Implications for statistical inference}
The focus of the present work has been on probabilistic aspects of SPAR representations, with particular consideration of the effects of the choice of coordinate system and margins. The motivation is that these theoretical considerations will inform an approach to inference, discussed in \textcite{MurphyBarltrop2023}. The general conclusion from this paper is that using Laplace margins and $L^1$ polar coordinates provides a variable space in which the SPAR model assumptions are satisfied for a wide range of copulas, with various tail dependence properties. As mentioned in the introduction, SPAR inference is equivalent to fitting a non-stationary peaks-over-threshold (POT) model, with angle as covariate. There are many examples of inference for non-stationary POT in the literature, \parencite[e.g.][]{chavez2005, Randell2016, Youngman2019, Zanini2020, Barlow2023}. In the examples considered here, it was found that when the SPAR assumptions are satisfied, angular densities are smooth functions of angle, which are readily amenable to semi-parametric estimation. However, although the GP scale functions were found to be continuous in many cases (assumption A2), they were not always smooth functions of angle. Appropriate applications and extensions of semi-parametric approaches will be necessary to represent scale functions with cusps, e.g. a piecewise-linear angular model \parencite{Barlow2023}, co-located knots in a spline angular model, or B\'ezier splines \parencite{majumder2023semiparametric}. 

\subsection{Future work}
For a SPAR model on Laplace margins with constant shape parameter $\xi(\mathbf{w}) = 0$, the tail order in corner $\mathbf{u}_0=(u_{0,1},...,u_{0,d})$ is linked to the GP scale parameter by $\sigma(\mathbf{w}_0) = 1/\kappa_{\mathbf{u}_0}$, where $d\mathbf{w}_0 = \left((-1)^{u_{0,1}+1}, ..., (-1)^{u_{0,d}+1} \right)$. However, general relations between SPAR model parameters and tail dependence coefficients have not been considered here. The general dependence properties of SPAR models can be derived from the marginal and joint exceedance probabilities, calculable from integrals of the SPAR density over angle. Whilst this is straightforward in principle, initial work indicates that the calculations involved are lengthy, and require consideration of numerous separate cases. These considerations are therefore deferred to future work.

Section \ref{sec:margin} considered the effect of margins on the validity of the SPAR assumptions. However, as discussed in Section \ref{sec:motivation}, conducting inference on prescribed margins is not a requirement of the SPAR approach. When applying the model to observational data, generally the marginal distribution is not known. In this case, working on prescribed margins requires first estimating the distribution for each margin and then applying a transformation. This will introduce further uncertainty to the estimated joint distribution. An alternative approach would be to apply the SPAR model to the observations directly, perhaps after a suitable normalisation for scale and location, and estimate both the extremal properties of the margins and copula in a single inference. This type of approach may be applicable when it is reasonable to assume that all variables have a finite upper and lower bounds; as discussed in Section \ref{sec:SPAR_shorttail}, we consider this is likely to be the case for most environmental applications \parencite{MurphyBarltrop2023}. 

A further consideration for the application of the SPAR model to real-world data, is appropriate location of the origin of the polar coordinate system. When working on prescribed margins, the natural choice is to define the polar origin to coincide with the Cartesian origin. For distributions which have support in all of $\mathbb{R}^d$, a finite shift in the origin has no effect on the asymptotic properties of the model, but can result in a faster convergence to the asymptotic form, as discussed in Example \ref{ex:EV_laplace}. However, on short-tailed margins, the choice of origin can have an impact on the SPAR model parameters, or even whether the model assumptions are satisfied. For instance, suppose that random vector $\mathbf{X}\in \mathbb{R}^d$ has bounded support, $\text{supp}(\mathbf{X}) = G \subset \mathbb{R}^d$. The bounded set $G$ can be viewed as analogous to a limit set for random vectors with unbounded support. Clearly, a minimum requirement for the SPAR representation when $G$ is bounded, is that $G$ is star-shaped with respect to the origin. The choice of origin for cases where random vector $\mathbf{X}\in \mathbb{R}^d$ has bounded support will be discussed in detail in future work.

\section*{Declarations}
\subsection*{Availability of supporting data}
The data shown in Figures \ref{fig:data_response} and \ref{fig:data_transformation} was provided as part of the Benchmarking Exercise for Environmental Contours \parencite{haselsteiner2021} and is available from \url{https://github.com/ec-benchmark-organizers}. The generic response curves shown in Figures \ref{fig:data_response}(a) and \ref{fig:data_response}(c) were reproduced from \textcite{deHauteclocque2022} and \textcite{haselsteiner2022}, respectively.

\subsection*{Competing interests}
The authors have no relevant financial or non-financial interests to disclose.

\subsection*{Funding}
This work was funded by the EPSRC, United Kingdom Supergen Offshore Renewable Energy Hub [EP/S000747/1] Flexible Fund
project ‘‘Improved Models for Multivariate Metocean Extremes (IMEX)’’. 

\subsection*{Acknowledgement}
We would like to thank Callum Murphy-Barltrop for useful comments on a draft of this paper. 

\printbibliography

\appendix
\renewcommand{\theequation}{\thesection.\arabic{equation}}
\setcounter{equation}{0}

\newpage
\begin{center}
{\huge Supplementary Material}    
\end{center}

\section{Proofs} \label{app:proofs}
\subsection*{Proof of Proposition \ref{prop:LpJacobian}}
Define $u=\cos_p(q)$ and $v=\sin_p(q)$. Then we can write
\begin{align*}
	\cos_p'(q) &= \frac{du}{dq},\\
	\sin_p'(q) &= \frac{dv}{dq} = \frac{du}{dq} \frac{dv}{du}.
\end{align*}
Differentiating (\ref{eq:ell_int}), the rate of change of pseudo-angle with $u$ is
\begin{equation*}
	\left|\frac{dq}{du} \right| = \frac{4}{\mathcal{C}_p}\left(1+ \left(\frac{dv}{du}\right)^2 \right)^{1/2},
\end{equation*}
where $\mathcal{C}_p$ is the circumference of the $L^p$ unit circle (see Definition \ref{def:angle}). Substituting these expressions into (\ref{eq:Jq}) gives 
\begin{equation} \label{eq:detJ_wx}
	J_p(q) = \frac{\mathcal{C}_p}{4} \left| u \frac{dv}{du} -v\right| \left(1+ \left(\frac{dv}{du}\right)^2\right)^{-1/2}.
\end{equation}
To simplify the presentation, we just consider the case for the first quadrant. Results for other quadrants follow by symmetry. For $L^p$ norms, in the first quadrant we have $v=(1-u^p)^{1/p}$. Therefore
\begin{equation*}
	\frac{dv}{du} = -u^{p - 1} \left(1 - u^p\right)^{\tfrac{1}{p} - 1} = - \left( \frac{u}{v}\right)^{p-1}.
\end{equation*}
Substituting this gives
\begin{equation*}
	\left|u \frac{d v}{d u} - v  \right| = \left|\frac{u^p}{v^{p-1}}  + v  \right| = v^{1-p} \left|u^p  + v^p\right| = v^{1-p}.
\end{equation*}
Then, using (\ref{eq:detJ_wx}) we have
\begin{equation*} 
	J_p(q) = \frac{\mathcal{C}_*}{4} v^{1-p} \left(1+ \left|\frac{u}{v}\right|^{2p-2}\right)^{-1/2} = \frac{\mathcal{C}_*}{4} \left(u^{2(p-1)} +  v^{2(p-1)}\right)^{-1/2}.
\end{equation*}
Substituting $u=\cos_p(q)$, $v=\sin_p(q)$ gives the first part of the result. The specific forms of the Jacobian for $p=1$, 2 and $\infty$ are straightforward to verify and follow directly from Example \ref{ex:Lp_angle}.

For the last part of the result, we again restrict attention to the first quadrant. We write $J_p$ as a function of $u$, substituting $v=(1-u^p)^{1/p}$, and consider which values of $p$ lead to zero derivative. We have 
\begin{equation*}
    \frac{dJ_p}{du} = 2(p-1) \left[u^{2p-3}-u^{p-1} (1-u^p)^{1-2/p} \right].
\end{equation*}
For $p\in (1,2)\cup(2,\infty)$ we have $dJ_p/du = 0$ implies $u^{p-2} = v^{p-2}$, which only occurs when $u=v$, and completes the proof.

\subsection*{Proof of Lemma \ref{lem:radius_change}}
For a vector $\mathbf{x} \in \mathbb{R}^2 \setminus \{\mathbf{0}\}$, define $r_a=\mathcal{R}_a(x,y)$, $r_b=\mathcal{R}_b(x,y)$ and $q = \mathcal{A}_*^{(-2,2]}((x,y)/\mathcal{R}_*(x,y))$. Denote $\mathbf{w}=(\cos_*(q),\sin_*(q))$. Then we can write
\begin{equation*}
	\mathbf{x} = r_a \frac{\mathbf{w}}{\mathcal{R}_a(\mathbf{w})} = r_b \frac{\mathbf{w}}{\mathcal{R}_b(\mathbf{w})}.
\end{equation*}
Hence 
\begin{equation} \label{eq:ra2rb}
	r_a = r_b \frac{\mathcal{R}_a(\mathbf{w})}{\mathcal{R}_b(\mathbf{w})}.
\end{equation}
Therefore $dr_b /dr_a = \mathcal{R}_b(\mathbf{w})/ \mathcal{R}_a(\mathbf{w})$, and the result follows by substituting the Jacobian.

\subsection*{Proof of Corollary \ref{cor:SPAR_radius_change}}
Denote the upper end point of $R|(Q=q)$ as $r_F(q)$, and denote $\mathbf{w}=(\cos_*(q),\sin_*(q))$. From Lemma \ref{lem:radius_change} we can use the change of variables (\ref{eq:ra2rb}) and write
\begin{align*}
    \int_0^{r_F(q)} f_{R_b,Q} (r_b,q) dr_b 
    &= \frac{\mathcal{R}_a(\mathbf{w})}{\mathcal{R}_b(\mathbf{w})} \int_0^{r_F(q)} f_{R_a,Q} \left(r_b \frac{\mathcal{R}_a(\mathbf{w})}{\mathcal{R}_b(\mathbf{w})}, q\right) dr_b = \int_0^{r_F(q)} f_{R_a,Q} \left(r_a , q\right) dr_a = f_Q(q).
\end{align*}
So the angular density $f_Q(q)$ remains unchanged by the change of gauge function and satisfies (A3) from the assumptions of the corollary. The relation between the conditional quantile functions $\mu_a$ and $\mu_b$ follows by noting that for given $\zeta$ and $q$, these must both refer to the same point in Cartesian space, so must be related by (\ref{eq:ra2rb}). Similarly, the conditional upper end points of the conditional radial distributions, $r_{a,F}(q) $ and $r_{b,F}(q)$ must also be related by (\ref{eq:ra2rb}).

To demonstrate $f_{R_b,Q}$ satisfies (A1) we have (writing $\mu_a$ and $\mu_b$ for brevity, instead of $\mu_a(\zeta,q)$, $\mu_b(\zeta,q)$)
\begin{align*} 
& \lim_{\zeta \to 1} \sup_{r_b\in[0,r_{b,F}(q)-\mu_b]} \left| \frac{\Bar{F}_{R_b|Q}(\mu_b + r_b |q)}{\Bar{F}_{R_b|Q}(\mu_b|q)} - \Bar{F}_{GP}(r_b;\xi(q),\sigma_b(\mu_b,q)) \right|\\
= 
& \lim_{\zeta \to 1} \sup_{r_b\in[0,r_{b,F}(q)-\mu_b]} \left| \frac{\int_{\mu_b+r_b}^{r_F(q)} f_{R_b|Q}(s |q) ds}{\int_{\mu_b}^{r_F(q)} f_{R_b|Q}(s|q) ds} - \left( 1+ \xi(q)\frac{r_b}{\sigma_b(\mu_b,q)} \right)^{-1/\xi(q)} \right|\\
= 
& \lim_{\zeta \to 1} \sup_{r_a\in[0,r_{a,F}(q)-\mu_a]} \left|
\frac{\int_{\mu_a+r_a}^{r_F(q)} f_{R_a|Q}(s |q) ds}{\int_{\mu_a}^{r_F(q)} f_{R_a|Q}(s|q) ds} - \left( 1+ \xi(q)\frac{r_a \frac{\mathcal{R}_b(\mathbf{w})}{\mathcal{R}_a(\mathbf{w})}}{\sigma_b(\mu_b,q)} \right)^{-1/\xi(q)} \right|\\
= 
& \lim_{\zeta \to 1} \sup_{r_a\in[0,r_{a,F}(q)-\mu_a]} \left|
\frac{\Bar{F}_{R_a|Q}(\mu_a + r_a |q)}{\Bar{F}_{R_a|Q}(\mu_a|q)} - \left( 1+ \xi(q)\frac{r_a}{\sigma_a(\mu_a,q)} \right)^{-1/\xi(q)} \right| = 0.
\end{align*}
Here we have tacitly assumed that $\xi(q)\neq0$. For values of $q$ where $\xi(q)=0$, the proof is identical to that above, substituting the appropriate form of the GP survivor function. Regarding assumption (A2), about the finiteness and continuity of the GP parameter functions, we note that the ration $\mathcal{R}_a(\mathbf{w})/\mathcal{R}_b(\mathbf{w})$ is finite and continuous, since it was stipulated that polar coordinate systems are defined in terms of gauge functions for continuous surfaces (see Definition \ref{def:polarR2}). Hence if $\sigma_a$ satisfies (A2), so does $\sigma_b$.

\subsection*{Proof of Theorem \ref{thm:SPAR_indep}}
Applying Corollary \ref{cor:SPAR_radius_change} to $f_{S,Q}$ shows that (i) implies (ii). For the second part, first note that $\mathcal{R}_\alpha$ must define a gauge function, since the set $\{\mathbf{x}\in\mathbb{R}^2 : \mathbf{x}\leq \sigma(q) (\cos_*(q),\sin_*(q)), q\in I\}$ is star-shaped and has continuous boundary. Then, applying Corollary \ref{cor:SPAR_radius_change} to $f_{R,Q}$ with $S=\mathcal{R}_\alpha(R)$ shows that (ii) implies (i).

\subsection*{Proof of Lemma \ref{lem:angle_change}}
The first part of the lemma is simply a statement of the general rule for transformation of coordinate systems for probability density functions. Note that the Jacobian is always positive, since any function $q_{a,b}(q)$ must be monotonically increasing. The particular forms of $q_{1,2}$ and $q_{2,1}$ follow immediately from Example (\ref{ex:Lp_angle}). For the Jacobians we just need to consider the case $q\in[0,1]$. The Jacobians in other quadrants follow by symmetry. For $q\in[0,1]$ we can write 
\begin{equation*}
    q_{1,2}(q) = \frac{\sin(\pi q/2)}{\cos(\pi q/2) + \sin(\pi q/2)}.
\end{equation*}
Differentiating this expression gives the first part of the result. For the second part, for $q\in[0,1]$ we can write $q_{2,1} = (2/\pi) \mathrm{atan}(q/(1-q))$, so that 
\begin{equation*}
	\frac{d q_{2,1}(q)}{dq} = \frac{2}{\pi} (2q^2-2q+1)^{-1} = \frac{2}{\pi} \left(\cos_1^2(q) + \sin_1^2(q)\right)^{-1}.
\end{equation*}

\subsection*{Proof of Corollary \ref{cor:SPAR_angle_change}}
From Lemma \ref{lem:angle_change}, for $q\in [q_{b,a}(s),q_{b,a}(t)]$ we have
\begin{align*}
    f_{Q_b}(q) &= \int_0^{r_F(q)} f_{R,Q_b}(r,q) dr = \left|\frac{d q_{a,b}(q)}{dq}\right| \int_0^{r_F(q)} f_{R,Q_a}(r,q_{a,b}(q)) dr =\left| \frac{d q_{a,b}(q)}{d q}\right| f_{Q_a}\left(q_{a,b}(q)\right),
\end{align*}
where $r_F(q)$ is the upper end point of $R|Q_b$. Since $f_{Q_a}$ is continuous and finite for $q\in [s,t]$ and hence satisfies assumption (A3), $f_{Q_b}(q)$ also satisfies assumption (A3). For the conditional radial density, we have 
\begin{align*}
    f_{R|Q_b}(r|q) = \frac{f_{R,Q_b}(r,q)}{f_{Q_b}(q)} = \frac{f_{R,Q_a}(r,q_{a,b}(q))}{f_{Q_a}(q_{a,b}(q))} = f_{R|Q_a}(r|q_{a,b}(q)).
\end{align*}
Therefore, the conditional radial density $f_{R|Q_b}(r|q)$ satisfies the SPAR assumptions (A1) and (A2) for the mapped angular range $q\in [q_{b,a}(s),q_{b,a}(t)]$.

\subsection*{Proof of Proposition \ref{prop:ARE2ARL}}
We shall start by considering the relation between the ARE and ARL models for the copula. The relation between the models for the copula density is derived in the same way. Under the assumptions of the proposition, for $\mathbf{w}=(1-w,w)$ and $w\in(0,1)$ we have
\begin{equation*}
    C_{\mathbf{u}_0}(r^{1-w},r^w) \sim \mathcal{L}_{\mathbf{u}_0}(r,\mathbf{w}) r^{\Lambda_{\mathbf{u}_0}(\mathbf{w})}, \quad r\to0^+,
\end{equation*}
where $\mathcal{L}_{\mathbf{u}_0}$ is slowly-varying in $r$ at $0^+$. We wish to derive a representation of the form (\ref{eq:ARL_dist}), for $\mathbf{z}=(1-z,z)$, $z\in(0,1)$, given by
\begin{equation*}
    C_{\mathbf{u}_0}(t\mathbf{z}) \sim B_{\mathbf{u}_0}(\mathbf{z}) \tilde{\mathcal{L}}_{\mathbf{u}_0}(t) t^{\kappa_{\mathbf{u}_0}}, \quad t\to0^+,
\end{equation*}
where $\Tilde{\mathcal{L}}_{\mathbf{u}_0}\in RV_0(0^+)$. Equating the variables of each model, $r^{1-w}=t(1-z)$, $r^w=tz$, and solving for $r$ and $w$ gives
\begin{align*}
    r&=t^2z(1-z),\\
    w&=\frac{\log(tz)}{\log(t^2z(1-z))} = \frac{1}{2} + \frac{\log(z)-\log(1-z)}{2\log(t^2z(1-z))}.
\end{align*}
For fixed $z$, $w\to1/2$ for $t\to0^+$. Hence $\mathcal{L}_{\mathbf{u}_0}(r,\mathbf{w}) \sim \tilde{\mathcal{L}}_{\mathbf{u}_0}(t)$ as $t\to0^+$ for some $\Tilde{\mathcal{L}}_{\mathbf{u}_0}\in RV_0(0^+)$. From the assumptions of the proposition, we can expand $\Lambda_{\mathbf{u}_0}(1-z,z)$ as a Taylor series about $z=1/2$:
\begin{align*}
    \Lambda_{\mathbf{u}_0}(1-z,z) = \Lambda_{\mathbf{u}_0}(\tfrac{1}{2},\tfrac{1}{2}) + \beta (z-\tfrac{1}{2}) + O(z-\tfrac{1}{2})^2,
\end{align*}
where $\beta = [\tfrac{d}{dz}\Lambda_{\mathbf{u}_0}(1-z,z)]_{z=1/2}$. From the discussion in Section \ref{sec:ARE_model}, we have $\Lambda_{\mathbf{u}_0}(\tfrac{1}{2},\tfrac{1}{2}) = \tfrac{1}{2} \kappa_{\mathbf{u}_0}$. Therefore, we have
\begin{align*}
    r^{\Lambda_{\mathbf{u}_0}(\mathbf{w})} &= \exp \left(\log(t^2z(1-z)) \left[\frac{1}{2} \kappa_{\mathbf{u}_0} + \beta \frac{\log(z)-\log(1-z)}{2\log(t^2z(1-z))} + O(\log(t))^{-2}\right] \right)\\
    &= (t^2 z(1-z))^{\tfrac{\kappa_{\mathbf{u}_0}}{2}} \left(\frac{z}{1-z}\right)^{\tfrac{\beta}{2}} O(1).
\end{align*}
This gives the required form of $B_{\mathbf{u}_0}(\mathbf{z})$, stated in the proposition. The required form of $B_{\mathbf{u}_0}$ is obtained in the same way. For the final claim, note that when $\kappa_{\mathbf{u}_0}<2$ and $\beta=0$, the tail density function tends to infinity for $z\to0^+$ and $z\to1^-$. Since $t^{\kappa_{\mathbf{u}_0}-2}\to\infty$ as $t\to0^+$, if the representation (\ref{eq:ARL_dens}) was valid for $t\to0^+$ and $z\to0^+$ (or $z\to1^-$) simultaneously, then this would imply that copula density is unbounded over a finite region, which is a contradiction, since the integral of the density over $(u,v)\in[0,1]^2$ must be equal to one. 

\subsection*{Proof of Proposition \ref{prop:ARL2ARE}}
This proof follows a similar reasoning to that for Proposition \ref{prop:ARE2ARL}. From the assumptions of the proposition we have $c_{\mathbf{u}_0}(t(1-z),tz) \sim b_{\mathbf{u}_0}(1-z,z) \mathcal{L}_{\mathbf{u}_0} (t)\, t^{\kappa_{\mathbf{u}_0}-2}$ for $t\to0^+$ and $z\in[0,1]$, and also $c_{\mathbf{u}_0}(r^{1-w},r^w) \sim \mathcal{M}_{\mathbf{u}_0} (r;w)\, r^{\lambda_{\mathbf{u}_0}(\mathbf{w})-1}$ for $r\to0^+$ and $w\in(0,1)$. Equating the variables of each model and solving for $t$ and $z$ gives
\begin{align*}
    t &= r^{1-w} + r^w \sim r^{\min(w,1-w)} \times 
    \begin{cases}
        1, & w\in(0,1/2)\cup(1/2,1),\\
        2, & w=1/2,\\
    \end{cases}, \qquad r\to 0^+,\\
    z &= \frac{r^w}{r^{1-w} + r^w} \sim \begin{cases}
        1-r^{1-2w}, & w<1/2,\\
        1/2, & w=1/2,\\
        r^{2w-1}, & w>1/2,
    \end{cases} \qquad r\to0^+.
\end{align*}
When $w=1/2$, we have 
\begin{align*}
    c_{\mathbf{u}_0}(r^{1/2},r^{1/2}) = c_{\mathbf{u}_0}(t/2,t/2) \sim b_{\mathbf{u}_0}(1/2,1/2) \mathcal{L}_{\mathbf{u}_0} (2r^{1/2}) \, (2r)^{\tfrac{\kappa_{\mathbf{u}_0}}{2}-1}.
\end{align*}
Since $b_{\mathbf{u}_0}(1-z,z)\in RV_{\beta_1}(0^+)$, for $w>1/2$ and $r\to0^+$ we have $b_{\mathbf{u}_0}(1-z,z) = \mathcal{M}_0(r,w) r^{\beta_1 (2w-1)}$ for some function $\mathcal{M}_0$ that is slowly-varying in $r$ at $0^+$. Also note that $\mathcal{L}_{\mathbf{u}_0}(t)$ is also slowly-varying in $r$. So, putting these together, for $w>1/2$ we have
\begin{align*}
    c_{\mathbf{u}_0}(r^{1-w},r^w) = c_{\mathbf{u}_0}(t,(1-z),tz) \sim \mathcal{M}_{\mathbf{u}_0}(r;w) r^{\beta_1 (2w-1)+ (1-w)(\kappa_{\mathbf{u}_0}-2)}, \quad r\to0^+,
\end{align*}
for some function $\mathcal{M}_{\mathbf{u}_0}(r;w)$ that is slowly-varying in $r$ at $0^+$. Similarly, for $w<1/2$, we have 
\begin{align*}
    c_{\mathbf{u}_0}(r^{1-w},r^w) = c_{\mathbf{u}_0}(t,(1-z),tz) \sim \mathcal{M}_{\mathbf{u}_0}(r;w) r^{\beta_2 (1-2w)+ w(\kappa_{\mathbf{u}_0}-2)}, \quad r\to0^+.
\end{align*}
Combining these results gives the desired form of $\lambda_{\mathbf{u}_0}$.

\subsection*{Proof of Proposition \ref{prop:delta_laplace_ARE}}
For $\mathbf{w} \in \{(w_1,...,w_d)\in\mathcal{U}_1 : (-1)^{1+u_{d,j}} w_j>0, \, j=1,...,d\}$, we can write
\begin{align*}
    \delta_L(r,\mathbf{w}) = c_{\mathbf{u}_0}(\tfrac{1}{2}\exp(-rw_1), ..., \tfrac{1}{2}\exp(-rw_d)) = c_{\mathbf{u}_0}(\exp(-tz_1), ..., \exp(-tz_d)).
\end{align*}
Equating marginal-scale and copula-scale coordinates for $j=1,...,d$, gives
\begin{equation*}
	\frac{1}{2}\exp(-rw_j) = \exp(-tz_j) \iff \log(2)+rw_j = tz_j.
\end{equation*} 
To solve for $t$ and $\mathbf{z}$, we assume $\mathbf{z}\in\mathcal{U}_1$ (this is always possible by rescaling), which gives
\begin{align*}
    t &= \sum_{j=1}^d t z_j = d\log(2) + r,\\
	z_j &= \frac{\log(2)+rw_j}{d\log(2) + r} = w_j + O(r^{-1}).
\end{align*}
When $|w_j|=1/d$ for $j=1,...d$, we have $z_j=w_j$ for $j=1,...d$, and 
\begin{align*}
    \delta_L(r,\mathbf{w}) &\sim \mathcal{M}_{\mathbf{u}_0}(\exp(-t),\mathbf{z}) \exp(-t(\lambda_{\mathbf{u}_0}(\mathbf{z})-1)), \qquad r\to\infty\\
    &= \mathcal{M}_{\mathbf{u}_0}(2^{-d} \exp(-r),\mathbf{w}) \exp(-(d\log(2)+r)(\lambda_{\mathbf{u}_0}(\mathbf{w})-1))\\
    &= \mathcal{M}_{\mathbf{u}_0}(2^{-d} \exp(-r),\mathbf{w}) 
    2^{-d(\lambda_{\mathbf{u}_0}(\mathbf{w})-1)} \exp(-r(\lambda_{\mathbf{u}_0}(\mathbf{w})-1)).
\end{align*}
For other values of $\mathbf{w} \in \{(w_1,...,w_d)\in\mathcal{U}_1 : (-1)^{1+u_{d,j}} w_j>0, \, j=1,...,d\}$, since $\lambda_{\mathbf{u}_0}$ is continuous and has bounded partial derivatives everywhere apart from the ray $\mathbf{z}=\mathbf{1}_d$, we have $\lambda_{\mathbf{u}_0}(\mathbf{z}) = \lambda_{\mathbf{u}_0}(\mathbf{w}+O(r^{-1})) = \lambda_{\mathbf{u}_0}(\mathbf{w})+O(r^{-1})$. Therefore, for $r\to\infty$ we can write
\begin{align*}
    \delta_L(r,\mathbf{w}) &\sim \mathcal{M}_{\mathbf{u}_0}(\exp(-t),\mathbf{z}) \exp(-t(\lambda_{\mathbf{u}_0}(\mathbf{z})-1))\\
    &= \mathcal{M}_{\mathbf{u}_0}(2^{-d} \exp(-r),\mathbf{w}+O(r^{-1})) \exp(-(d\log(2)+r)(\lambda_{\mathbf{u}_0}(\mathbf{w})-1+O(r^{-1})))\\
    &= \mathcal{M}_{\mathbf{u}_0}(2^{-d} \exp(-r),\mathbf{w}+O(r^{-1})) 
    O(1) \exp(-r(\lambda_{\mathbf{u}_0}(\mathbf{w})-1)).
\end{align*}
The terms outside the exponential function are slowly-varying in $\exp(r)$ at $\infty$, which completes the proof.

\subsection*{Proof of Proposition \ref{prop:delta_GP_ARL}}
For this case we write
\begin{align*}
    \delta_{GP}(r,\mathbf{w}) = c_{\mathbf{1}_d}((1+\xi_m r w_1)^{-1/\xi_m},...,(1 + \xi_m r w_d)^{-1/\xi_m}) = c_{\mathbf{1}_d}(t\mathbf{z}).
\end{align*}
We can relate the copula-scale angular-radial coordinates $(t,\mathbf{z})$ with the marginal-scale angular-radial coordinates $(r,\mathbf{w})$ as follows. If we assume $\mathbf{z}\in\mathcal{U}_1$ and denote $s_{\mathbf{w}} = \xi_m^{-1/\xi_m} \sum_{j=1}^d w_j^{-1/\xi_m}$ then 
\begin{align*}
    t&= \sum_{j=1}^d tz_j = \sum_{j=1}^d (1 + \xi_m r w_j)^{-1/\xi_m} \sim  r^{-1/\xi_m} s_{\mathbf{w}}, \quad r\to\infty \\
    z_j &= t^{-1} (1 + \xi_m r w_j)^{-1/\xi_m} \sim w_j^{-1/\xi_m} s_{\mathbf{w}}^{-1}, \quad r\to\infty.
\end{align*}
Therefore $\mathbf{z}$ is asymptotically constant along rays of constant $\mathbf{w}$. Substituting this into the assumed asymptotic form of the copula gives
\begin{equation*}
    \delta_{GP}(r,\mathbf{w}) \sim s_{\mathbf{w}}^{\kappa_{\mathbf{1}_d}-d} b_{\mathbf{1}_d} \left(s_{\mathbf{w}}^{-1} \mathbf{w}^{-1/\xi_m} \right) \mathcal{L}_{\mathbf{1}_d} \left(r^{-1/\xi_m} \right) r^{\tfrac{d-\kappa_{\mathbf{1}_d}}{\xi_m}}, \quad r\to\infty.
\end{equation*}
Above we have used the fact that since $\mathcal{L}_{\mathbf{1}_d}$ is slowly-varying at $0^+$, $\mathcal{L}_{\mathbf{1}_d} \left(r^{-1/\xi_m} s_{\mathbf{w}}\right) \sim \mathcal{L}_{\mathbf{1}_d} \left(r^{-1/\xi_m} \right)$ as $r\to\infty$.    

\subsection*{Proof of Proposition \ref{prop:fq_convergence_HJ}}
To simplify the presentation, we use the notation $\mathbf{w}=(\cos_1(q), \sin_1(q))$ and $\mathbf{u}=(F_*(r\cos_1(q)), F_*(r\sin_1(q)))$ throughout this proof. The general strategy for proving these results is as follows. We have
\begin{equation*}
	f_{Q_{*}}(q)=\int_0^{r_F(q)} f_{R_{*},Q_{*}}(r,q) dr,
\end{equation*}
where $r_F(q)$ is the upper end point of $R_*|(Q_*=q)$. We need to examine the behaviour of the integrand $f_{R_{*},Q_{*}}(r,q)$ for $r\in[0,r_F(q)]$. First, we write the joint angular-radial density as
\begin{align*}
	f_{R_*,Q_*} (r,q) &= r\, m_*(r,\mathbf{w}) c(\mathbf{u})\\
	&= r\, m_*(r,\mathbf{w}) \delta_*(r,\mathbf{w}).
\end{align*}
To examine the behaviour of $f_{R_*,Q_*}$ for $\mathbf{u}$ close to corner $(u_0,v_0)$, we write $\delta_*(r, \mathbf{w}) = c_{(u_0,v_0)}(t(1-z),tz)$, where
\begin{align*}
	t(1-z) &= u_0+(-1)^{u_0} F_*(r\cos_1(q)),\\
	tz &= v_0 + (-1)^{v_0} F_*(r\sin_1(q))).
\end{align*}
We then find asymptotic relations between copula-scale polar coordinates $(t,z)$ and marginal-scale polar coordinates $(r,q)$, so that the assumed asymptotic ARL form (\ref{eq:ARL_dens}) can be applied. To check whether the integral is finite, we use the assumed regular variation property, together with Karamata's theorem (see e.g. \cite{Resnick1987}) which states that
\begin{enumerate}[(i)]
	\item if $\alpha>-1$ then $g\in RV_{\alpha}(0^+)$ implies $G(x)=\int_0^x g(t) dt \in RV_{\alpha+1}(0^+)$ and is finite;
	\item if $\alpha<-1$ then $g\in RV_{\alpha}(\infty)$ implies $G(x)=\int_x^{\infty} g(t) dt \in RV_{\alpha+1}(\infty)$ and is finite.
\end{enumerate}

(a) For GP margins, we have $q\in[0,1]$ and hence $\mathbf{w}=(1-q,q)$. The rays originate in the lower left corner of the copula. Write
\begin{align*}
	f_{R_{GP},Q_{GP}} (r,q) &= r\, [(1+\xi_m r (1-q))(1+\xi_m r q)]^{-1-\tfrac{1}{\xi_m}} c_{(0,0)}(t(1-z),tz),
\end{align*}
where
\begin{align*}
	t(1-z) &= 1 - (1+\xi_m r(1-q))^{-\tfrac{1}{\xi_m}} \\
	tz &= 1 - (1+\xi_m rq)^{-\tfrac{1}{\xi_m}}.
\end{align*}
Solving for $t$ and $w$ and expanding as Taylor series about $r=0$ gives $t\sim r$ and $z\sim q$ for $r\to 0^+$. Therefore, we have
\begin{align*}
	f_{R_{GP},Q_{GP}} (r,q) &\sim b_{(0,0)}(1-q,q) \mathcal{L}_{(0,0)} (r) \, r^{\kappa_{(0,0)}-1}, \quad r\to0^+.
\end{align*}
Therefore, the angular radial density is bounded as $r\to0^+$ since either $\kappa_{(0,0)}=1$ and $\lim_{r\to0^+} \mathcal{L}_{(0,0)} (r)=\Upsilon_{(0,0)}\leq 1$, or $\kappa_{(0,0)}>1$. 

For the behaviour as $r\to\infty$, we need to consider three cases, $\xi_m<0$, $\xi_m=0$ and $\xi_m>0$. When $\xi_m<0$, we have $r \leq r_F(q) = \min \left(-(\xi_m(1-q))^{-1}, -(\xi_mq)^{-1}\right)$. When $q<1/2$ the path terminates on the right hand edge of the copula and when $q>1/2$ the path terminates on the upper edge. As the copula density is bounded away from the corners, $f_{R_{GP},Q_{GP}}$ is also bounded for $r\in[0,r_F(q)]$. Therefore $f_{Q_{GP}}(q)$ is also finite, since the integral is over a finite range. For $q=1/2$, substitute $r=r_F(q)-s = -2/\xi_m -s$, so that
\begin{align*}
	f_{R_{GP},Q_{GP}} (r,1/2) &= \left(-\frac{2}{\xi_m} -s\right)\, (-\xi_m s/2)^{-2-\tfrac{2}{\xi_m}} c_{(1,1)}(t/2,t/2),
\end{align*}
where $t=2(-\xi_m s/2)^{-\tfrac{1}{\xi_m}}$. For $s\to0^+$ we have
\begin{align*}
	f_{R_{GP},Q_{GP}} (r,1/2) &\sim -\frac{2}{\xi_m}\, (-\xi_m s/2)^{-2-\tfrac{2}{\xi_m}} b_{(1,1)}(1/2)\mathcal{L}_{(1,1)}(t) t^{\kappa_{(1,1)}-2}\\
	&\propto \mathcal{L}_{(1,1)}\left(s^{-\tfrac{1}{\xi_m}}\right)   s^{-2-\tfrac{\kappa_{(1,1)}}{\xi_m}}.
\end{align*}
In the second line we have used the fact that since $\mathcal{L}_{(1,1)}$ is slowly-varying at $0^+$, for all $a>0$, $\mathcal{L}_{(1,1)}(at) \sim \mathcal{L}_{(1,1)}(t)$ as $t\to0^+$. So $f_{R_{GP},Q_{GP}} (r,1/2)$ is in $RV_\alpha(0^+)$, with $\alpha = -2 -(\kappa_{(1,1)}/\xi_m)$. Applying Karamata's theorem, the integral is finite when $\alpha>-1$ or $\xi_m>-\kappa_{(1,1)}$.

When $\xi_m>0$ we have
\begin{align*}
f_{R_{GP},Q_{GP}} (r,q) &= r\, [(1+\xi_m r (1-q))(1+\xi_m r q)]^{-1-\tfrac{1}{\xi_m}} c_{(1,1)}(t(1-z),tz),
\end{align*}
where
\begin{align*}
t(1-z) &= (1+\xi_m r(1-q))^{-\tfrac{1}{\xi_m}} \\
tz &= (1+\xi_m rq)^{-\tfrac{1}{\xi_m}}.
\end{align*}
Solving for $t$ and $z$ gives for $r\to\infty$
\begin{align*}
t &\sim r^{-\tfrac{1}{\xi_m}} \left[(\xi_m(1-q))^{-\tfrac{1}{\xi_m}} + (\xi_m q)^{-\tfrac{1}{\xi_m}}\right],\\
z &\sim \frac{q^{-\tfrac{1}{\xi_m}}}{(1-q)^{-\tfrac{1}{\xi_m}} + q^{-\tfrac{1}{\xi_m}}}.
\end{align*}
Therefore, rays of constant $q$ asymptote to rays of constant $z$ on the copula scale. So for the joint density, we get
\begin{align*}
f_{R_{GP},Q_{GP}} (r,q) &\sim (\xi_m^2 q(1-q))^{-1-\tfrac{1}{\xi_m}} b_{(1,1)}(z) \left[(\xi_m(1-q))^{-\tfrac{1}{\xi_m}} + (\xi_m q)^{-\tfrac{1}{\xi_m}} \right]^{\kappa_{(1,1)}-2} \mathcal{L}_{(1,1)}(t) r^{-1-\tfrac{\kappa_{(1,1)}}{\xi_m}}.
\end{align*}
This is in $RV_{\alpha}(\infty)$ with $\alpha=-1-(\kappa_{(1,1)}/\xi_m)<-1$. Hence, by Karamata's theorem, this integral is finite for any $\xi_m>0$.

Finally, in the case that $\xi_m=0$, we have
\begin{align*}
	f_{R_{GP},Q_{GP}} (r,q) &= r \exp(-r) c_{(1,1)}(t(1-z),tz),
\end{align*}
where $t(1-z) = \exp(-r(1-q))$ and $tz = \exp(-rq)$, so that 
\begin{align*}
    t &= \exp(-r(1-q))+\exp(-rq),\\
    z &= \frac{\exp(-rq)}{\exp(-r(1-q))+\exp(-rq)}.
\end{align*}
Note that for $q<1/2$, $z\sim \exp(-r(2q-1))$ as $r\to\infty$. In some cases, the assumed asymptotic for of the copula is not valid for $t,w\to0^+$ simultaneously. So in these cases, we note that $t c_{(1,1)}(t(1-z),tz)$ is bounded for $t\to0^+$ for all $z\in[0,1]$. (If $t c_{(1,1)}(t(1-z),tz)$ is not bounded for $z\to0^+$, then this contradicts the fact that $\int_0^1 c_{(1,1)}(t,0) dt =1$). Since $t c_{(1,1)}(t(1-z),tz)$ is bounded, there exists $\beta\in(0,\infty)$ and $\tau>0$ such that $t c_{(1,1)}(t(1-z),tz)<\beta$ for $t<\tau$. Then for $t<\tau$
\begin{align*}
	f_{R_{GP},Q_{GP}} (r,q) &\leq r \exp(-r) \frac{\beta}{t}\\
	&= r \exp(-r) \frac{\beta}{\exp(-r(1-q))+\exp(-rq)}\\
	&\sim \beta r \begin{cases}
		\exp(-r(1-\min(q,1-q))), & q\in(0,1/2)\cup(1/2,1),\\
		2\exp(-r/2), & q=1/2.
	\end{cases}
\end{align*}
So by the dominated convergence theorem, $f_{Q_{GP}} (q)$ is finite for $\xi_m=0$, which completes the proof for part (a).

(b) For GP margins, as $q\to0^+$ the paths $\mathbf{u}$ through the copula pass closer to the lower right corner of the copula (see \autoref{fig:Copula_paths}). To examine the behaviour of $f_{R_{GP},Q_{GP}}$ for $\mathbf{u}$ close to corner $(1,0)$, we need to determine the range of $r$ for which $\mathbf{u}=(F_{GP}(r(1-q)),F_{GP}(rq))\to(1^-, 0^+)$ as $q\to0^+$. For part (i), we treat the cases $\xi_m=0$ and $\xi_m>0$ separately. For the case of exponential margins we have
\begin{align*}
	f_{R_E,Q_E}(r,q) = r \exp(-r) c_{(1,0)}(t(1-z),tz),
\end{align*}
where $t(1-z)=\exp(-r(1-q))$ and $tz=1-\exp(-rq)$. Solving for $t$ and $z$ gives
\begin{align*}
	t &= \exp(-r(1-q))+ 1 - \exp(-rq),\\
	z &= \frac{1-\exp(-rq)}{\exp(-r(1-q))+ 1 - \exp(-rq)}.
\end{align*}
Consider the behaviour of the joint density in the range $r\in J=[-\tfrac{1}{2} \log(q),-2\log(q)]$. For $q\to0^+$, the lower limit of $J$ tends to infinity, but $rq\to 0$ for all $r\in J$. Therefore, for $r\in J$ and $q\to0^+$ we have 
\begin{align*}
	t &\sim\exp(-r)+rq,\\
	z &\sim \frac{rq}{\exp(-r)+rq}.
\end{align*}
For $r\in J$ and $q\to0^+$, when $r q = \exp(-r)$, we have $z\to1/2$. The solution of $r q = \exp(-r)$ is $r = W_0(1/q)$, where $W_0$ is the zeroth branch of the Lambert W function \parencite[see e.g.][\S4.13]{NIST_DLMF}. Note that $W_0(1/q)\sim -\log(q) - \log(-\log(q))$ as $q\to0^+$. Define $s=r-W_0(1/q)$, then for $r\in J$ and $q\to0^+$ we have 
\begin{align*}
	t &\sim \exp(-s -W_0(1/q))+(s+W_0(1/q))q\\
	&= (s+W_0(1/q))q(e^{-s}+1)\\
	&\sim W_0(1/q) q (e^{-s}+1)\\
	z &\sim (e^{-s}+1)^{-1}.
\end{align*}
Substituting this into the assumed form of the copula gives for $q\to0^+$ and $r\in J$,
\begin{align*}
	&c_{(1,0)}(t(1-z),tz) \sim \Upsilon_{(1,0)} b_{(1,0)}\left(\frac{e^{-s}}{e^{-s}+1},\frac{1}{e^{-s}+1}\right) \left[W_0(1/q) q (e^{-s}+1)\right]^{-1},
\end{align*}
and
\begin{align*}
	f_{R_E,Q_E}(r,q) &= (W_0(1/q)+s) W_0(1/q) q \exp(-s) c_{(1,0)}(t(1-z),tz)\\
	&\sim \Upsilon_{(1,0)} \frac{\exp(-s)}{\exp(-s)+1} b_{(1,0)}\left(\frac{e^{-s}}{e^{-s}+1},\frac{1}{e^{-s}+1}\right) W_0(1/q),
\end{align*}
Therefore, $f_{R_E,Q_E}$ is unbounded for $r\in J$ as $q\to0^+$, and hence $f_{Q_E}(q)\to\infty$ as $q\to0^+$.

For the case $\xi_m>0$ we take a similar approach to above and consider the behaviour of the joint density for $r$ in the interval $J=[-\tfrac{1}{2} \log(q),-2\log(q)]$. In this case we have
\begin{align*}
	f_{R_{GP},Q_{GP}}(r,q) = r (1+\xi_m r(1-q))^{-1-\tfrac{1}{\xi_m}} (1+\xi_m rq)^{-1-\tfrac{1}{\xi_m}} c_{(1,0)}(t(1-z),tz),
\end{align*}
where $t(1-z)=(1+\xi_m r(1-q))^{-\tfrac{1}{\xi_m}}$ and $tz=1-(1+\xi_m rq)^{-\tfrac{1}{\xi_m }}$. For $r\in J$ and $q\to0^+$ we have 
\begin{align*}
	t(1-z)&\sim (\xi_m r)^{-\tfrac{1}{\xi_m}},\\
	tz&\sim rq.
\end{align*}
Solving for $t$ and $z$, gives for $r\in J$ and $q\to0^+$,
\begin{align*}
	t &\sim (\xi_m r)^{-\tfrac{1}{\xi_m}} + rq,\\
	z &\sim \frac{rq}{(\xi_m r)^{-\tfrac{1}{\xi_m}} + rq}.
\end{align*}
Note that $z\to1/2$ when $(\xi_m r)^{-\tfrac{1}{\xi_m}} = rq$ for $r\in J$ and $q\to0^+$. If we substitute $r=sq^{-1}(q/\xi_m)^{\tfrac{1}{\xi_m+1}}$ then for $r\in J$ and $q\to0^+$,
\begin{align*}
	t &\sim \left(\frac{q}{\xi_m}\right)^{\tfrac{1}{\xi_m+1}}\left[s^{-\tfrac{1}{\xi_m}} + s\right],\\
	z &\sim \frac{s}{s^{-\tfrac{1}{\xi_m}} + s}.
\end{align*}
Substituting this for the joint density, and recalling that for $r\in J$ and $q\to0^+$ we have $r\to\infty$ and $rq\to0^+$, and hence
\begin{align*}
	f_{R_{GP},Q_{GP}}(r,q) &= r (1+\xi_m r(1-q))^{-1-\tfrac{1}{\xi_m}} (1+\xi_m rq)^{-1-\tfrac{1}{\xi_m}} c_{(1,0)}(t(1-z),tz)\\
	&\sim \xi_m^{-1-\tfrac{1}{\xi_m}} r^{-\tfrac{1}{\xi_m}} c_{(1,0)}(t(1-z),tz)\\
	&\propto q^{\tfrac{1}{\xi_m+1}} s^{-\tfrac{1}{\xi_m}} c_{(1,0)}(t(1-z),tz)\\
	&\sim q^{\tfrac{1}{\xi_m+1}} s^{-\tfrac{1}{\xi_m}} \Upsilon_{(1,0)} b_{(1,0)}(1-z,z) t^{-1}\\
	&\propto b_{(1,0)}(1-z,z) \frac{s^{-\tfrac{1}{\xi_m}}}{s^{-\tfrac{1}{\xi_m}} + s}.
\end{align*}
Therefore, for $r\in J$ and $q\to0^+$ the joint density tends to a fixed form in terms of $s$. However, since $dr/ds \propto q^{-\tfrac{\xi_m}{\xi_m+1}}$, the integral of the joint density over $r$ diverges as $q\to0^+$, which completes the proof for part (i).

For (ii) we proceed in the same way as for part (i), but using a different change of variables. As with part (a), we substitute $r=r_F(q)-s = a(1-s)/(1-q)$, where $a=-1/\xi_m$, to get $t(1-z)=s^a$ and $tz=a(1-s)q+O(q^2)$. Therefore, for $q\to0^+$ we have
\begin{align*}
	t&\sim s^a + aq(1-s),\\
	z&\sim \frac{aq(1-s)}{s^a+aq(1-s)}.
\end{align*}
As before, we are interested in the behaviour of the joint density around $z=1/2$, which occurs when $aq=s^a/(1-s)$. The value $s^a/(1-s)$ is monotonically increasing in $s$ for $s\in[0,1)$, so when $q$ is small, the solution of $aq=s^a/(1-s)$ is approximately $s=(aq)^{1/a}$. Therefore, we substitute $s=m(aq)^{\tfrac{1}{a}}$, with $m\in[0,(aq)^{-\tfrac{1}{a}}]$. As we are interested in the behaviour of the joint density for $\mathbf{u}$ close to $(1,0)$, we restrict our interest to $m\in J = [0,-\log(q)]$, so that the upper end point of this interval tends to infinity as $q\to0^+$, but $mq^b\to0^+$ for any $b>0$ and $m\in J$. So for $q\to0^+$ and $m\in J$ we have $t\sim aq \left[m^a + 1\right]$ and $z\sim (m^a+1)^{-1}$. We now write the joint density as
\begin{align*}
	f_{R_{GP},Q_{GP}}(r,q) &= r (1+\xi_m r(1-q))^{-1-\tfrac{1}{\xi_m}} (1+\xi_m rq)^{-1-\tfrac{1}{\xi_m}} c_{(1,0)}(t(1-z),tz)\\
	&=a \frac{1-s}{1-q}\left[\left(1-q\frac{1-s}{1-q}\right) s \right]^{a-1} c_{(1,0)}(t(1-z),tz)\\
	&\sim a s^{a-1} c_{(1,0)}(t(1-z),tz), \qquad q\to0^+, m\in J\\
	&= a m^{a-1} (aq)^{\tfrac{a-1}{a}} c_{(1,0)}(t(1-z),tz)\\
	&\sim a m^{a-1} (aq)^{\tfrac{a-1}{a}} \Upsilon_{(1,0)} b_{(1,0)}(1-z,z) t^{-1}\\
	&\sim a (aq)^{-\tfrac{1}{a}} \Upsilon_{(1,0)} b_{(1,0)}(1-z,z) \frac{m^{a-1}}{1+m^a}.
\end{align*}
Noting that $ds/dm=(aq)^{\tfrac{1}{a}}$, the integral of $f_{R_{GP},Q_{GP}}(r,q)$ over $r$ converges to a fixed form in terms of $m$ for $\qquad q\to0^+$ and $m\in J$.
For $m\to0^+$, $z\sim 1-m^a$, under the assumptions above,
\begin{align*}
	f_{R_{GP},Q_{GP}}(r,q) &\sim a (aq)^{-\tfrac{1}{a}} \Upsilon_{(1,0)} \mathcal{M}_1(1-m^a) m^{a(\beta_1+1)-1},
\end{align*}
for some $\mathcal{M}_1(z)\in RV_0(1^-)$. The integral $\int_0^{m_0} f_{R_{GP},Q_{GP}}(r,q) dm$ for small $m_0$ has an integrand that is independent of $q$ and is regularly-varying at $0^+$ with index $\alpha = a(\beta_1+1)-1 >-1$. Hence by Karamata's theorem, this part of the integral is finite. For the upper limit of the interval $J$, $m=-\log(q)$ and $z\sim m^{-a}$ as $q\to0^+$. So we can write
\begin{align*}
	f_{R_{GP},Q_{GP}}(r,q) &\sim a (aq)^{-\tfrac{1}{a}} \Upsilon_{(1,0)} \mathcal{M}_0(m^{-a}) m^{a\beta_0-1},
\end{align*}
for some $\mathcal{M}_0(z)\in RV_0(0^+)$. As above, Karamata's theorem implies that the integral is finite as $q\to0^+$ and $m\to\infty$. For the remaining part of the integral, we note that $c$ is finite away from the corners, so $f_{R_{GP},Q_{GP}}(r,q)$ is also finite and the range of the integral is also finite. Hence $f_{Q_{GP}}(q)$ is finite as $q\to0^+$.

(c) For symmetric margins, the rays of constant angle on the copula scale, emanate from the centre of the copula $\mathbf{u}=(0.5,0.5)$ and pass close to at most one corner of the copula -- see \autoref{fig:Copula_paths}. Since the copula is assumed to be continuous and bounded away from the corners, we need only consider the behaviour of the joint density for $\mathbf{u}$ close to a corner of the copula. The same steps used in part (a) then show that the density is convergent when $q\notin \mathbb{Z}$, and the ray terminates in a corner. For the cases where $q\in\mathbb{Z}$, the ray terminates on the edge of the copula. From the assumed form of the copula, there is a $\beta>0$ such that $c(\mathbf{u})\leq\beta$ for all values of $\mathbf{u}$ along the path and hence $f_{R_{SGP},Q_{SGP}}\leq \beta \, r \, m_{SGP}(r,\mathbf{w})$. When $\xi_m<0$, the range of the integral is finite and the joint density $f_{R_{SGP},Q_{SGP}}$ is finite, so the integral is finite. When $\xi_m=0$, we have $f_{R_{SGP},Q_{SGP}}\leq \tfrac{1}{4} \beta \, r \, \exp(-r)$, which has finite integral over $r\in[0,\infty)$. Finally, when $\xi_m>0$ we have $f_{Q_{SGP}}\leq \tfrac{1}{4} \beta \int_0^\infty r (1+\xi_m r)^{-1-1/\xi_m} dr$. This integral is finite when $\xi_m<1$, which completes this part of the proof.

(d) For simplicity, consider the case $q=0$ -- other cases are treated identically. In this case, the path terminates at $\mathbf{u}=(1,1/2)$. Since $c$ is positive and finite on the edges $c(1,1/2)=\beta$ for some $\beta\in(0,\infty)$. Since $c$ is continuous, for any $\epsilon\in(0,\beta)$ there exists $\Delta>0$ such that $c(1-\Delta,1/2)\geq\beta-\epsilon$. Therefore for $\tfrac{1}{2}(1+\xi_m r)^{-1/\xi_m}<\Delta$, we have $f_{R_{SGP},Q_{SGP}}(r,0)\geq \tfrac{1}{4} (\beta-\epsilon) r(1+\xi_m r)^{-1-1/\xi_m}$. Define $r_0=((2\Delta)^{-\xi_m}-1)/\xi_m$. Then 
\begin{align*}
    f_{Q_{SGP}}(0) > \int_{r_0}^\infty f_{R_{SGP},Q_{SGP}}(r,0) dr \geq \frac{1}{4} (\beta-\epsilon) \int_{r_0}^\infty r(1+\xi_m r)^{-1-1/\xi_m} dr.
\end{align*}
The integral on the RHS is infinite, so since this inequality holds for any $\epsilon>0$, $f_{Q_{SGP}}(0)$ is also infinite.

(e) For this case, we equate the copula-scale coordinate with the marginal-scale coordinate, to give $t = (1+\xi_m r)^{-1/\xi_m} \sim (\xi_m r)^{-1/\xi_m}$ for $r\to\infty$. Then, since $c(1-t,1/2)\in RV_\alpha(0^+)$, we have $c(1-t,1/2) = \mathcal{L}(t)t^\alpha$ as $t\to 0^+$, for some function $\mathcal{L}\in RV_0(0^+)$. Therefore 
\begin{align*}
    f_{R_{SGP},Q_{SGP}}(r,0) &= \frac{1}{4} r (1+\xi_m r)^{-1-\tfrac{1}{\xi_m}} \mathcal{L}\left(r^{-\tfrac{1}{\xi_m}}\right) (\xi_m r)^{-\tfrac{\alpha}{\xi_m}}\\
    &\sim \frac{1}{4} \xi_m^{-1-\tfrac{\alpha+1}{\xi_m}} \mathcal{L}\left(r^{-\tfrac{1}{\xi_m}}\right) r^{-\tfrac{\alpha+1}{\xi_m}}, \quad r\to\infty,
\end{align*}
Therefore $f_{R_{SGP},Q_{SGP}}(r,0)\in RV_\beta(\infty)$ with $\beta=-(\alpha+1)/\xi_m$. Therefore, by Karamata's theorem, $f_{Q_{SGP}}(0)$ is finite when $\xi_m<\alpha+1$.

(f) For this case the joint density for $q=1/2$ is
\begin{align*}
    f_{R_{P},Q_{P}}(r,1/2) &= r (r/2)^{-4} c(t/2,t/2), \quad r\geq 2
\end{align*}
where $t/2=1-(2/r)$. Substitute, $r=2+s$ for $s\geq0$, then $t \sim s$ for $s\to0^+$. Then we have
\begin{align*}
    f_{R_{P},Q_{P}}(2+s,1/2) &= (2+s) ((2+s)/2)^{-4} c(t/2,t/2)\\
    &\sim 2 \Upsilon_{(0,0)} b_{(0,0)}(1/2) s^{-1}, \quad s\to0^+.
\end{align*}
The integral of this function from 0 to any $s_0>0$ is infinite, and hence $f_{Q_{P}}(1/2)$ is also infinite. 

\subsection*{Proof of Proposition \ref{prop:fw_convergence_WT}}
We have two cases to consider, depending on whether the path $\mathbf{u}=(F_L(rw_1),...,F_L(rw_d))$ terminates in a corner of the copula or not, as $r\to\infty$. First, assume that $w_j\neq 0$ for $j=1,...,d$, so that the path $\mathbf{u}$ terminates in the corner $\mathbf{u}_0=(H(w_1),...,H(w_d))$ as $r\to\infty$, where $H$ is the Heaviside step function, $H(x)=1$ for $x\geq0$ and $H(x)=0$ for $x<0$. By Proposition \ref{prop:delta_laplace_ARE}, there exists a function $\mathcal{M}_{\mathbf{u_0}}(t,\mathbf{w})$ that is slowly-varying in $t$ at $0^+$, such that 
\begin{equation*}
    c_{\mathbf{u}_0}(\exp(-t\mathbf{z})) \sim \mathcal{M}_{\mathbf{u_0}} (\exp(-r),\mathbf{w}) \exp(-r (\lambda_{\mathbf{u}_0} (\mathbf{w}) - 1)), \quad r\to\infty.
\end{equation*}
Therefore, we can write 
\begin{align*}
    f_{R,\mathbf{W}} (r,\mathbf{w}) &= 2^{-d} r^{d-1} \exp(-r) c_{\mathbf{u}_0}(\exp(-t\mathbf{z})),\\
    &\sim 2^{-d} r^{d-1} \mathcal{M}_{\mathbf{u_0}} (\exp(-r),\mathbf{w}) \exp(-r \lambda_{\mathbf{u}_0} (\mathbf{w})), \quad r\to\infty\\
    &= 2^{-d} (\log(m))^{d-1} \mathcal{M}_{\mathbf{u_0}} (1/m,\mathbf{w}) m^{-\lambda_{\mathbf{u}_0} (\mathbf{w})},
\end{align*}
where $r=\log(m)$. Noting that $dr/dm = m^{-1}$, the integrand for the angular density is regularly-varying at infinity with index less than $-1$, and hence the integral is finite by Karamata's theorem. 

In the second case, assume that some component of $\mathbf{w}=(w_1,...,w_d)$ is zero, then the path $\mathbf{u}$ does not terminate at a corner of the copula. Therefore the copula density is bounded everywhere along the path and hence $f_{R,\mathbf{W}} (r,\mathbf{w}) \leq \beta r^{d-1} \exp(-r)$ for some $\beta>0$. Hence the angular density is finite in this case as well.

\subsection*{Proof of Proposition \ref{prop:fw_continuity}}
Suppose that $f_{R,\mathbf{W}}(r,\mathbf{w})$ is defined on $[0,\infty)\times \Sigma$, where $\Sigma \subseteq \mathcal{U}_1$. By definition, for any $r_1\in(0,\infty)$ we have $f_{\mathbf{W}}(\mathbf{w}) = \int_{0}^{r_1} f_{R,\mathbf{W}}(r,\mathbf{w}) dr + \int_{r_1}^{\infty} f_{R,\mathbf{W}}(r,\mathbf{w}) dr$. Since $f_{\mathbf{W}}(\mathbf{w})$ is finite and $f_{R,\mathbf{W}}$ is bounded (no point masses), $\int_{r_1}^{\infty} f_{R,\mathbf{W}}(r,\mathbf{w}) dr \to 0$ as $r_1\to\infty$. Therefore, for all $\mathbf{w}\in \Sigma$ and any $\varepsilon_1>0$ there exists $r_1$ such that $\int_{r_1}^\infty f_{R,\mathbf{W}}(r,\mathbf{w}) dr <\varepsilon_1$. Since $f_{R,\mathbf{W}}$ is continuous and finite for $r\in[0,r_1]$, for any $\mathbf{w}\in\Sigma$ and $\varepsilon_2>0$ there exists a $\Delta>0$ and neighborhood $N_{\Delta}=\{\mathbf{z}\in\Sigma : \|\mathbf{w}-\mathbf{z}\|_1<\Delta \}$ such that $|f_{R,\mathbf{W}}(r,\mathbf{w})-f_{R,\mathbf{W}}(r,\mathbf{z})|<\varepsilon_2$ for all $(r,\mathbf{z})\in[0,r_1]\times N_{\Delta}$. Now choose $\varepsilon_1$ and $r_1$ such that $\int_{r_1}^\infty f_{R,\mathbf{W}}(r,\mathbf{z}) dr <\varepsilon_1$ for all $\mathbf{z}\in N_{\Delta}$. Then we have
\begin{align*}
    \left| f_{\mathbf{W}}(\mathbf{w})-f_{\mathbf{W}}(\mathbf{z}) \right| &= \left| \int_{0}^{r_1} f_{R,\mathbf{W}}(r,\mathbf{w}) dr - \int_{0}^{r_1} f_{R,\mathbf{W}}(r,\mathbf{z}) dr + \int_{r_1}^{\infty} f_{R,\mathbf{W}}(r,\mathbf{w}) dr - \int_{r_1}^{\infty} f_{R,\mathbf{W}}(r,\mathbf{z}) dr \right|\\
    &< \left| \int_{0}^{r_1} f_{R,\mathbf{W}}(r,\mathbf{w}) dr - \int_{0}^{r_1} f_{R,\mathbf{W}}(r,\mathbf{z}) dr \right| + \varepsilon_1\\
    &< r_1 \varepsilon_2 + \varepsilon_1
\end{align*}
Note that $r_1\to\infty$ as $\varepsilon_1\to0^+$. However, for any given $\varepsilon_1$ and $r_1$, we can set $\varepsilon_2=r_1^{-2}$, so that $r_1 \varepsilon_2\to0^+$ as $\varepsilon_1\to0^+$. Hence $f_{\mathbf{W}}(\mathbf{w})$ is continuous for all $\mathbf{w}\in\Sigma$. If the upper end point of $R$ is finite, then the continuity follows in the same way.

\subsection*{Proof of Lemma \ref{lemma:frw_conditions}}
The angular-radial joint density is given by $f_{R,\mathbf{W}}(r,\mathbf{w}) = 2^{-d} r^{d-1} \exp(-r) c(\mathbf{u})$, where $\mathbf{u}=(F_L(rw_1), ..., F_L(rw_d))$. For any $r\in[0,\infty)$, we have $\mathbf{u} \in (0,1)^d$. Hence $c(\mathbf{u})$ is continuous and therefore $f_{R,\mathbf{W}}(r,\mathbf{w})$ is also continuous. The bounds on $f_{R,\mathbf{W}}(r,\mathbf{w})$ can be established in the same way as in the proof of Proposition \ref{prop:fw_convergence_WT}. For the case that the path $\mathbf{u}$ terminates at an edge of the copula, the copula density is assumed to be bounded everywhere along the path. In the case that $\mathbf{u}$ terminates at a corner of the copula, the assumed asymptotic ARE form (\ref{eq:AREmod_dens}) multiplied by the marginal product function $m_L(r,\mathbf{w})=2^{-d}\exp(-r)$, ensures that the angular-radial joint density is bounded as well. Hence $f_{R,\mathbf{W}}(r,\mathbf{w})$ is bounded everywhere and hence satisfies the assumptions of Proposition \ref{prop:fw_continuity}.

\subsection*{Proof of Proposition \ref{prop:SPAR_Laplace}}
From the assumptions of the proposition, the angular density is finite at all angles. So for $r \to \infty$, the conditional radial density has asymptotic form
\begin{equation*}
    f_{R|\mathbf{W}}(r|\mathbf{w}) \sim \frac{2^{-d} r^{d-1}}{f_{\mathbf{W}}(\mathbf{w})} \mathcal{M}(\exp(-r),\mathbf{w}) \exp( -r \lambda(\mathbf{w})) \coloneqq \mathcal{M}^*(\exp(-r),\mathbf{w}) \exp( -r \lambda(\mathbf{w})).
\end{equation*}
Note that $\mathcal{M}^*(t,\mathbf{w})$ is also slowly-varying in $t$ at $0^+$. Since $f_{R|\mathbf{W}}(r|\mathbf{w})$ is ultimately monotone, from the monotone convergence theorem for $r \to \infty$ we have
\begin{align*}
    \Bar{F}_{R|\mathbf{W}}(r|\mathbf{w}) &= \int_r^\infty f_{R|\mathbf{W}}(s|\mathbf{w}) ds \sim  \int_r^\infty \mathcal{M}^*(\exp(-s),\mathbf{w}) \exp( - s \lambda(\mathbf{w})) ds = \int_{\exp(r)}^\infty \mathcal{M}^*(1/t,\mathbf{w}) t^{-\lambda(\mathbf{w}) - 1} dt,
\end{align*}
where we have made the substitution $t=\exp(s)$ in the last step. The integrand on the RHS is regularly-varying in $t$ with index $-\lambda(\mathbf{w}) - 1 < -1$ and hence by Karamata's theorem, the integral is finite and regularly-varying with index $-\lambda(\mathbf{w})$. Therefore, there exists a function $\mathcal{L}(t,\mathbf{w})$ that is slowly-varying in $t$ at $0^+$, such that $\Bar{F}_{R|\mathbf{W}}(r|\mathbf{w}) = \mathcal{L}(\exp(-r),\mathbf{w}) \exp(-r \lambda(\mathbf{w}))$. So for all $r>0$ we have
\begin{align*}
    \frac{\Bar{F}_{R|\mathbf{W}}(\mu + r |\mathbf{w})}{\Bar{F}_{R|\mathbf{W}}(\mu|\mathbf{w})} 
    =
    \frac{\mathcal{L}(\exp(-(\mu+r)),\mathbf{w})}{\mathcal{L}(\exp(-\mu),\mathbf{w})} \frac{\exp(-(\mu+r) \lambda(\mathbf{w}))}{\exp(-\mu \lambda(\mathbf{w}))}
    \sim 
    \exp(-r\lambda(\mathbf{w})), \quad \mu\to\infty.
\end{align*}
Therefore, assumptions (A1) and (A2) are satisfied taking $\xi(\mathbf{w})=0$ and $\sigma(\mu,\mathbf{w})=1/\lambda(\mathbf{w})$. 

\subsection*{Proof of Proposition \ref{prop:SPAR_Pareto}}
To demonstrate that the angular density is finite, we need to consider the behaviour of the joint density in the lower and upper tails. Taking a similar approach to the proof of Proposition \ref{prop:fq_convergence_HJ}, in the lower tail we have
\begin{equation*}
    f_{R,\mathbf{W}}(r,\mathbf{w}) \sim b_{\mathbf{0}_d} \left(\mathbf{w} \right) \mathcal{L}_{\mathbf{0}_d} \left(r \right) r^{\kappa_{\mathbf{0}_d}-1}, \quad r\to 0^+.
\end{equation*}
Hence $f_{R,\mathbf{W}}(r,\mathbf{w})$ is bounded for $r\to0^+$. Moreover, since the copula density is bounded away from the corners, $f_{R,\mathbf{W}}(r,\mathbf{w})$ is bounded for any finite value of $r$. For the upper tail, we note that when the polar origin is defined at $\mathbf{x}=\mathbf{1}_d$ on Pareto margins and $\xi_m=1$, $\delta_{P}(r,\mathbf{w})=\delta_{GP}(r,\mathbf{w})$. Hence from Proposition \ref{prop:delta_GP_ARL}, for $\mathbf{w}\in \mathcal{U}_1\cap(0,1]^d$ we have, 
\begin{equation*}
    \delta_{P}(r,\mathbf{w}) \sim s_{\mathbf{w}}^{\kappa_{\mathbf{1}_d}-d} b_{\mathbf{1}_d} \left((s_{\mathbf{w}} \mathbf{w})^{-1} \right) \mathcal{L}_{\mathbf{1}_d} \left(r^{-1} \right) r^{d-\kappa_{\mathbf{1}_d}}, \quad r\to\infty.
\end{equation*}
For the marginal product function we have
\begin{equation*}
    m_P(r,\mathbf{w}) = \prod_{j=1}^d (1+rw_j)^{-2} \sim r^{-2d} \prod_{j=1}^d w_j^{-2}, \quad r\to\infty.
\end{equation*}
Combining this we get
\begin{equation*}
    f_{R,\mathbf{W}}(r,\mathbf{w}) \sim \left[\prod_{j=1}^d w_j^{-2} \right] s_{\mathbf{w}}^{\kappa_{\mathbf{1}_d}-d} b_{\mathbf{1}_d} \left((s_{\mathbf{w}} \mathbf{w})^{-1} \right) \mathcal{L}_{\mathbf{1}_d} \left(r^{-1} \right) r^{-1-\kappa_{\mathbf{1}_d}}, \quad r\to\infty.
\end{equation*}
So $f_{R,\mathbf{W}}\in RV_{-1-\kappa_{\mathbf{1}_d}}(\infty)$. Hence by Karamata's theorem, $f_{\mathbf{W}}(\mathbf{w})$ is finite and $\Bar{F}_{R|\mathbf{W}}\in RV_{-\kappa_{\mathbf{1}_d}}(\infty)$. Thus from the regular variation property
\begin{equation*}
    \lim_{\mu\to\infty} \frac{\Bar{F}_{R|\mathbf{W}} (\mu+r|\mathbf{w})}{\Bar{F}_{R|\mathbf{W}} (\mu|\mathbf{w})} = \left(1+\frac{r}{\mu}\right)^{-\kappa_{\mathbf{1}_d}} = \Bar{F}_{GP} \left(r, \frac{1}{\kappa_{\mathbf{1}_d}} , \frac{\mu}{\kappa_{\mathbf{1}_d}}\right).
\end{equation*}
Therefore, assumptions (A1) and (A2) are satisfied, taking $\xi(\mathbf{w})=1/\kappa_{\mathbf{1}_d}$ and $\sigma(\mu,\mathbf{w})=\mu/\kappa_{\mathbf{1}_d}$. 

\subsection*{Proof of Proposition \ref{prop:SPAR_Pareto_AI}}
From the discussion at the end of Section \ref{sec:pdf_coords_multivar}, we know that for two-dimensional angular-radial densities in $L^1$ polar coordinates, we have $f_{R,\mathbf{W}}(r,(1-q,q))) = f_{R,Q}(r,q)$ for $q\in[0,1]$ and $r\in[\max(q^{-1},(1-q)^{-1}),\infty)$. Using Proposition \ref{prop:SPAR_Pareto} we have
\begin{equation*}
    f_{R,Q}(r,q) \sim (q(1-q))^{-\kappa_{\mathbf{1}_d}} b_{\mathbf{1}_d} \left(1-q,q\right) \mathcal{L}_{\mathbf{1}_d} \left(r^{-1} \right) r^{-1-\kappa_{\mathbf{1}_d}}, \quad r\to\infty.
\end{equation*}
From Proposition \ref{prop:ARE2ARL} we have $b_{\mathbf{1}_d}(1-q,q) = \gamma (q(1-q))^{\tfrac{\kappa_{\mathbf{1}_d}}{2} -1}$ for some $\gamma>0$. Substituting this into the equation above completes the proof.

\section{Examples of angular-radial models for copulas} \label{app:copula_calcs}
\setcounter{equation}{0}
In this appendix we present examples of the various angular-radial models for copulas introduced in Section \ref{sec:copula_tail_model}. Where applicable, we also confirm the relations between the various models defined in Propositions \ref{prop:ARE2ARL} and \ref{prop:ARL2ARE}. 

\subsection{Extreme value copulas} \label{app:EVcopula}
Extreme value (EV) copulas can be written in the form
\begin{equation*}
    C(\mathbf{u}) = \exp(-A(\mathbf{x})),
\end{equation*}
where $\mathbf{x}=(x_1,...,x_d)=(-\log(u_1),...,-\log(u_d))$, and $A:[0,\infty)^d \to [0,\infty)$ is a convex homogeneous function of order one, known as the stable tail dependence function \parencite{Beirlant2004}, that satisfies $\max(x_1,...x_d)\leq A(x_1,...,x_d) \leq x_1 + \cdots + x_d$. The lower and upper bounds in the inequality correspond to the cases of complete dependence and independence respectively. In general we will rule out these cases and assume that the inequality is strict. The homogeneity property of $A$ suggests an angular-radial representation. Define $\tau=\|\mathbf{x}\|_1$ and $\boldsymbol{\psi} =\mathbf{x}/\mathbf{\tau}$. The variables $\tau$ and $\boldsymbol{\psi}$ are referred to as the Pickands coordinates \parencite{Falk2005}, and are the $L^1$ polar coordinates of $\mathbf{x}$. EV copulas can then be expressed as
\begin{equation*}
    C(\mathbf{u}) = \exp(-\tau A(\boldsymbol{\psi})).
\end{equation*}

The general expression for the density involves many products of partial derivatives of $A$. However, the following general form is useful for asymptotic calculations in the upper and lower tail. Progressively taking partial derivatives shows that
\begin{align*}
    \frac{\partial}{\partial u_1} C(\mathbf{u}) &= -\frac{1}{u_1} \frac{\partial}{\partial x_1} \exp(-A(\mathbf{x})) = \frac{1}{u_1} \exp(-A(\mathbf{x})) A^{(1,0,0,...)}(\mathbf{x}) \\
    \frac{\partial^2}{\partial u_1 \partial u_2} C(\mathbf{u}) &= -\frac{1}{u_1 u_2} \frac{\partial}{\partial x_2} \left[\exp(-A(\mathbf{x})) A^{(1,0,0,...)}(\mathbf{x})\right] \\
    &= \frac{1}{u_1 u_2} \exp(-A(\mathbf{x})) \left[A^{(1,0,0,...)}(\mathbf{x})A^{(0,1,0,...)}(\mathbf{x}) - A^{(1,1,0,...)}(\mathbf{x})\right]\\
    &\;\;\vdots\\
    \frac{\partial^d}{\partial u_1 \cdots \partial u_d} C(\mathbf{u}) &= \left( \prod_{j=1}^d u_j^{-1} \right) \exp(-A(\mathbf{x})) \left[\left( \prod_{j=1}^d A_j(\mathbf{x}) \right) + ... + (-1)^{d-1} A^{(1,...,1)}(\mathbf{x})\right],
\end{align*}
where $A_j=\partial A/\partial x_j$ and the notation $A^{(n_1,...,n_d)}$ denotes partial differentiation $n_j$ times in the $j^{\text{th}}$ variable. The omitted terms on the final line involve products of partial derivatives of $A$ of mixed order. Note that $n^{\text{th}}$-order partial derivatives of $A$ are homogeneous of order $1-n$, and that $\prod_{j=1}^d u_j^{-1} = \exp(\tau)$. We can therefore write
\begin{equation} \label{eq:EV_density}
    c(\mathbf{u}) = \exp(-\tau (A(\boldsymbol{\psi})-1)) \left[\left(\prod_{j=1}^d A_j(\boldsymbol{\psi}) \right) + \tau^{-1} [...] +  \tau^{-2} [...] + ... + (-\tau)^{1-d} A^{(1,...,1)}(\boldsymbol{\psi})\right],
\end{equation}
where the omitted terms involve partial derivatives of $A$ of orders between $2$ and $d-1$, evaluated at $\boldsymbol{\psi}$. 

\subsubsection{ARL model}
\textbf{Lower tail:} In the lower tail we have
$C(t,...,t) = t^{A(\mathbf{1}_d)}$. Hence EV copulas have lower tail order $\kappa_{\mathbf{0}_d}=A(\mathbf{1}_d)$ with $\mathcal{L}_{\mathbf{0}_d}(t) \equiv 1$. \textcite{Hua2011} noted that in the lower tail 
\begin{align*}
    C(t\mathbf{w}) &= \exp\left(\log(t) A\left(1+\frac{\log(w_1)}{\log(t)},...,1+\frac{\log(w_d)}{\log(t)}\right)\right)\\
    &\sim \exp\left(\log(t) \left[A(\mathbf{1}_d)+A_1(\mathbf{1}_d)\frac{\log(w_1)}{\log(t)} + \cdots+ A_d(\mathbf{1}_d)\frac{\log(w_d)}{\log(t)}\right]\right), \quad \log(t)\to0^+\\
    &=t^{A(\mathbf{1}_d)}\prod_{j=1}^d w_j^{A_j(\mathbf{1}_d)}.
\end{align*}
Hence EV copulas have lower tail order function $B_{\mathbf{0}_d}(\mathbf{w})=\prod_{j=1}^d w_j^{A_j(\mathbf{1}_d)}$. However, note that the asymptotic equivalence above assumes that $\log(w_j)/\log(t)\to0^+$ as $t\to0^+$ for $j=1,...,d$, so this form of the copula does not hold if $w_j$ decreases at a similar rate to $t$. This is discussed further in Section \ref{sec:SPAR_longtail}. Using (\ref{eq:tail_dens_pdif}) and differentiating the expression for $C(t\mathbf{w})$, we obtain for the density 
\begin{equation} \label{eq:EV_lower_tail_density}
    c(t\mathbf{w}) \sim \left[\prod_{j=1}^d A_j(\mathbf{1}_d)\right] t^{A(\mathbf{1}_d)-d} \prod_{j=1}^d w_j^{A_j(\mathbf{1}_d)-1}.
\end{equation}
Alternatively, we can work directly with the density, using the same expansion as above to obtain
\begin{equation*} 
    \exp(-\tau (A(\boldsymbol{\psi})-1)) \sim t^{A(\mathbf{1}_d)-d} \prod_{j=1}^d w_j^{A_j(\mathbf{1}_d)-1}.
\end{equation*}
For the terms in the square brackets in (\ref{eq:EV_density}), we note that in the lower tail
\begin{align*}
    \tau &= -\log\left(t^d \prod_{j=1}^d w_j\right) \to \infty, \quad t\to 0^+\\
    \boldsymbol{\psi} &\to \left(\tfrac{1}{d}, ..., \tfrac{1}{d}\right), \quad t\to 0^+.
\end{align*}
Therefore, the terms in the square brackets in (\ref{eq:EV_density}) which are proportional to $\tau^{-1}$, $\tau^{-2}$, ..., $\tau^{1-d}$, tend to zero. The remaining terms are all homogeneous of order zero, so $\prod_{j=1}^d A_j(\mathbf{1}_d) =  \prod_{j=1}^d A_j\left(\tfrac{1}{d}, ..., \tfrac{1}{d}\right)$, which agrees with the asymptotic expression obtained from the tail order function.

\textbf{Upper tail:} In the upper tail, for $\mathbf{u}=\mathbf{1}_d - t\mathbf{w}$ and $\mathbf{w}\in\mathcal{U}_1 \cap [0,1]^d$, for $t\to0^+$ we have
\begin{align*}
    \tau &= -\sum_{j=1}^d\log\left(1-t w_j\right) = \sum_{j=1}^d (t w_j + O(t^2)) = t + O(t^2),\\
    \boldsymbol{\psi} &= \tau^{-1}(-\log(1-tw_1),...,-\log(1-tw_d)) = \mathbf{w} + O(t).
\end{align*}
The bounds on $A$ together with the convexity property imply that $A(\mathbf{w}+O(t)) = A(\mathbf{w}) + O(t)$. Hence
\begin{equation} \label{eq:EV_uppertail}
    C(\mathbf{1}_d-t\mathbf{w}) = \exp(-tA(\mathbf{w})+O(t^2)) = 1 -t A(\mathbf{w})+O(t^2).
\end{equation}
The general relation between asymptotic forms of $C(\mathbf{1}_d-t\mathbf{w})$ and $C_{\mathbf{1}_d}(t\mathbf{w})$ is discussed in \textcite{Joe2010}. For the two-dimensional case, for any copula we have $C_{(1,1)}(t(1-w),tw) = t - 1 + C(1-t(1-w),1-tw)$, and hence for EV copulas $C_{(1,1)}(t\mathbf{w}) \sim t(1-A(\mathbf{w}))$ for $t\to0^+$. For the copula density, since the partial derivatives of $C$ are ultimately monotone, we can differentiate (\ref{eq:EV_uppertail}) to obtain
\begin{equation*}
    c_{\mathbf{1}_d}(t\mathbf{w}) \sim t^{1-d} \left|A^{(1,...,1)}(\mathbf{w})\right|.
\end{equation*}
Hence EV copulas have upper tail order $\kappa_{\mathbf{1}_d}=1$, although the expression above does not give separate values for $\Upsilon_{\mathbf{1}_d}$ and $b_{\mathbf{1}_d}(\mathbf{w})$. As with the lower tail order function, this asymptotic form can be obtained directly from the density. Since $\tau = t+O(t^2)$, $\exp(-\tau (A(\boldsymbol{\psi})-1)) \to 1$ as $t\to0^+$. Similarly, the dominant terms in the square brackets in (\ref{eq:EV_density}) are those proportional to $\tau^{1-d}$, giving the same asymptotic expression obtained by differentiating (\ref{eq:EV_uppertail}). 

\subsubsection{ARE model} \label{sec:EV_ARE}
\textbf{Lower tail:} In the lower tail we have $u_j = \exp(-rw_j)$ and hence for $\mathbf{w}\in\mathcal{U}_1\cap[0,1]^d$ we have $\tau=r$ and $\boldsymbol{\psi}=\mathbf{w}$. Therefore 
\begin{equation*}
    C(\exp(-r\mathbf{w}))=\exp(-r A(\mathbf{w})),
\end{equation*}
and hence $\Lambda_{\mathbf{0}_d} (\mathbf{w}) = A(\mathbf{w})$, and $\mathcal{L}_{\mathbf{0}_d} (t,\mathbf{w})\equiv 1$. For the density, we note that the terms in the square brackets of (\ref{eq:EV_density}) that are proportional to $\tau^{-1}$, $\tau^{-2}$, etc., all tend to zero for $r\to\infty$. Therefore
\begin{equation} \label{eq:EV_lower_tail_exponent_density}
    c(\exp(-r\mathbf{w})) \sim \left(\prod_{j=1}^d A_j(\mathbf{w}) \right) \exp(-r (A(\mathbf{w})-1)), \quad r\to \infty.
\end{equation}
Hence we have $\lambda_{\mathbf{0}_d} (\mathbf{w}) = A(\mathbf{w})$, and $\mathcal{M}_{\mathbf{0}_d} (t,\mathbf{w}) = \prod_{j=1}^d A_j(\mathbf{w})$.

The copula and copula density exponent functions satisfy the assumptions of Proposition \ref{prop:ARE2ARL}. We can confirm that the expressions for the tail order and tail density functions derived in the previous subsection are in agreement with the forms given in Proposition \ref{prop:ARE2ARL}. As noted above, $\kappa_{(0,0)}=A(1,1)$. Also note that, $\beta \coloneqq [\tfrac{d}{dw}\Lambda_{(0,0)}(1-w,w)]_{w=1/2} = A_2(\tfrac{1}{2},\tfrac{1}{2}) - A_1(\tfrac{1}{2},\tfrac{1}{2}) = A_2(1,1) - A_1(1,1)$, where the final equality is obtained by using the fact that since $A$ is homogeneous of order one, its first partial derivatives are homogeneous of order zero. Next note that by Euler's theorem on homogeneous functions $A(1,1)=A_1(1,1)+A_2(1,1)$. Substituting these expressions into the form of the tail order and tail density functions given in Proposition \ref{prop:ARE2ARL}, shows that this agrees with the expressions derived above.

\textbf{Upper tail:} Since EV copulas have an upper tail order of 1, $\Lambda_{\mathbf{1}_d}(\mathbf{w})=\max(w_1,...,w_d)$. The case for the density exponent function is more complicated. Examples of $\delta_L(r,\mathbf{w})$ (which is related to $\lambda_{\mathbf{1}_d}(\mathbf{w})$ by Proposition \ref{prop:delta_laplace_ARE}) for two-dimensional cases are considered in the following subsection.

\subsubsection{AR copula function for Laplace margins}
\begin{figure}[!t]
	\centering
	\includegraphics[scale=0.6]{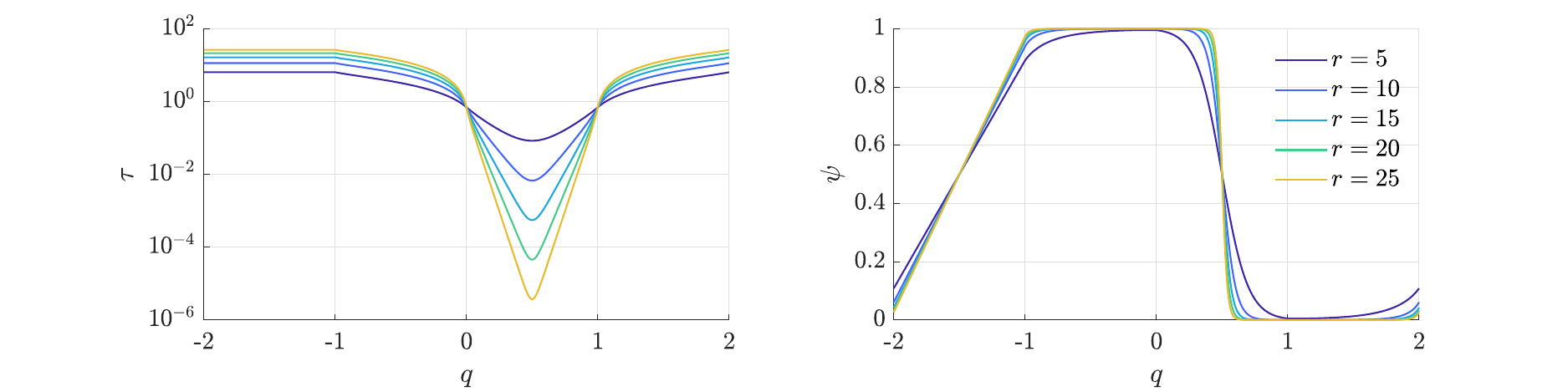}
	\caption{Pickands coordinates $(t,\psi)$ for two-dimensional EV copulas as a function of $L^1$ polar coordinates $(r,q)$ on Laplace margins.} 
	\label{fig:Pickands_coords}
\end{figure}

For the two-dimensional case we write
\begin{equation} \label{eq:EV_density_2D}
    c(u,v) = T_1 (T_2 + T_3),
\end{equation}
where 
\begin{align*}
    T_1 &= \exp(-\tau (A(\boldsymbol{\psi})-1)),\\
    T_2 &= A^{(1,0)}(\boldsymbol{\psi}) A^{(0,1)}(\boldsymbol{\psi}),\\
    T_3 &= -\tau^{-1} A^{(1,1)}(\boldsymbol{\psi}).
\end{align*}
To understand the asymptotic behave of the copula density, we consider the asymptotic behaviour of the Pickands coordinates. On Laplace margins with $(u,v)=(F_L(rw_1),F_L(rw_2))$,  $\mathbf{w}=(w_1,w_2)=(\cos_1(q),\sin_1(q))$, and $\boldsymbol{\psi}=(\psi,1-\psi)$, we have
\begin{equation*}
    \tau = 
    \begin{cases}
    \log(4) + r & q\in[-2,-1],\\
    \log(2) + r|w_2| + O(\exp(-r|w_1|) & q\in(-1,0],\\
    \tfrac{1}{2}\exp(-r|w_2|) + O(\exp(-r\min(2|w_2|,|w_1|)) & q\in(0,1/2),\\
    \exp(-r/2) + O(\exp(-r)) & q=1/2,\\
    \tfrac{1}{2}\exp(-r|w_1|) + O(\exp(-r\min(2|w_1|,|w_2|)) & q\in(1/2,1),\\
    \log(2) + r|w_1| + O(\exp(-r|w_2|) & q\in[1,2),
    \end{cases}
\end{equation*}
\begin{equation*}
    \psi = 
    \begin{cases}
     |w_2| + O(r^{-1}) & q\in[-2,-1],\\
     1 - O((r|w_2|)^{-1}\exp(-r|w_1|)) & q\in(-1,0),\\
     1 - O(\exp(-r(|w_1|-|w_2|)) & q\in[0,1/2),\\
     1/2 & q=1/2,\\
     O(\exp(-r(|w_2|-|w_1|)) & q\in(1/2,1],\\
     O((r|w_1|)^{-1}\exp(-r|w_2|)) & q\in(1,2).
    \end{cases}
\end{equation*}
The variation of $(\tau,\psi)$ with $(r,q)$ is illustrated in \autoref{fig:Pickands_coords}. From Section \ref{sec:EV_ARE} and Proposition \ref{prop:delta_laplace_ARE}, we know the asymptotic behaviour of $\delta_L(r,\mathbf{w})$ in the third quadrant. In the second and fourth quadrants, $\psi$ is exponentially close to zero or one, so since $A(0,1)=A(1,0)=1$ for all stable tail dependence functions, $T_1\to1$ as $r\to\infty$. Similarly, in the first quadrant $\tau\to0$, and hence $T_1\to1$ as $r\to\infty$ in this quadrant as well. Therefore, in the first, second and fourth quadrants, the asymptotic behaviour of $\delta_L(r,\mathbf{w})$ is determined by the terms $T_2$ and $T_3$. To illustrate various possibilities for how $T_2$ and $T_3$ behave in these quadrants, and hence the asymptotic behaviour of $\delta_L(r,\mathbf{w})$, we consider three examples for the stable tail dependence function $A$: the symmetric and asymmetric logistic models, and the H{\"u}sler-Reiss model.

\textbf{Symmetric logistic:} The symmetric logistic model with parameter $\alpha\geq 1$ is given by
\begin{equation*}
    A(\boldsymbol{\psi}) = z^{\tfrac{1}{\alpha}},
\end{equation*}
with $z=\psi^\alpha + (1-\psi)^{\alpha}$. The case $\alpha=1$ corresponds to the independence copula. When $\alpha>1$, we have
\begin{align*}
    T_2 &= z^{\tfrac{2}{\alpha}-2} (\psi (1-\psi))^{\alpha-1}\\
    T_3 &= (\alpha-1) z^{-\tfrac{1}{\alpha}} \tau^{-1} T_2.
\end{align*}
After substituting the expressions for $\tau$ and $\psi$ above we find that $\delta_L(r,\mathbf{w})$ has asymptotic form (\ref{eq:delta_Laplace}) with
\begin{align*}
    \beta(\mathbf{w}) &= 
    \begin{cases}
        0, & q\in(-2,-1)\cup[0,1],\\
        1-\alpha, & q\in[-1,0)\cup(1,2],
    \end{cases}\\
    \lambda(\mathbf{w}) &= 
    \begin{cases}
    A(\mathbf{w}), & q\in(-2,-1],\\
    1+(\alpha-1)|w_1|, & q\in(-1,0],\\
    1 - \alpha + (2\alpha - 1) \max(|w_1|,|w_2|), & q\in(0,1]\\
	1 + (\alpha - 1) |w_2| & q\in(1,2].
    \end{cases}
\end{align*}
Note that in this case $A^{(1,1)}(\psi,1-\psi) = (1-\alpha) z^{\tfrac{1}{\alpha}-2} (\psi (1-\psi))^{\alpha-1}$, so as $\psi\to0^+$ we have $A^{(1,1)}(\psi,1-\psi) \sim (1-\alpha) \psi^{\alpha-1}$. Therefore, the assumptions of Proposition \ref{prop:ARL2ARE} are satisfied, taking $\beta_1 = \beta_2 = \alpha-1$. It can then be verified that Proposition \ref{prop:ARL2ARE} in combination with Proposition \ref{prop:delta_laplace_ARE} gives the same form of $\lambda(\mathbf{w})$ obtained above for the range $q\in(0,1)$.

\textbf{Asymmetric logistic:} The asymmetric logistic model proposed by \textcite{Tawn1988}, is given by
\begin{equation*} \label{eq:asymlog}
    A(\boldsymbol{\psi}) = (1-\gamma_1)\psi + (1-\gamma_2)(1-\psi) + z^{\tfrac{1}{\alpha}},
\end{equation*}
where $z=(\gamma_1 \psi)^{\alpha} + (\gamma_2 (1-\psi))^{\alpha}$ and $\alpha\geq 1$ and $\gamma_1,\gamma_2\in[0,1]$. When $\gamma_1=\gamma_2=1$ this reduces to the symmetric logistic model. Independence occurs when $\alpha=1$ or $\gamma_1=0$ or $\gamma_2=0$. In the case that $\gamma_1,\gamma_2\in(0,1)$, we have
\begin{align*}
    T_2 &= \left(1-\gamma_1 + \gamma_1^\alpha \psi^{\alpha-1} z^{\tfrac{1}{\alpha}-1}\right) \left(1-\gamma_2 + \gamma_2^\alpha (1-\psi)^{\alpha-1} z^{\tfrac{1}{\alpha}-1} \right)\\
    T_3 &= (\alpha-1) (\gamma_1 \gamma_2 )^\alpha (\psi(1-\psi))^{\alpha-1} \tau^{-1} z^{\tfrac{1}{\alpha}-2}.
\end{align*}
For $q\in[-1,1/2)$, $\psi\to1$ as $r\to\infty$ and hence $T_2\to 1 - \gamma_2$ as $r\to\infty$. Similarly, for $q\in(1/2,2]$, $T_2\to 1 - \gamma_1$ as $r\to\infty$. For $q$ in the second and fourth quadrants, $\tau\to\infty$ as $r\to\infty$, hence $T_3\to0$. In the first quadrant for $r\to\infty$ we have 
\begin{equation*}
    T_3 \sim 2(\alpha-1)\exp(r[1-\alpha+(2\alpha-1)\min(w_1,w_2)]) \times \begin{cases}
        \gamma_1^\alpha \gamma_2^{1-\alpha}, & q<1/2,\\
        \gamma_2^\alpha \gamma_1^{1-\alpha}, & q>1/2.
    \end{cases}
\end{equation*}
 The terms in the exponential function are positive when $\min(w_1,w_2) > (\alpha-1)/(2\alpha-1)$. Overall, we find that $\delta_L(r,\mathbf{w})$ has asymptotic form (\ref{eq:delta_Laplace}) with $\beta(\mathbf{w})=0$ for all $q$ and 
\begin{align*}
    \lambda(\mathbf{w}) &= 
    \begin{cases}
    A(\mathbf{w}), & q\in(-2,-1),\\
    1 & q\in\left[-1,\dfrac{\alpha-1}{2\alpha-1}\right],\\[10pt]
    1 - \alpha + (2\alpha - 1) \max(|w_1|,|w_2|), & q\in\left(\dfrac{\alpha-1}{2\alpha-1}, \dfrac{\alpha}{2\alpha-1}\right)\\[10pt]
	1 & q\in\left[\dfrac{\alpha}{2\alpha-1},2\right].
    \end{cases}
\end{align*}
Note that although $A^{(1,1)}$ has the same form as for the symmetric logistic model, the assumptions of Proposition \ref{prop:ARL2ARE} are not satisfied, since $T_3\to0^+$ for $q\to0^+$ and $q\to1^-$, but $T_2\to1-\gamma_1$ or $1-\gamma_2$, respectively. Hence the ARL model is not valid for $q\to0^+$ or $q\to1^-$.

In the first quadrant $\lambda(\mathbf{w})$ is symmetric about $q=1/2$ and is not dependent on the values of $\gamma_1$ and $\gamma_2$. However, at finite levels, the corresponding isodensity contours are not symmetric about $q=1/2$, as illustrated in \autoref{fig:SPAR_Laplace_EV}, for a case with $\alpha=5$, $\gamma_1=0.9$ and $\gamma_2=0.1$. To improve the agreement between the SPAR approximation and true joint density at finite levels, we can define the angular-radial coordinate system relative to a different origin. Note that a finite shift in the origin does not affect the asymptotic behaviour of $\delta_L(r,\mathbf{w})$, and hence the values of $\beta(\mathbf{w})$ and $\lambda(\mathbf{w})$ remain unchanged.

Define new angular-radial variables relative to origin $(x_0,y_0)$, $r^* = (x-x_0)+(y-y_0)$ and $q^*=(y-y_0)/r^*$, for $x>x_0$ and $y>y_0$. Then, in terms of the original angular-radial coordinates,
\begin{align*}
    r&=r^*+x_0+y_0,\\
    q&=\frac{r^* q^* +y_0}{r}.
\end{align*}
We can find $(x_0,y_0)$ such that $T_2\sim T_3$ as $r\to\infty$ at constant angle $q^*$. This gives two sets of equations
\begin{align*}
    1-\gamma_1 &= 2(\alpha-1) \gamma_1^{\alpha} \gamma_2^{1-\alpha} \exp(r[1-\alpha+(2\alpha-1)q]), \quad q\in(0,1/2)\\
    1-\gamma_2 &= 2(\alpha-1) \gamma_2^{\alpha} \gamma_1^{1-\alpha} \exp(r[\alpha+(1-2\alpha)q]), \quad q\in(1/2,1).
\end{align*}
Substituting the new coordinates and rearranging gives
\begin{align*}
    \beta_1 &= r^* (1-\alpha +(2\alpha-1)q^*) + (1-\alpha)x_0 + \alpha y_0, \quad q\in(0,1/2)\\
    \beta_2 &= r^* (\alpha +(1-2\alpha)q^*) + \alpha x_0 + (1-\alpha)y_0, \quad q\in(1/2,1),
\end{align*}
where
\begin{align*}
\beta_1 &= \log \left( \frac{1-\gamma_1}{2(\alpha-1)\gamma_2} \left(\frac{\gamma_2}{\gamma_1}\right)^{\alpha} \right),\\
\beta_2 &= \log \left( \frac{1-\gamma_2}{2(\alpha-1)\gamma_1} \left(\frac{\gamma_1}{\gamma_2}\right)^{\alpha} \right).
\end{align*}
For these equations to hold approximately for $r^*\to\infty$, the terms multiplying $r^*$ must be zero. Then solving for $(x_0,y_0)$ gives
\begin{align*}
x_0 = \frac{(\alpha-1)\beta_1+\alpha\beta_2}{2\alpha-1}, \quad 
y_0 = \frac{\alpha\beta_1+(\alpha-1)\beta_2}{2\alpha-1}.
\end{align*}

\begin{figure}[!t]
	\centering
    \includegraphics[scale=0.6]{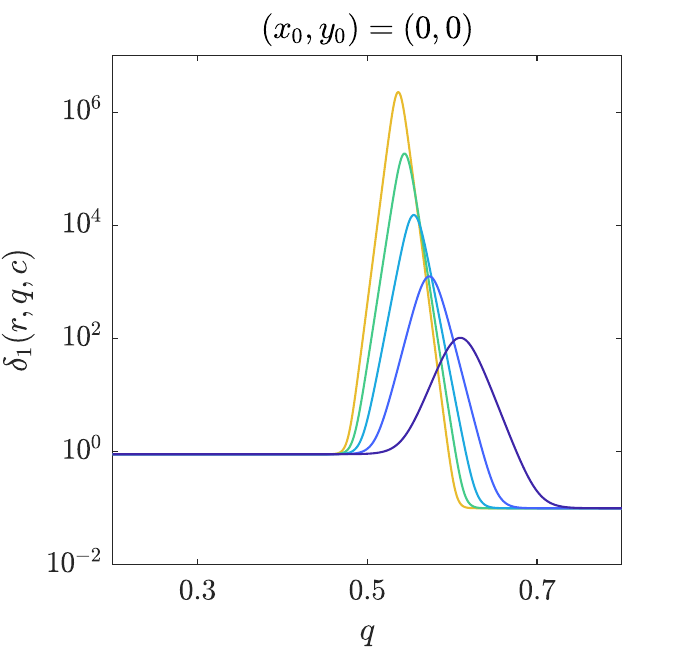}
    \includegraphics[scale=0.6]{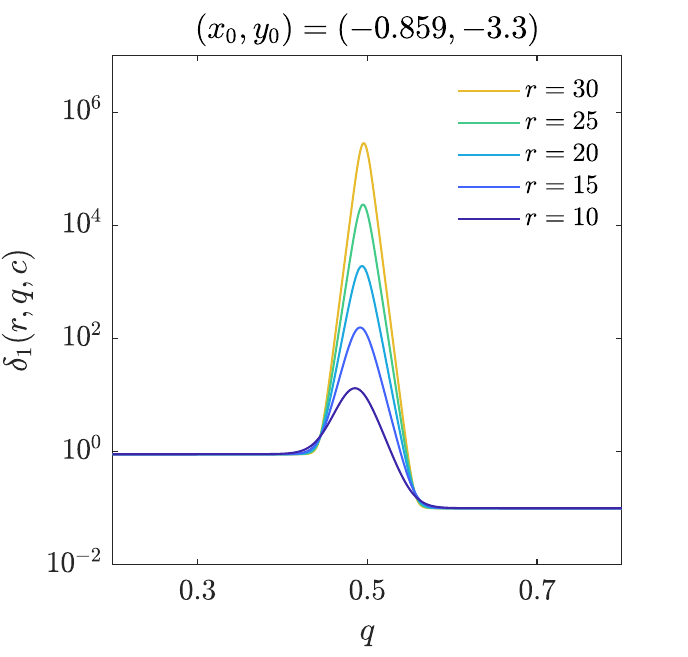}
	\caption{Effect of choice of origin for polar coordinates on speed of convergence of angular-radial dependence function to asymptotic form, for EV copula with asymmetric logistic dependence with $\alpha=0.5$, $\gamma_1=0.9$ and $\gamma_2=0.1$. Left plot has polar origin at $(x_0,y_0)=(0,0)$, right plot has origin selected such that $T_2=T_3$ at fixed angle -- see (\ref{eq:EV_density_2D}).}
	\label{fig:Asym_origin}
\end{figure}

The effect of the choice of origin on the speed of convergence of $\delta_L(r,\mathbf{w})$ to its asymptotic form is illustrated in \autoref{fig:Asym_origin} for the case $\alpha=5$, $\gamma_1=0.9$ and $\gamma_2=0.1$. The plots show that placing the origin at $(0,0)$ results in slow convergence, whereas placing the origin so that $T_2\sim T_3$ at constant angle in the first quadrant improves the rate of convergence.

\textbf{H{\"u}sler-Reiss:} The H{\"u}sler-Reiss dependence model \parencite{Husler1989}, with parameter $\alpha>0$, is given by
\begin{equation*}
    A(\boldsymbol{\psi}) = \psi z_1 + (1-\psi) z_2,
\end{equation*}
where
\begin{align*}
    z_1 &= \Phi \left(\frac{1}{\alpha} + \frac{\alpha}{2}\log\left(\frac{\psi}{1-\psi}\right) \right),\\
    z_2 &= \Phi \left(\frac{1}{\alpha} - \frac{\alpha}{2}\log\left(\frac{\psi}{1-\psi}\right) \right),
\end{align*}
and $\Phi$ is the standard normal distribution function. In this case we have
\begin{align*}
    T_2 &= z_1 z_2,\\
    T_3 &= \frac{\alpha}{2\tau\psi} \, \phi \left(\frac{1}{\alpha} - \frac{\alpha}{2}\log\left(\frac{\psi}{1-\psi}\right) \right),
\end{align*}
where $\phi$ is the standard normal density function. In this case we have 
\begin{align*}
    \left|A^{(1,1)}(\psi,1-\psi)\right| &= \frac{\alpha}{2\psi} \, \phi \left(\frac{1}{\alpha} - \frac{\alpha}{2}\log\left(\frac{\psi}{1-\psi}\right) \right)\\
    &= \phi \left(\frac{1}{\alpha}\right) \frac{\alpha}{2\psi} \, \exp \left(\log\left(\frac{\psi}{1-\psi}\right) - \frac{\alpha^2}{8}\log^2\left(\frac{\psi}{1-\psi}\right) \right)\\
    &= \phi \left(\frac{1}{\alpha}\right) \frac{\alpha}{2\psi} \left(\frac{\psi}{1-\psi}\right) ^ {1 - \tfrac{\alpha^2}{8}\log\left(\tfrac{\psi}{1-\psi}\right)}.
\end{align*}
This is rapidly-varying for $\psi\to0^+$ and $\psi\to1^-$ with index $+\infty$. The assumptions of Proposition \ref{prop:ARL2ARE} are therefore not satisfied, so we cannot obtain the copula density exponent function from the tail density function. Instead, we can calculate the AR copula function directly.

To calculate asymptotic expressions for $T_2$ we utilise the asymptotic Mill's ratio for the standard normal distribution. Combining this with the other terms, we obtain the following results
\begin{equation*}
    \delta_L(r,\mathbf{w}) \sim 
    \begin{cases}
    A_1 \exp\left(-\dfrac{1}{2} \left[\left(\dfrac{\alpha}{2} r \right)^2 + \left(\dfrac{\alpha^2}{2} \log(\log(4)) - 1\right) r\right] \right), & q=0,1\\
    A_2 \exp\left(-\dfrac{1}{2} \left[\left(\dfrac{\alpha}{2}(w_1-w_2) r\right)^2 - r \right] \right), & q\in(0,1)\\
    A_3 \sqrt{\dfrac{2}{r}} \exp\left(-\dfrac{1}{2} \left[ \left(\dfrac{\alpha}{2} r|w_2|\right)^2 +\left(\dfrac{\alpha^2}{2} \log(2r|w_1|) - 1\right) r|w_2| + \left(\dfrac{\alpha}{2} \log(2r|w_1|)\right)^2 \right] \right), & q\in(1,2)\\
    A_4 \, (\log(r))^{-1} \, r^{\tfrac{1}{4} \alpha^2 \log(\log(2)) + \tfrac{1}{2}}\, \exp\left(-\dfrac{1}{2}\left(\dfrac{1}{2} \alpha \log(r)\right)^2 \right), & q=2,3\\
    A_5 \exp(-r (A(\mathbf{w})-1)), & q\in(2,3),\\
    \end{cases}
\end{equation*}
where the terms $A_j$ are constants dependent on $q$ only, given by
\begin{align*}
    A_1 &= \alpha \phi\left(\frac{1}{\alpha}\right) (\log(4))^{-\tfrac{1}{2} -\tfrac{\alpha^2}{8}\log(\log(4))} \\  
    A_2 &= \alpha \phi\left(\frac{1}{\alpha}\right)\\
    A_3 &= \phi\left(\frac{1}{\alpha}\right) 2^{-\tfrac{\alpha^2}{4} \tfrac{|w_1|}{|w_2|}} \left[\frac{2}{\alpha} \frac{\sqrt{|w_1|}}{|w_2|} + \frac{\alpha}{2}  \frac{1}{\sqrt{|w_1|}}\right] \\
    A_4 &= \frac{4}{\alpha} \phi\left(\frac{1}{\alpha}\right) (\log(2))^{-\tfrac{1}{2} -\tfrac{\alpha^2}{8}\log(\log(2))} \\
    A_5 &= 4^{1-A(\mathbf{w})} \Phi \left(\frac{1}{\alpha} + \frac{\alpha}{2}\log\left(\frac{|w_1|}{|w_2|}\right) \right) \Phi \left(\frac{1}{\alpha} - \frac{\alpha}{2}\log\left(\frac{|w_1|}{|w_2|}\right) \right).
\end{align*}
Clearly, this does not have asymptotic form (\ref{eq:delta_Laplace}) for all $q$. Moreover, outside the third quadrant, the limit set is the line segment $\{(x,x),x\in[0,1]\}$. 
However, after multiplying by $m_L(r,\mathbf{w}) = \tfrac{1}{4}\exp(-r)$, the terms multiplying the leading order contributions within the exponential are continuous in $q$, and hence assumptions (A1) and (A2) are satisfied. 

\subsection{Gaussian copula} \label{app:Gaussian}
For the Gaussian copula we focus on models for the copula density, since the copula does not admit a simple closed form expression. The bivariate Gaussian copula density with Pearson correlation coefficient $\rho\in(-1,1)$ is given by
\begin{align*}
c(u,v) &= \frac{1}{\sqrt{1-\rho^2}} T_1 T_2,\\
T_1 &= \exp(-\tfrac{1}{2}\rho \beta(x^2+y^2)),\\
T_2 &= \exp(\beta xy),
\end{align*}
where $x=\Phi^{-1}(u)$, $y=\Phi^{-1}(v)$, where $\Phi$ is the standard normal cdf $\beta = \rho/(1-\rho^2)$. From the symmetry of the copula, we need only consider the asymptotic behaviour of the copula in the lower left corner. For $u\to 0^+$, $x\to -\infty$, and utilising the asymptotic Mill's ratio we have $\Phi(x)\sim -\phi(x)/x$, where $\phi$ is the standard normal pdf. Solving for $x$ gives
\begin{align*}
x^2 \sim W_0\left( \frac{1}{2\pi (1-u)^2}\right) \sim L_u-\log(L_u), \quad u\to0^+,
\end{align*}
where $W_0$ is the zeroth branch of the Lambert W function, and $L_u = -\log(2\pi u^2)$. Substituting this approximation shows that
\begin{align} \label{eq:Normal_T1}
	T_1 \sim \left(2\pi (1-u)(1-v) \sqrt{L_uL_v}\right)^{\rho\beta}, \quad u,v\to0^+.
\end{align}
For the product terms, for $u,v\to0^+$ we have
\begin{align}
\nonumber xy & \sim \sqrt{(L_u-\log(L_u))(L_v-\log(L_v))}\\
\nonumber &= \sqrt{L_u L_v} \sqrt{1 - \frac{\log(L_u)}{L_u} - \frac{\log(L_v)}{L_v} + \frac{\log(L_u)\log(L_v)}{L_u L_v}}\\
\nonumber &\sim \sqrt{L_u L_v} \left[1 - \frac{\log(L_u)}{2 L_u} - \frac{\log(L_v)}{2 L_v} + \frac{\log(L_u)\log(L_v)}{2 L_u L_v}\right]\\
\label{eq:gaussian_xy} &\sim \sqrt{L_u L_v} - \frac{1}{2} \left[ \sqrt{\frac{L_v}{L_u}} \log(L_u) + \sqrt{\frac{L_u}{L_v}} \log(L_v)\right].
\end{align}
These expressions can be used to derive the tail density function and copula density exponent functions, as discussed below.

\subsubsection{ARL model}
In the lower left corner we have $(u,v)=(t(1-w),tw)$, so that
\begin{align*}
L_u L_v = -4\left(\log^2(\sqrt{2\pi} t) + \log(\sqrt{2\pi} t)\log(w(1-w)) + \log(w)\log(1-w)\right)
\end{align*}
The terms which dominate depend on the relative sizes of $t$ and $w$. For fixed $w\in(0,1)$ and $t\to0^+$, we have
\begin{align*}
	\sqrt{L_u L_v} &= -2 \log(\sqrt{2\pi}t) 
	\left(1 + \frac{\log(w(1-w))}{\log(\sqrt{2\pi} t)} + \frac{\log(w)\log(1-w)}{\log^2(\sqrt{2\pi} t)}\right)^{1/2}\\
	&\sim -2 \log(\sqrt{2\pi}t) 
	\left(1 + \frac{\log(w(1-w))}{2\log(\sqrt{2\pi} t)} + \frac{\log(w)\log(1-w)}{2\log^2(\sqrt{2\pi} t)}\right)\\
	&\sim - \log(2\pi t^2 w(1-w)).
\end{align*}
However, for $w\to0^+$ with $t$ fixed, we have
\begin{align*}
	\sqrt{L_u L_v} &\sim 2  
	\left(\log\left(\sqrt{2\pi}t\right)\log(w) \right)^{1/2}.
\end{align*}
The other terms in (\ref{eq:gaussian_xy}) also behave differently depending on the relative sizes of $w$ and $t$. For $t\to0^+$ with $w$ fixed we have $L_u / L_v \to 1$. In contrast, when $w\ll t$, $L_u/L_v\to0$. Gathering all the terms together, for fixed $w\in(0,1)$ and $t\to 0^+$ we get
\begin{equation*}
c(t(1-w),tw) \sim \frac{1}{1-\rho^2} \left(-4\pi t^2 w(1-w) \log(t) \right)^{-\tfrac{\rho}{1+\rho}}.
\end{equation*}
Comparing this with (\ref{eq:ARL_dens}), we see that $\kappa_{(0,0)}=2/(1+\rho)$. A similar analysis for the other corners shows that $\kappa_{(1,1)}=2/(1+\rho)$ and $\kappa_{(1,0)}=\kappa_{(0,1)}=2/(1-\rho)$. 

For small $t$ and $w\to0^+$ with $w\ll t$, combining the terms above gives
\begin{align*}
c(t(1-w),tw) \sim \frac{1}{1-\rho^2} \left(2 \sqrt{2} \pi t^2 w (\log(2\pi t^2)\log(w))^{\tfrac{1}{4}} \right)^{\tfrac{\rho^2}{1-\rho^2}} \exp\left(\frac{2\rho}{1-\rho^2} \sqrt{\log(2\pi t^2)\log(w)}\right).
\end{align*}

\subsubsection{ARE model}
For the lower left corner we have $(u,v)=(\exp(-r(1-w), \exp(-rw))$, so that $L_u=r(1-w)-\log(2\pi)$ and $L_v=rw-\log(2\pi)$, and $x^2\sim 2r(1-w)-\log(2\pi L_u)$, $y^2\sim 2rw-\log(2\pi L_v)$ as $r\to\infty$. Substituting this back shows that
\begin{align*} 
	T_1 \sim (4\pi r \sqrt{w(1-w)})^{2\alpha} \exp(-2\alpha r), \quad r\to\infty.
\end{align*}
Similarly, from (\ref{eq:gaussian_xy}) we have
\begin{align*} 
	xy \sim 2 r \sqrt{w(1-w)} - \frac{1}{2} \sqrt{\frac{w}{1-w}} \log(2\pi L_u) - \frac{1}{2} \sqrt{\frac{1-w}{w}} \log(2\pi L_v), \quad r\to\infty.
\end{align*}
Therefore
\begin{align*} 
	T_2 \sim \exp(2\beta r \sqrt{w(1-w)}) (4\pi r(1-w))^{ - \tfrac{\beta}{2} \sqrt{\tfrac{w}{1-w}}} (4\pi r w)^{-\tfrac{\beta}{2} \sqrt{\tfrac{1-w}{w}}}, \quad r\to\infty.
\end{align*}
Combining these terms gives
\begin{align*}
c(\exp(-r\mathbf{w})) \sim \mathcal{M}_{(0,0)}(\exp(-r),\mathbf{w}) \exp (-r (\lambda_{(0,0)}(\mathbf{w})-1)), \quad r\to\infty,
\end{align*}
where
\begin{align*}
    \mathcal{M}_{(0,0)}(\exp(-r),(1-w,w))) &= 
    \frac{1}{1-\rho^2} (4\pi r)^{a_1} w^{a_2} (1-w)^{a_3},\\
    \lambda_{(0,0)}(1-w,w) &= \frac{1 -2\rho\sqrt{w(1-w)}}{1-\rho^2},
\end{align*}
and
\begin{align*}
    a_1 &= \frac{\rho}{2-2\rho^2} \left[2\rho - \frac{1}{\sqrt{w(1-w)}} \right],\\
    a_2 &= \frac{\rho}{2-2\rho^2} \left[ \rho - \sqrt{\frac{1-w}{w}}\right],\\
    a_3 &= \frac{\rho}{2-2\rho^2} \left[ \rho - \sqrt{\frac{w}{1-w}}\right].
\end{align*}
Note that this asymptotic form is only valid for $w\in(0,1)$, since it assumes that both $u$ and $v$ go to zero as $r\to\infty$. 

The copula density exponent function satisfies the assumptions of Proposition \ref{prop:ARE2ARL} with $\tfrac{d}{dw} \lambda_{(0,0)}(w,1-w)=0$. Recalling that $\kappa_{(0,0)}=2/(1+\rho)$, we see that the form of $b_{(0,0)}$ obtained from Proposition \ref{prop:ARE2ARL} agrees with that obtained in the preceding subsection.

\subsubsection{AR copula function for Laplace margins}
From Proposition \ref{prop:delta_laplace_ARE}, we just need to check that $\lambda(-1)=\lim_{w\to1^-} \lambda_{(0,0)}(w)$. On Laplace margins, for $q=-1$ we have $u=\tfrac{1}{2}\exp(-r)$ and $v=1/2$. In this case $y=0$ and $L_u = 2r -\log(\pi/2)$, so that $x^2 \sim 2r-\log(\pi L_u/2)$. Therefore
\begin{align*}
    \delta_L(r,(0,-1),c) &\sim \frac{1}{\sqrt{1-\rho^2}} \exp(-\tfrac{1}{2}\rho\beta(2r-\log(\pi L_u/2)))\\
    &\sim \frac{1}{\sqrt{1-\rho^2}} (\pi r)^{\tfrac{1}{2}\rho\beta} \exp\left(-r \frac{\rho^2}{1-\rho^2}\right).
\end{align*}
Hence, $\lambda(-1)=\lim_{w\to1^-} \lambda_{(0,0)}(w)$.

\subsection{T copula} \label{app:t_copula}
As with the Gaussian copula, the t-copula does not admit a simple closed form expression. We therefore focus on the copula density. The bivariate t copula density function with parameters $\rho\in(-1,1)$, $\nu>0$ is given by
\begin{equation*}
c(u,v) = \frac{g^2 \nu}{2\sqrt{1-\rho^2}} \, \left(1+\frac{x^2 +y^2 - 2\rho xy}{\nu(1-\rho^2)}\right)^{-\tfrac{\nu}{2}-1} \, \left(\left(1+\frac{x^2}{\nu}\right) \left(1+\frac{y^2}{\nu}\right)\right)^{\tfrac{\nu+1}{2}} , 
\end{equation*}
where $g = \Gamma(\nu/2) / \Gamma((\nu+1)/2)$, $x=F_t^{-1}(u,\nu)$, $y=F_t^{-1}(v,\nu)$ and $F_t^{-1}(\cdot,\nu)$ is the inverse cdf of the univariate t distribution on $\nu$ degrees of freedom. The quantiles of the univariate t distribution asymptote to \parencite{shaw2006}
\begin{align*}
    F_t^{-1}(p,\nu) &\sim -\sqrt{\nu}(pg\nu\sqrt{\pi})^{-1/\nu}, \quad p\to 0,\\
    F_t^{-1}(p,\nu) &\sim \sqrt{\nu}((1-p)g\nu\sqrt{\pi})^{-1/\nu}, \quad p\to 1.
\end{align*}

\subsubsection{ARL model}
Substituting these asymptotic approximations shows that 
\begin{equation*}
c(t\mathbf{w})  \sim \Upsilon_{(0,0)} b_{(0,0)}(\mathbf{w}) t^{-1}, 
\end{equation*}
where
\begin{align*}
    \Upsilon_{(0,0)} b_{(0,0)}(\mathbf{w}) = \frac{g}{2\sqrt{\pi}} (1-\rho^2 )^{\tfrac{\nu+1}{2}} (w_1 w_2)^{\tfrac{1}{\nu}} \left(w_1^{\tfrac{2}{\nu}}+ w_2^{\tfrac{2}{\nu}} -2\rho (w_1 w_2)^{\tfrac{1}{\nu}} \right)^{-1-\tfrac{\nu}{2}}.
\end{align*}
So the copula density has asymptotic form (\ref{eq:ARL_dens}) with $\kappa_{(0,0)}=1$. The same tail order is obtained in other corners, whereas the tail density function has the opposite sign of $\rho$ in the lower right and upper left corners. Separating $\Upsilon_{(0,0)}$ and $b_{(0,0)}$ requires consideration of the copula, which does not have a simple closed form. However, the tail dependence coefficient can be expressed in terms of the cdf of the univariate t-distribution (see e.g. \cite{schmidt2002tail, frahm2003elliptical, chan2008tcopula, Joe2015}).

\subsubsection{ARE model}
A similar analysis shows that the copula density also has ARE form with 
\begin{align*}
\lambda_{(0,0)}(1-w,w) &= \left(1+\frac{2}{\nu}\right)\max(w,1-w) - \frac{1}{\nu}\\
\mathcal{M}_{(0,0)}(t,w) &= \frac{g}{2\sqrt{\pi}} (1-\rho^2)^{\tfrac{\nu+1}{2}} \times
\begin{cases}
    (2-2\rho)^{-1-\tfrac{\nu}{2}}, & w=1/2\\
    1, & w\in[0,1/2)\cup(1/2,1].
\end{cases}
\end{align*}
As with the tail order function, the same functions are obtained in the other corners, but with the sign of $\rho$ changed in the lower right and upper left corners.

Since the t-copula satisfies the assumptions of Proposition \ref{prop:ARL2ARE}, we can confirm that the copula density exponent function obtained above is the same as that given in Proposition \ref{prop:ARL2ARE}. The tail density function $b_{(0,0)}(1-z,z)$ has regularly-varying tails with order $1/\nu$ for $z\to0^+$ and $z\to1^-$. Also note that $|w-\tfrac{1}{2}|=\max(w,1-w)-\tfrac{1}{2}$. Substituting these into the expression given in Proposition \ref{prop:ARL2ARE} gives the desired result.

\subsubsection{AR copula function for Laplace margins}
As before, we just need to check the behaviour for $\mathbf{w}=(1,0)$. In this case, we have $y=0$ and $u=1-\tfrac{1}{2}\exp(-r)$, so that $x\sim \sqrt{\nu}(\tfrac{1}{2}\exp(-r)g\nu\sqrt{\pi})^{-1/\nu}$. Substituting this shows that 
\begin{align*}
\delta_L(r,(1,0),c) &\sim \frac{g^2 \nu}{2} \, \left(\frac{g\nu\sqrt{\pi}}{2}\right)^{1/\nu} \,
\left(1-\rho^2\right)^{(\nu+1)/2} \, 
\exp(-r/\nu).
\end{align*}
Hence $\lambda(0)=\lim_{w\to0^+} \lambda_{(1,1)}(w)$.

\end{document}